%% file: SBC_Background_ISR_2.tex
\RequirePackage[l2tabu, orthodox]{nag}
\documentclass{stsci_report}

\usepackage{booktabs}
\usepackage{rotating}
\usepackage{natbib}
\usepackage{url}
\usepackage{graphicx}
\usepackage{caption}
\usepackage[T1]{fontenc}
\usepackage[font=footnotesize,labelfont=bf]{caption}
\usepackage{refcount} 
\usepackage[breaklinks,colorlinks=true,allcolors=blue]{hyperref}
\usepackage{color}
\usepackage{amsmath,amssymb}
\usepackage{makecell}
\usepackage[yyyymmdd]{datetime}

\usepackage[dvipsnames]{xcolor}
\definecolor{AquaGreen}{HTML}{12e193}
\definecolor{BrightSkyBlue}{HTML}{02ccfe}
\definecolor{LightRed}{HTML}{ff474c}
\definecolor{Melon}{HTML}{ff7855}
\definecolor{ShockingPink}{HTML}{fe02a2}
\definecolor{Perrywinkle}{HTML}{8f8ce7}

\newcommand{\degr}{$^{\circ}$}
\newcommand{\cpspsqarc}{${\rm counts\,sec^{-1}\,arcsec^{-2}}$}

\renewcommand{\th}{\textsuperscript{th}}

\makeatletter 
\newcommand\footnoteref[1]{\protected@xdef\@thefnmark{\ref{#1}}\@footnotemark}
\makeatother

\copyrighttext{Copyright \copyright \the\year\ The Association of Universities for Research in Astronomy, Inc. All Rights Reserved.}

\title{\textbf{Using 23 Years of ACS/SBC Data to Understand Backgrounds}:\\Explaining \&\ Predicting\\Background Variations}
\presubtitle{\flushleft{\includegraphics[width=5cm]{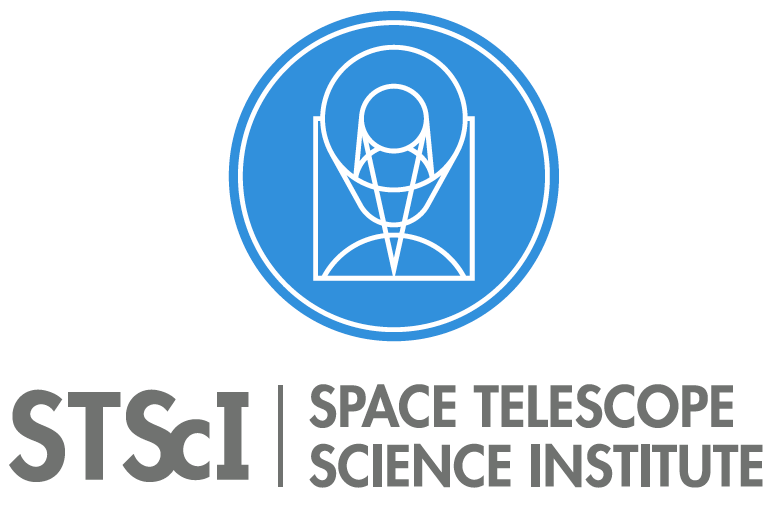}} \\ \hfill Instrument Science Report ACS 2026-01}
\author{Christopher J. R. Clark, Roberto J. Avila, Alyssa M. Guzman, Norman A. Grogin}
\date{2026-01-08}

\begin{document}

\maketitle

\abstract{Recent analysis of 23 years of Hubble Space Telescope ACS/SBC data has shown that background levels can vary considerably between observations, with most filters showing over an order of magnitude variation. For the shorter-wavelength filters, the background is understood to be dominated by airglow; however, what precisely drives background variations is not well constrained for any filter. Here, we explore the causes of the background variation. Using over 8,000 archival SBC observations, we developed a machine learning model that can accurately predict the background for an observation, based upon a set of 23 observational parameters. This model indicates that, depending on filter, the SBC background is generally dominated by Solar elevation, Solar separation angle, Earth limb angle of observation, SBC temperature, and target Galactic latitude.}

\newpage

\section{Introduction \&\ Motivation} \label{Section:Introduction}

The Solar Blind Channel (SBC; \citealp{Tran2003B}) of the Advanced Camera for Surveys (ACS; \citealp{Clampin2000B}) on the Hubble Space Telescope (HST; \citealp{Bahcall1982C}) is a Multi-Anode Microchannel Array (MAMA), a photon-counting detector used for Far-UltraViolet (FUV) observations. The SBC observes via a suite of 6 imaging filters and 2 low-spectral-resolution prisms, with spectral sensitivty over a range of approximately 1150--1700\,\AA. Although they do exhibit dark current, MAMAs like the SBC do not produce any read noise; as a result, the instrumental/astrophysical background is the main source of noise in SBC observations. 

However, presented in \citet{CJRClark2025C}, the background levels in SBC data can vary considerably -- by well over an order of magnitude in several filters. This is especially true for the filters affected by FUV geocoronal airglow lines, which are the dominant source of background for observations with several SBC filters. 
 
 The primary lines that can effect SBC observations are Ly-$\alpha$ at 1216\,\AA\ , and O{\sc i} at 2471\,\AA, 1304\,\AA, and 1356\,\AA, all of which are excited by Solar photons. Other airglow lines present in the SBC wavelength range are Lyman--Birge--Hopfield system of N$_{2}$ lines, which emit at 1250--2800\,\AA\ (and which can dominate at 1300--1900\AA), which are electronically-excited by both Solar and geocoronal electrons \citep{Torr1994F,Eastes2000C,Cantrall2021A}. There are also weaker airglow lines that are minor contributors to the total airglow luminosity; for instance, the Schumann--Runge O$_{2}$ lines at 1750--2000\,\AA\  \citep{Hedin2009A}. 

The impact of airglow lines upon the SBC background for a given observation will therefore be affected by time of day, Solar altitude, Earth limb angle, Solar activity variations throughout (and between) Solar cycles, geomagnetic field strength, and more \citep{Waldrop2013A,Putis2018A}. Understanding which factors are the primary drivers of variations in SBC backgrounds is valuable for helping users design observing programs to achieve optimal backgrounds, and therefore benefit from shorter integration times and/or stronger detections.

In this ISR, we take advantage of the 23 years of archival data available for the SBC, consisting of many thousands of exposures, to constrain what observational parameters actually drive variations in the background.

This ISR is primarily concerned with imaging observations with SBC; however, as spectral observations with the SBC prisms use the same detector, background levels are considered in the same manner -- however, the consequent implications for variations in {\it spectral} sensitivity are not explored. 

\subsection{Report Structure} \label{Subsection:Report_Structure}

This ISR is laid out as follows:

In Section~\ref{Section:Archival_Backgrounds}, we summarise the archival observations considered in this report, and how the background levels in those observations were measured. 

In Section~\ref{Section:Observational_Parameters}, we describe the set of ephemeris, pointing, and other observational parameters we assembled. 

In Section~\ref{Section:Background_Modeling}, we present how we modelled the relationships between the measured SBC background levels, and the various observational parameters. 

In Section~\ref{Section:Background_Causes}, we explore the model results, and how different observational parameters affect the background levels experienced by the SBC. 

In Section~\ref{Section:Conclusion}, we summarize our findings, and describe the primary considerations that users should bear in mind when designing SBC observations.

\section{Background Measurements from Archival Observations} \label{Section:Archival_Backgrounds}

In this ISR, we use the background measurements presented in the companion report, \citet{CJRClark2025C}. We refer the reader to that report for a full description of the archival data used, and how the background measurements were made. 

To briefly summarize, \citet{CJRClark2025C} identified 8,640 suitable archival SBC exposures. The number of suitable exposures in a given filter vary from 261 in F115LP to 2,933 in F150LP.  A procedure was developed to mask both compact and diffuse sources of emission in the exposures. Each exposure was then divided into a 5$\times$5 grid of 25 square regions, and the background was measured in each. The faintest 5 of those square regions were then used to make the final background measurement (to exclude areas where marginally-detected sources avoided being masked).

It is important to note that measuring the background empirically from the data in this manner means that the SBC dark rate cannot be distinguished from other sources of background. This aligns with the objectives of this study, because various observational parameters that can drive the SBC dark rate are also likely to impact other sources of background. For instance SAA proximity can increase the luminosity of atmospheric airglow lines \citep{He2020B}, which will  drive up the observed background. Even the small temperature effect from Earth being at perihelion versus aphelion is known to cause changes to the dark rate for the MAMA aboard HST/COS \citep{Johnson2024C}.

For a discussion of the implications of considering dark current together with other sources of background, see Section 1.1 of \citet{CJRClark2025C}.

\section{Observational Parameters} \label{Section:Observational_Parameters}

The distributions of measured SBC backgrounds presented in \citet{CJRClark2025C} show that the background level can vary significantly between observations. The median difference between the 16\th\ and 84\th\ percentiles is a factor of 9.3, and ranges from a factor of 5.3 for F165LP, to a factor of 20 for F125LP. We wish to understand what drives this variation, especially if this suggests practicable ways in which future observations could be planned to reduce backgrounds.

To do this, we identified a range of observational parameters that could be reasonably expected to affect the background level encountered by the SBC. We found the value for each of these observational parameters for every archival exposure in our analysis, to enable us to explore if/how they correlate with background level (see Section~\ref{Section:Background_Modeling}). Here we describe each of these observational parameters, 23 in total, and how they were obtained. We will then use these parameters in Section~\ref{Section:Background_Modeling} to explore what affects SBC background levels. The methods used to obtain many of these parameters are illustrated with worked examples in a \texttt{jupyter} notebook available on the STScI \texttt{hst\_notebooks} GitHub repository\footnote{\url{https://github.com/spacetelescope/hst_notebooks/blob/main/notebooks/ACS/hst_orbits_ephem/hst_orbits_ephem.ipynb}}.

\paragraph{Date}

To capture whether any background-affecting phenomena are evolving over time, we record the date of the exposure. We use the mid-point of each exposure for this purpose, taken as the mean of the \texttt{EXPSTART} and \texttt{EXPEND} header keywords, quantified in terms of decimal Modified Julian Date. 

\paragraph{Day of Year}

There are various phenomena that could affect SBC background levels, which would be expected to dictated by time of year. For instance, it has already been found that the dark rate of the SBC tends to be elevated in the spring of most years \citep{Guzman2024I}. Similarly, the dark rate of the NUV MAMA on HST/COS varies annually due to the slight average temperature increase when the Earth is at perihelion \citep{Johnson2024C}. To capture any similar kinds of annually-periodic effects, we record the day of year of each SBC exposure (without rounding).

\paragraph{Instrument Temperature}

As previously discussed, it is well established that dark current increases as the SBC heats up from the operation of the electronics during an observation \citep{Avila2017B}. The header of SBC \texttt{\_flt.fits} records the instrument temperature at the start and end of the exposure (under header keywords \texttt{MDECODT1} and \texttt{MDECODT2}). We record both of these temperatures (in \degr C).

\paragraph{Exposure Duration}

Exposure time will certainly impact the background, through the increase in instrument temperature as it heats up under operation. However, it could additionally be the case that the background is increased by transient/intermittent phenomena, which would be more likely to occur during longer exposures. Exposure duration is provided in seconds in the header of the \texttt{\_flt.fits}, under the keyword \texttt{EXPTIME}.

\paragraph{Earth Limb Angle}

Earth limb angle describes the angular separation between the coordinates observed by HST, and the nearest part of the Earth's limb. Naturally, observations with a lower Earth limb angle observe through a greater column of atmosphere, and so are more vulnerable to airglow and scattered light. Earth limb angle changes over the course of an exposure, due to HST's orbital motion; this is recorded in each exposure's \texttt{\_jit.fits} file, under \texttt{LimbAng}. We therefore record the minimum, maximum, and mean Earth limb angle (in degrees) for each exposure.

\paragraph{Solar Time}

Air glow in the FUV is primarily driven by Solar photons exciting atoms high in the atmosphere; the strength of airglow therefore varies with time of day. Specifically, it varies with local Solar time (ie, the time frame where the Sun is precisely at zenith at noon, and at nadir at midnight). Because of HST's orbital motion, its local Solar time does change significantly during the course of an exposure. To compute local Solar time, we need to know the latitude and longitude of HST during the observation. To retrieve this information, we use public `two line element' orbital information provided by the United States Space Command (USSPACECOM\footnote{\label{Footnote:USSPACECOM}\url{https://www.space-track.org/documentation\#odr}}), from which orbital ephemeris was computed using the Simplified General Perturbations 4 (SGP4; \citealp{Vallado2006A}) model, as implemented by the \texttt{skyfield}\footnote{\url{https://github.com/skyfielders/python-skyfield/}} package for \texttt{Python}. We thereby computed the local Solar time of HST at the start, midpoint, and end of each exposure, which we recorded in hours.

\paragraph{Solar Altitude}

Solar altitude is the angle between the Sun, and the nearest part of the Earth's limb. As with Solar time, this relates to atmospheric excitation by FUV Solar photons. Solar altitude is provided (in degrees) in the header of each exposure's \texttt{\_flt.fits}, under the keyword \texttt{SUN\_ALT}.

\paragraph{Solar Separation Angle}

The Solar separation angle is the angle between the Sun, and the direction in which HST is pointing, recorded in degrees. This is most likely to relate to any contribution of scattered light to the background. In particular, interplanetary Ly-$\alpha$ emission, arising from Solar Ly-$\alpha$ photons scattered by interplanetary neutral hydrogen atoms, should be more significant at smaller Solar separation angles \citep{Lallement1993C}. The Solar separation angle is provided in degrees in the header of the \texttt{\_flt.fits}, under the keyword \texttt{SUNANGLE}.

\paragraph{Orbital Height}

In general, the line-of-sight of an observation will intersect a greater column of Earth's atmosphere at lower orbital height -- thereby increasing the average potential exposure to airglow, and to atmospherically-scattered light. It should be noted, however, that observations at greater orbital height can intersect a {\it greater} atmospheric column, due to the horizon being lower -- ie, there will be a greater atmospheric column at a given Earth limb angle. We the USSPACECOM/SGP4 ephemerides\hyperref[Footnote:USSPACECOM]{\textsuperscript{\getrefnumber{Footnote:USSPACECOM}}} to find the orbital height (in km) for each exposure.

\paragraph{Geomagnetic Field Strength}

\begin{figure}
\centering
\includegraphics[width=0.975\textwidth]{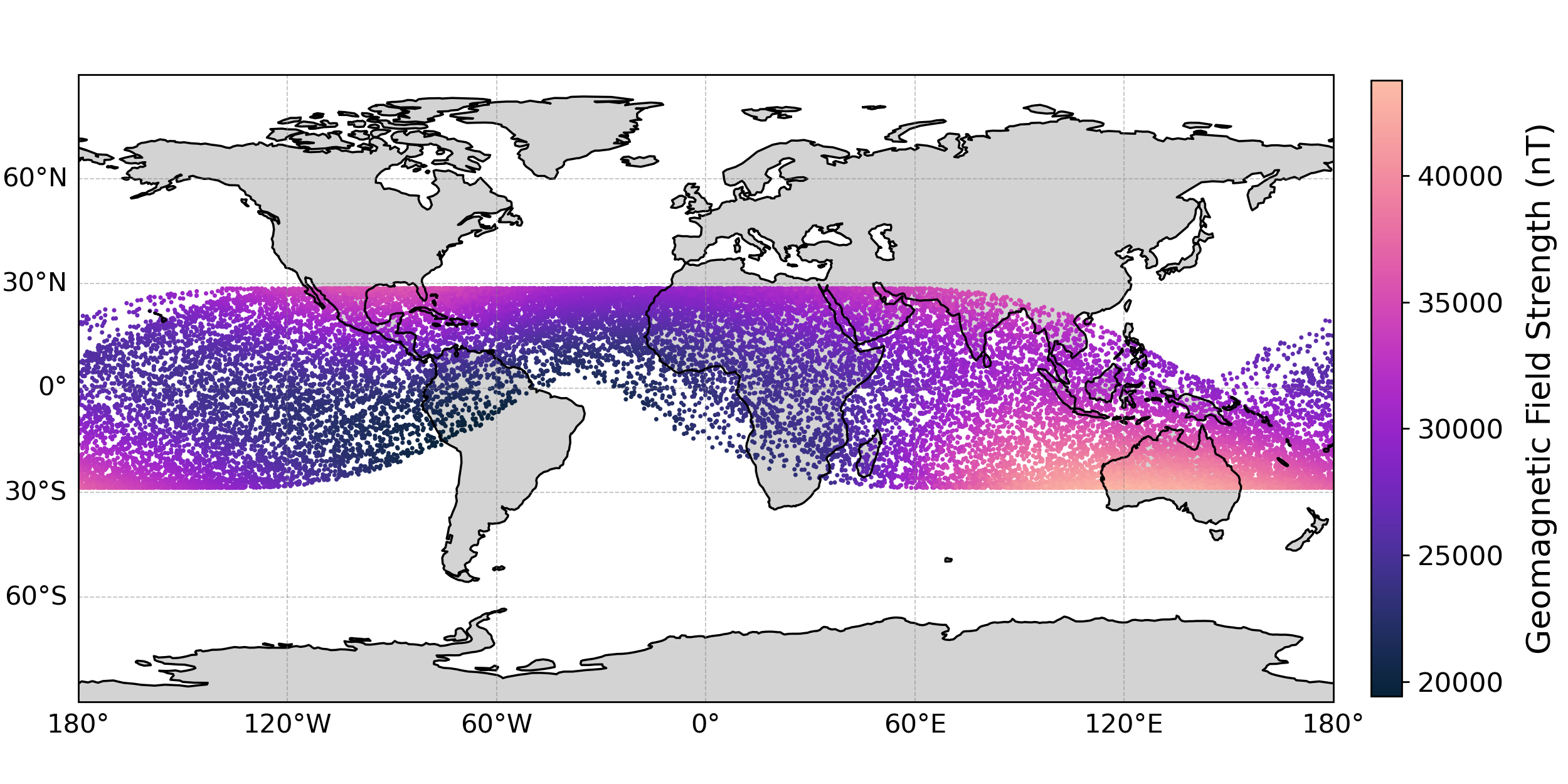}
\caption{Map showing the position of HST during every SBC exposure considered in this study. Each exposure is plotted with 3 points, corresponding to its start, mid-point, and end. All points are color-coded according to the strength of the geomagnetic field at that position, as per IGRF-14. Orbits that would intersect the SAA, and where no SBC exposures were therefore taken, are neatly illustrated by the gaps in this map. The region around the SAA still has a notably weaker geomagnetic field than elsewhere.}
\label{Fig:SBC_Geomag_Coords}
\end{figure}

Earth's geomagnetic field mediates the interactions between the Solar wind and Earth's atmosphere. This is particularly significant in the vicinity of the South Atlantic Anomaly (SAA), where the relative weakness of the geomagnetic field allows one of the Van Allen belts to descend to an altitude of only 200\,km \citep{Pinto1992A}, intersecting HST's orbit. The trapped Solar protons \&\ electrons of the Van Allen belt cause a major increase in airglow in the atmosphere above the SAA, exciting emission well above the usual level caused by FUV Solar photons.  There can also be elevated airglow where the geomagnetic field is {\it stronger}, due to more effective funneling of photoelectrons (produced by Solar ionization) to the emitting layers of the thermosphere; this effect is especially known for the 1356\,\AA\ O{\sc i} and Lyman-Birge-Hopfield N$_{2}$ airglow lines \citep{Solomon2020A}. The many impacts of the SAA on HST mean that the observatory does not observe within defined contours around the anomaly \citep{Lupie2002A} -- however, the SAA does not have a discrete `edge'. We therefore expect higher background levels in exposures taken when HST is closer to the SAA. We also expect a similar, albeit smaller, impact for exposures taken in other areas where the geomagnetic field is weaker (as observed in cosmic ray impacts for other instruments; \citealp{Miles2021A}). We therefore determine the geomagnetic field strength during each exposure. For this, we use 14\th\ version of the International Geomagnetic Reference Field (IGRF-14;  \citealp{Alken2021C}); accessed via the \texttt{Python} package \texttt{ppigrf}\footnote{\url{https://github.com/IAGA-VMOD/ppigrf}}. We evaluate the magnitude of the field strength at the start, mid-point, and end of each exposure -- for the latitude, longitude, and orbital height corresponding to each of those three points in time, as per the USSPACECOM/SGP4 ephemeris values. This allows us to capture variation in the field strength during each exposure, whilst still keeping later modelling computationally reasonable. The geomagnetic field strengths are recorded in units of nT. The position of HST at the start, mid-point, and end of each SBC exposure, and the corresponding strength of the geomagnetic field at that location, is plotted in Figure~\ref{Fig:SBC_Geomag_Coords}.

\paragraph{Sunspot Count}

Variations in Solar activity levels can significantly impact the backgrounds observed by the SBC. The FUV emission from the Sun (the primary driver of the airglow emission detected by the SBC) varies by over a factor of 2 across the course of the 11-year Solar cycle, peaking at Solar maximum \citep{Lean2003A}. Conversely, the
contraction of the thermosphere at Solar {\it minimum} can also increase the background, by making the airglow-emitting atmospheric layers thinner and lower, thereby increasing the airglow-emitting column observed by the SBC for sightlines closer to the Earth's limb. Sunspot counts are a standard measure of the level of Solar activity at a given time \citep{Hoyt1998A}. We obtained daily historical sunspot numbers from the Royal Observatory of Belgium's Sunspot Index and Long-term Solar Observations (SILSO\footnote{\url{http://www.sidc.be/silso/DATA/SN_d_tot_V2.0.csv}}) database. Day-to-day variation in sunspot numbers introduces noise to their use as a tracer of Solar activity -- therefore, for each SBC exposure, we take the mean of the SILSO daily number of sunspots reported over the 7-day period centered on the date of the observation.

\paragraph{Lunar Phase Angle}

The Moon has a FUV albedo of $3.9\pm0.2$\%\ in the 1250--1750\,\AA\ wavelength range \citep{Henry1995F}. This means that, depending on phase, the Moon is one of the brightest FUV sources in the sky. As a result, the phase of the Moon can potentially affect SBC background levels, due to  added excitation of atmospheric airglow lines, and possible also via scattered light. We used the \texttt{astroplan} package \citep{Morris2018D} for \texttt{Python} to determine the phase of the Moon at the mid-point of each SBC exposure. We record Lunar phase in degrees, from 0\degr\ (full Moon) to 180\degr\ (new Moon).

\paragraph{Lunar Separation Angle}

As with Lunar phase, the Lunar separation angle also has scope to impact SBC background levels (moreso from potential scattered light effects than from boosted airglow). The Lunar separation angle is provided in degrees in the header of the \texttt{\_flt.fits}, under the keyword \texttt{MOONANGL}. 

\paragraph{Ecliptic Coordinates}

Zodiacal light (being Solar light scattered from the interplanetary dust cloud) is very minimal in the FUV, being $\sim$3 orders of magnitude fainter than it is in {\it V}-band \citep{Leinert1998A}. Instead, in the FUV, the Solar Ly-$\alpha$ photons scattered by interplanetary neutral hydrogen dominate over zodiacal light \citep{Lallement1993C,Leinert1998A}. This `interplanetary' neutral hydrogen is in fact hydrogen from the local interstellar medium passing through the Solar system (\citealp{Gladstone2025B}, and references therein). As such, the density of this material, and therefore the brightness of the interplanetary scattered Ly-$\alpha$, is generally greatest in the direction of the incoming interstellar gas\footnote{Approximate ecliptic coordinates 230\degr\ E, 0\degr\ N \citep{Ostgaard2003D}.}. Conversely, the interplanetary Ly-$\alpha$ is weakest towards the opposite side of the Solar system\footnote{Approximate ecliptic coordinates 80\degr\ E, 10\degr\ S \citep{Ostgaard2003D}.}, where less of the interstellar hydrogen penetrates, and where a larger fraction of it has been ionized by the Sun, and hence less able to scatter Solar Ly-$\alpha$ \citep{Pryor1996A,Pryor1998D,Ostgaard2003D}. Therefore, to capture the impact of interplanetary Ly-$\alpha$ light, and any contribution from FUV zodiacal light, we record the ecliptic latitude and longitude (in degrees) of the targets of all the SBC exposures.

\paragraph{Absolute Galactic Latitude}

As shown in \citet{CJRClark2025C}, diffuse emission from large numbers of undetected and/or unresolved sources contribute to the effective background in SBC observations. The greater density of stellar sources, especially young stars, towards the Galactic plane, mean that this contribution to SBC background levels should generally be greater at lower Galactic latitude. We therefore record the absolute Galactic latitude (ie, ranging form 0\degr--180\degr) of the target coordinates of all SBC observations.

\section{Modeling the SBC Background Variation} \label{Section:Background_Modeling}

With the background level measurement and 23 observational parameters obtained for each of the 8,640 suitable SBC exposures, we were able to explore the relationships between them. Specifically, we wished to establish to what degree these observational parameters dictate the background levels encountered, and their variation. In others words: How accurately can we predict the background level that will be in an SBC observation {\it a priori}, if all we know about it are the 23 observational parameters?

This is a task well-suited to a Machine Learning (ML) approach. We tried applying a number of ML regression methods, including Gaussian process regression, Bayesian ridge regression, support vector regression, gradient boost regression, and simple neural networks (all using the \texttt{scikit-learn}\footnote{\url{https://scikit-learn.org/stable/index.html}} library for  \texttt{Python}; \citealp{Pedregosa2011A}), along with deep neural networks (via the \texttt{keras}\footnote{\url{https://keras.io/}} implementation of \texttt{tensorflow}; \citealp{Chollet2015A}). However, we found that all of these methods tended to be vulnerable to models over-fitting the data, whilst often simultaneously struggling yield accurate predictions for the highest- and lowest-background observations (although Gaussian process regression was somewhat less vulnerable to these problems). The vulnerability to over-fitting makes sense, due to the degeneracies found in the data -- for instance, if a particular object is the target of a multiple-orbit SBC observing campaign in a given cycle, then that specific part of the parameter space will be very densely populated, easily skewing models.

Ultimately, we found that random forest regression was by far the most successful technique. Specifically, we employed the Quantile Random Forest (QRF) regression method, as presented in the \texttt{Python} package \texttt{quantile-forest}\footnote{\label{Footnote:QRF_Docs}\url{https://zillow.github.io/quantile-forest/index.html}} by \citet{Johnson2024A}. Random forest regression constructs many individual decision trees, each of which is trained based upon only a subset of the features (ie, the observational parameters), and using a bootstrap-resampled set of the observations. This makes the process relatively robust against over-fitting. The {\it quantile} forest implementation uses this information to furthermore provide uncertainties on the predicted outputs (in our case, the background level).

\begin{figure}
\centering
\includegraphics[width=0.325\textwidth]{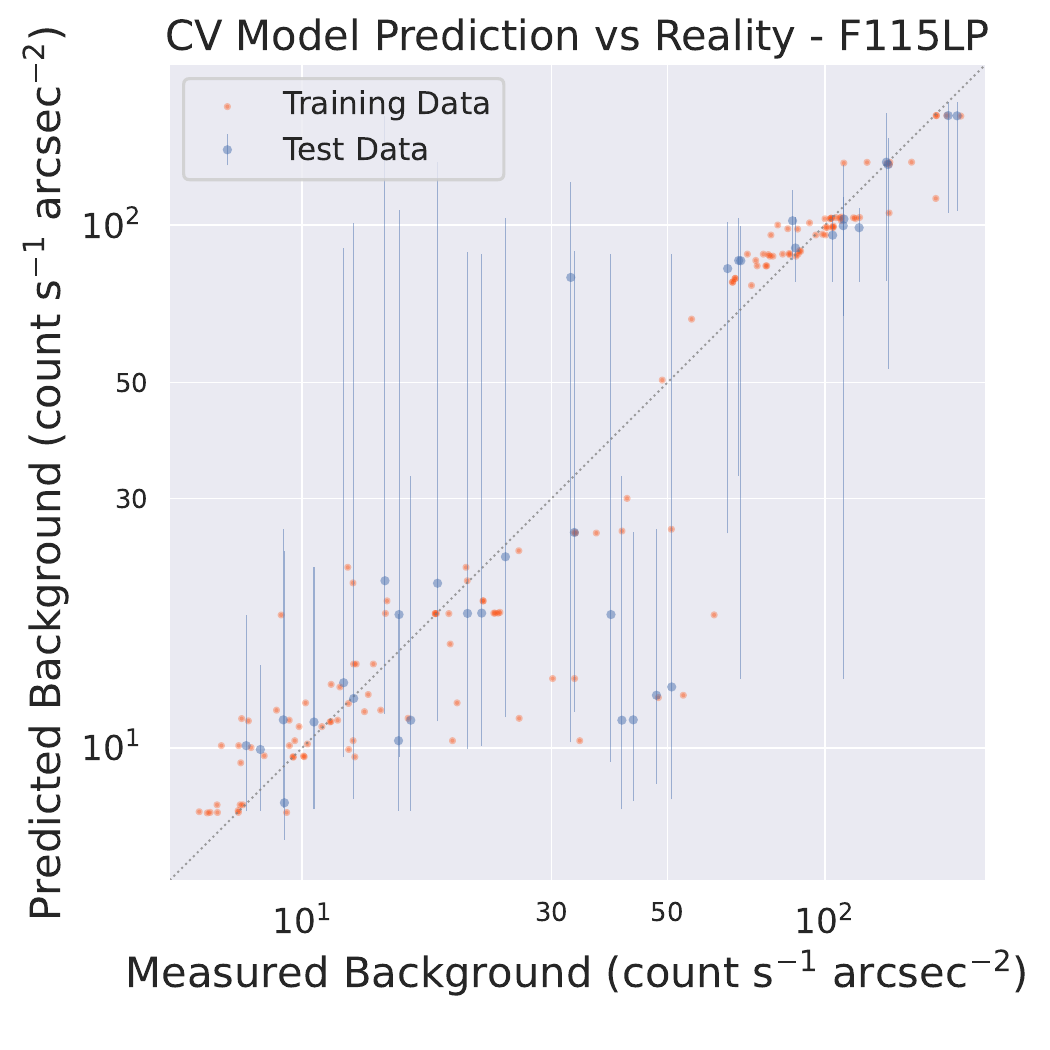}
\includegraphics[width=0.325\textwidth]{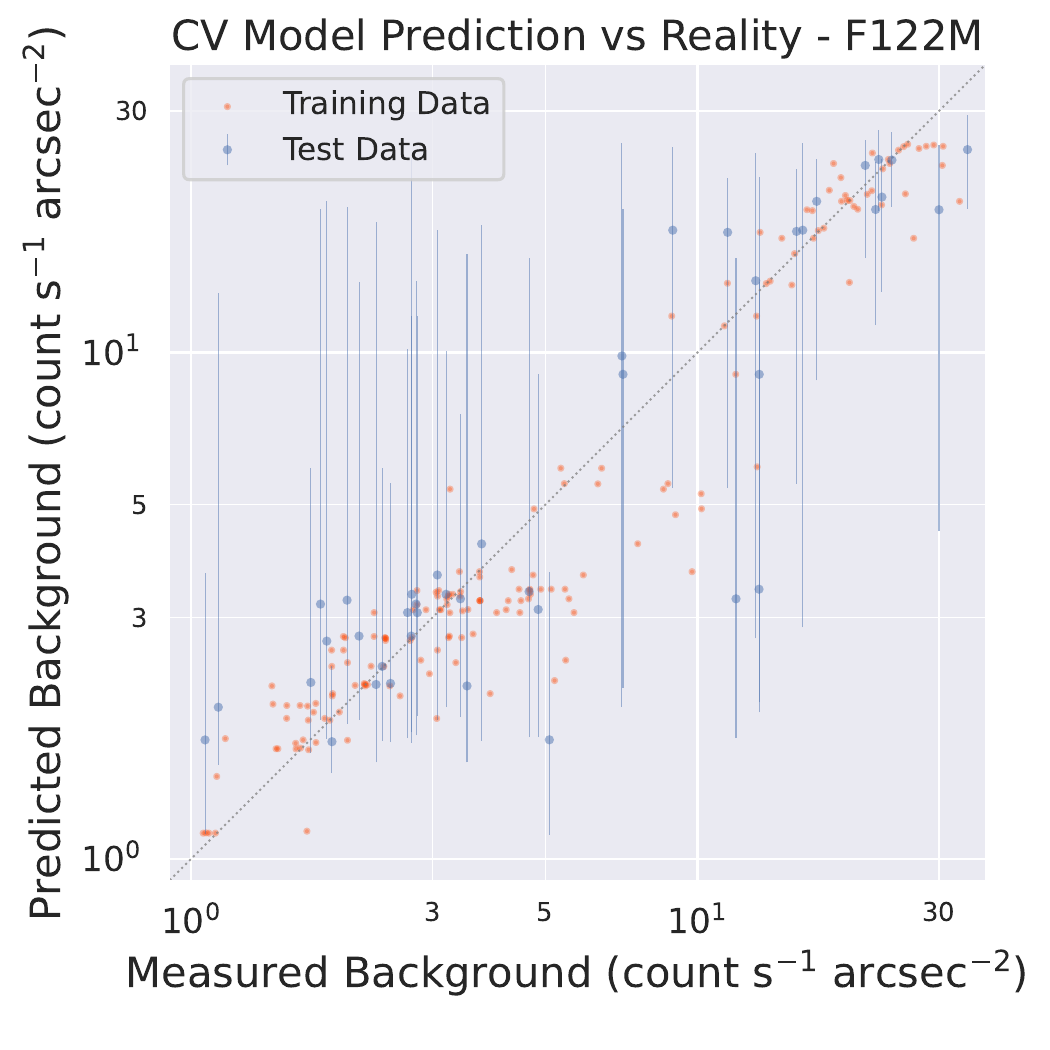}
\includegraphics[width=0.325\textwidth]{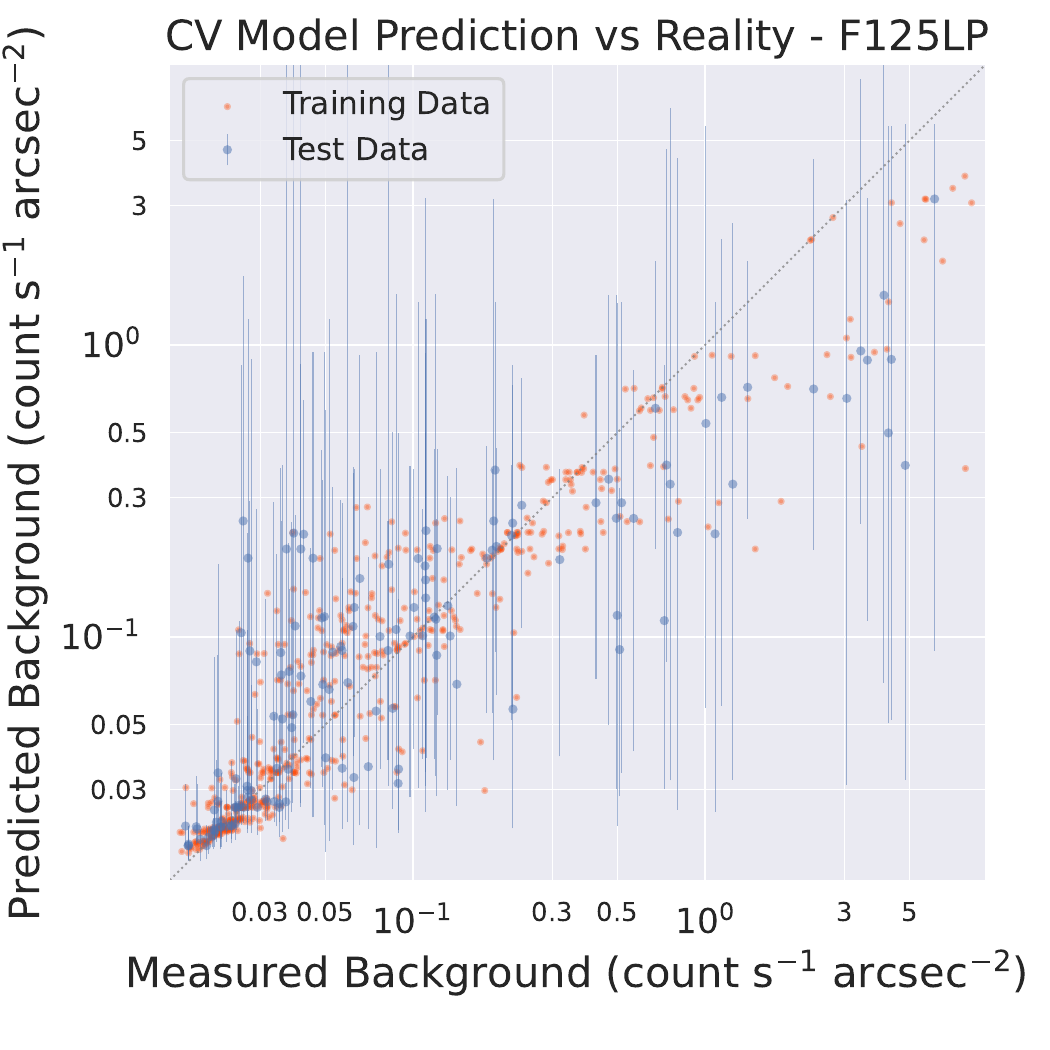}
\includegraphics[width=0.325\textwidth]{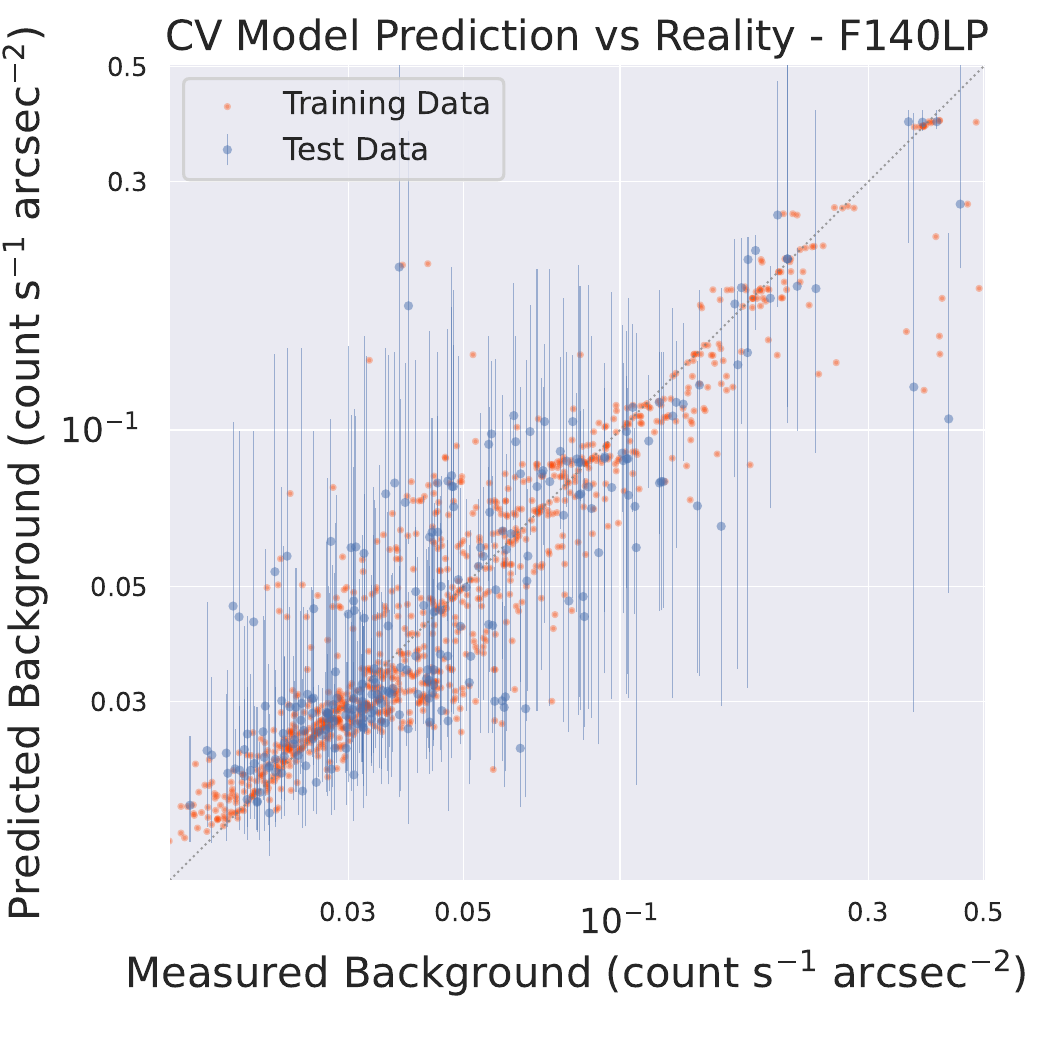}
\includegraphics[width=0.325\textwidth]{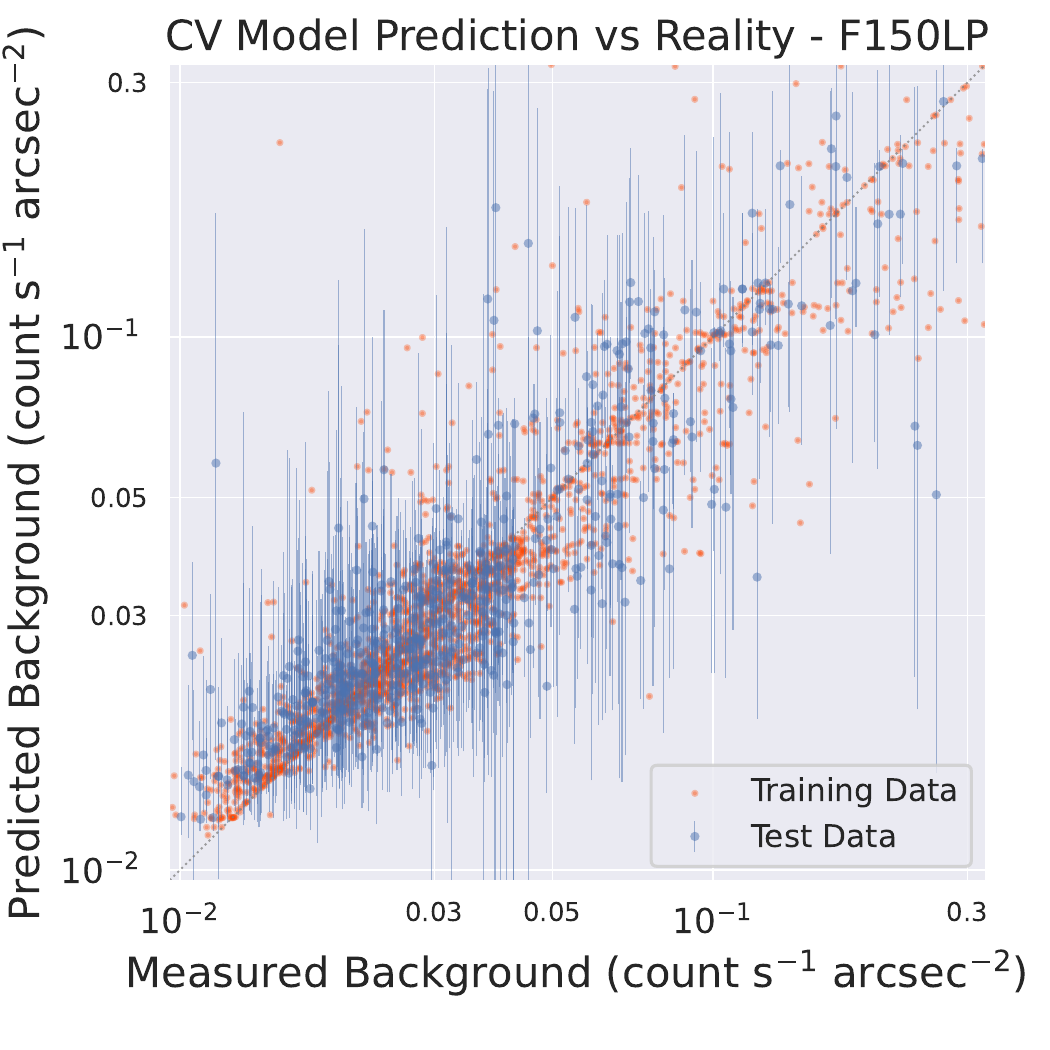}
\includegraphics[width=0.325\textwidth]{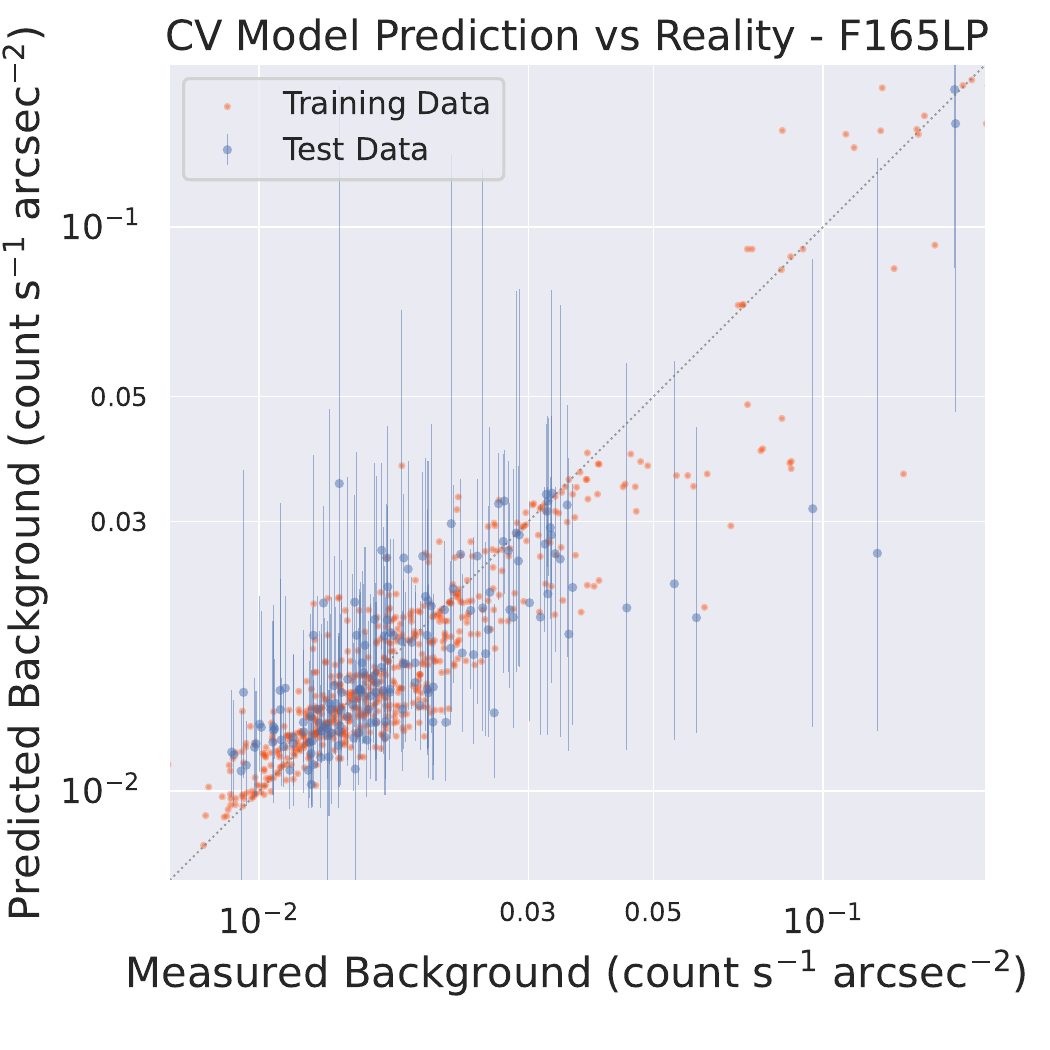}
\includegraphics[width=0.325\textwidth]{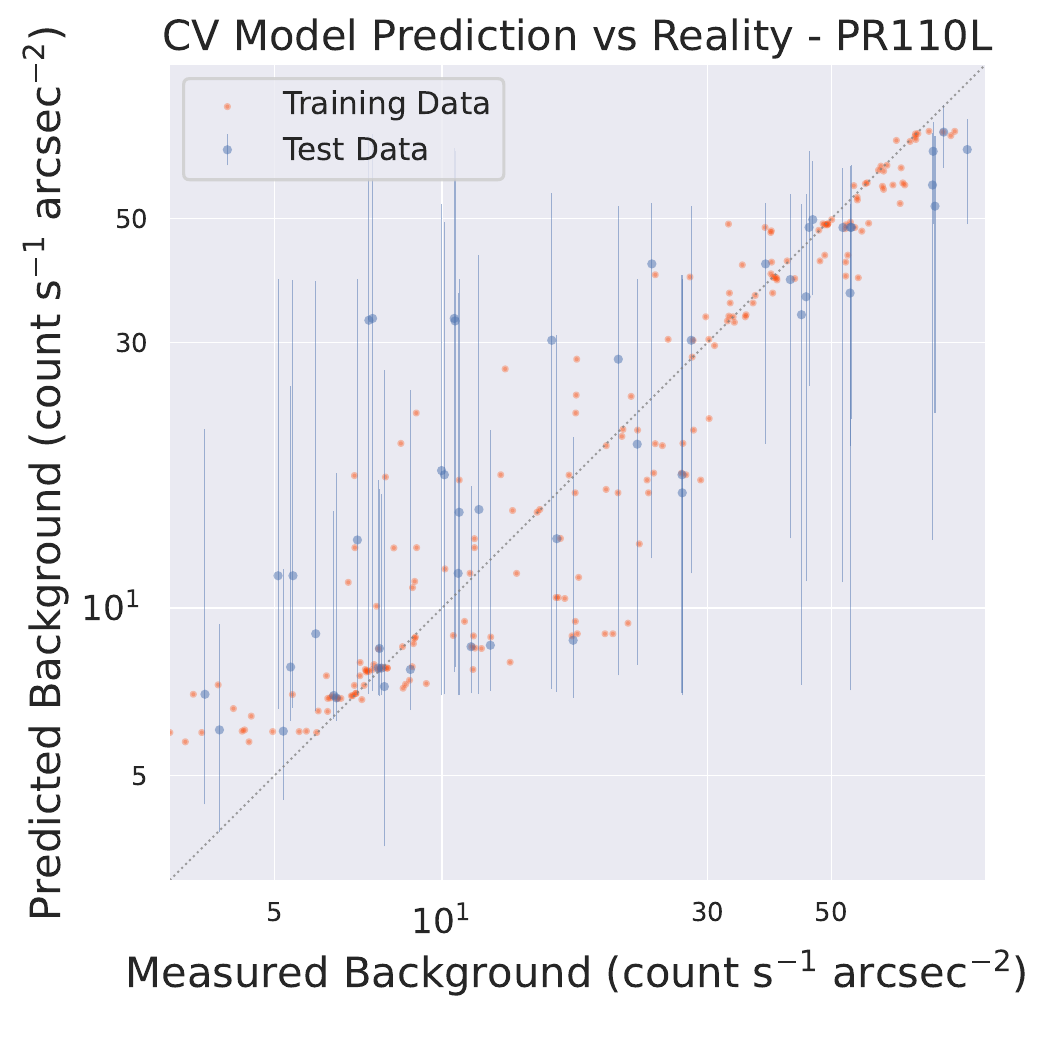}
\includegraphics[width=0.325\textwidth]{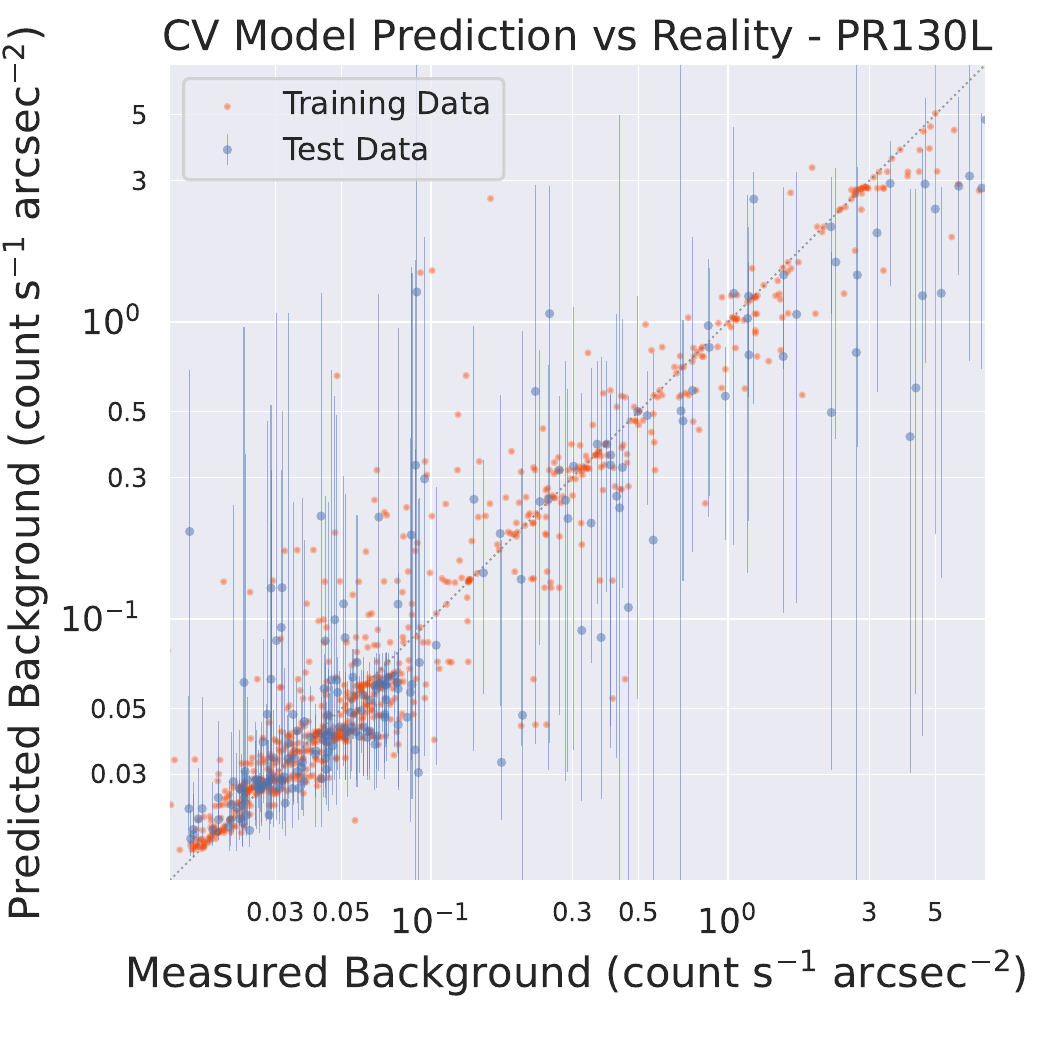}
\caption{Comparison of predicted versus true background levels, for each SBC filter, from our quantile forest regression modelling. For each filter's data, we trained using 80\%\ of the observations, and withheld the remaining 20\%\ for validation testing. For each filter, the training data is plotted in orange, and the test data is plotted in blue (along with the uncertainties on the predictions, as provided by the model). Prism data is shown only for completeness.}
\label{Fig:Model_Accuracy}
\end{figure}

Following standard ML best practice, we only trained the model using 80\%\ of the data, withholding the remaining 20\%\ for validation testing. The results of the modelling are shown in Figure~\ref{Fig:Model_Accuracy}, where for each SBC filter we plot the true background levels for each exposure, versus the background levels predicted by the model, for both the training and test data. When tuning the fitting, we found we were unable to avoid a degree of overfitting for the filters where there are fewer observations available for training (specifically F115LP and PR110L), without compromising the accuracy of the trained models. In-depth descriptions of the training and validation methodologies are provided in Appendix~\ref{AppendixSection:Model_Training}.

Summary statistics for the accuracy of the model in each filter are given in Table~\ref{Table:Model_Accuracy}. We consider the prediction relative accuracy, which we define as the ratio of the model-predicted background to the measured background, for the 20\% of the data held back for validation testing (ie, where a prediction relative accuracy of 1 indicates a perfectly accurate prediction, and a prediction relative accuracy of 1.1 indicates a 10\% over-prediction). 

Additionally, we report the median relative error for each filter (ie, the median of the relative difference between the predicted and measured backgrounds, among the validation data). We use the {\it median} error, as opposed to the mean, to avoid unrepresentative skew from a small number of major outliers (see Figure~\ref{Fig:Model_Accuracy})

We also compute what fraction of the variation in the observed SBC backgrounds is explained by our models. We assume that the residuals between each observation's background value, and the median background value for that filter, represent the `null hypothesis' residuals -- this quantifies the overall scatter in the observed backgrounds when no model is applied. We then compare this to the residuals versus the backgrounds predicted by our model. Or, in other words: for each observation, if one were to predict the background using our model, and compare that to the background `predicted' by just using the median historical background for that filter, how much smaller is the error in the predicted background when using our model? 

For each observation, we therefore compute the ratio of these two residuals. The median of these ratios tells us the average relative decrease in residual when using our model – ie, the fraction of the variation in observed backgrounds explained by our model. For instance, for F140LP, the residuals when using our model are 0.237$\times$ the residuals just using the historical median background as the prediction. As such, our model accounts for $1 - 0.237 = 76.3\%$ of the variation in observed F140LP background levels (ie, our model reduced the scatter by a factor of 4.22). We report these explained scatter fractions for each filter, in Table~\ref{Table:Model_Accuracy}.

For 6 of the 8 filters, our model accounts for 73--81\% of the observed variation in background levels observed. The exceptions to this are F125LP and PR130L, where we find that 97.2\%\ and 96.0\%\ of the variation is explained by our model, respectively. This is because observed backgrounds in F125LP and F130LP span very large ranges -- over 2.5 orders of magnitude (whereas the measured backgrounds for all other filters have a range of \textless\,1.5 orders of magnitude). Therefore, even though the model accuracy at F125LP and PR130L is similar to the other filters, the decrease in scatter is much larger.

\begin{table*}
\centering
\caption{Summary statistics for the relative accuracy of our background-predicting quantile forest regression model. We report the median, 16\th\ percentile, and 84\th\ percentile values for the relative accuracy for each filter. We also report the median of the relative error (in percent), and the fraction of the scatter in the measured background levels that is explained by the model.}
\footnotesize
\label{Table:Model_Accuracy}
\begin{tabular}{lrrrrrrrr}
\toprule \toprule
\multicolumn{1}{c}{} &
\multicolumn{1}{c}{F115LP} &
\multicolumn{1}{c}{F122M} &
\multicolumn{1}{c}{F125LP} &
\multicolumn{1}{c}{F140LP} &
\multicolumn{1}{c}{F150LP} &
\multicolumn{1}{c}{F165LP} &
\multicolumn{1}{c}{PR110L} &
\multicolumn{1}{c}{PR130L} \\
\cmidrule(lr){2-9}
Median Accuracy & 1.049 & 1.003 & 0.958 & 1.006 & 1.022 & 0.982 & 1.134 & 1.060 \\
16\th Percentile & 0.742 & 0.796 & 0.570 & 0.784 & 0.808 & 0.780 & 0.811 & 0.561 \\
84\th Percentile & 1.278 & 1.443 & 1.614 & 1.463 & 1.344 & 1.287 & 1.552 & 1.491 \\
\cmidrule(lr){2-9}
Median Error & 21.3\% & 23.1\% & 27.6\% & 19.2\% & 16.7\% & 15.8\% & 22.7\% & 28.4\% \\
 \cmidrule(lr){2-9}
Explained Scatter & 77.3\% & 80.8\% & 97.2\% & 77.0\% & 80.2\% & 73.8\% & 74.6\% & 96.0\% \\
\bottomrule
\end{tabular}
\end{table*}

\section{Causes of the SBC Background Variation} \label{Section:Background_Causes}

We have verified that our QRF regression model, using the 23 observational parameters laid out in Section~\ref{Section:Observational_Parameters}, is able to accurately predict the background levels encountered by the SBC. We want to understand specifically which observational parameters are driving the observed background level, and how they do so. This can be especially useful if it indicates under what circumstances users could achieve lower backgrounds.

\subsection{Model Analysis with SHapely Additive exPlanations} \label{Subsection:SHAP_Method}

To determine how each observational parameters affects the background level, we used the analysis technique of SHapely Additive exPlanations (SHAP; \citealp{Lundberg2020A}). The SHAP technique is drawn from cooperative game theory, and is designed to describe how the inputs of a machine learning model affect the outputs. SHAP values are computed by measuring the average marginal contribution of each feature to the model’s predictions across all possible combinations of features, by substituting individual feature values for an observation with values drawn at random from other observations, and then repeating this across large numbers of observations. This provides model-agnostic explanations of feature importance, and their effect on model outputs. 

A key value of SHAP analysis is that {\bf it can reveal relationships that can not be seen in the raw data} (due to the confounding interactions between parameters), by isolating the marginal impacts of each parameter.

It should be noted, however, that the SHAP method is technically finding causal relationships in the {\it model}. In this sense, the SHAP approach effectively assumes that a model {\it fully} explains the corresponding observations; whereas, as per Table~\ref{Table:Model_Accuracy}, we know that 3--27\% of the scatter in our observations remains unexplained by our model. Nonetheless, SHAP is a powerful tool for interpreting flexible-but-complex modelling tools, such as the QRF regression model we employ.

We use the implementation provided by the \texttt{SHAP}\footnote{\url{github.com/shap/shap}} package for \texttt{Python}. Specifically, we employed the \texttt{shap.Explainer} function, using the \texttt{permutation} algorithm, and passed a background distribution (ie, the set of points whose values are substituted in to gauge marginal effects) constructed from 250 randomly-sampled observations.

As per the \texttt{SHAP} documentation, a background distribution of 100--1000 samples is generally expected to be adequate. We found that using more than 250 samples for the background distribution did not improve the reliability or convergence of the results, but did require considerably more computational resources. Moreover, for F115LP, the SBC filter with the least archival observations available, there are only 261 observations suitable for our analysis. We therefore opted to use 250 samples for each filter.

\subsection{SHAP Results} \label{Subsection:SHAP_Results}

\begin{figure}
\centering
\includegraphics[width=0.32\textwidth]{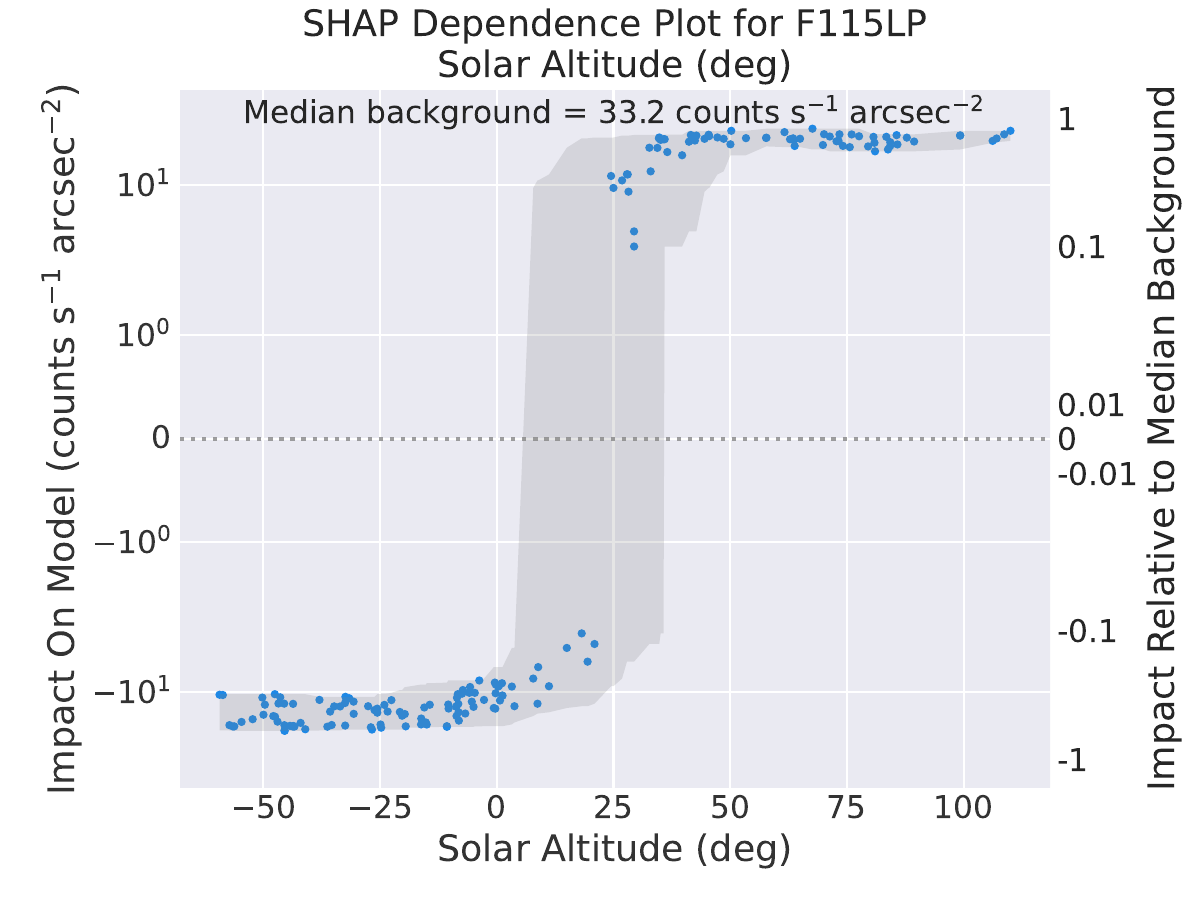}
\includegraphics[width=0.32\textwidth]{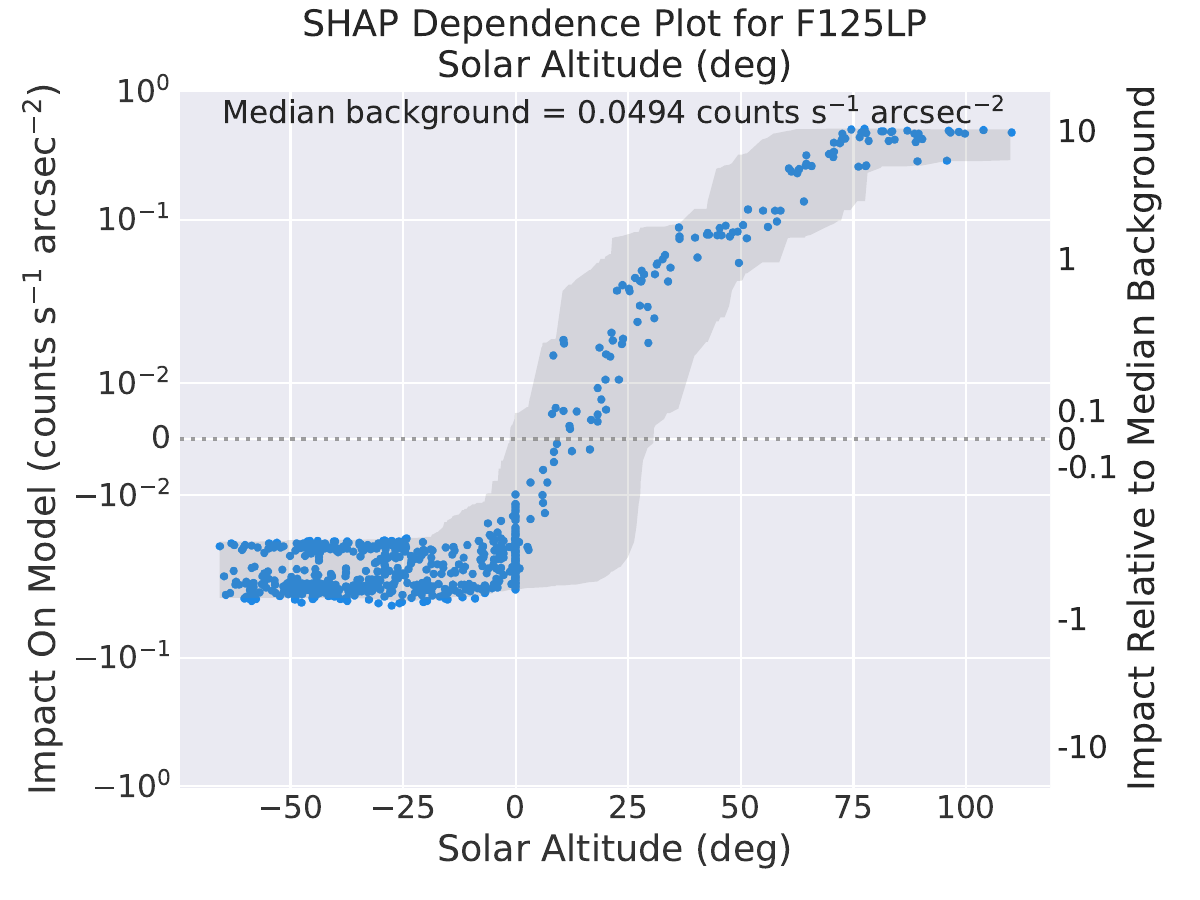}
\includegraphics[width=0.32\textwidth]{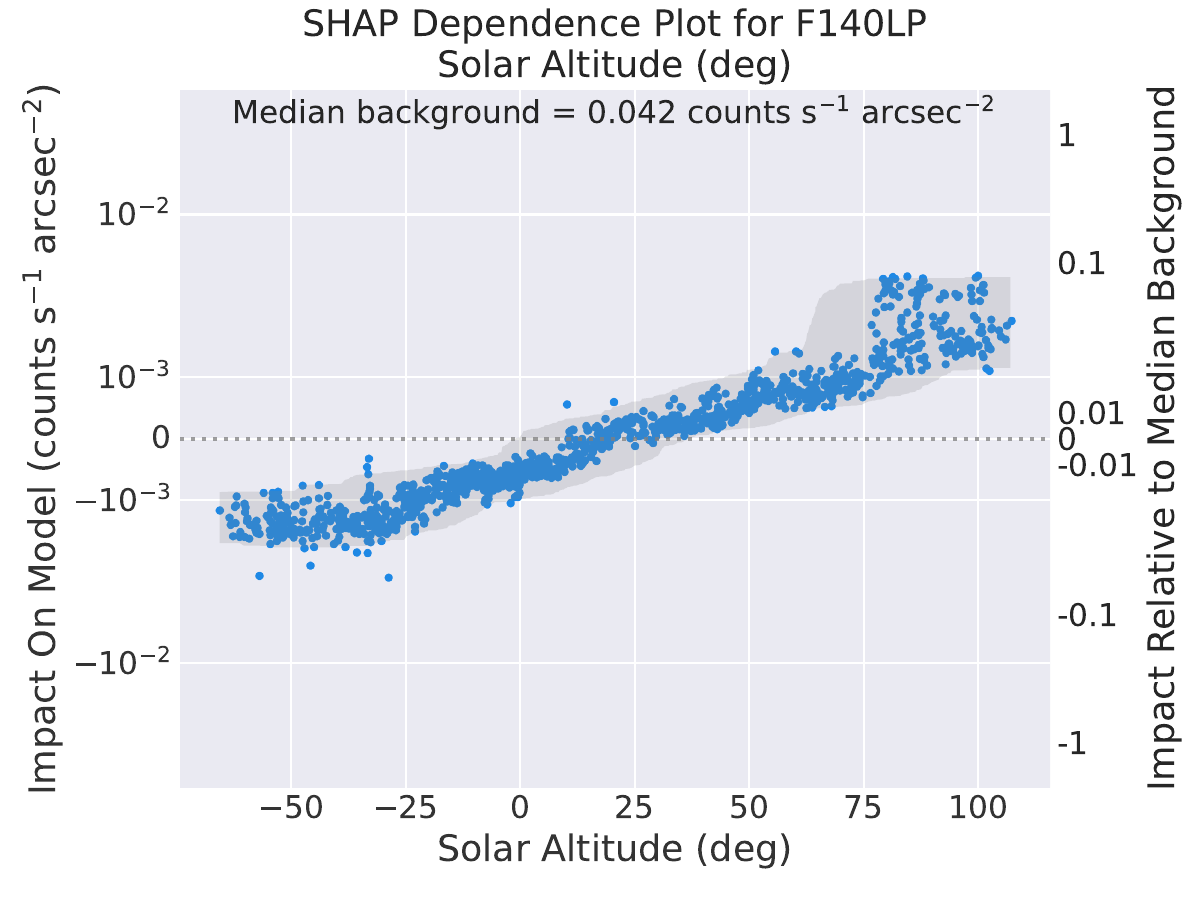}
\includegraphics[width=0.32\textwidth]{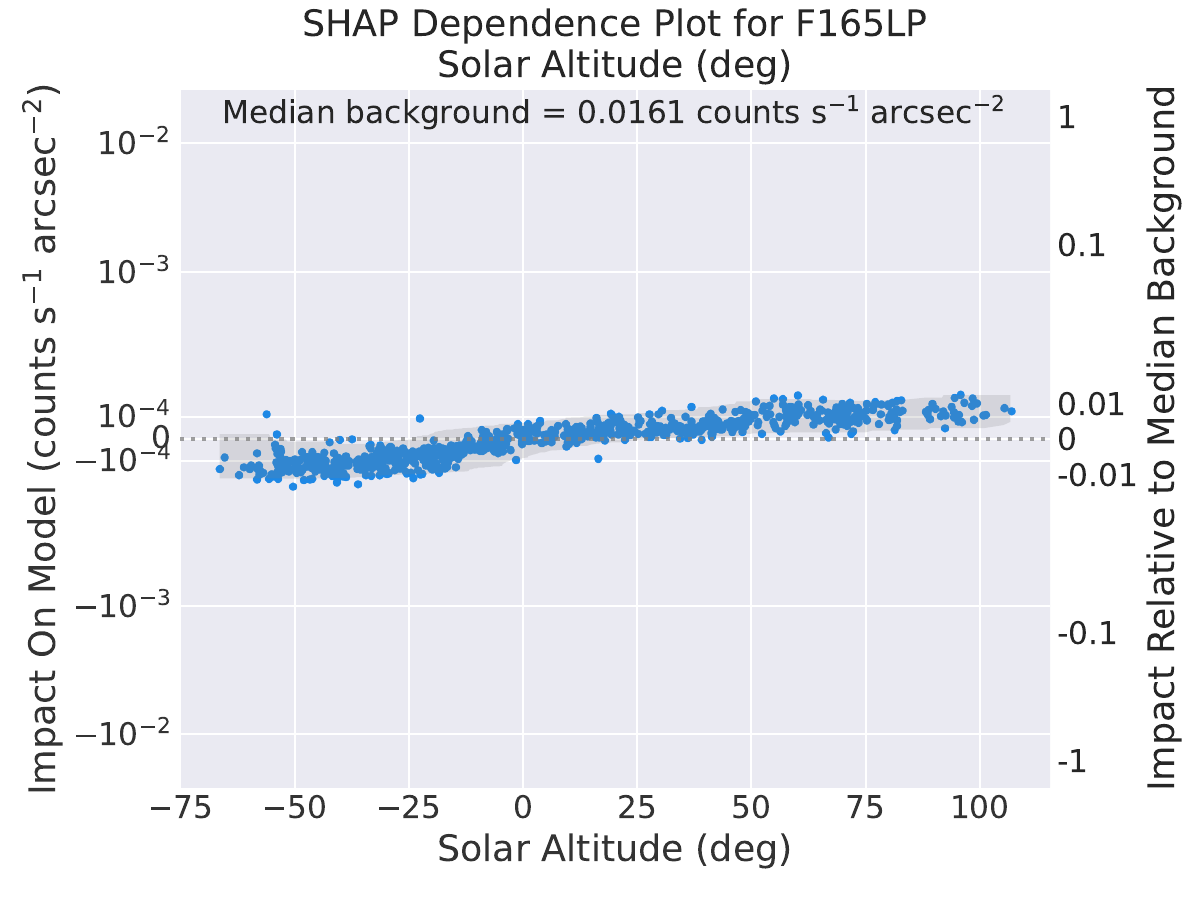}
\includegraphics[width=0.32\textwidth]{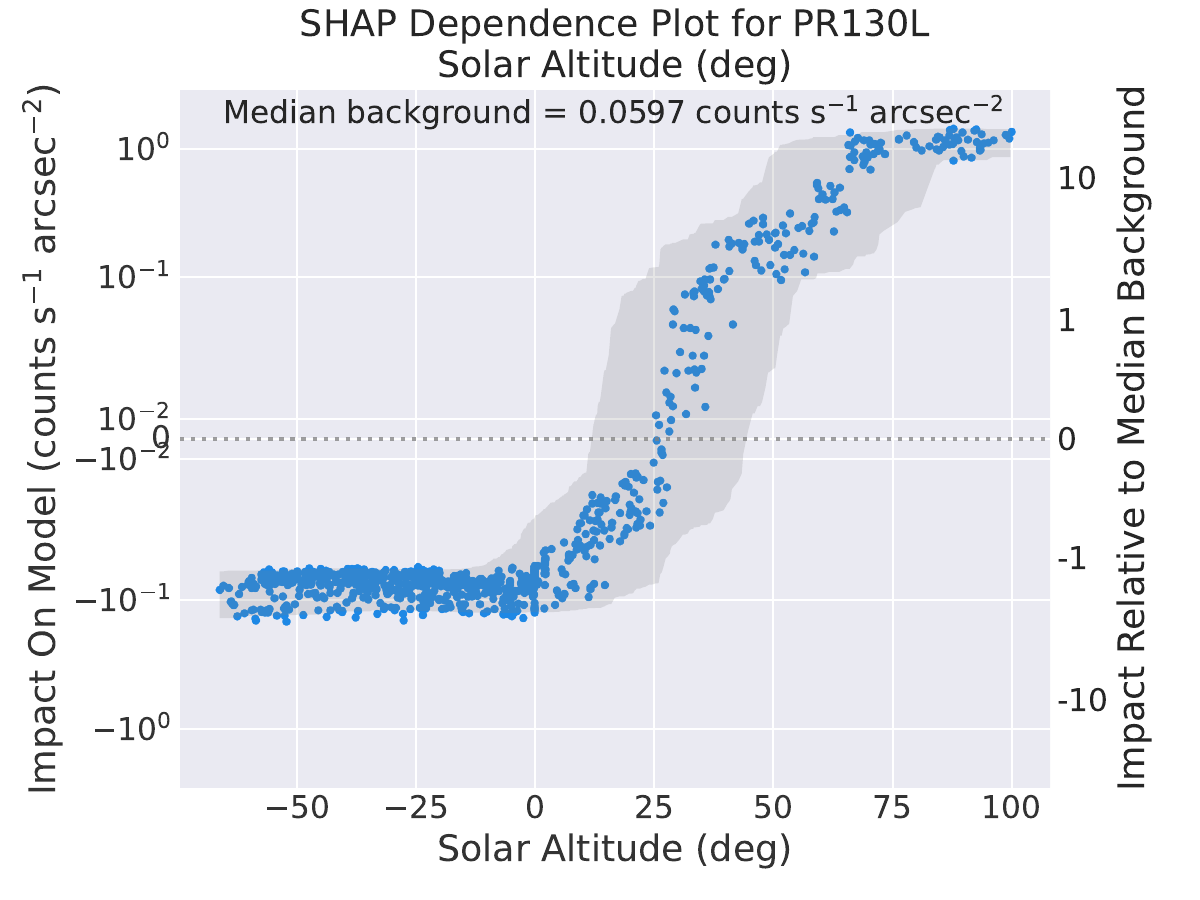}
\includegraphics[width=0.32\textwidth]{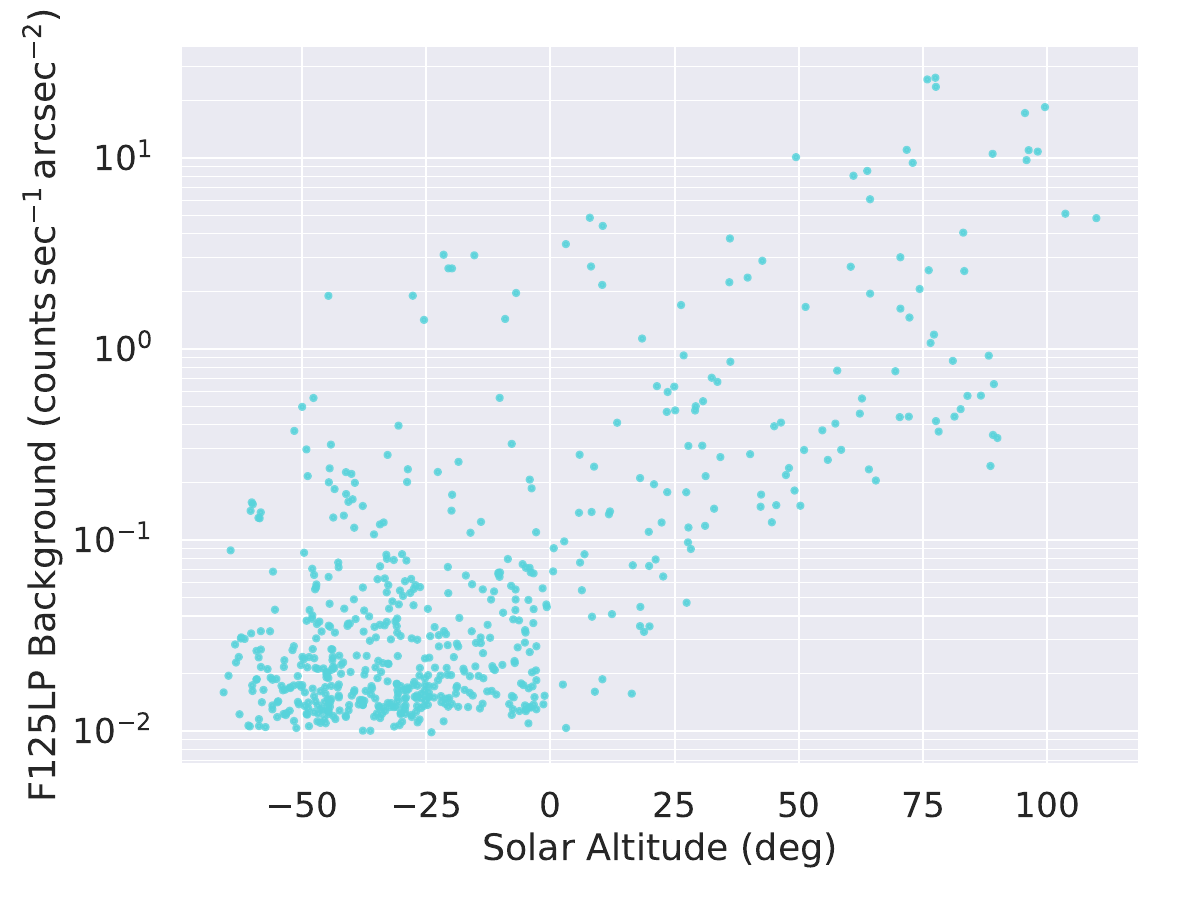}
\caption{SHAP dependence plots for Solar altitude for a representative subset of SBC filters. The left-hand y-axis shows the marginal impact of the parameter on the model prediction in \cpspsqarc, whilst the right-hand y-axis shows the impact on the model normalized to the median background for that filter (the median background for the filter in question is noted on each plot). The y-axes are plotted using a symlog scale (linear either side of 0, and logarithmic elsewhere). The shaded region shows the 90\th\ percentile range in SHAP dependence, averaged within a moving Gaussian window with sigma equal to 5\% of the parameter range. Shown for comparison is a plot of measured background versus Solar altitude for F125LP observations, to illustrate a noisier raw relation, as compared to the marginal impact derived by SHAP analysis.}
\label{Fig:Solar_Alt_Dependence_Plots}
\end{figure}

The SHAP analysis of the QRF regression model indicates that many of the observational parameters in our modeling do indeed affect the background -- but that only a few of them have a major impact. We examine these relationships below, in Section~\ref{Subsection:SHAP_Main_Drivers}; for this examination, we make use of dependence plots and `beeswarm' plots. Here, we first briefly explain these plots, for readers who may be unfamiliar with them.

\subsubsection{SHAP Dependence Plots} \label{Subsubsection:SHAP_Dependence_Plots}

Dependence plots show the marginal impact a given observational parameter had on the final model prediction. Each point on a dependence plot corresponds to one observation, indicating the impact the parameter in question had on the final predicted background for that observation, as determined by the SHAP analysis. Note that the y-axis values a dependence plot do {\it not} show the actual predicted background for each observation; rather they show only the marginal contribution of the parameter in question to the background predictions. Example dependence plots, for the Solar altitude parameter, are shown in Figure~\ref{Fig:Solar_Alt_Dependence_Plots}.

\begin{figure}
\centering
\includegraphics[width=0.32\textwidth]{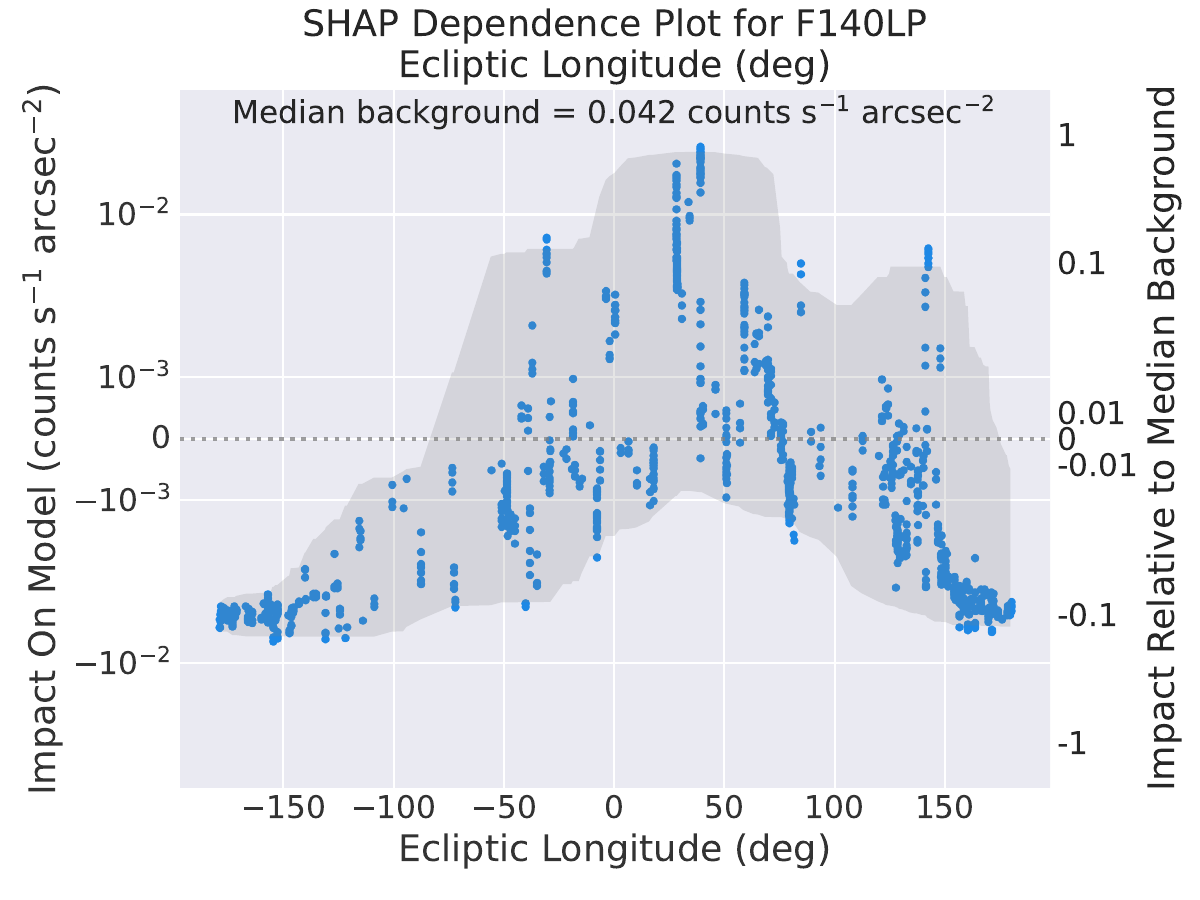}
\includegraphics[width=0.32\textwidth]{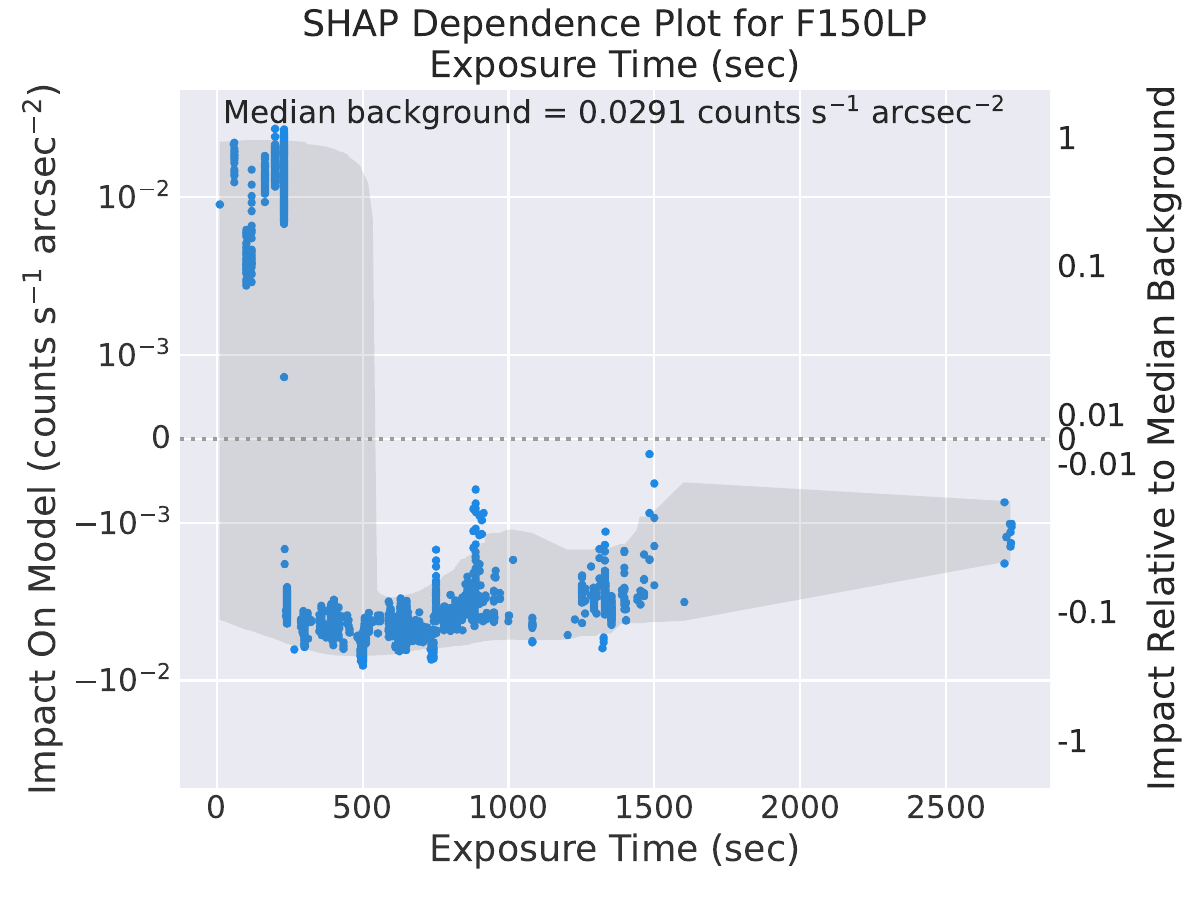}
\includegraphics[width=0.32\textwidth]{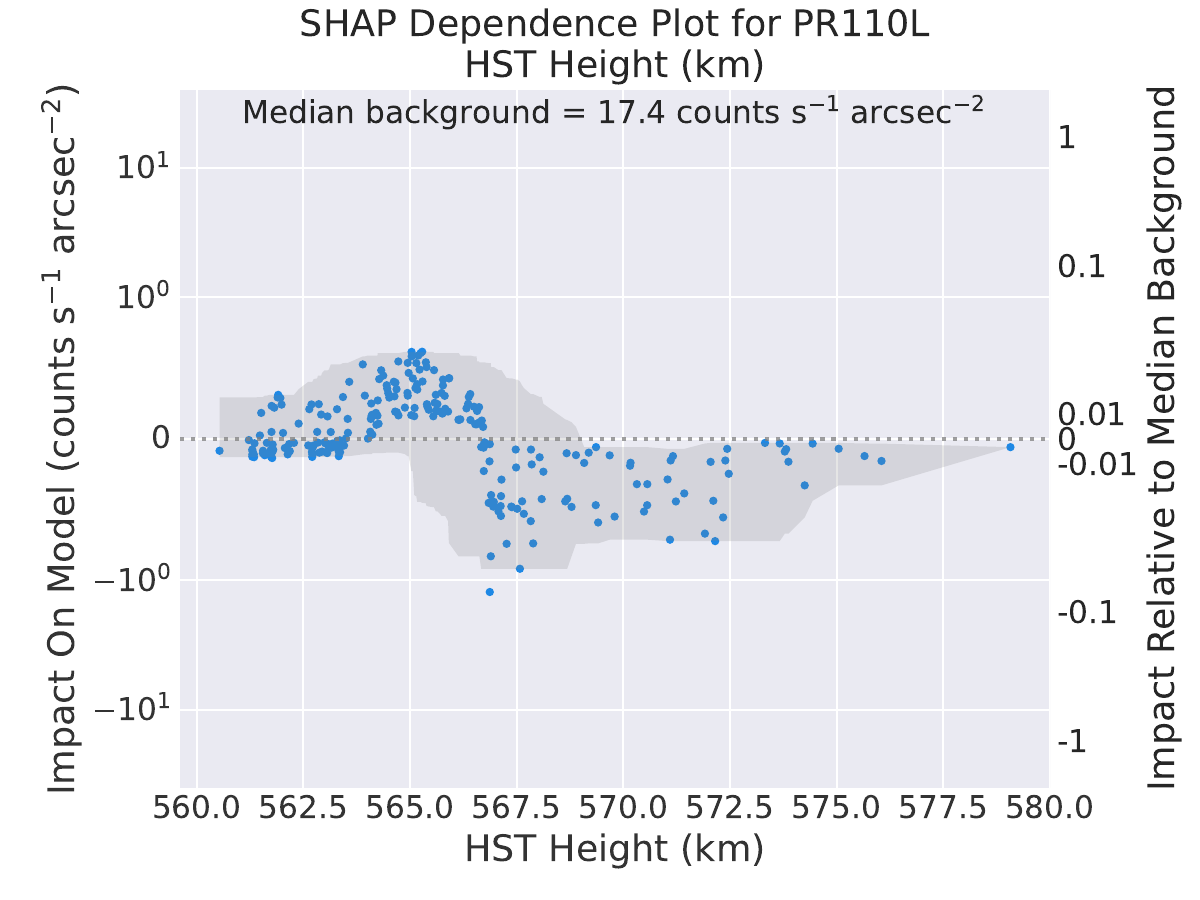}
\caption{SHAP dependence plots for several relations where our QRF regression model has not constrained how a parameter impacts the background, as indicated by large scatter, either in localized spikes and jumps, or throughout larger portions of the parameter range. Details otherwise as per Figure~\ref{Fig:Solar_Alt_Dependence_Plots}.}
\label{Fig:Bad_Dependence_Plots}
\end{figure}

For parameters whose marginal impact on the model prediction is not well-constrained, the dependence plot will show conspicuous scatter. This scatter can be present across the full parameter range, in certain regions, or as `spikes' limited to certain parts of the range. Any apparent trends in dependence plots with considerable scatter should be treated with a high degree of caution -- even if there appears to be a significant overall trend, the scatter indicates that the model does not have a stable grasp on the relationship between that parameter at the resulting background. Dependence plots often have more scatter at the lower and upper extremes of the parameter range, where behaviour is less well constrained. Examples of likely-unreliable dependence relations are shown in Figure~\ref{Fig:Bad_Dependence_Plots}. 

In general, if our QRF regression model has established a clear reliable correlation between a given parameter and the resulting background, then the marginal impact determined by the SHAP analysis should result in a tight trend. In all our dependence plots, we show a shaded region corresponding to the 90\th\ percentile range in SHAP dependence averaged within a moving Gaussian window with sigma equal to 5\% of the parameter range. This serves to illustrate the range of plausible dependence relations; any line (straight or otherwise) that could be drawn though this shaded region should be considered compatible with the SHAP dependence.

The full set of dependence plots, for all observational parameters, for all filters, is provided in Appendix~\ref{AppendixSection:SHAP_Dependence_Plots}.

\subsubsection{SHAP Beeswarm Plots} \label{Subsubsection:SHAP_Beeswarm_Plots}

\begin{figure}
\centering
\includegraphics[width=0.9\textwidth]{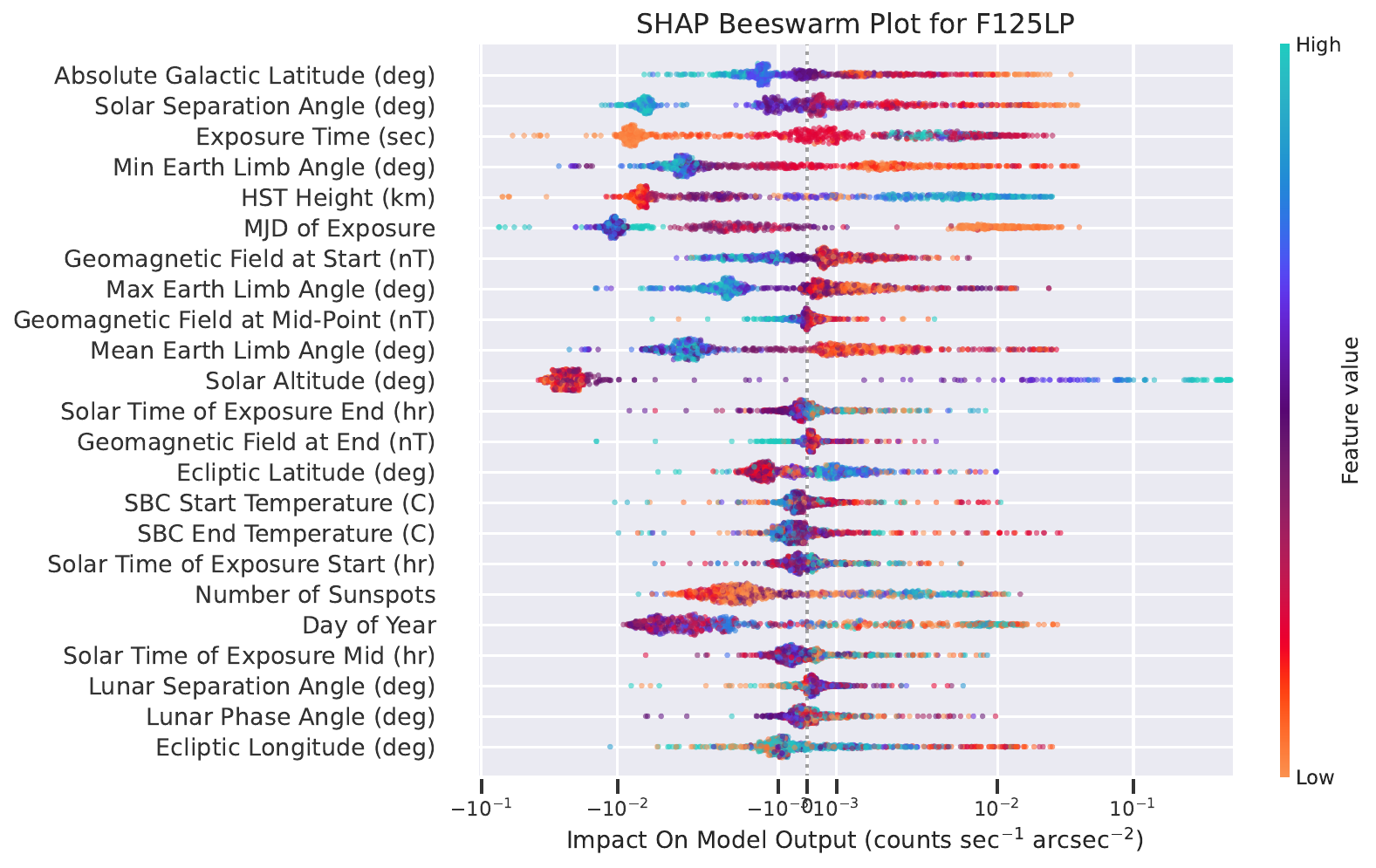}
\includegraphics[width=0.9\textwidth]{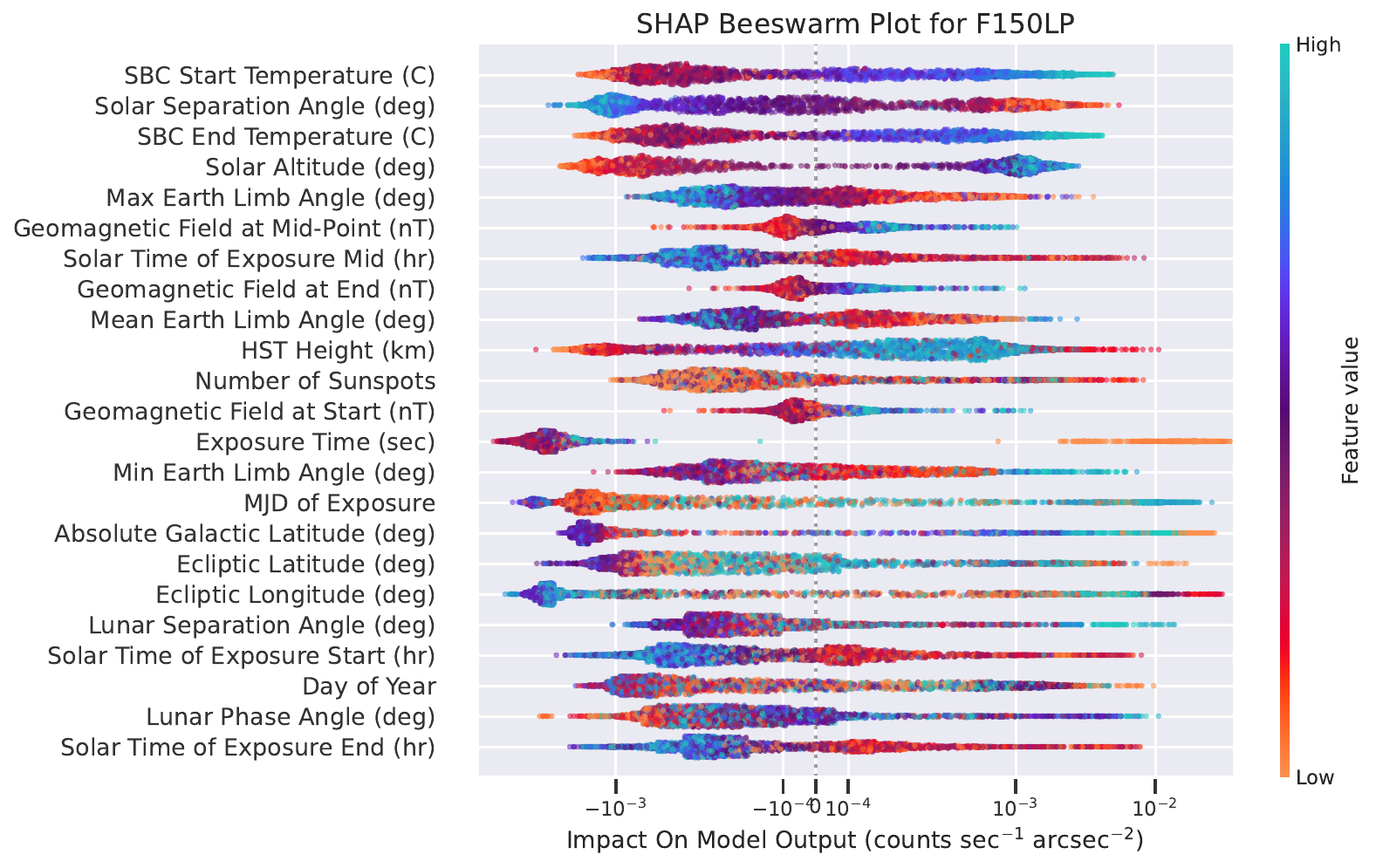}
\caption{Beeswarm plot for SHAP analysis of our QRF regression model for F125LP and F150LP. The rows are ordered according to the tightness of the correlation between the parameter in question, and the impact on the model (calculating using the Kendall's tau correlation coefficient; \citealp{Kendall1990}). This means that parameters located lower in the plot are more likely to have poorly-constrained or spurious impacts on the model predictions.}
\label{Fig:Example_Beeswarm_Plots}
\end{figure}

To display the dependence information of all observational parameters at once, we use `beeswarm' plots. Beeswarm plots summarize the marginal impact of the variation of every parameter on the model prediction. This aids in identifying the observational parameters that have the largest impact on the predicted background.

Each row in a beeswarm plot displays the SHAP results for one parameter, with each point representing one observation. The horizontal position of a point indicates the marginal contribution of that parameter's value to the total model prediction (ie, the predicted background) for that observation, relative to the average prediction. Points are color-coded according the value the observational parameter in question had for each observation. 

Within a row, the specific vertical position of a point does not connote any information; rather, at horizontal positions where there are more observations, the points simply become `bunched up' to indicate this fact.

We plot our beeswarm plots with the rows ordered according to the tightness of the correlation between the parameter in question, and the impact on the model (calculating using the Kendall's tau correlation coefficient; \citealp{Kendall1990}). As such, rows located higher in each plot indicate to parameters where the relationship between the parameter and the model impact is stronger. Conversely, the lower a row is, the weaker/noisier the correlation between the parameter, and its model impact.

Figure~\ref{Fig:Example_Beeswarm_Plots} shows the beeswarm plot of our SHAP analysis for F125LP and F150LP. As F150LP this is the SBC filter with the largest number of observations, it is the filter where the QRF regression model, and the SHAP analysis of it, are best able to capture the underlying relations.  

Using the F150LP plot in Figure~\ref{Fig:Example_Beeswarm_Plots} as an example of interpreting beeswarm plots: The rows for SBC start and end temperature show that higher temperatures are strongly associated with larger marginal increases to the predicted background -- whilst lower temperatures are associated with larger marginal {\it decreases} in the background (as expected). In contrast, the row for Earth limb angle has points spanning a reduced range, showing that this parameter has less impact in predicted background. The direction of the color gradient shows that higher Earth limb angles are generally associated with larger reductions in predicted background. The color gradient for Earth limb angle is also `noiser' than the color gradient for SBC temperature, indicating that the correlation is less tight. 

The full set of beeswarm plots, for all filters, is provided in Appendix~\ref{AppendixSection:SHAP_Beeswarm_Plots}.

\subsection{Main Drivers of Variation in the SBC Background} \label{Subsection:SHAP_Main_Drivers}

The SHAP analysis reveals that a relatively small subset of the observational parameters we incorporated into our QRF regression modelling appear to be the primary drivers of the variation in background levels encountered in SBC observations. 

\subsubsection{Solar Position} \label{Subsubsection:SHAP_Solar_Position}

The position of the Sun during SBC observations is often the observational parameter that has the largest impact on background levels. Figure~\ref{Fig:Solar_Alt_Dependence_Plots} shows the SHAP dependence plots for Solar altitude for several filters. For filters that can be strongly affected by the Ly-$\alpha$ and O{\sc i} airglow lines -- such as F112LP, F125LP, and, PR130L -- the predicted background increases dramatically if the Sun is high above the horizon, by a difference corresponding to 2--20$\times$ the median background level.

We note that this is an excellent validation our QRF regression model -- the model does not {\it a priori} `know' that a Solar altitude of 0\,$^{\circ}$ corresponds to the Sun coming above the horizon (note that this relationship is comparatively noisy in the raw data; see the lower-right panel of Figure~\ref{Fig:Solar_Alt_Dependence_Plots}). Nor does the model know which filters are particularly sensitive to the brightest FUV airglow lines. Nonetheless, the model precisely identified 0\,$^{\circ}$ as a sharp transition point in expected background levels in these filters. This also confirms the ability of the SHAP analysis method to extract the relationships learned by the QRF regression model.

Whilst the F150LP and F165LP filters are not sensitive to the brightest airglow lines, their backgrounds still appear to show clear dependence on Solar altitude (albeit a small one, relative to the median backgrounds). This could be due to the weaker airglow lines these filters do cover (see Section~\ref{Section:Introduction}), or potentially a scattered light effect. 

In most filters the background also shows a dependence on Solar separation angle. This may be due to the fact that there is more airglow in the direction of sunrise in the morning, and sunset in the evening. Therefore, if HST is pointing in the direction of increased airglow, it will necessarily also be observing in a direction with a reduced Solar separation angle.

In a similar vein, the filters sensitive to  Ly-$\alpha$ (F115LP, F122M, and PR110L) show background dependence on Solar time, with background peaking in the afternoon or early evening. This suggests that this effect is due to the increase in Solar excitation of atmospheric Ly-$\alpha$ throughout the day, as would be expected.

\begin{figure}
\centering
\includegraphics[width=0.32\textwidth]{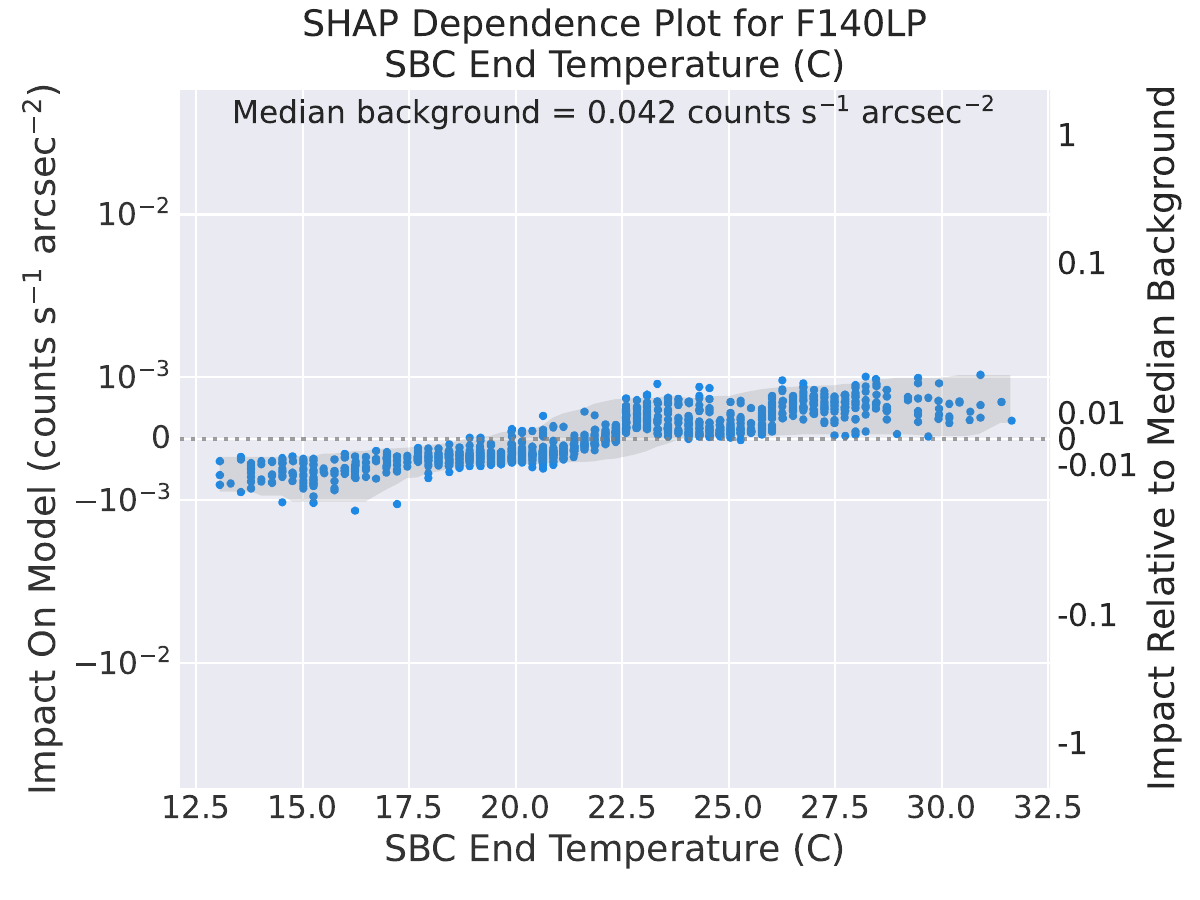}
\includegraphics[width=0.32\textwidth]{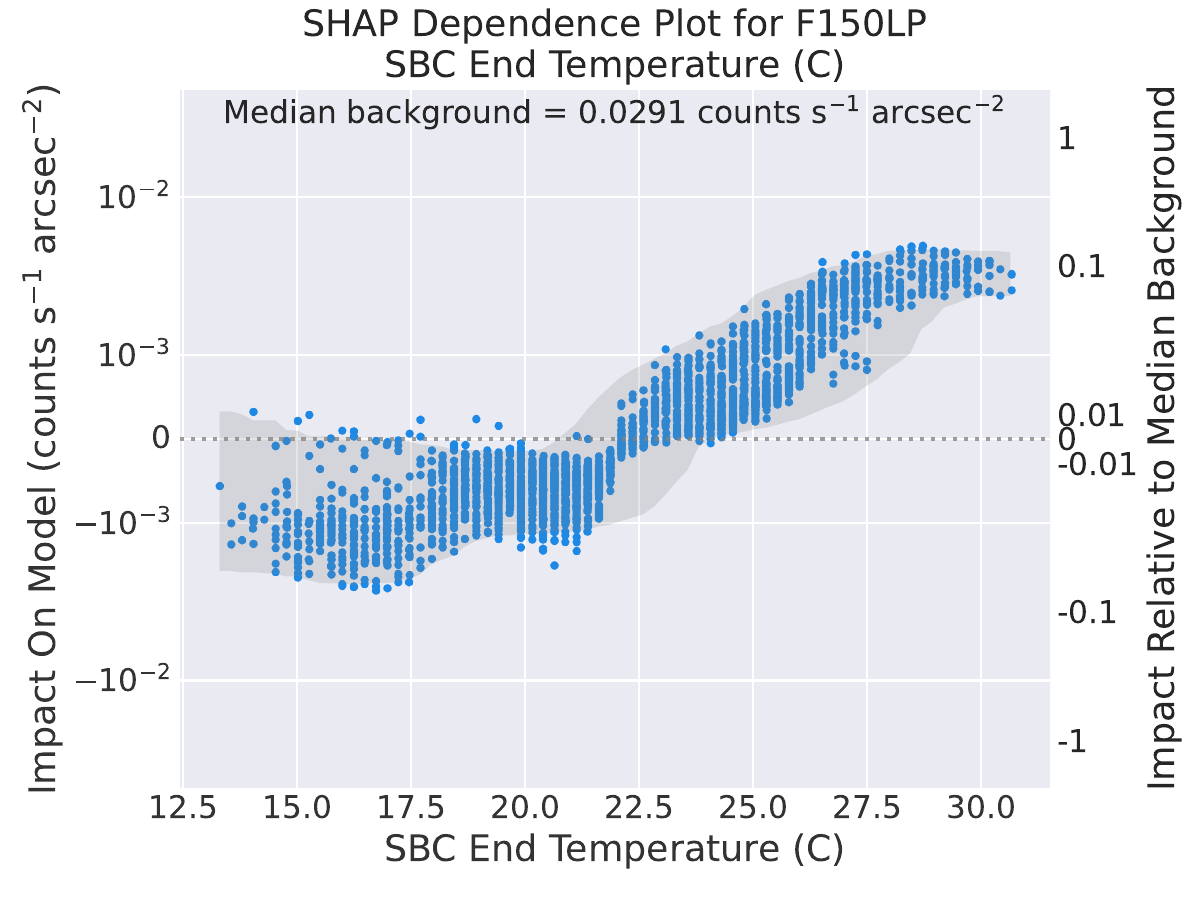}
\includegraphics[width=0.32\textwidth]{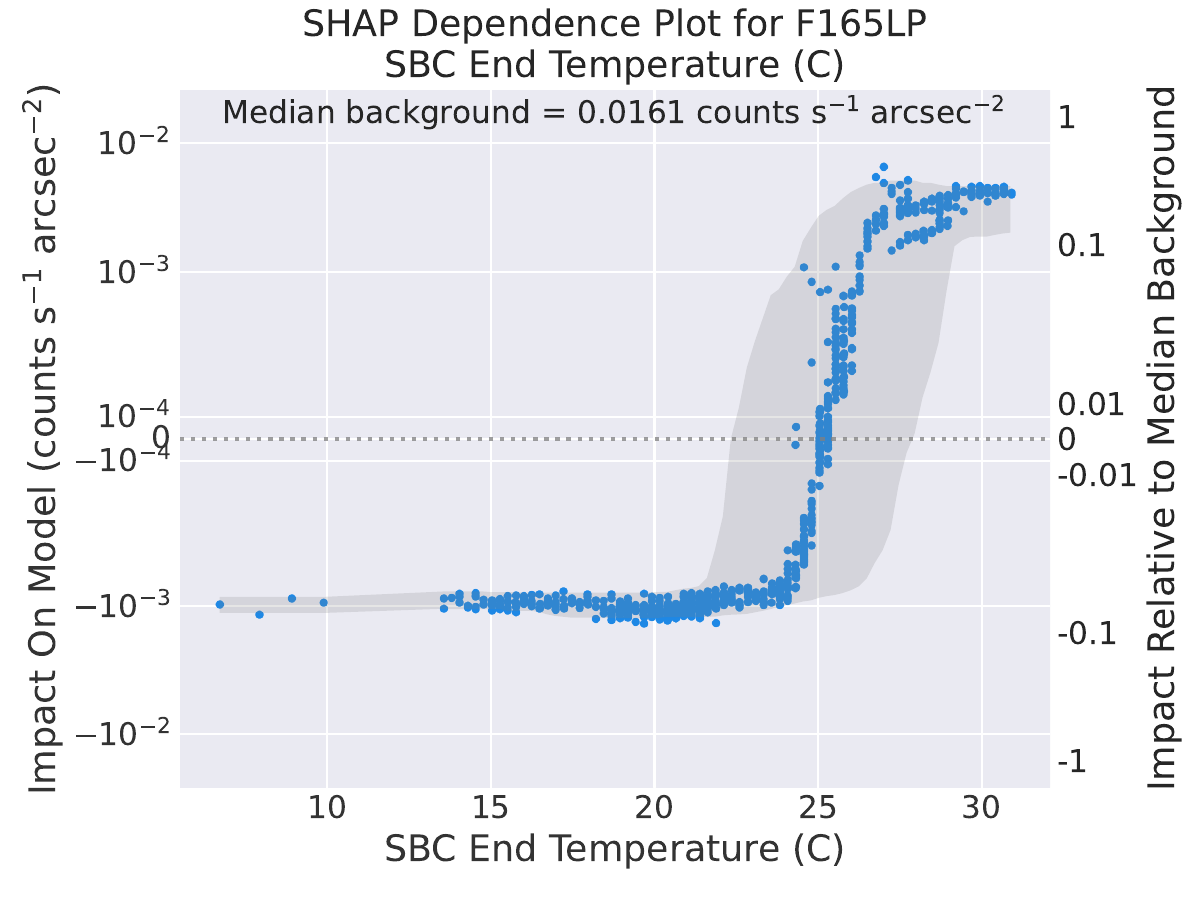}
\caption{SHAP dependence plots for SBC temperature at the end of the exposure, for several filters. Details otherwise as per Figure~\ref{Fig:Solar_Alt_Dependence_Plots}.}
\label{Fig:SBC_Temp_Dependence_Plots}
\end{figure}

\subsubsection{SBC Temperature} \label{Subsubsection:SHAP_Temperature}

As expected, the temperature of the SBC is the dominant driver of background in the longer-wavelength F150LP and F165LP filters (see Figure~\ref{Fig:SHAP_Beeswarm_F150LP_F165LP}). In these filters, where airglow is not a major contributor, the dark rate will be the main source of non-astrophysical counts. This is compounded by the fact that the longer-wavelenth long-pass filters have successively narrower bandpasses, with reduced throughput. Therefore, longer exposures are generally required, increasing the instrument temperature and thence the dark rate.

Figure~\ref{Fig:SBC_Temp_Dependence_Plots} shows the SHAP dependence plots for the SBC temperature as recorded at the end of the exposure\footnote{The SBC temperature at the start of each exposure is also included as one of our model parameters. Whilst both start and end temperature had similar impacts on the predicted background, the impact of the end temperature tended to be of a slightly larger magnitude on average, as would be expected.} for F140LP, F150LP, and F165LP.  The relationship for F165LP, the most-affected filter, is especially strong, with a sharp increase in background at $\approx$24\,$^{\circ}$C. The difference between an end temperature of 22\,$^{\circ}$C vs 27\,$^{\circ}$C in F165LP corresponds to 0.0025\,\cpspsqarc, or about 25\% of the filter median background. 

Even for most of the filters at shorter wavelengths than F140LP, the SHAP analysis also finds that the SBC temperature is still well-correlated with background (except for F125LP). However, the magnitude of the impact on the model is far smaller, \textless4\%\ of the median background in each filter.

We note that when measuring the background level in exposures, our source-masking procedure will mask the high-dark portion of the detector in exposures where the dark rate is significant, as can be seen in Figure~2 of \citet{CJRClark2025C}. The effect of temperature on the background shown in Figure~\ref{Fig:SBC_Temp_Dependence_Plots} therefore corresponds to impact expected over the rest of the detector, where the dark rate is lower (and/or the impact for exposure where the high-dark region is less conspicuous).

\begin{figure}
\centering
\includegraphics[width=0.32\textwidth]{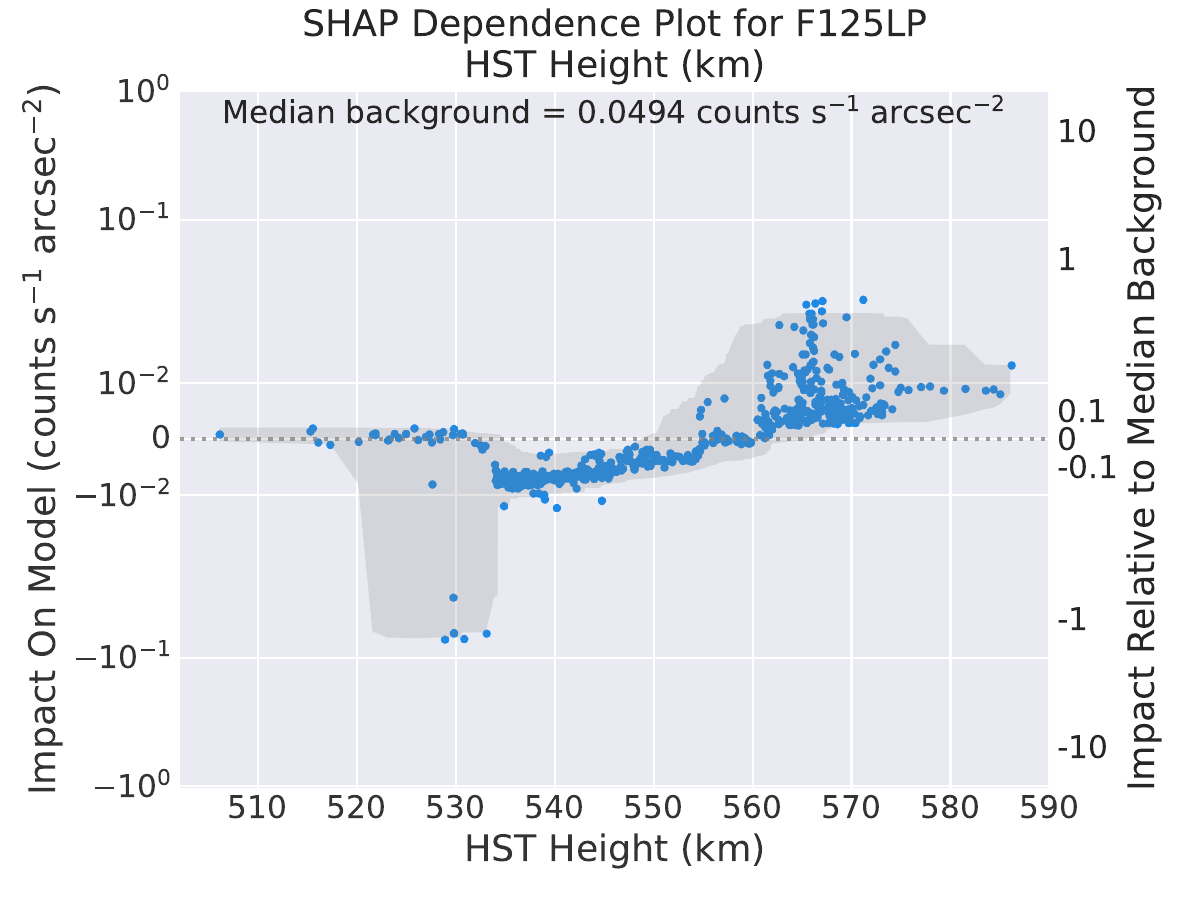}
\includegraphics[width=0.32\textwidth]{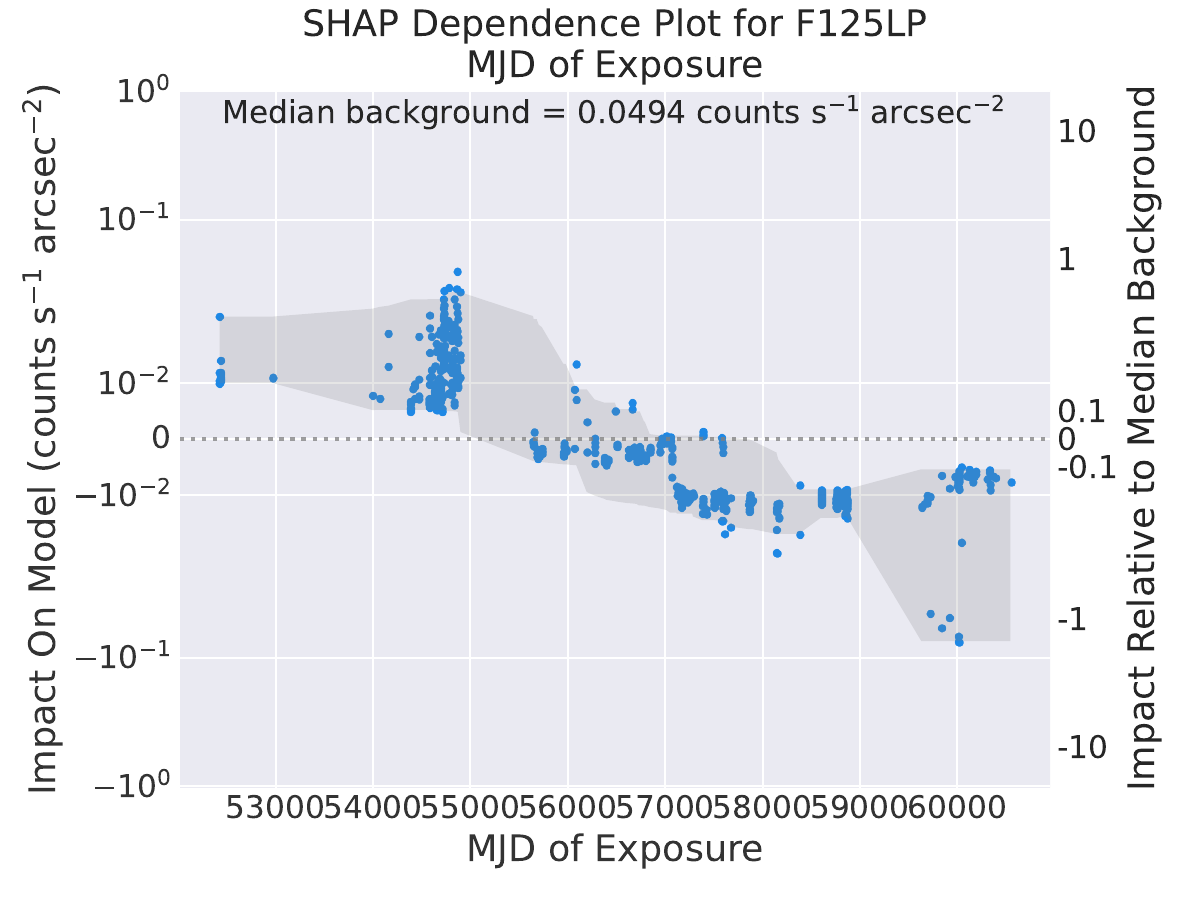}
\includegraphics[width=0.32\textwidth]{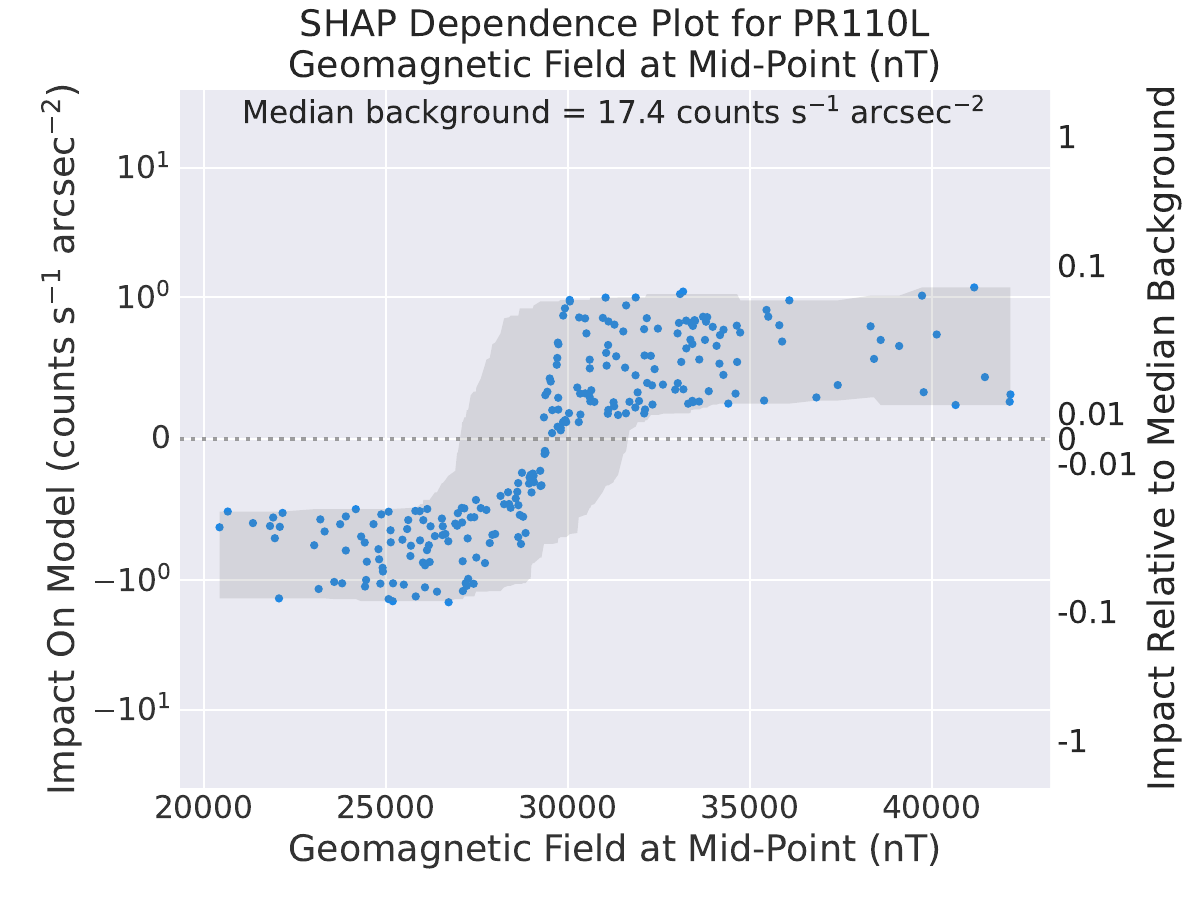}
\includegraphics[width=0.32\textwidth]{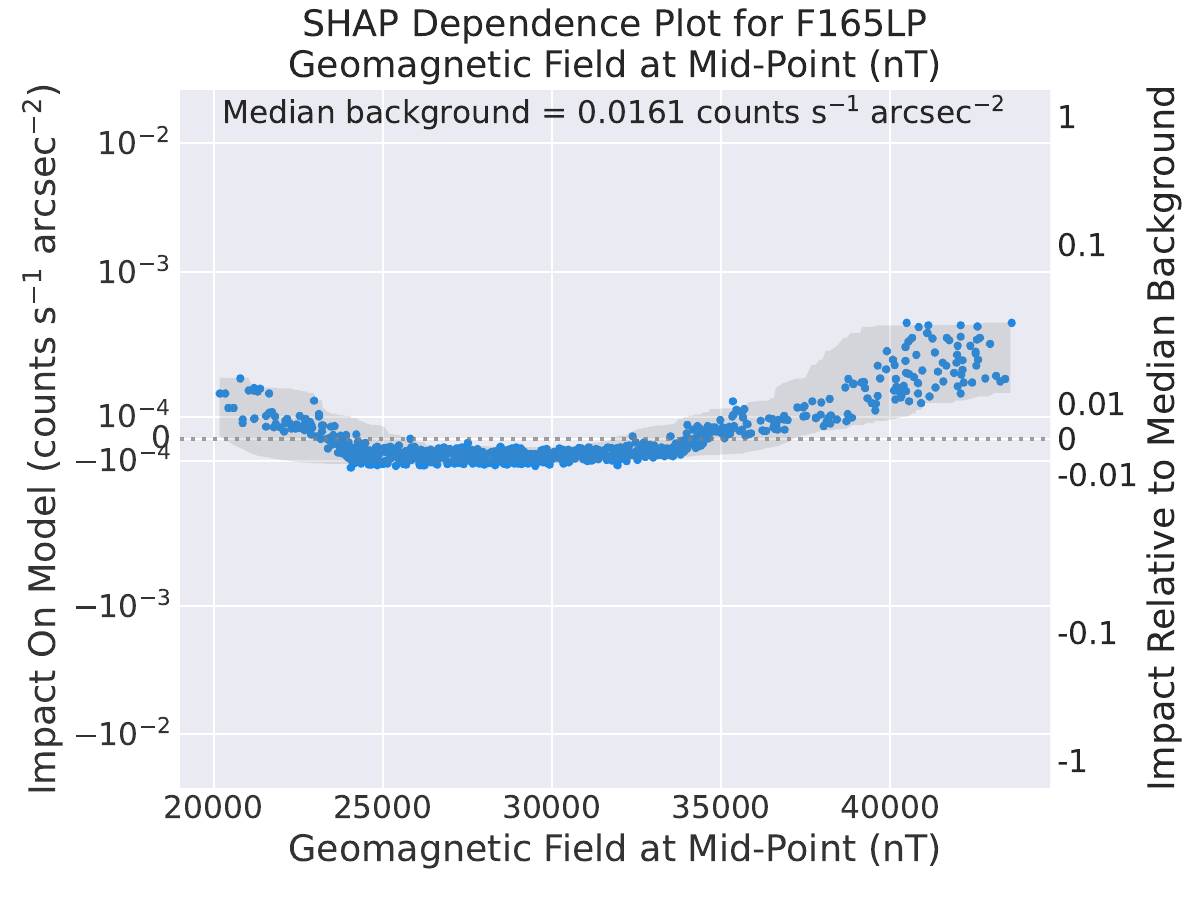}
\includegraphics[width=0.32\textwidth]{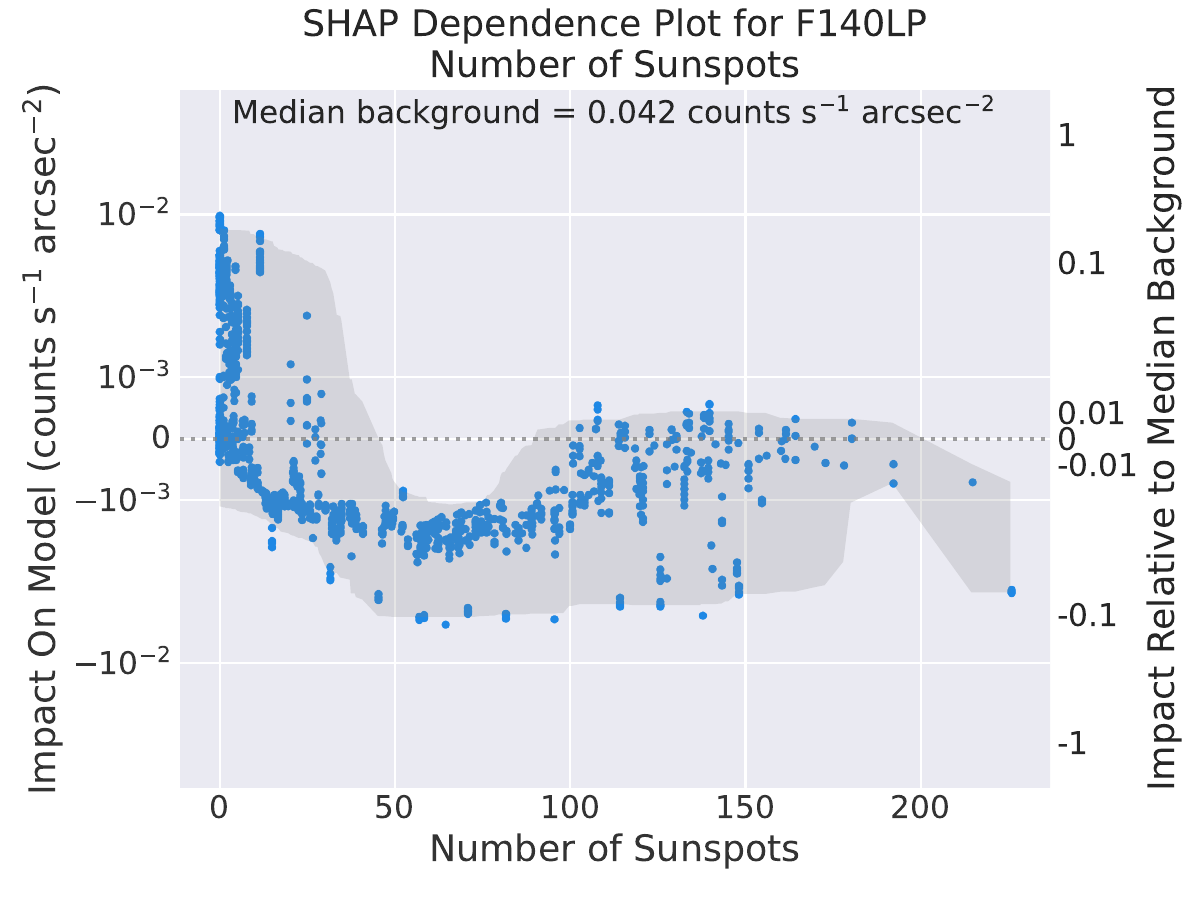}
\includegraphics[width=0.32\textwidth]{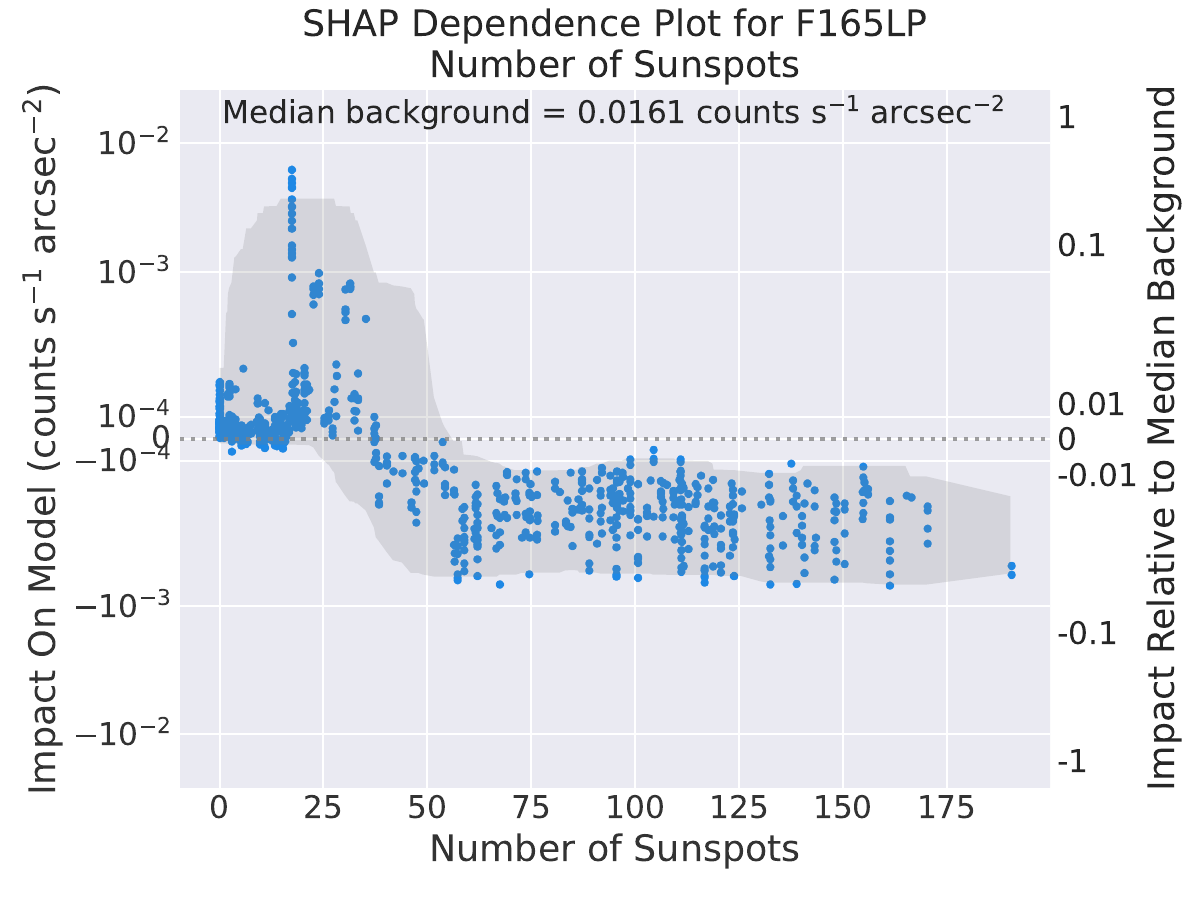}
\caption{SHAP dependence plots for observational parameters relating to geocoronal effects, for several filters. Details otherwise as per Figure~\ref{Fig:Solar_Alt_Dependence_Plots}.}
\label{Fig:Geocoronal_Dependence_Plots}
\end{figure}

Several filters also show a relationship between day of year and the predicted background, with background elevated at the start/end of the year (eg, see Figures~\ref{Fig:SHAP_Dependence_F125LP} and \ref{Fig:SHAP_Dependence_F140LP}). This suggests that the SBC may indeed experience the same annual variation in dark rate as the NUV MAMA on HST/COS, caused by the higher average telescope temperature when the Earth is at perihelion in January \citep{Johnson2024C}.

\subsubsection{Geocoronal Effects} \label{Subsubsection:SHAP_Geogoronal}

As described in Section~\ref{Section:Observational_Parameters}, the potential relationships between geocoronal effects and the SBC background can be complex. For instance, there are mechanisms by which both higher and lower geomagnetic field strength can increase airglow. Figure~\ref{Fig:Geocoronal_Dependence_Plots} show sthe SHAP dependence plots for several pertinent parameters. We indeed see that whilst some filters show the predicted background increasing with field strength (eg, PR110L), others show a `U'-shaped relation with elevated background predicted at both higher and lower field strength (eg, F165LP). The different relations at different wavelengths may be due to the different excitation mechanisms for different species. 

Interestingly, several filters show that the model expects a {\it greater} background for observations conducted when HST was at higher orbital altitudes (see Figure~\ref{Fig:Geocoronal_Dependence_Plots}). This also manifests as an anti-correlation between predicted background and date of observation -- as HST's orbital altitude is very tightly correlated with time, these seem hard for the model to disentangle. In general, this relation is not what we would expect. 

However, as discussed in Section~\ref{Section:Observational_Parameters}, observations at greater orbital height can intersect a greater airglow emitting column, due to the greater angular depression of the horizon -- ie, there will be a greater atmospheric column at a given Earth limb angle. 

Another surprising relationship is with sunspot count (a proxy for Solar activity). For a number of filters, the SHAP analysis finds that our model predicts higher background when there are fewer sunspots (see Figure~\ref{Fig:Geocoronal_Dependence_Plots}). It is unclear what might be causing this, as airglow should generally increase with more Solar activity. The CCDs on HST show more cosmic rays events during solar minimum, due to contraction of the thermosphere \citep{Miles2021A}, but this should not affect the SBC.

\subsubsection{Other Effects} \label{Subsubsection:SHAP_Other}

\begin{figure}
\centering
\includegraphics[width=0.32\textwidth]{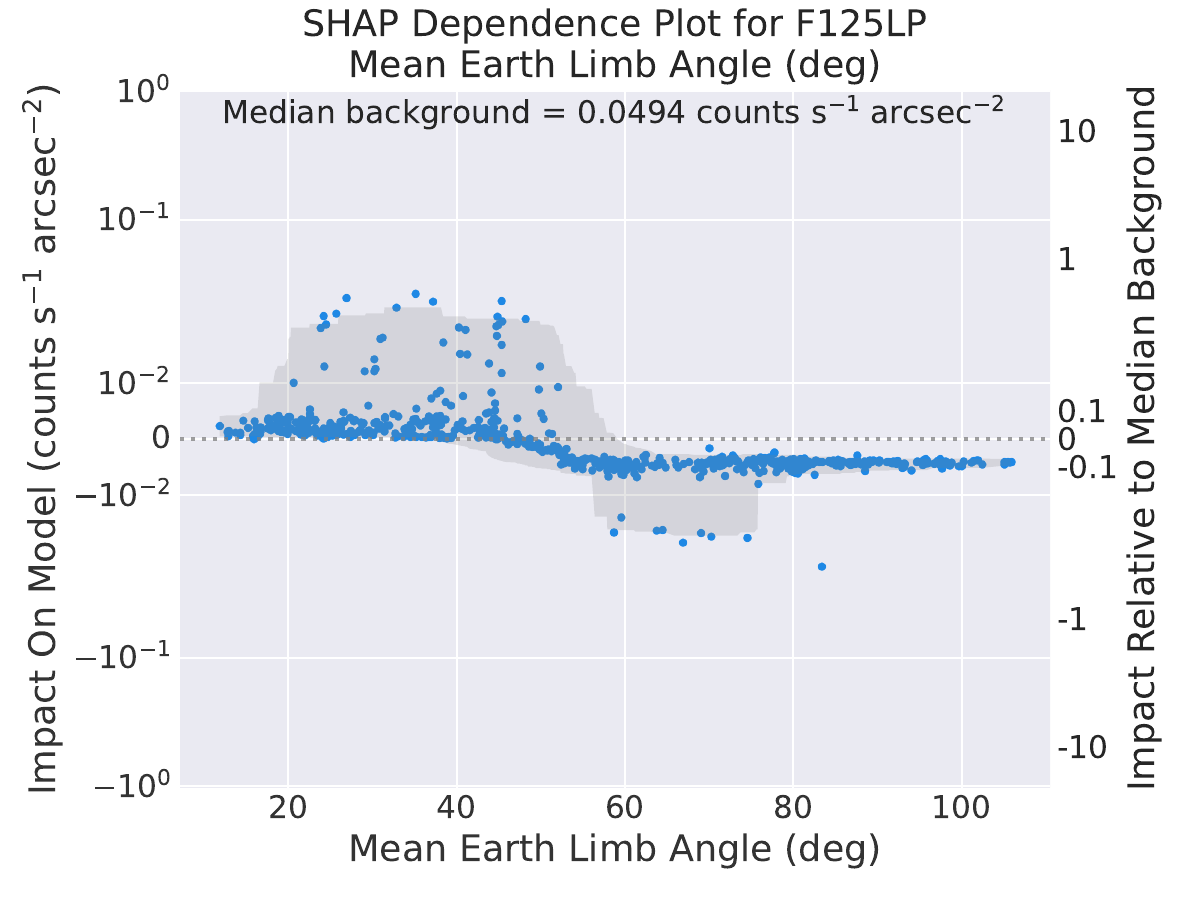}
\includegraphics[width=0.32\textwidth]{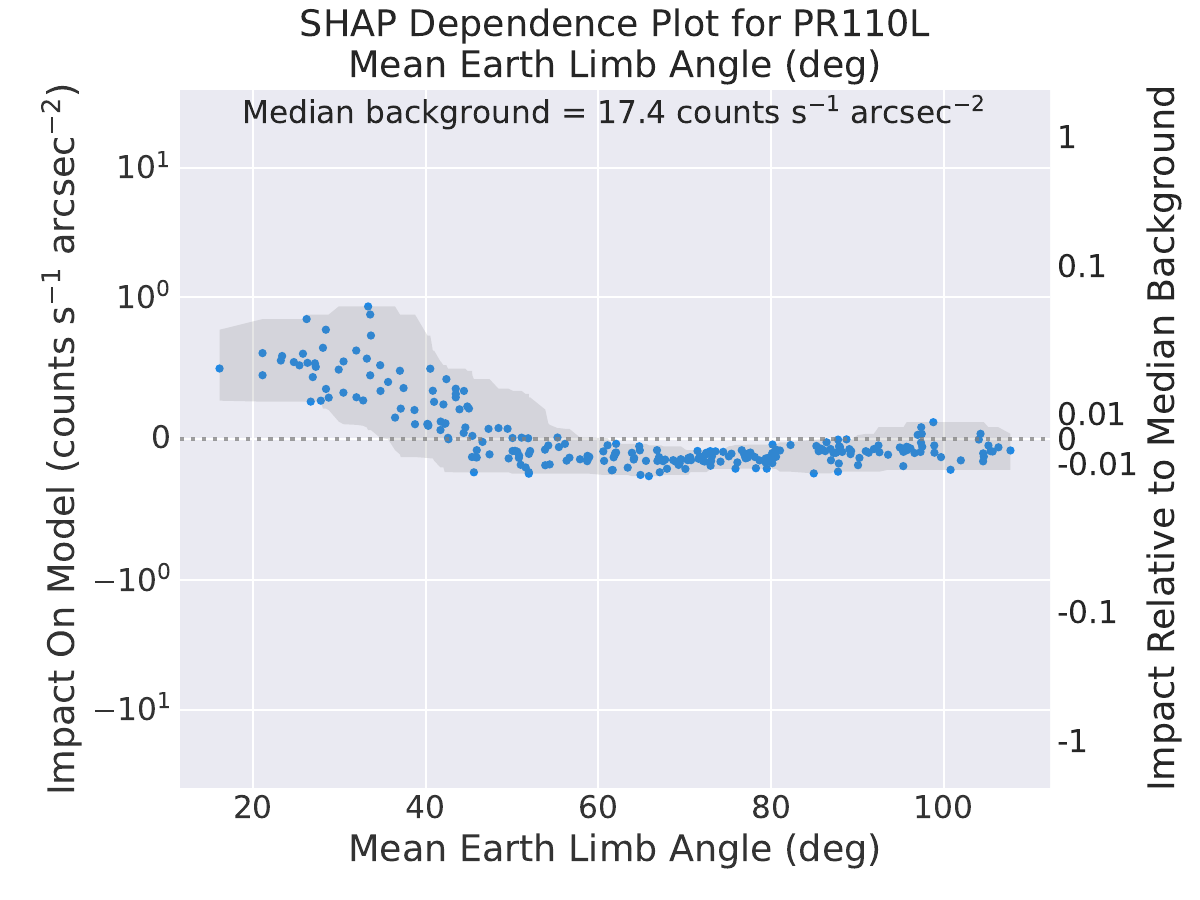}
\includegraphics[width=0.32\textwidth]{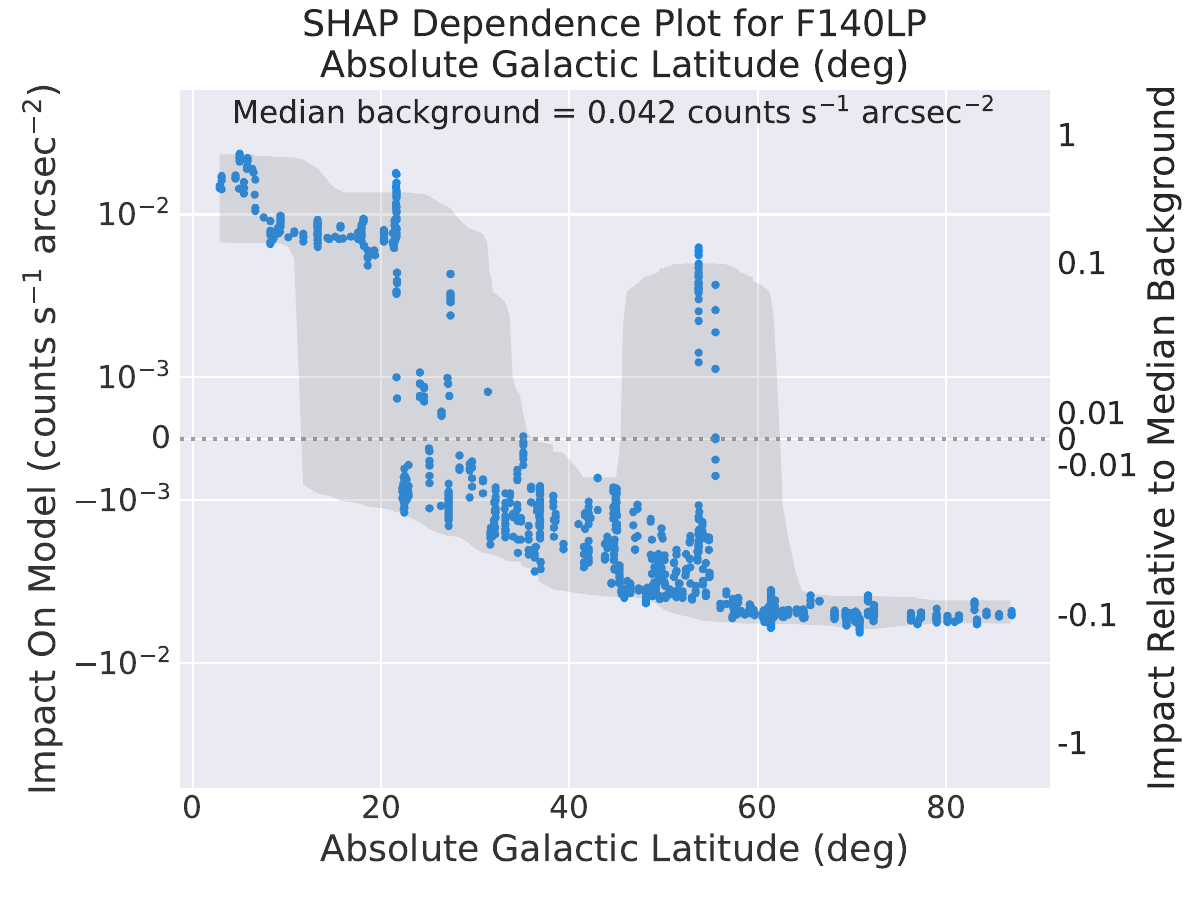}
\caption{SHAP dependence plots for Earth limb angle and Galactic latitude, for several filters. Details otherwise as per Figure~\ref{Fig:Solar_Alt_Dependence_Plots}.}
\label{Fig:SBC_Other_Dependence_Plots}
\end{figure}

Unsurprisingly, Earth limb angle has a clear impact on predicted background for most filters, at the \textgreater10\%\ level for the shorter wavelength filters (see left and central panels of Figure~\ref{Fig:SBC_Other_Dependence_Plots}). Earth limb angles of 50$^{\circ}$--60$^{\circ}$ appear to be the threshold below which predicted background begins to increase.

We also see significantly increased predicted background for observations targeting lower Galactic latitudes. As described in \citet{CJRClark2025C}, our masking procedure cannot remove the contribution of low-S/N or confused sources. Such sources will contribute to the background we measure. This is significant at low Galactic latitude, due to the greater number UV sources towards the Galactic plane. The right panel of Figure~\ref{Fig:SBC_Other_Dependence_Plots} shows the impact of this can amount to as much as 100\% of the median background in a given filter. 

\section{Conclusions \&\ Guidance for Users} \label{Section:Conclusion}

We have used 8,640 SBC exposures to try to identify what factors drive the large variation in background levels. For every observation, we assembled a consistent set of 23 observational parameters, quantifying properties such as Solar position, geocoronal conditions, instrument temperature, etc (see Section~\ref{Section:Observational_Parameters}). We have used quantile random forest regression to construct a machine learning model able to use these observational parameters to accurately predict the background for a given observation (Section~\ref{Section:Background_Modeling}). 

We then applied the technique of SHapley Additive eXplanations (SHAP) to extract from the model the relationships it learned to make its predictions (Section~\ref{Section:Background_Causes}). Our SHAP analysis of the predictive model indicates a few primary drivers of background levels encountered in SBC observations. 

For observations in filters where airglow is a primary source of the background (F115LP, F122M, F125LP, F140LP, PR110L, and PR130L), we find that the expected background level is significantly increased by observing when the Sun is above the horizon, or by observing targets at an Earth limb angle $\lesssim$\,50$^{\circ}$. Users  can use the \texttt{SHADOW} special requirement to request night-time observations, whilst the \texttt{LOW-SKY} special requirement imposes a minimum Earth limb angle of \textgreater\,40$^{\circ}$ (see the ACS Instrument Handbook; \citealp{Hubble-ACS2024}, Section 9.4\footnote{\url{https://hst-docs.stsci.edu/acsihb/chapter-9-exposure-time-calculations/9-4-detector-and-sky-backgrounds}}). Note, however, that these special requirements are a limited resource that constrain schedulability, and require specific justification in proposals.

For F150LP and F165LP, the background is often dominated by the contribution of the instrumental dark rate. The dark rate increases with temperature, especially above 25\,$^{\circ}$C \citep{Avila2017B}, and SBC temperature increases steadily during continuous operation. This analysis reaffirms the impact of this effect; for F165LP in particular, the dark rate is by far the dominant source of background. The SBC typically reaches 25\,$^{\circ}$C after $\approx$\,2 hours of operation; users should consider if their observations can be planned in such way that F150LP and F165LP observations are not conducted after this duration of continuous SBC operation.

Geocoronal parameters do seem to have some impact on background, but these are consistently at a level corresponding $\lesssim$10\% of the typical background level. This is relatively small impact is presumably thanks to the strict SAA avoidance requirements already imposed for SBC observations. There does therefore not seem to be good justification for incorporating additional geocoronal considerations when planning observations (not that this would even be practical in the case of, eg, number of sunspots).

Lastly, backgrounds are typically higher at low Galactic latitude, presumably due to the increased contribution of individually-undetected UV sources closer to the Galactic plane. As such, there is no way for users to avoid it for a given target, but users can use the data presented in this report to estimate the likely impact of this effect on their observations.

\section{Acknowledgements}

This research made use of \texttt{Astropy}\footnote{\url{https://www.astropy.org/}}, a community-developed core \texttt{Python} package for Astronomy \citep{astropy2013,astropy2019}. This research made use of Photutils\footnote{\url{https://photutils.readthedocs.io}}, an \texttt{Astropy}-affiliated package for
detection and photometry of astronomical sources\citep{Bradley2020C}. This research made use of \texttt{NumPy}\footnote{\url{https://numpy.org/}} \citep{VanDerWalt2011B,Harris2020A}, \texttt{SciPy}\footnote{\url{https://scipy.org/}} \citep{SciPy2001,SciPy2020}, and \texttt{Matplotlib}\footnote{\url{https://matplotlib.org/}} \citep{Hunter2007A}. This research made use of the \texttt{pandas}\footnote{\url{https://pandas.pydata.org/}} data structures package for \texttt{Python} \citep{McKinney2010}. This research made use of \texttt{corner}\footnote{\url{https://corner.readthedocs.io}}, a python package for the display of multidimensional samples \citep{ForemanMackey2016D}. This research made use of \texttt{iPython}, an enhanced interactive \texttt{Python} \citep{Perez2007A}. This research made use of sequential color-vision-deficiency-friendly colormaps from \texttt{cmocean}\footnote{\url{https://matplotlib.org/cmocean/}} \citep{Thyng2016A} and \texttt{CMasher}\footnote{\url{https://cmasher.readthedocs.io}} \citep{VanDerVelden2020A}. This research made use of \texttt{TOPCAT}\footnote{\url{http://www.star.bris.ac.uk/~mbt/topcat/}} \citep{Taylor2005A}, an interactive graphical viewer and editor for tabular data.

\vspace{3.5ex plus 1ex minus 0.2ex}
\appendix

{\Large \bf Appendix}

\section{Model Training \&\ Validation} \label{AppendixSection:Model_Training}

Here, we provide a full description of how we applied an ML model to predict SBC backgrounds based on observational parameters. It is not necessary to understand these granular details to interpret the outputs of the ML modelling. Rather, this information is provided here so any interested readers can check our approach, and/or replicate our methodology.

As described in Section~\ref{Section:Background_Modeling}, we used Quantile Random Forest (QRF) regression method, as implemented by the \texttt{Python} package \texttt{quantile-forest}\citep{Johnson2024A}. For each filter, we trained using 80\% of the observations, and kept back 20\% of the observations for validation testing, with the training and testing observations selected at random. 

QRF regression has a numbers of settings that can be tuned, to optimize the quality of the model fit. Our final setup used parameters (provided here for replicability) of: \texttt{criterion=`squared\_error'}, \texttt{n\_estimators=250}, \texttt{max\_features=`sqrt'}, \texttt{min\_samples\_split=10}, \texttt{min\_samples\_leaf=20}, \texttt{max\_samples\_leaf=0.75}, and \texttt{max\_depth=20}. For a full explanation of each of these settings, see the \texttt{quantile-forest} documentation\footnoteref{Footnote:QRF_Docs}. When tuning these model settings, we needed to strike a balance between two undesirable effects:

The first undesirable effect was under-sensitivity -- this is when the trained model over-predicts the true value for faint backgrounds, and under-predicts the true value for bright backgrounds. In other words, the model undergoes `regression towards the mean'. 

The second undesirable effect was over-fitting. Over-fitting is when the model `learns' the specific observations being fitted, as opposed to the general relations -- an issue that becomes apparent when the test data (which the model wasn't trained on) is much worse-fit than the training data.

\subsection{Over-Sampling} \label{AppendixSubsection:Over-Sampling}

 In general, model settings that reduced under-fitting tended to increase over-sensitivity. To combat this conflict, we trained our model for each filter on an {\it over-sampled} version of the training data, which had been weighted to draw more observations from the low-background and high-background tails of the observations. 

Specifically, we over-sampled our training data by a factor of 3. In other words, the over-sampled training data contains 3 times more observations than the original training data, meaning each individual observation can (and often will) appear in the over-sampled training data multiple times. Observations were drawn from the training data at random (with replacement), but with the probability of drawing a given observation being weighted according to the background percentile each observation lies in, according to:

\begin{equation}
w_{i} = \left(1 + \frac{|P_{i} - 50 |}{50}\right)^{O}
\label{Equation:Oversamp_Weight}
\end{equation}

\noindent where $w_{i}$ is the weight applied to the probability of the $i$\th\ observation being drawn during oversampling; $P_{i}$ is the percentile the $i$\th\ observation lies in amongst the distribution of backgrounds measured in for all observations in that filter (ie, the observation with the faintest background lies in the 0\th\ percentile, and the observation with the brightest background lies in the 100\th\ percentile); and $O$ is the over-sampling index, for which we found through experimentation that a value of $O = 2.5$ works best\footnote{The purpose of dividing by 50 in Equation~\ref{Equation:Oversamp_Weight} is to normalize the percentiles so that the observations with the brightest and faintest backgrounds have a value of 1, whilst the observation with the median background has a value of 0. We then add 1 to apply a constant offset to shift these values from the 0--1 range to the 1--2 range, as otherwise the median observation would have weighting of 0.}.

The result of this weighting is that an observation with the faintest (or brightest) background for a given filter is 5--6 times more likely to be drawn during over-sampling than the observation with the median background. More heavily over-sampling for the brightest and faintest backgrounds means that the QRF regression is much less able to `ignore' extremal observations at the edges of the parameter space. This does a good job of addressing the under-sensitivity problem, thereby allowing us to use settings for the QRF regression that reduce over-fitting (as otherwise, settings that reduce over-fitting tend to lead to more severe under-sensitivity). 

\begin{figure}
\centering
\includegraphics[width=0.475\textwidth]{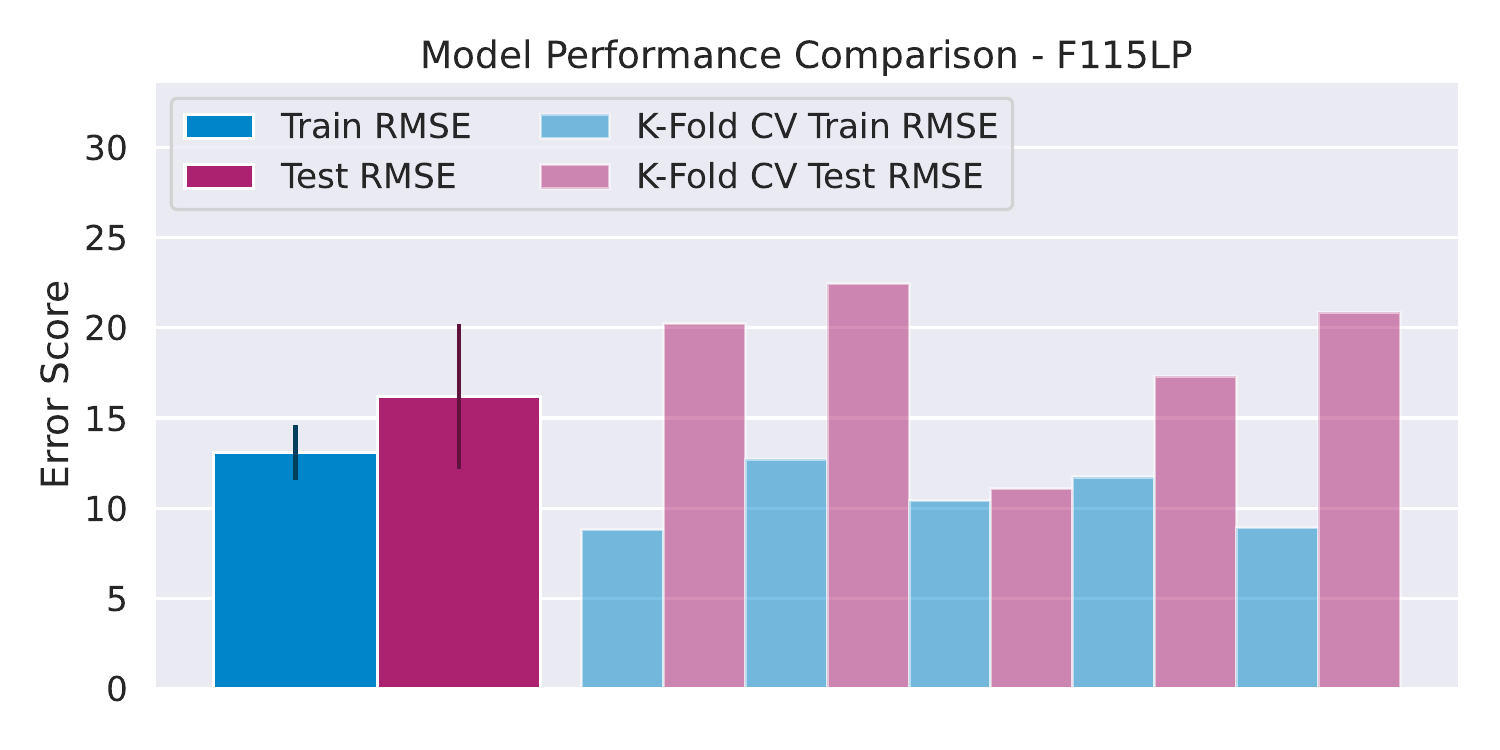}
\includegraphics[width=0.475\textwidth]{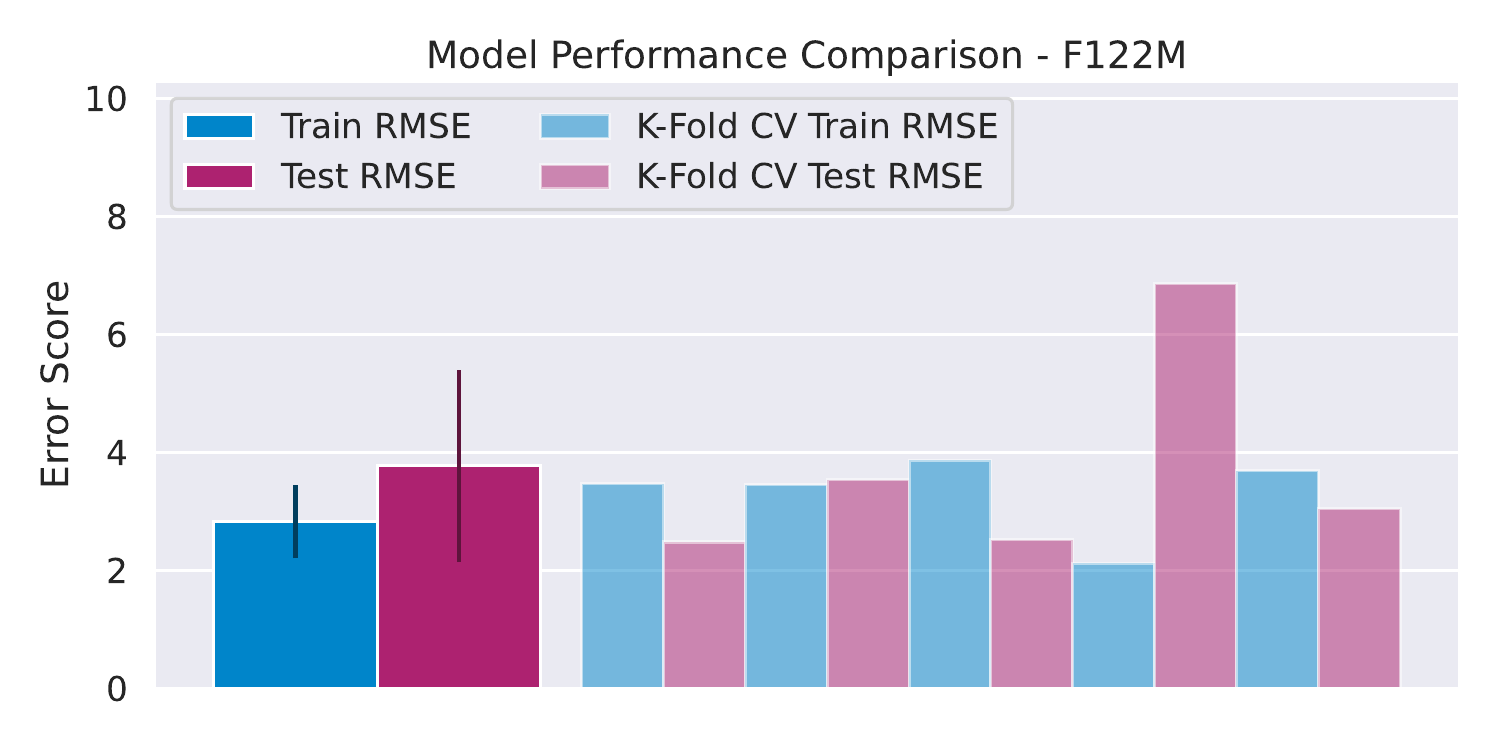}
\includegraphics[width=0.475\textwidth]{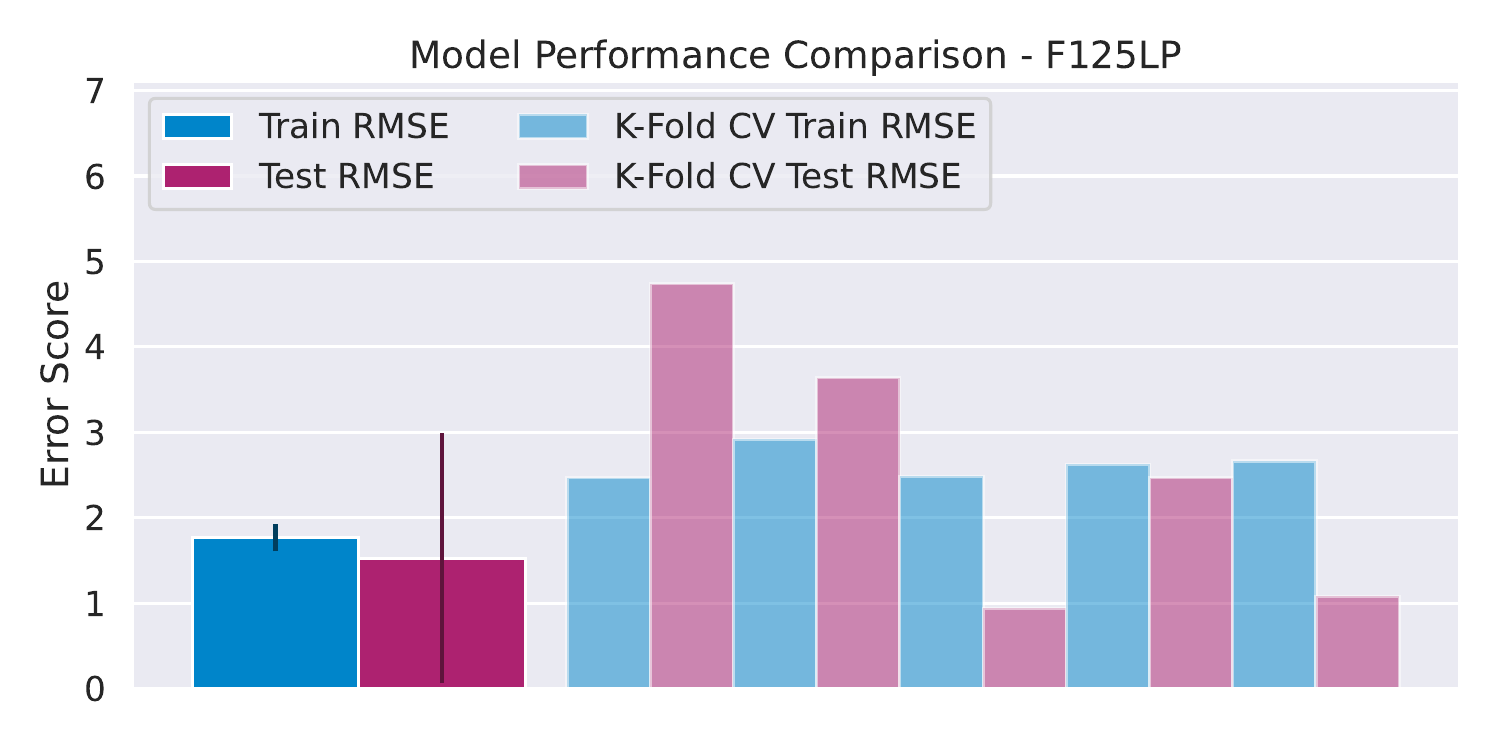}
\includegraphics[width=0.475\textwidth]{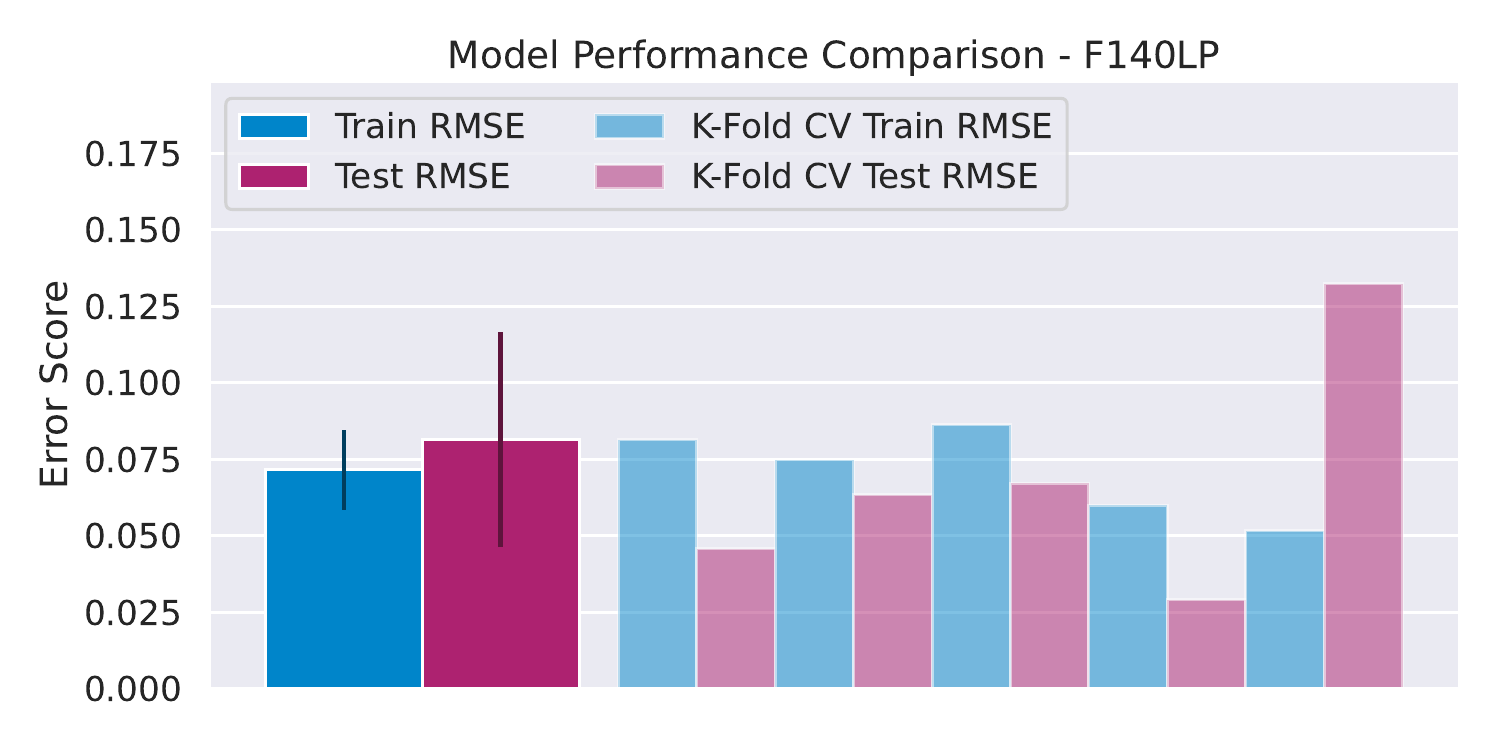}
\includegraphics[width=0.475\textwidth]{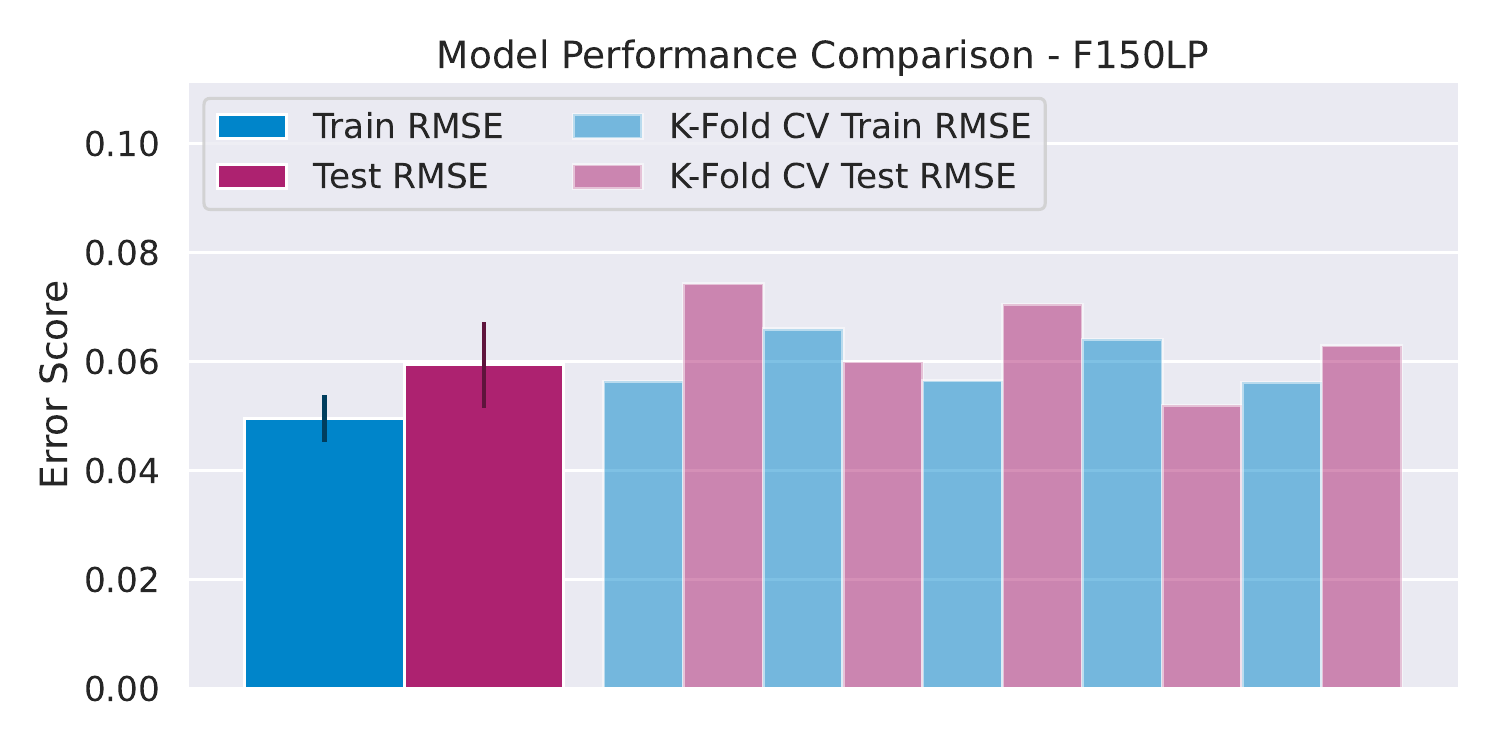}
\includegraphics[width=0.475\textwidth]{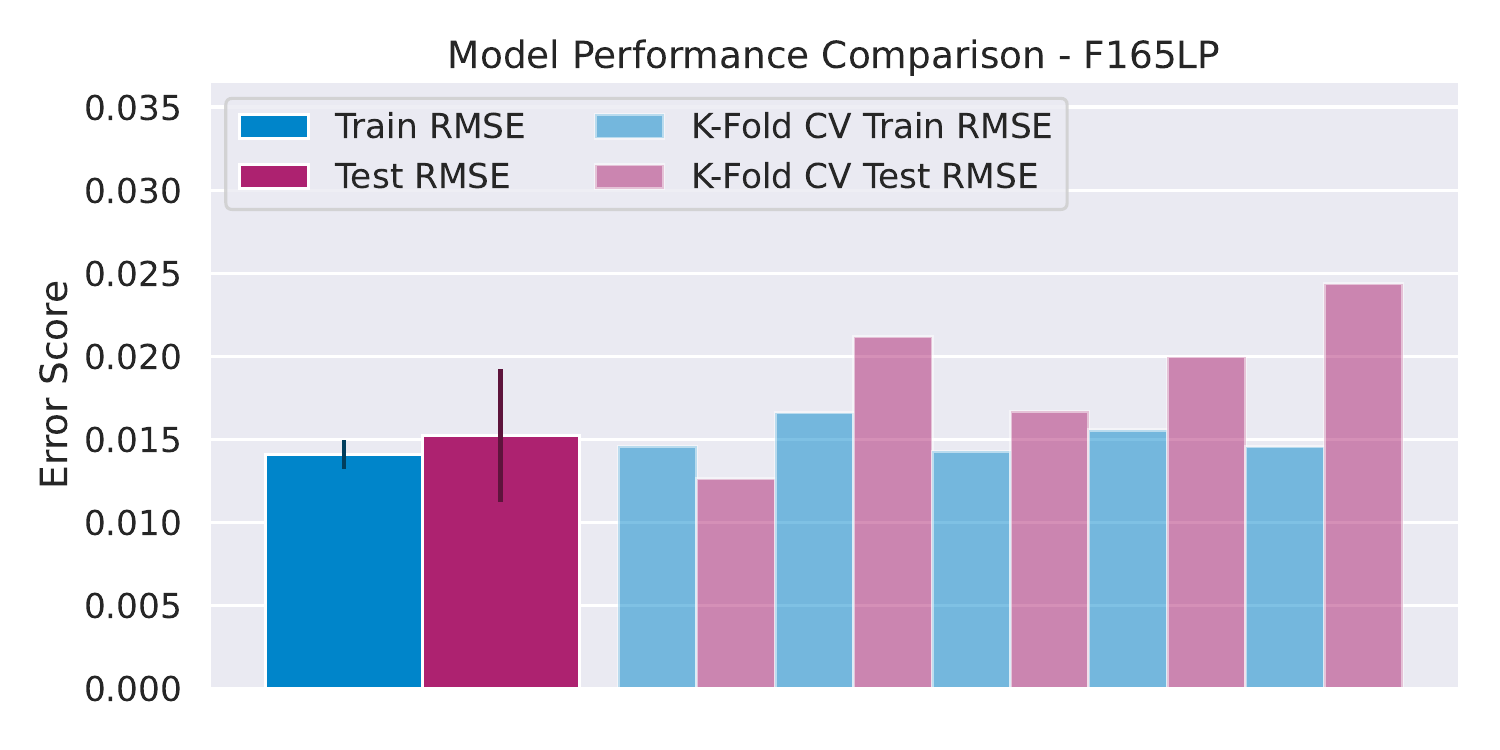}
\includegraphics[width=0.475\textwidth]{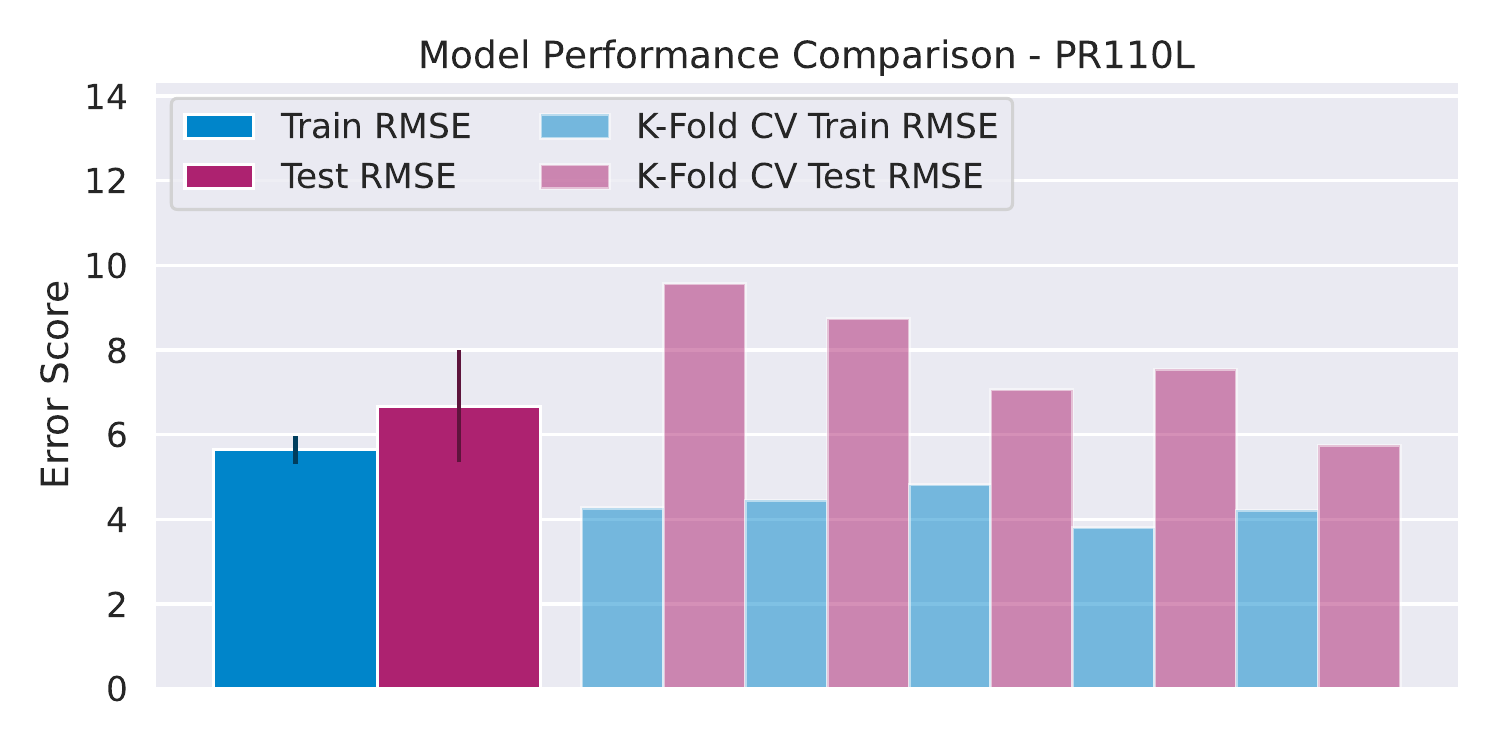}
\includegraphics[width=0.475\textwidth]{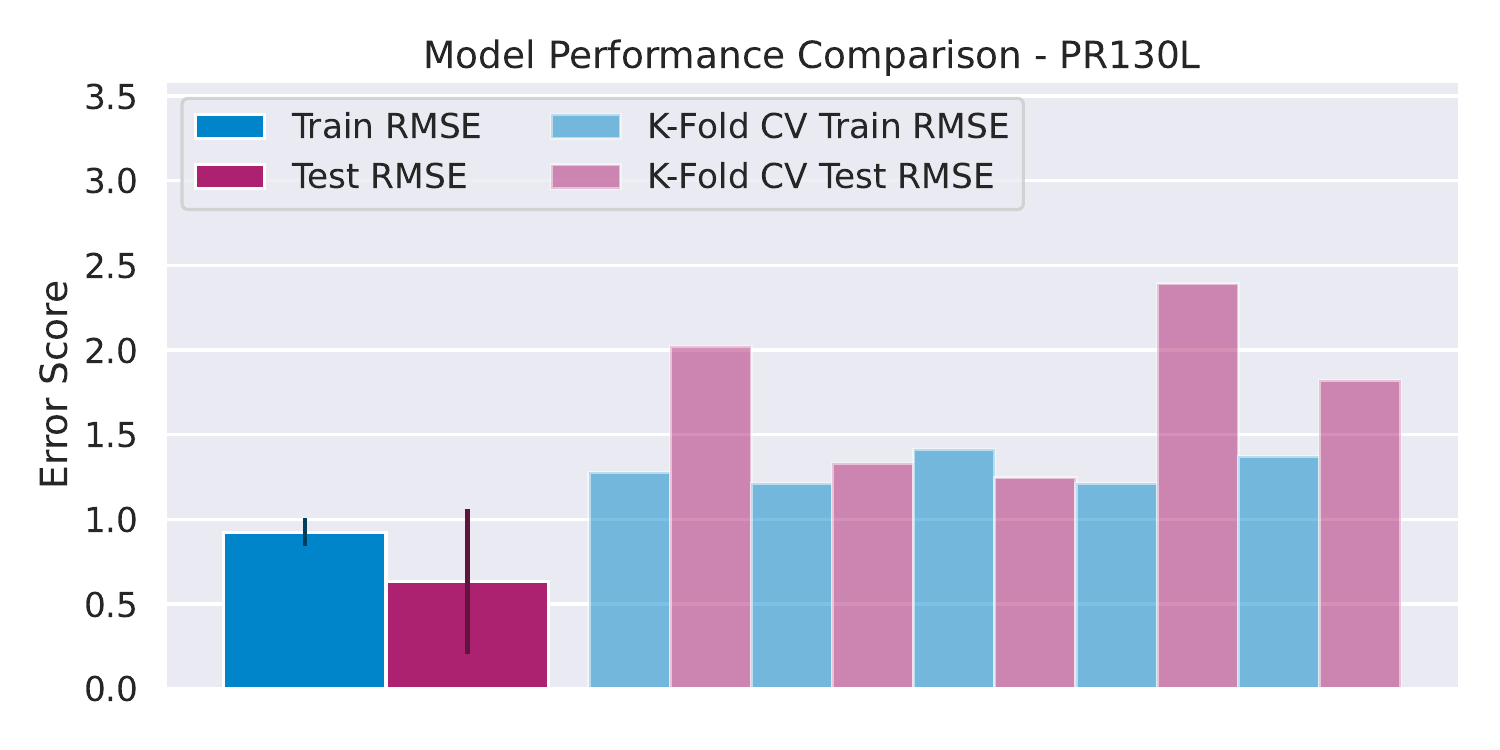}
\caption{Cross-validation plots for our QRF regression model fitting, for every filter. We plot the Root-Mean-Squared Error (RMSE) for the test and training samples for the fiducial model fit used throughout the rest of the analysis (wider bars, on the left of each plot), and of the 5-fold cross-validation (lighter thinner bars, on the right of each plot). We also plot error bars on the for the test and training RMSE values of the fiducial model fit; these error bars are set by calculating the standard deviation of the RMSEs of the 5 cross-validation values.}
\label{Fig:Model_CV}
\end{figure}

Every time an observation is drawn during the over-sampling, we also apply small random perturbations to its background value and all of its observational parameter values, by multiplying them each by a random number between 0.99 and 1.01. This ensures that the model won't `lock on' and over-fit to any specific observation that gets drawn multiple times in the over-sampling, whilst also keeping the perturbations small enough to not bias the final model.

We note that we only apply this over-sampling to the training data -- not the test data. The test data therefore still provides a `clean' way to assess the quality of the final model fit (and whether the over-sampling itself introduces any new issues).

\subsection{Cross-Validation} \label{AppendixSubsection:Cross-Validation}

To assess the quality of our final models for each filter, we not only used our validation data, but also performed $k$-fold cross validation. This is when the full dataset is split into $k$ randomly-assigned equally-sized portions. The fitting and validation process is then repeated $k$ times, where for each instance one $k$\th\ portion is used as the test data, and the other portions are used as the training data. We did 5-fold cross validation; ie, where a different 5\th\ of the data was used as the test set each time. 

This $k$-fold cross validation is useful as we don't have an especially large number of observations in each filter (by machine learning standards). As such, the 20\% of observations randomly assigned to a test may be unrepresentative. The difference in fit quality between folds gauges any impact from this, and lets us assess if any difference between the test and training errors is significant. 

The cross-validation results of our QRF regression model fit for each filter are plotted in Figure~\ref{Fig:Model_CV}, for both the 5-fold cross-validation, and for the fiducial model fit used throughout the rest of this analysis. Comparing the Root-Mean-Squared Errors (RMSE) of the test and training samples for each filter indicates that the QRF regression may be systematically over-fitting for F115LP and PR110L. This is suggested mainly by the consistently larger RMSE for the test vs training data in the 5-fold cross-validations. We attribute this to the fact that F115LP and PR110L have the smallest number of archival observations of all the SBC filters, possibly compromising the ability of the QRF regression to model the underlying relations.\footnote{As previously mentioned, we model the prism data for completeness, but do not extrapolate the implications of the analysis to the impacts upon {\it spectral} sensitivity}.

For all the other filters, however, the cross-validation results indicate no evidence for over-fitting, with the test and training samples exhibiting no significant differences in RMSE.

\section{SHAP Dependence Plots} \label{AppendixSection:SHAP_Dependence_Plots}

\begin{figure}
\centering
\includegraphics[width=0.24\textwidth]{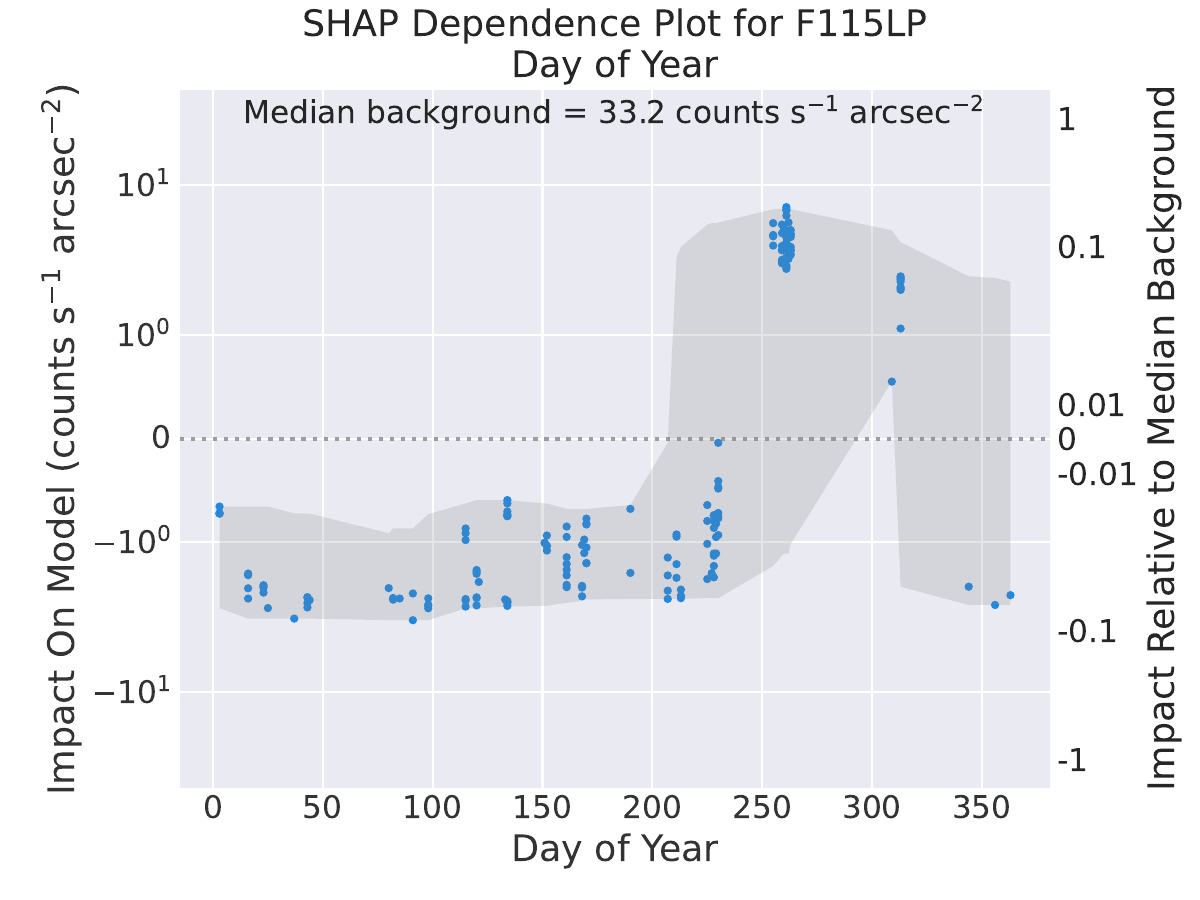}
\includegraphics[width=0.24\textwidth]{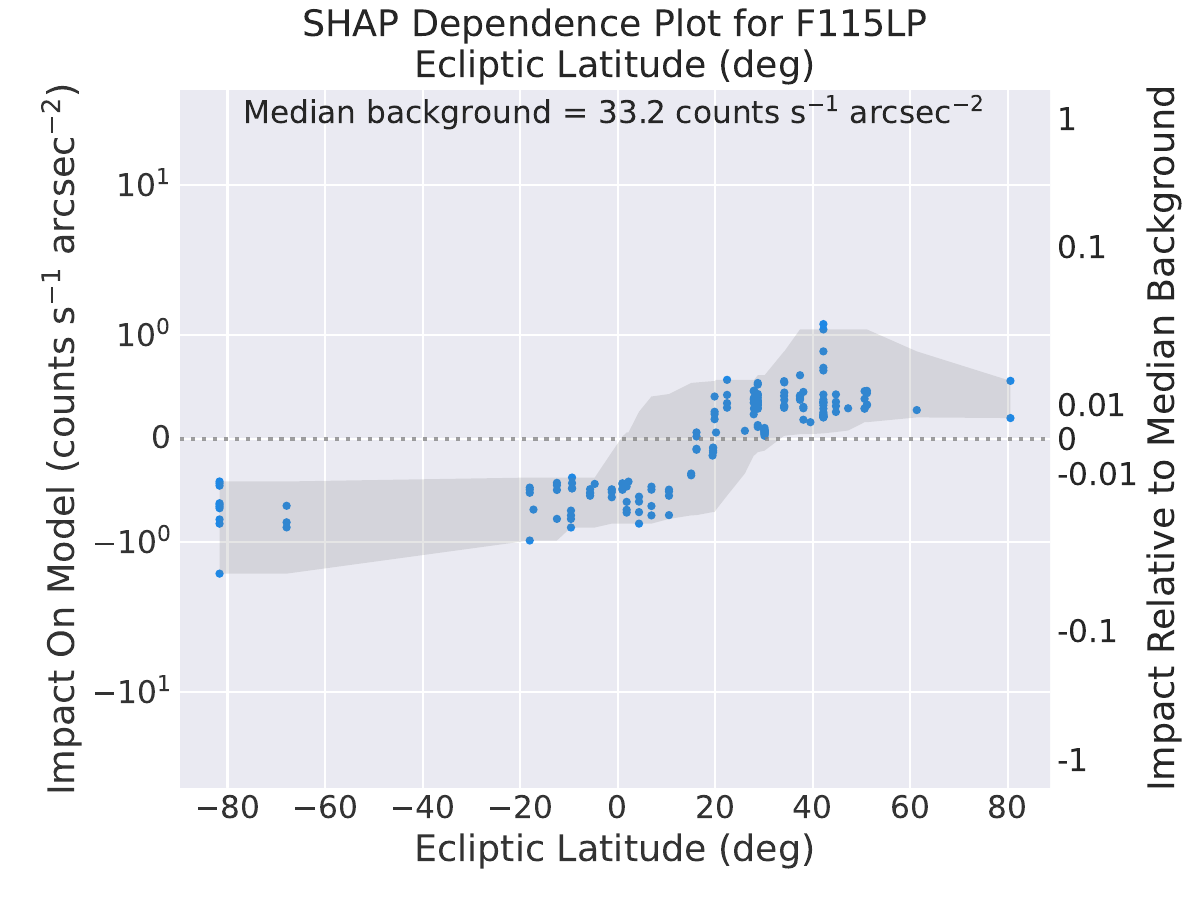}
\includegraphics[width=0.24\textwidth]{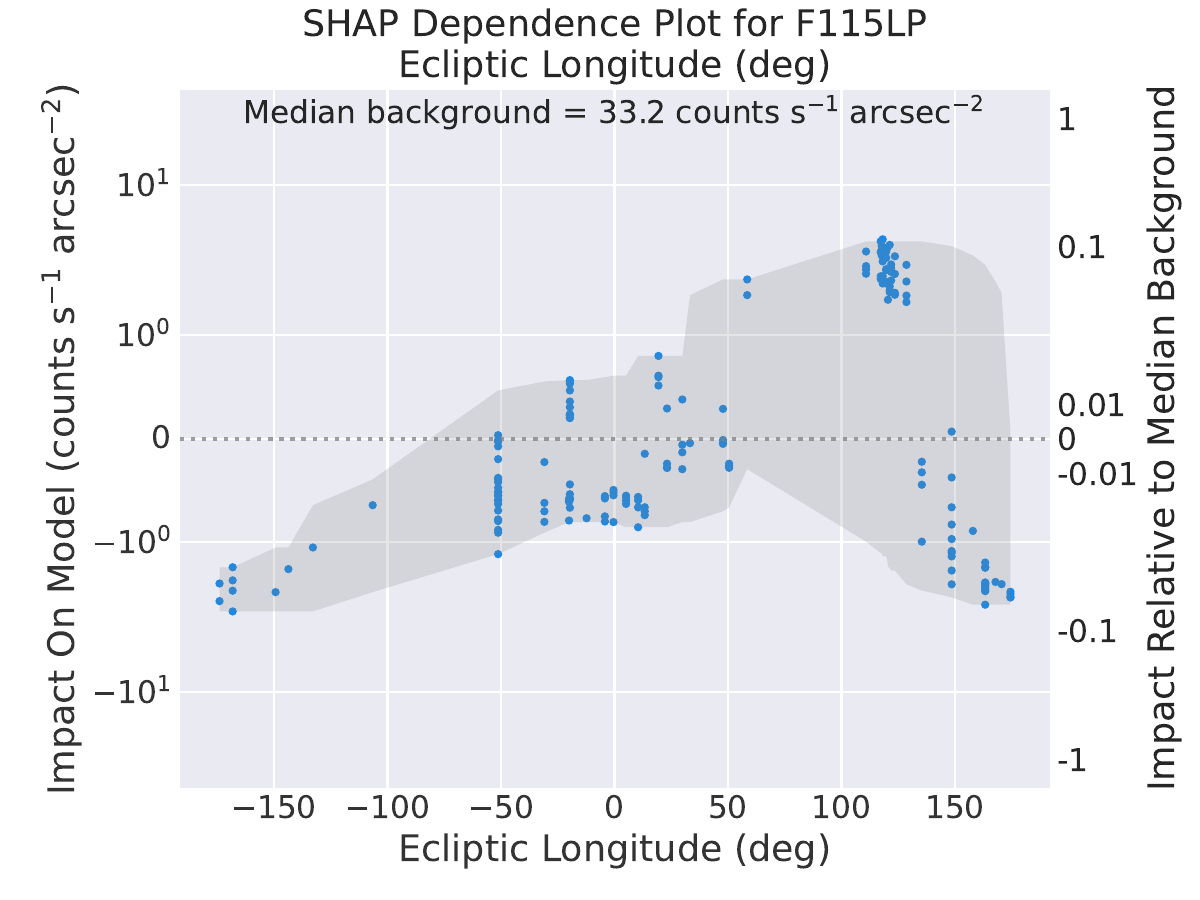}
\includegraphics[width=0.24\textwidth]{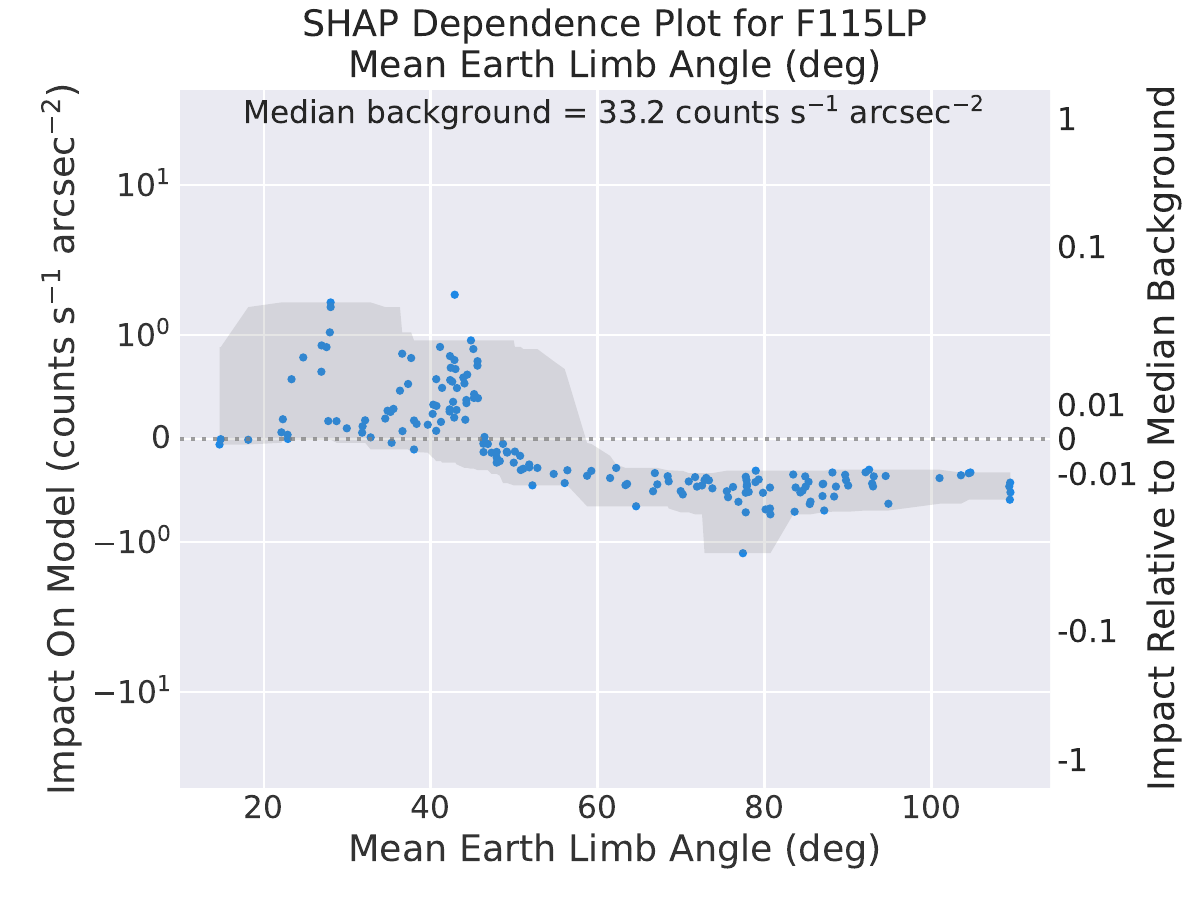}
\includegraphics[width=0.24\textwidth]{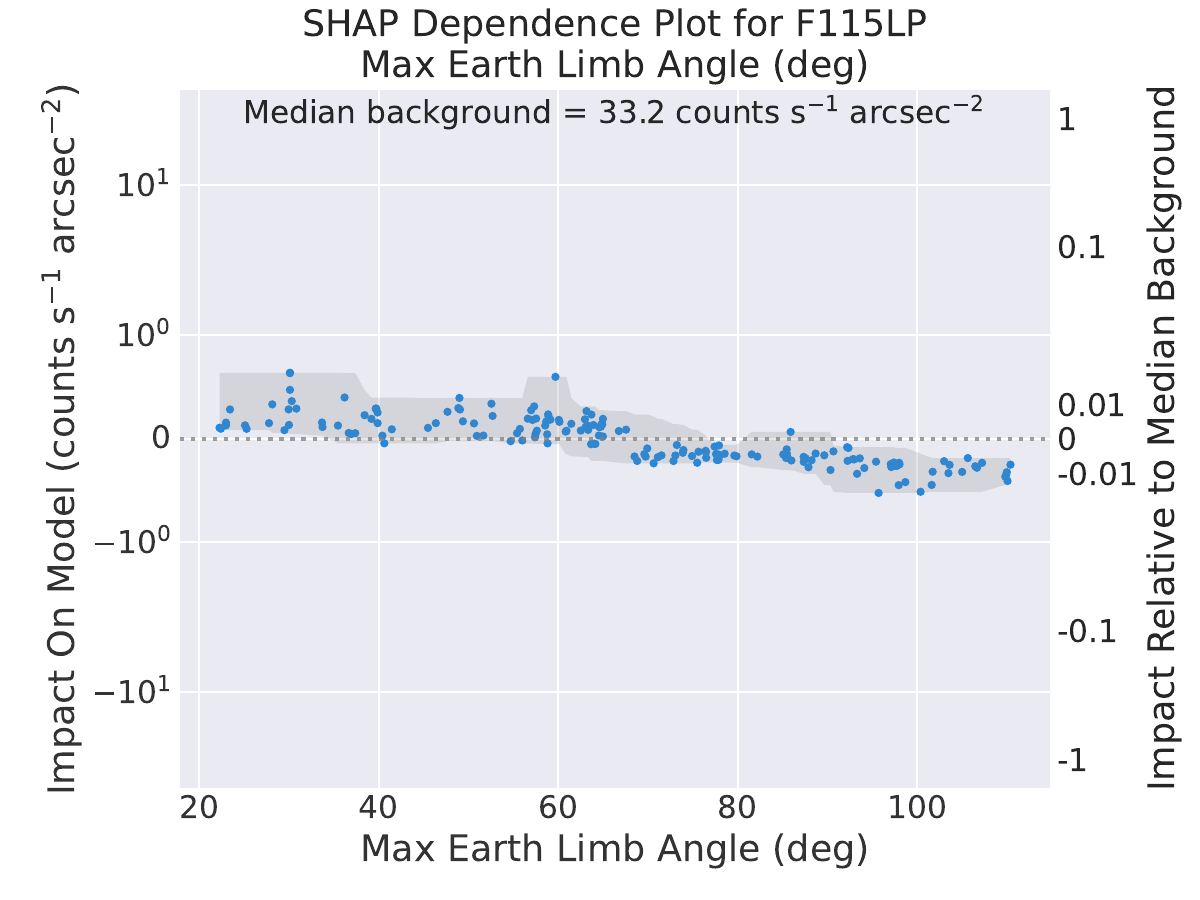}
\includegraphics[width=0.24\textwidth]{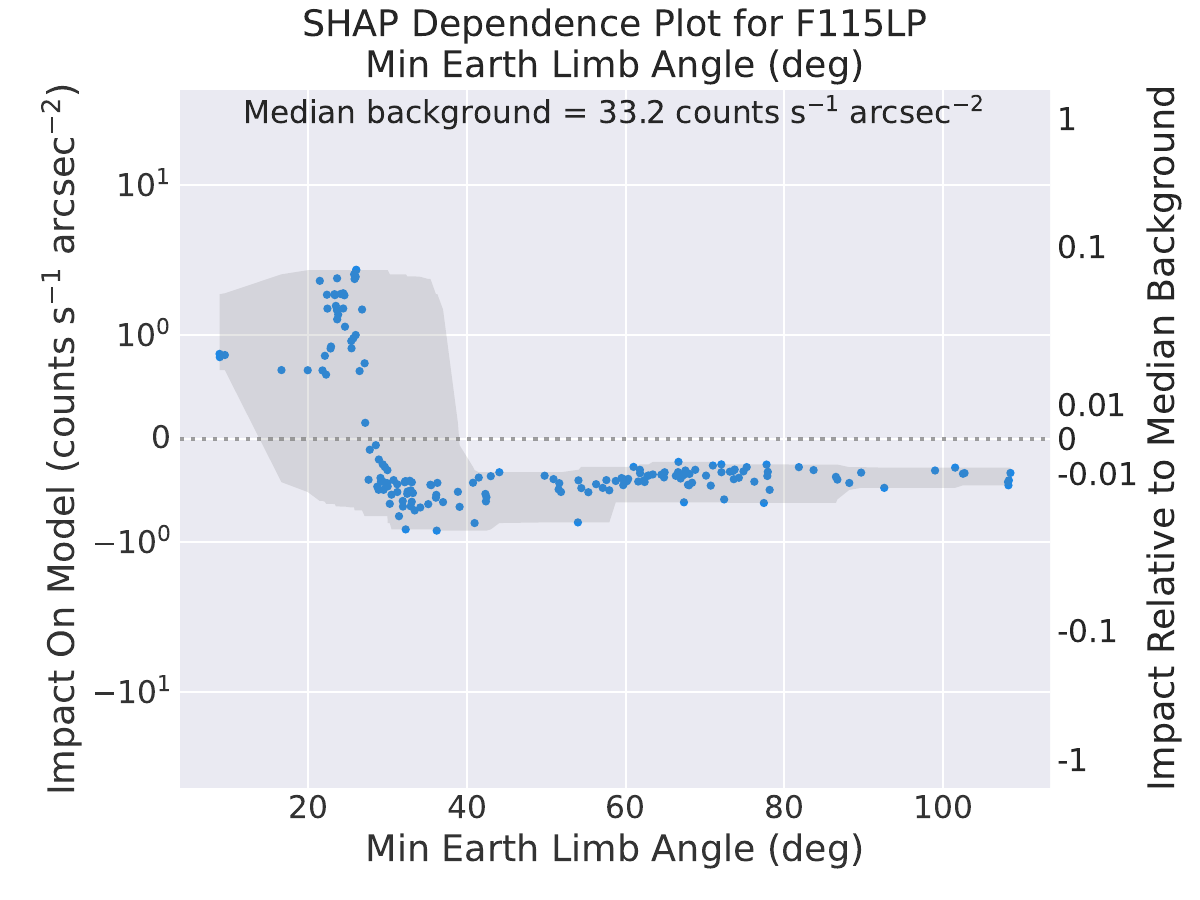}
\includegraphics[width=0.24\textwidth]{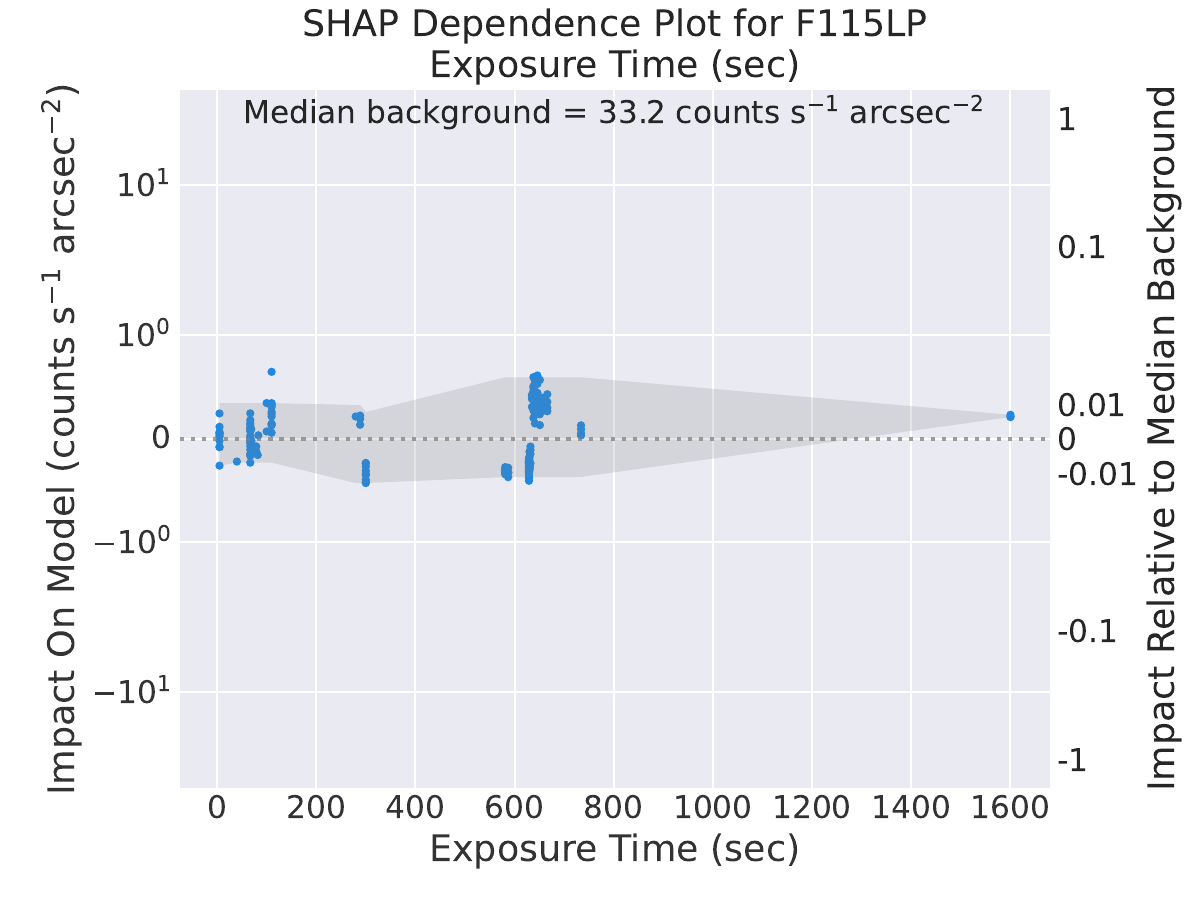}
\includegraphics[width=0.24\textwidth]{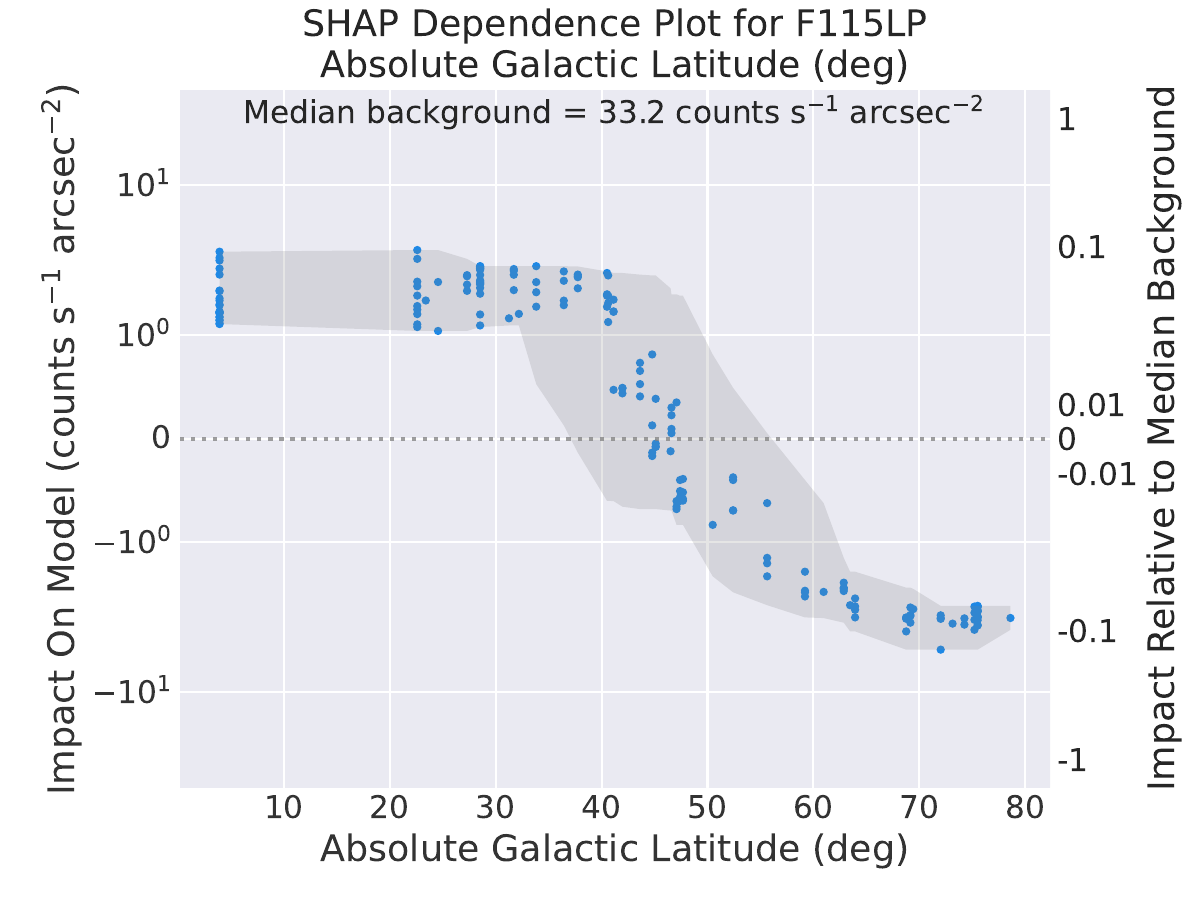}
\includegraphics[width=0.24\textwidth]{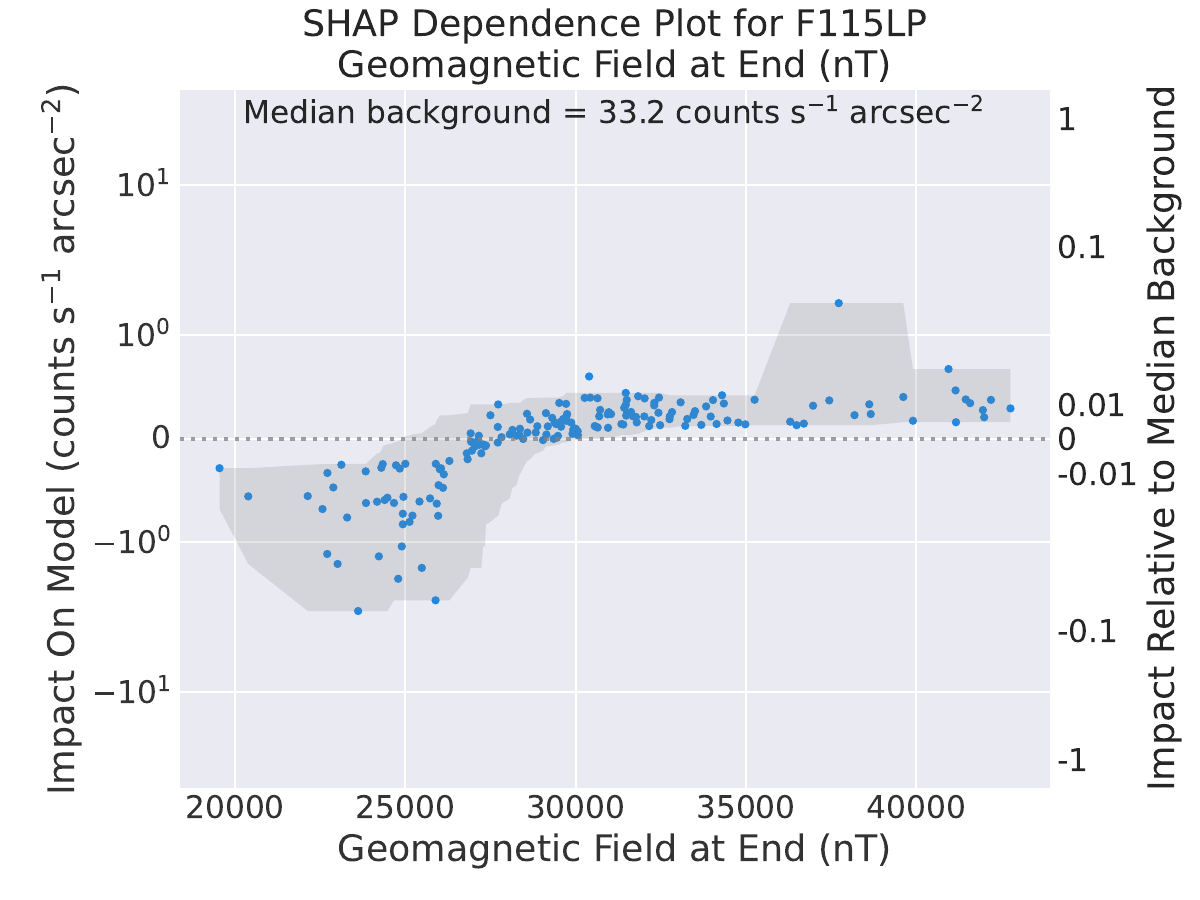}
\includegraphics[width=0.24\textwidth]{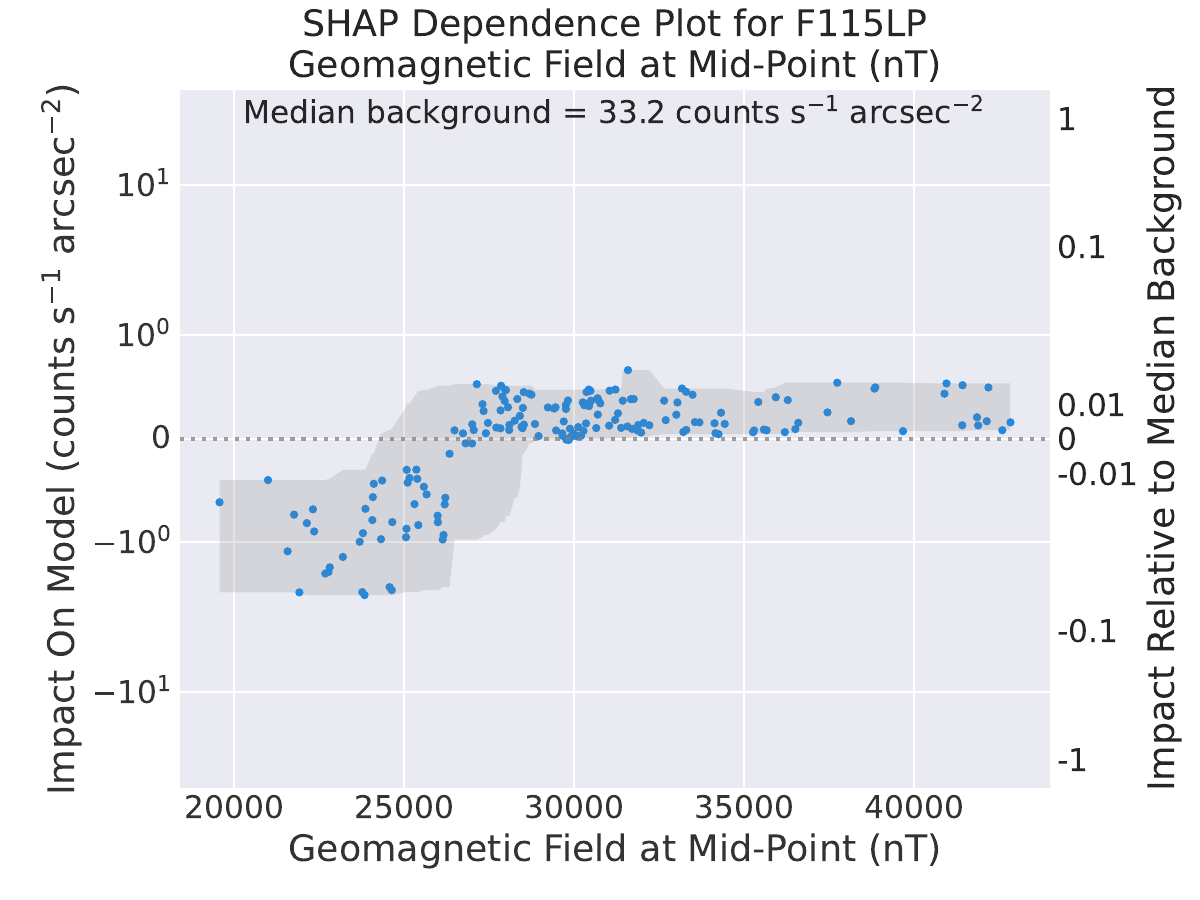}
\includegraphics[width=0.24\textwidth]{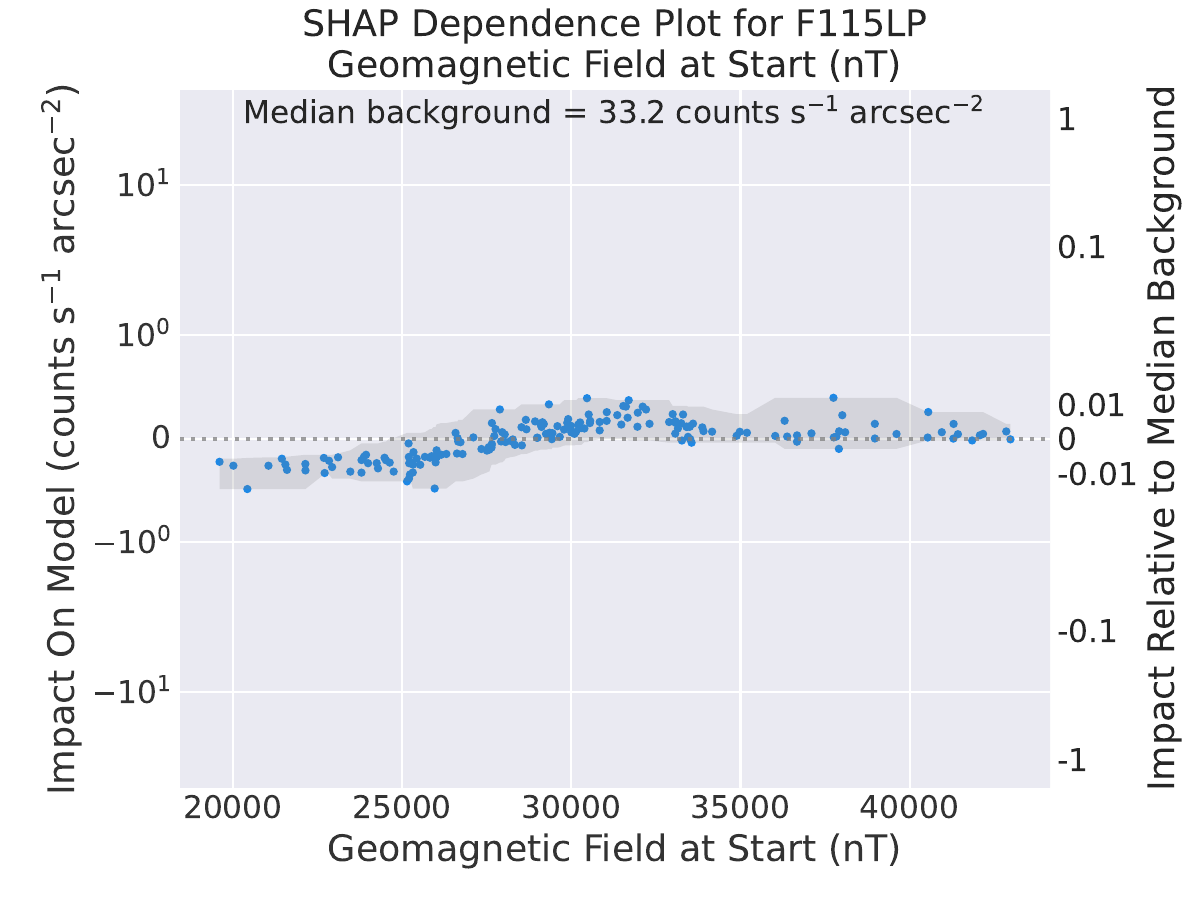}
\includegraphics[width=0.24\textwidth]{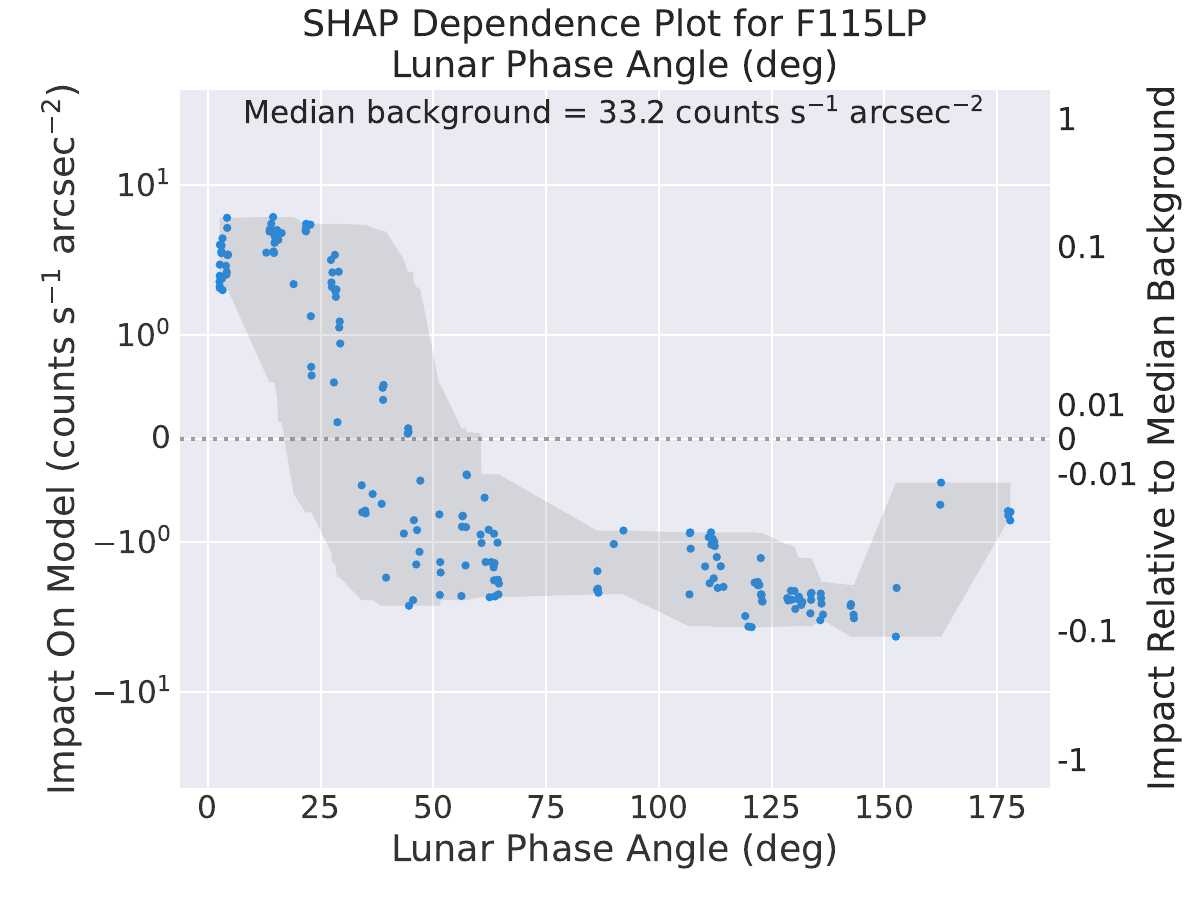}
\includegraphics[width=0.24\textwidth]{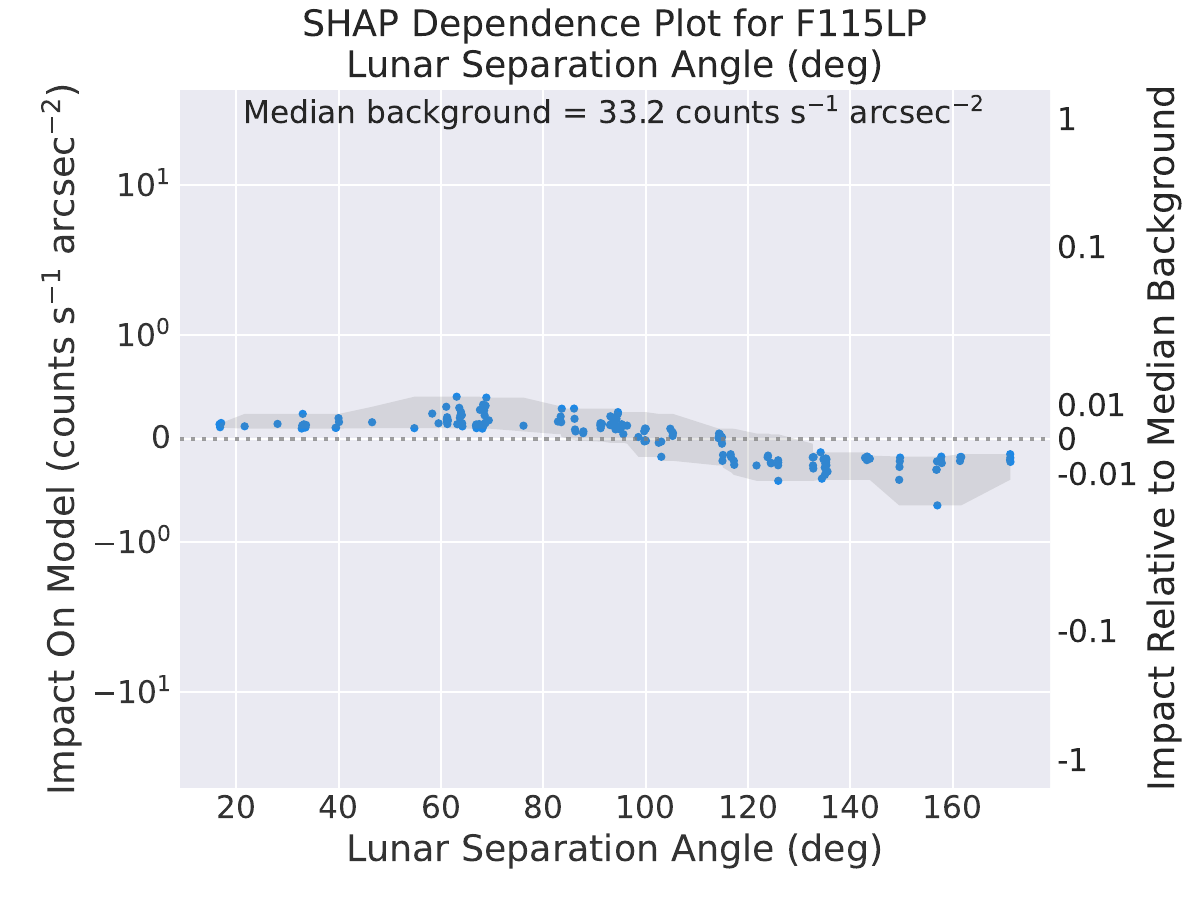}
\includegraphics[width=0.24\textwidth]{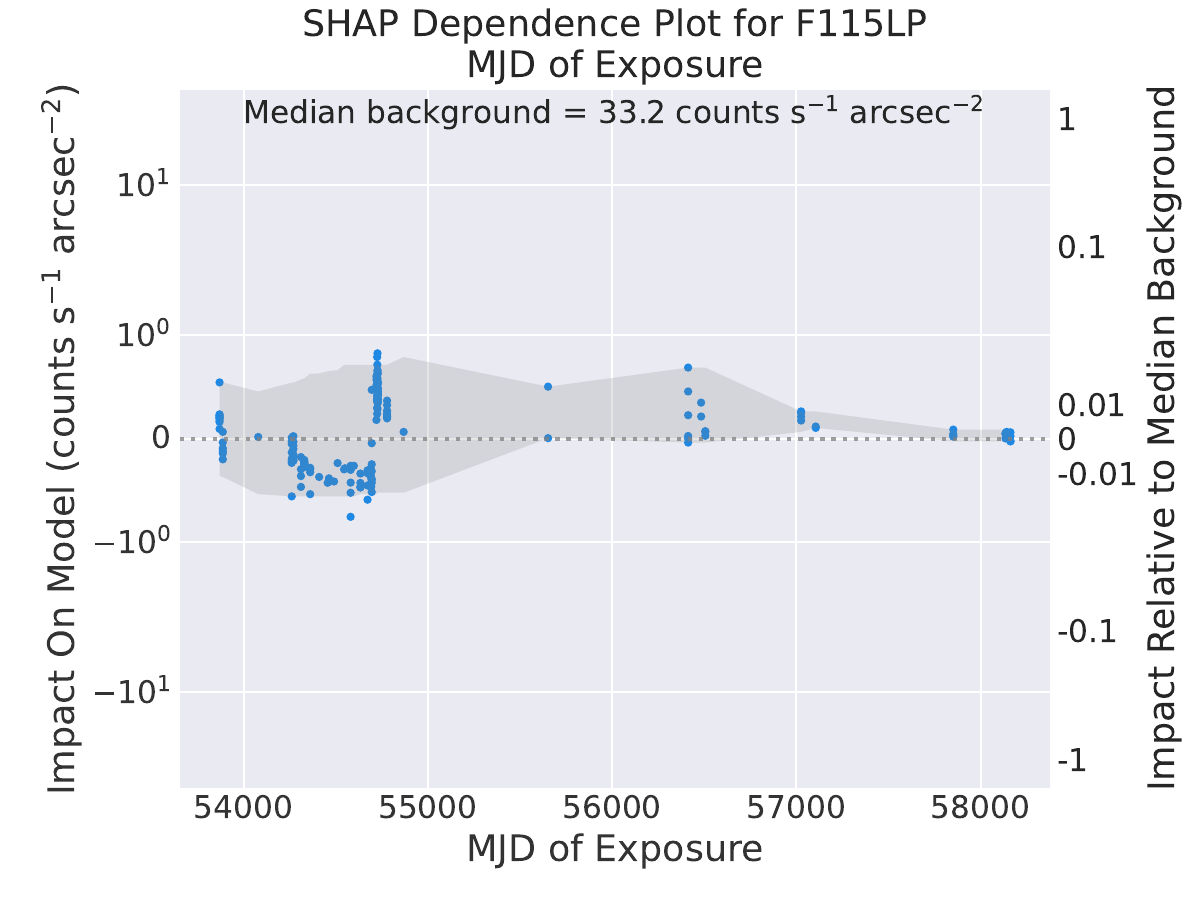}
\includegraphics[width=0.24\textwidth]{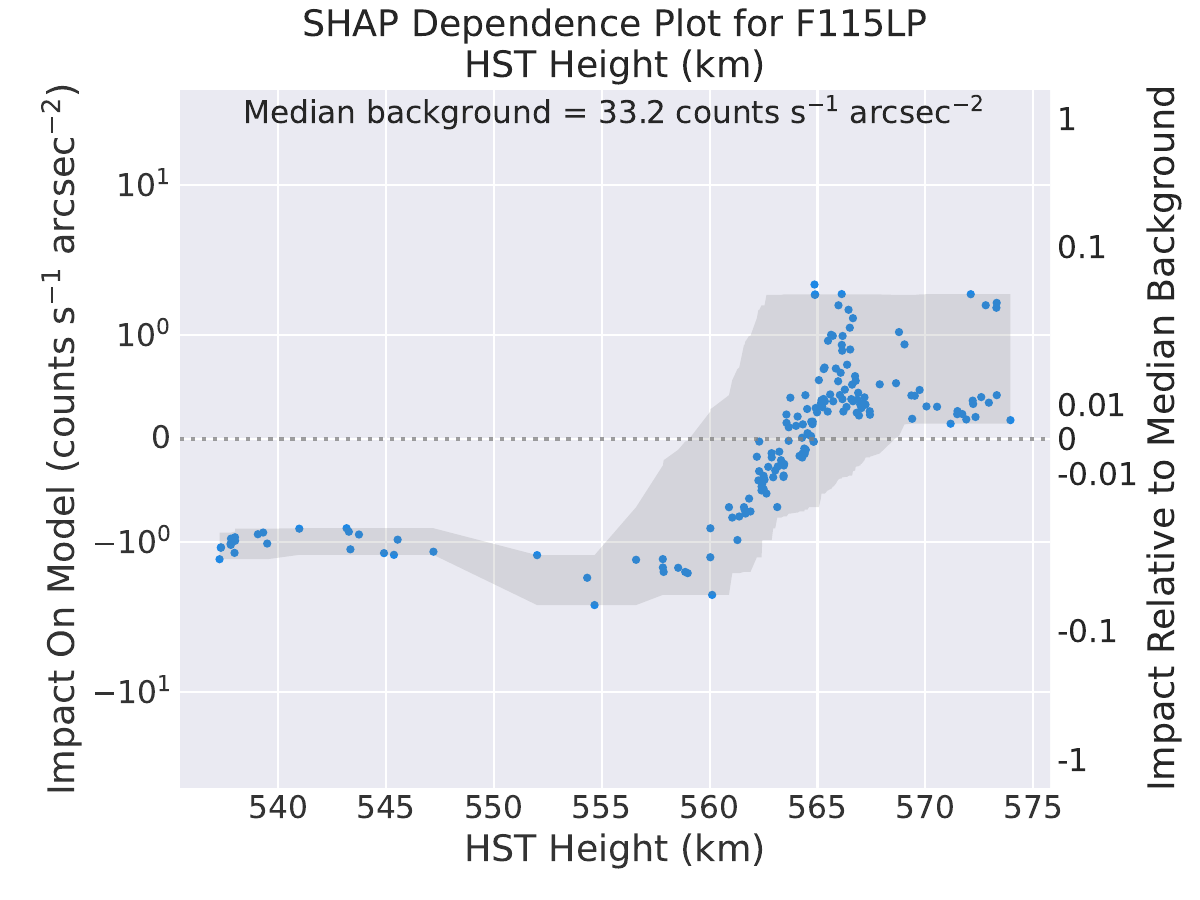}
\includegraphics[width=0.24\textwidth]{SHAP_Plots_QuantileForestRegr/F115LP_solar_alt_SHAP_Dependence.pdf}
\includegraphics[width=0.24\textwidth]{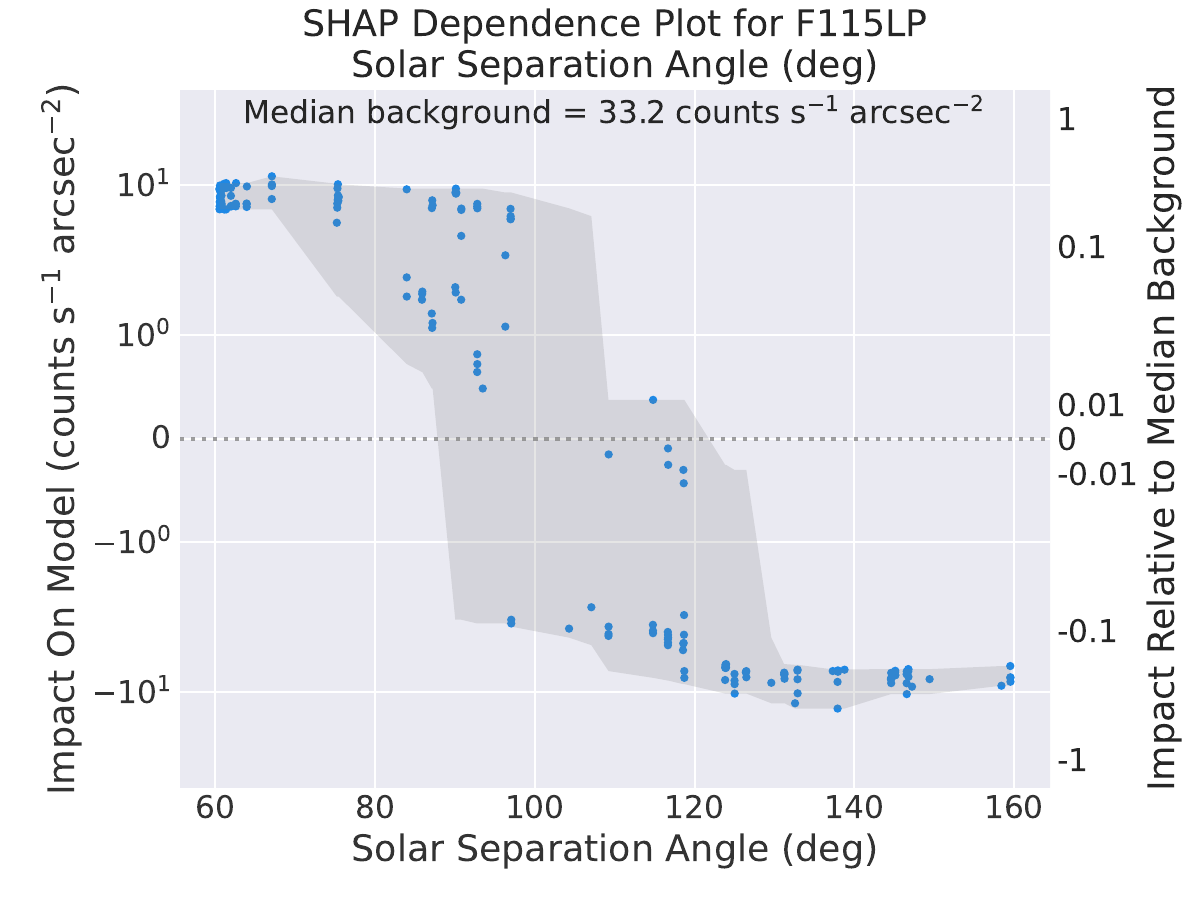}
\includegraphics[width=0.24\textwidth]{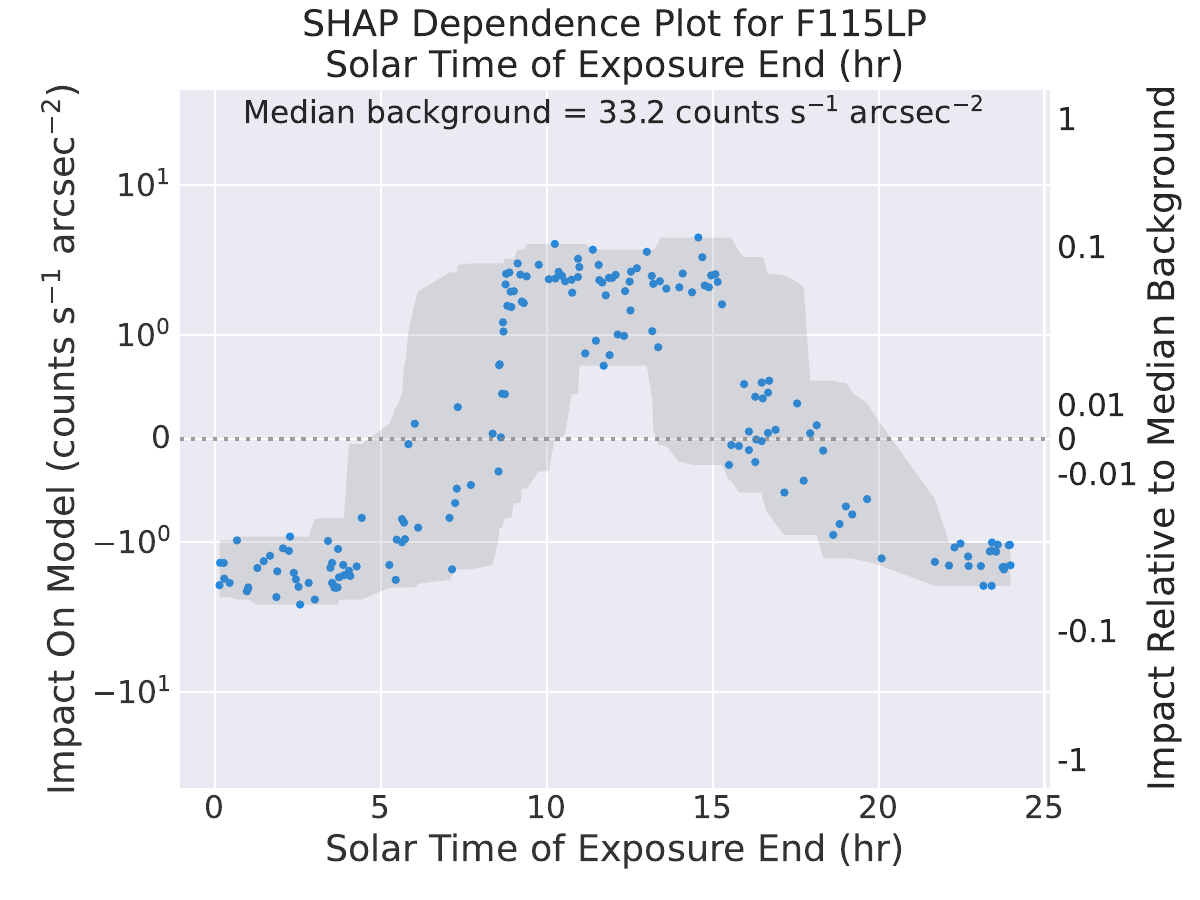}
\includegraphics[width=0.24\textwidth]{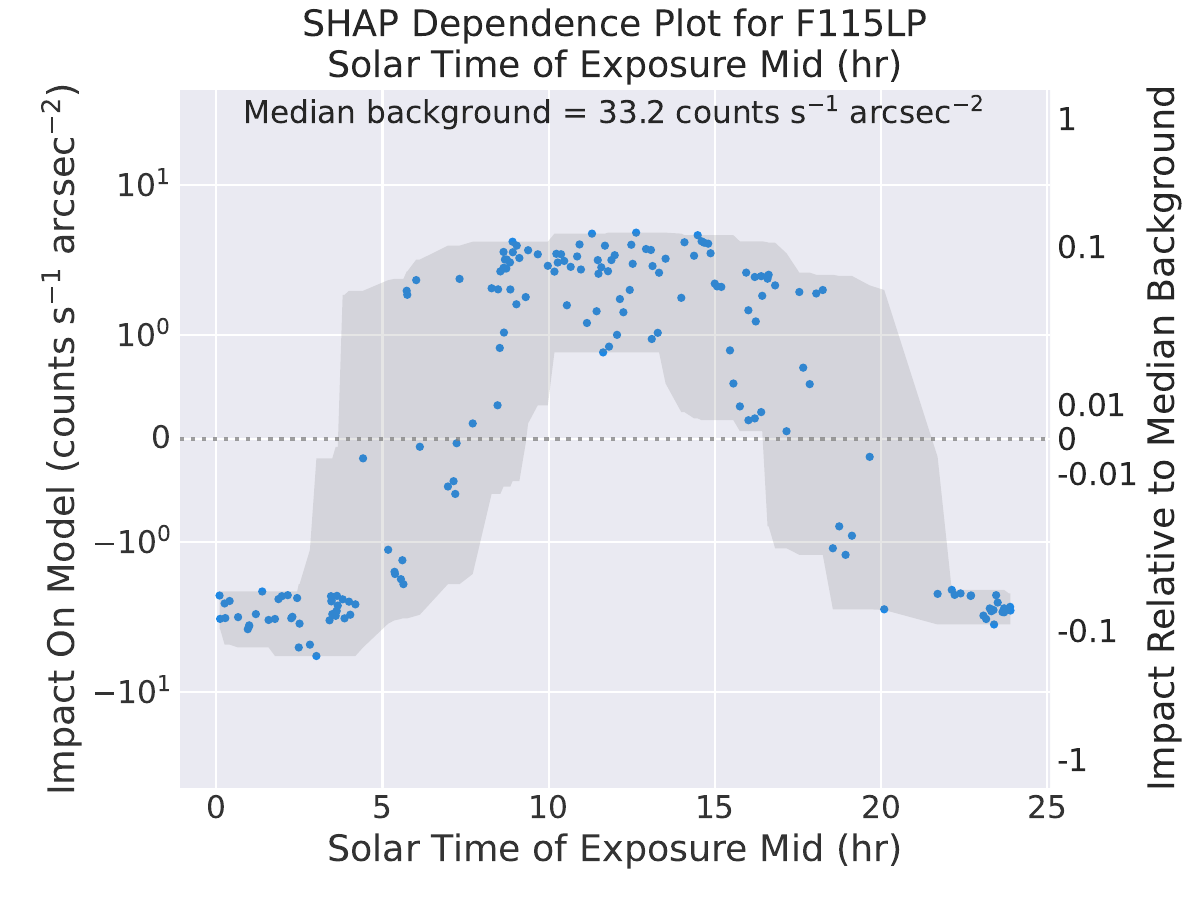}
\includegraphics[width=0.24\textwidth]{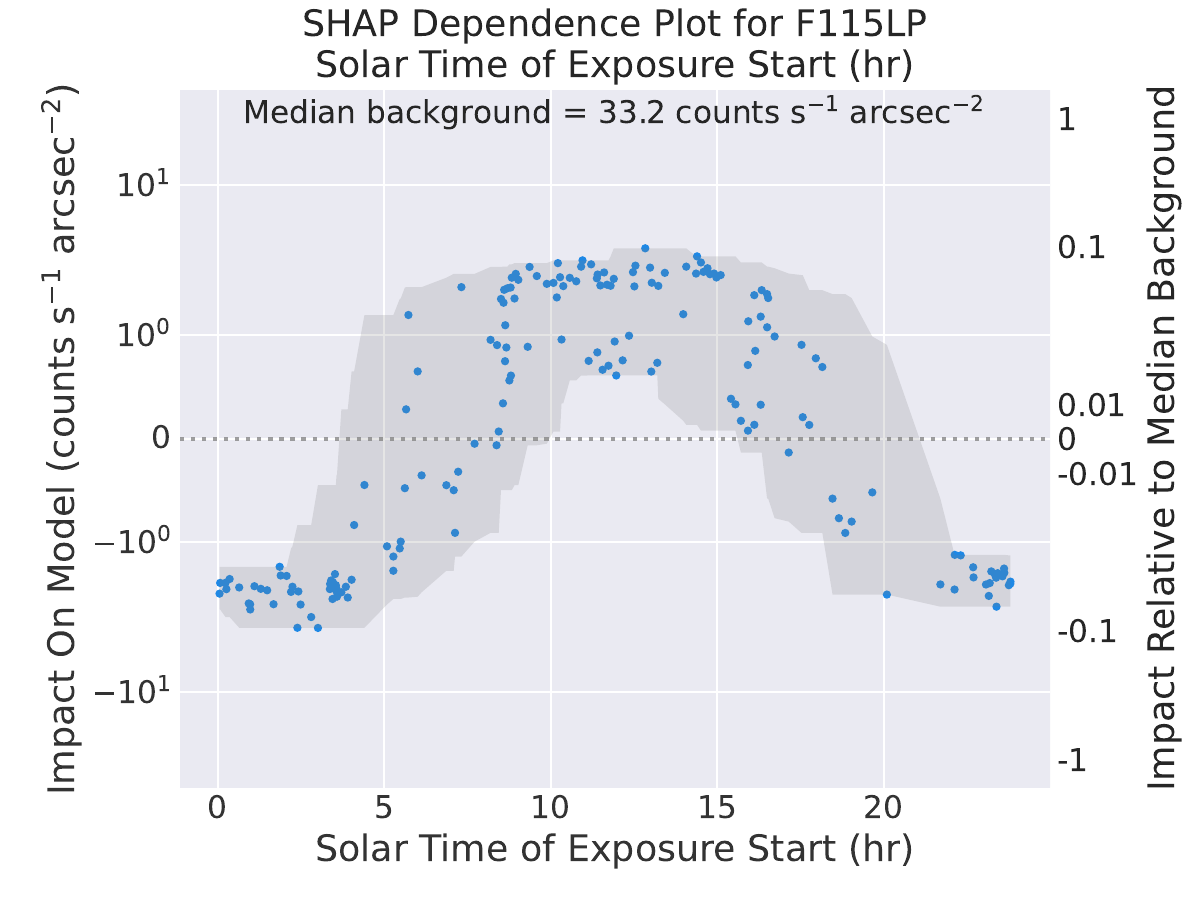}
\includegraphics[width=0.24\textwidth]{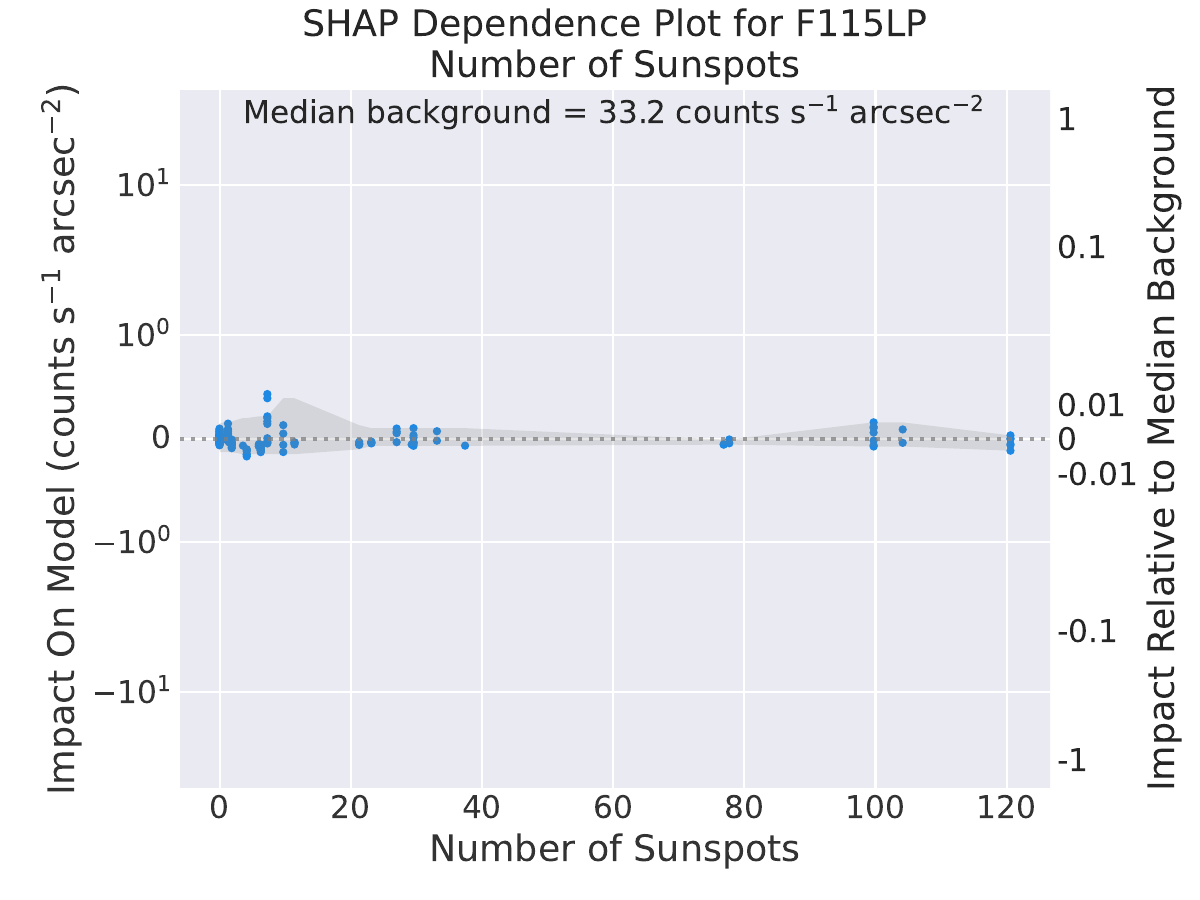}
\includegraphics[width=0.24\textwidth]{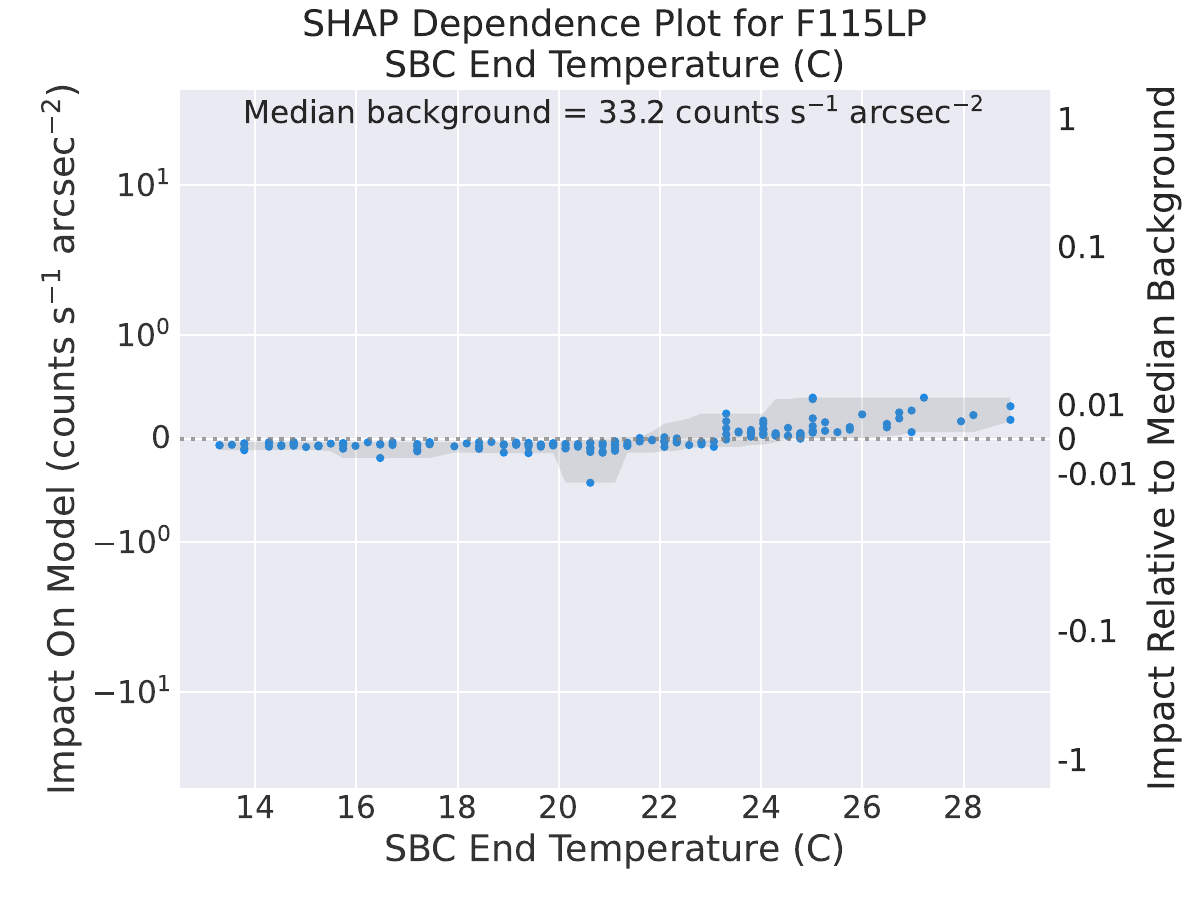}
\includegraphics[width=0.24\textwidth]{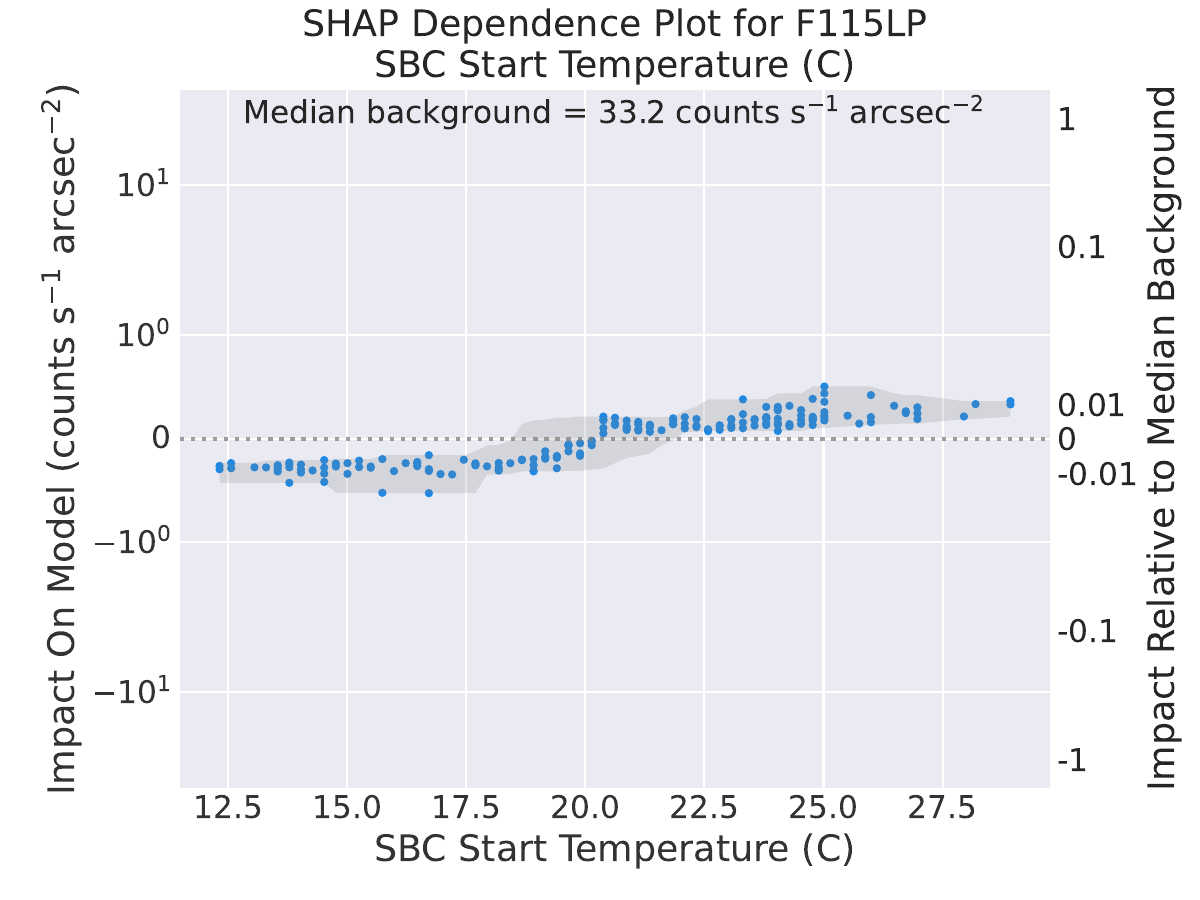}
\caption{SHAP dependence plots for QRF regression modeling of F115LP. SHAP dependence plots for Solar altitude, for several filters. The left-hand y-axis shows the marginal impact of the parameter on the model prediction in \cpspsqarc, whilst the right-hand y-axis shows the impact on the model normalized to the median background for that filter (the median background for the filter in question is noted on each plot). The y-axes are plotted using a symlog scale (linear either side of 0, and logarithmic elsewhere). The bars show the density of points.}
\label{Fig:SHAP_Dependence_F115LP}
\end{figure}

\begin{figure}
\centering
\includegraphics[width=0.24\textwidth]{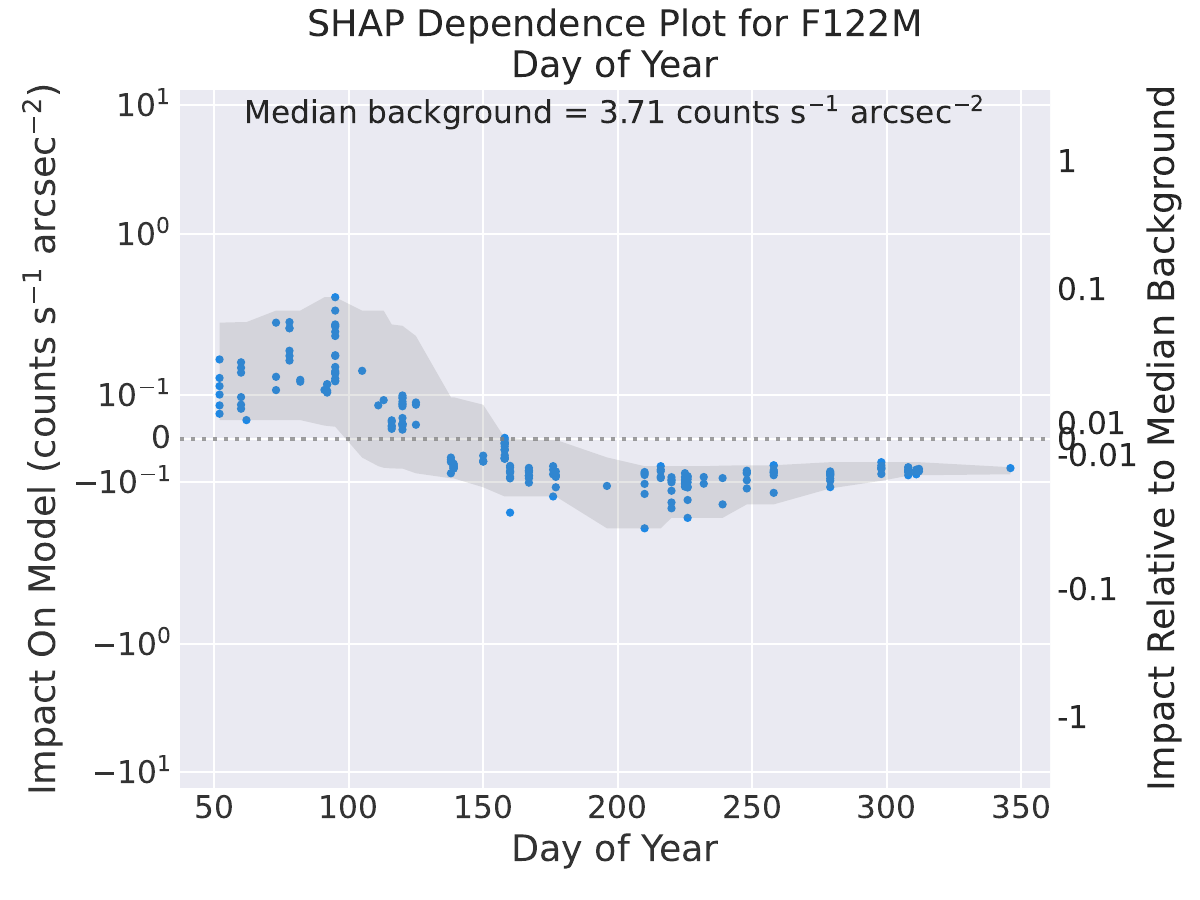}
\includegraphics[width=0.24\textwidth]{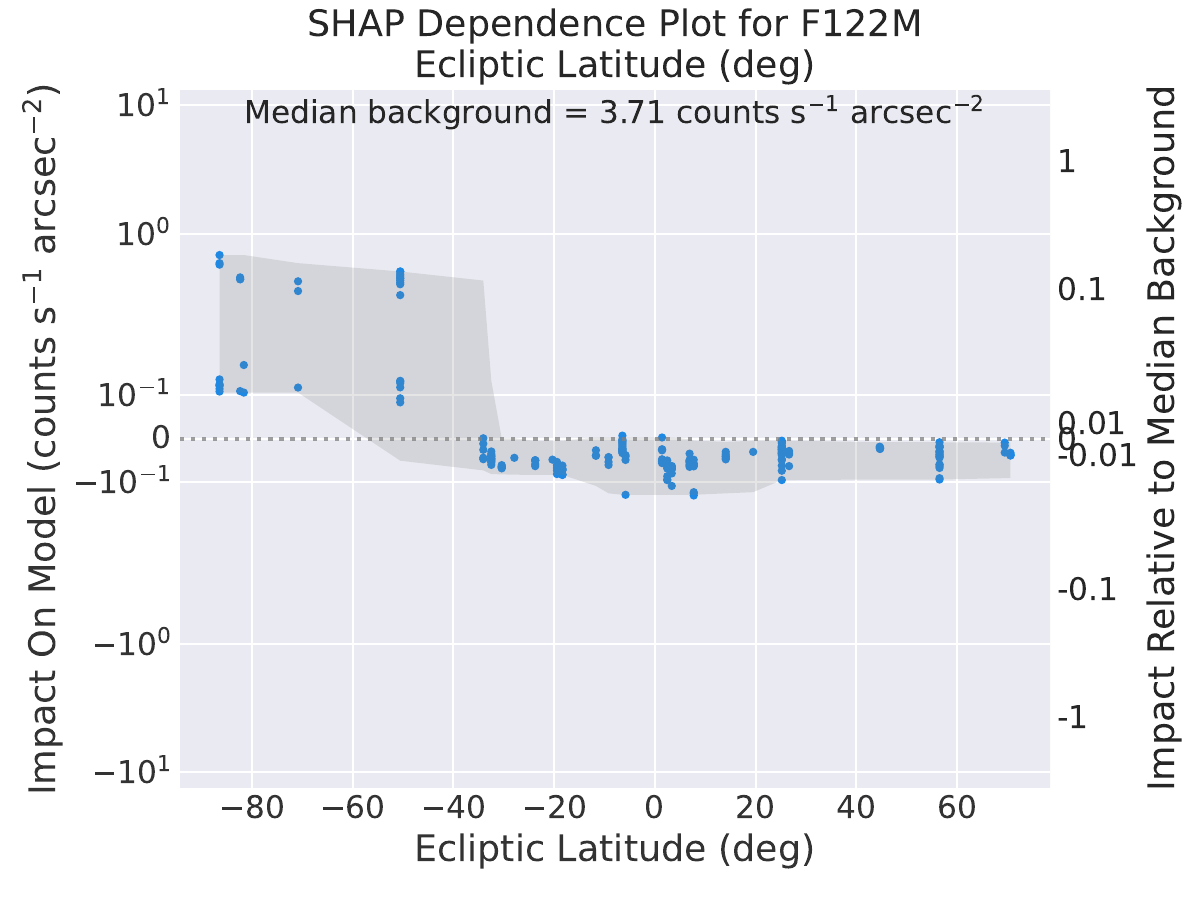}
\includegraphics[width=0.24\textwidth]{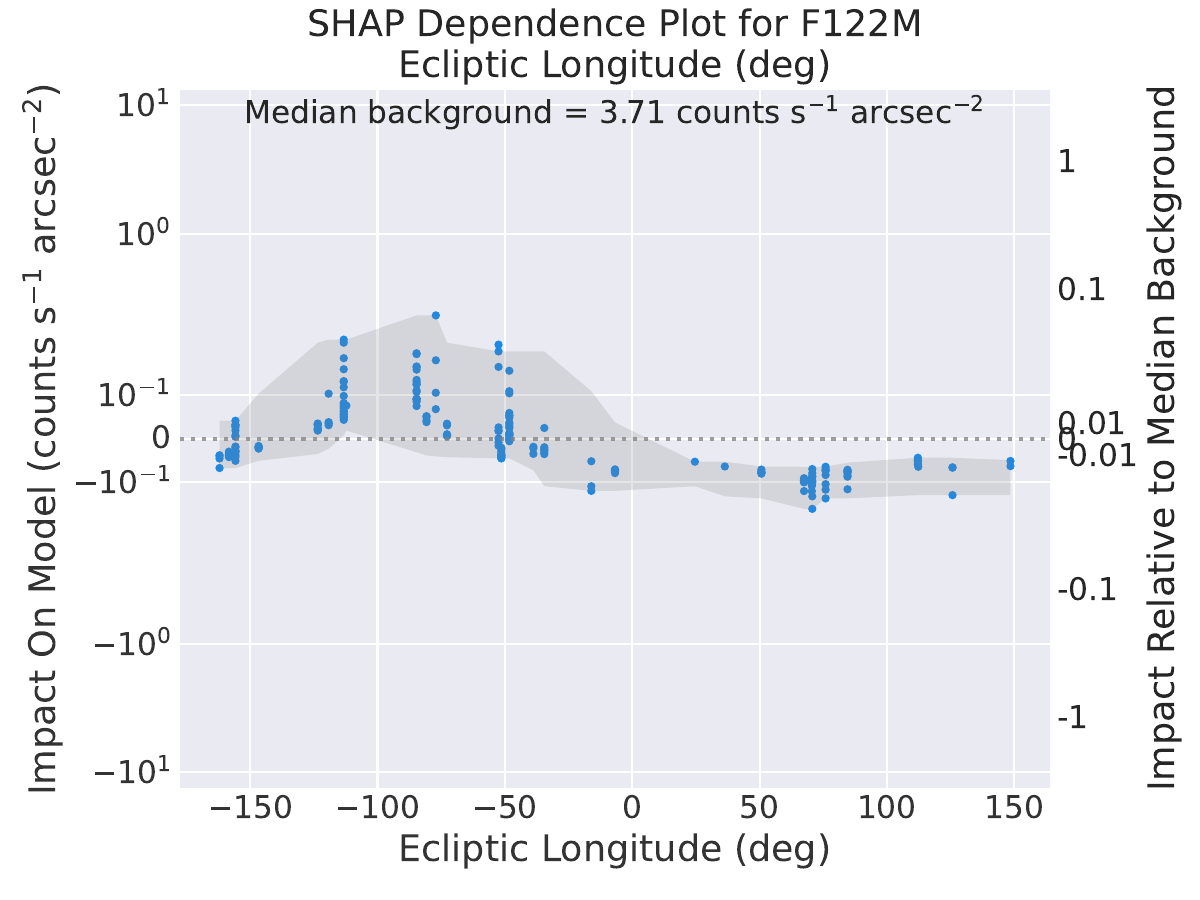}
\includegraphics[width=0.24\textwidth]{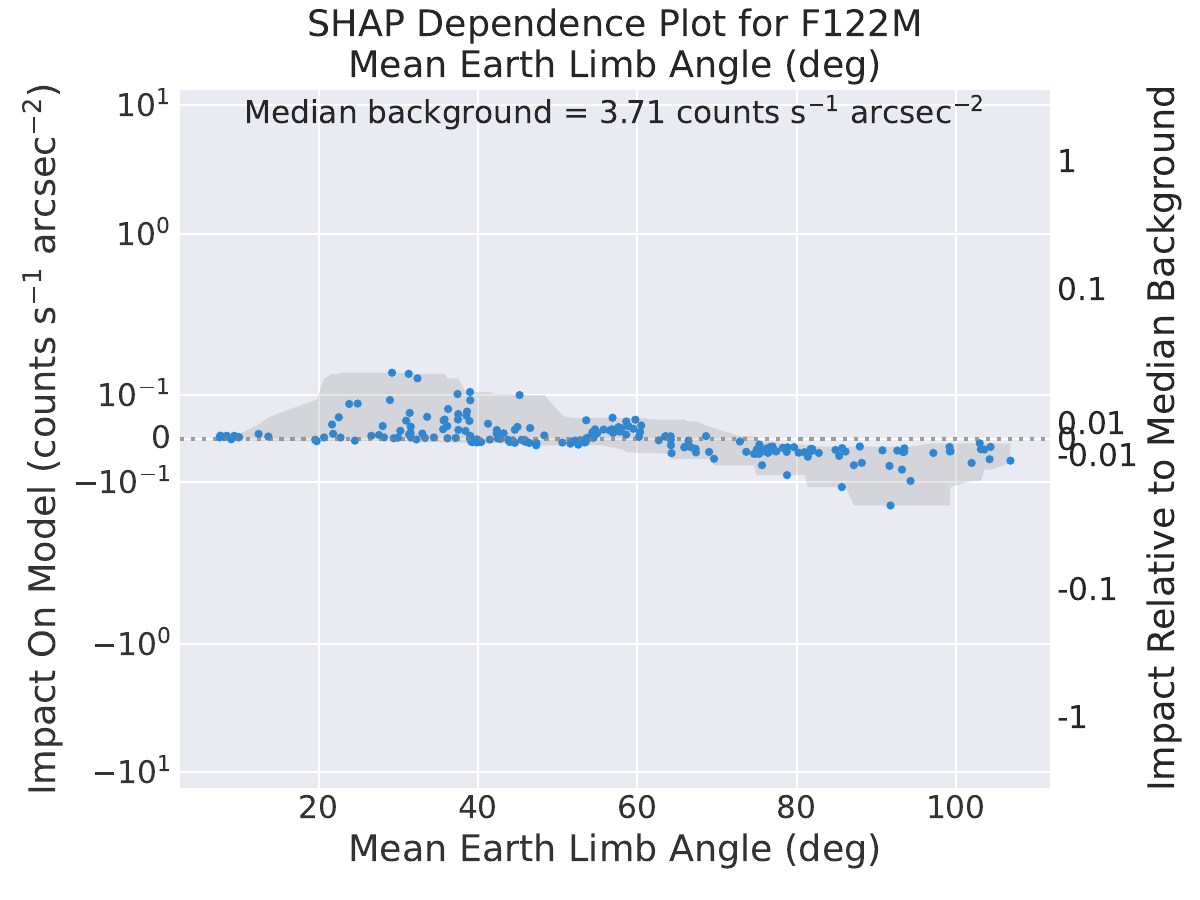}
\includegraphics[width=0.24\textwidth]{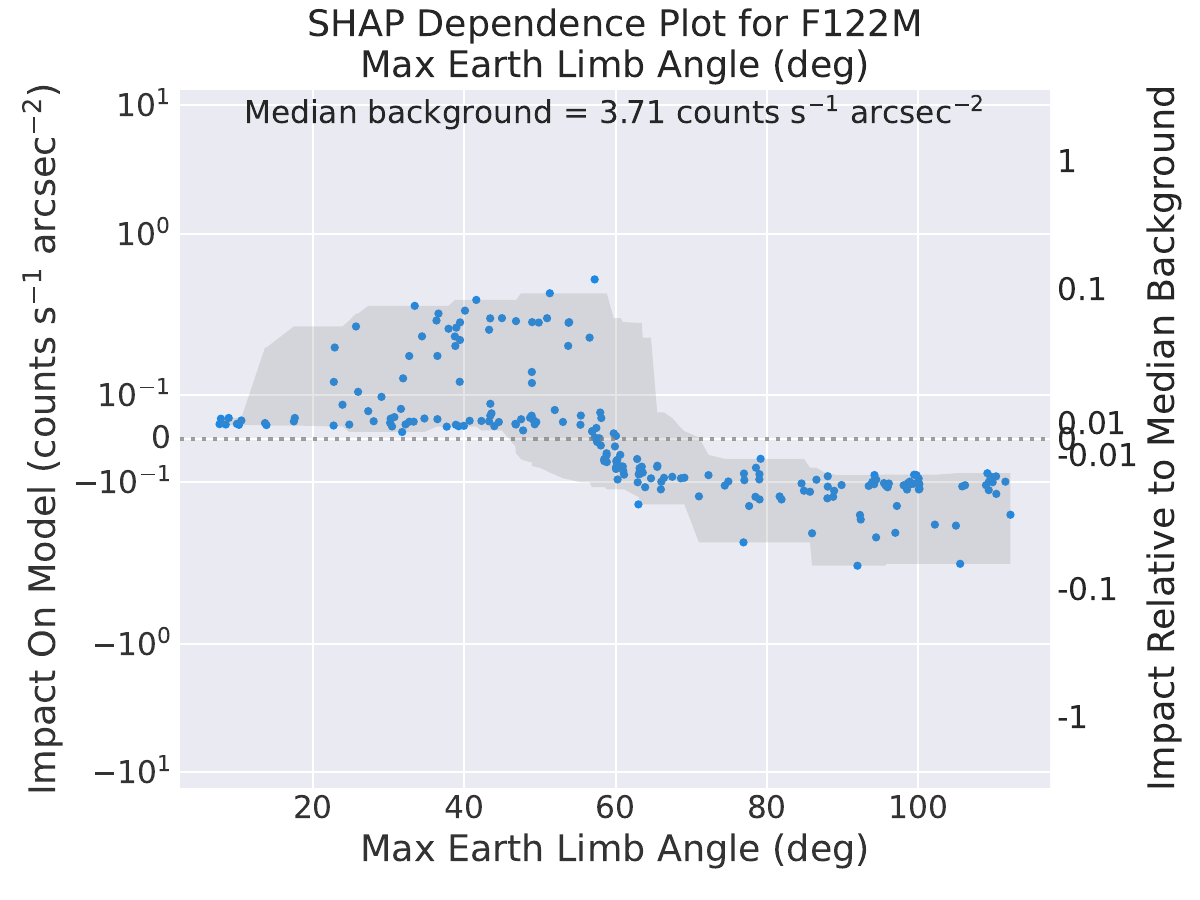}
\includegraphics[width=0.24\textwidth]{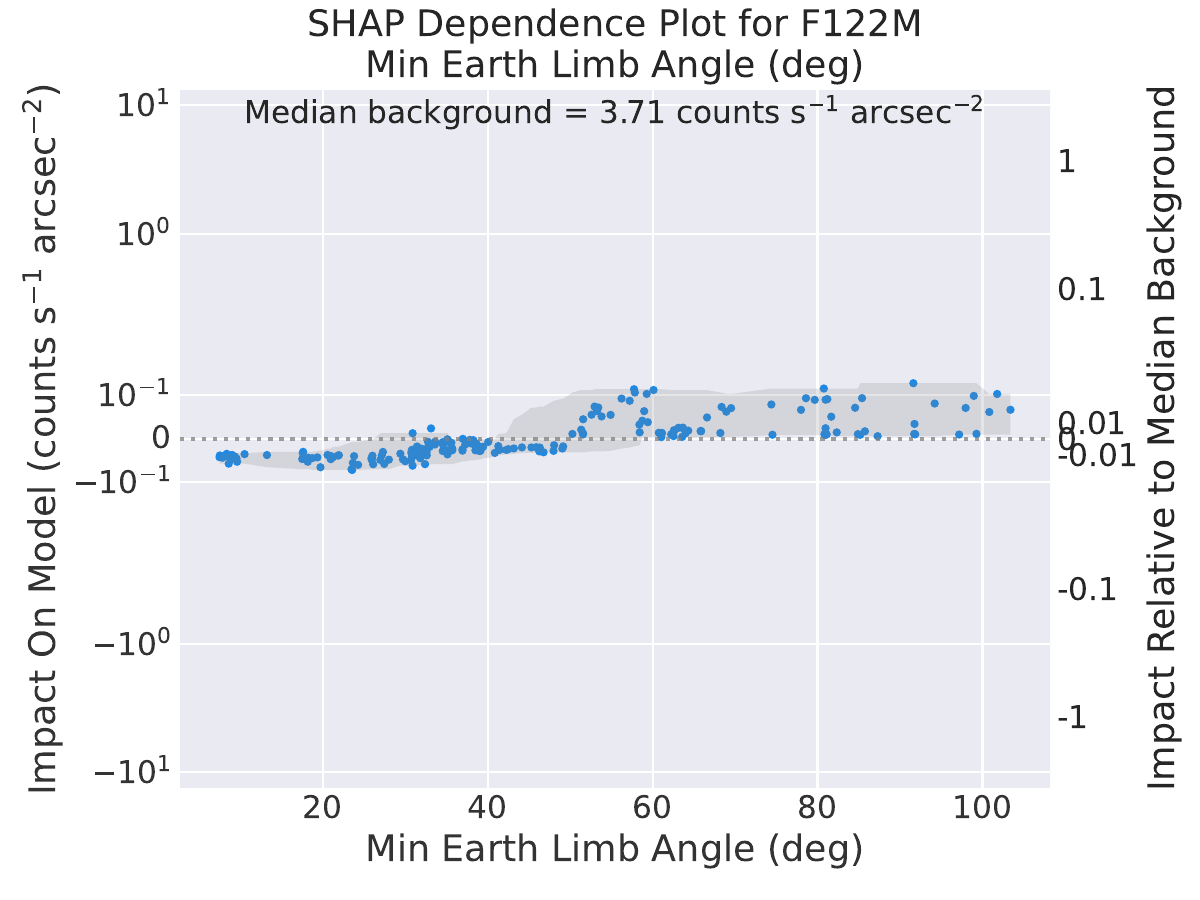}
\includegraphics[width=0.24\textwidth]{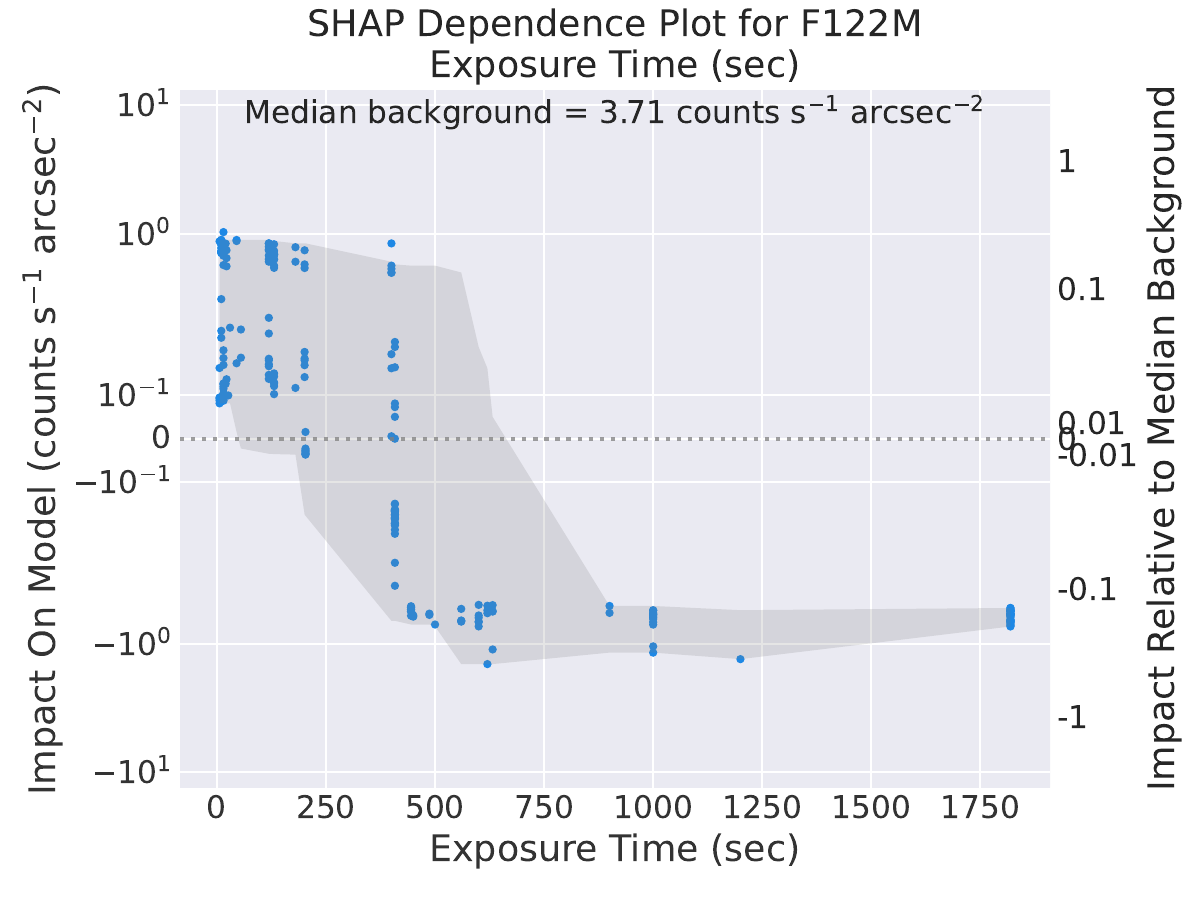}
\includegraphics[width=0.24\textwidth]{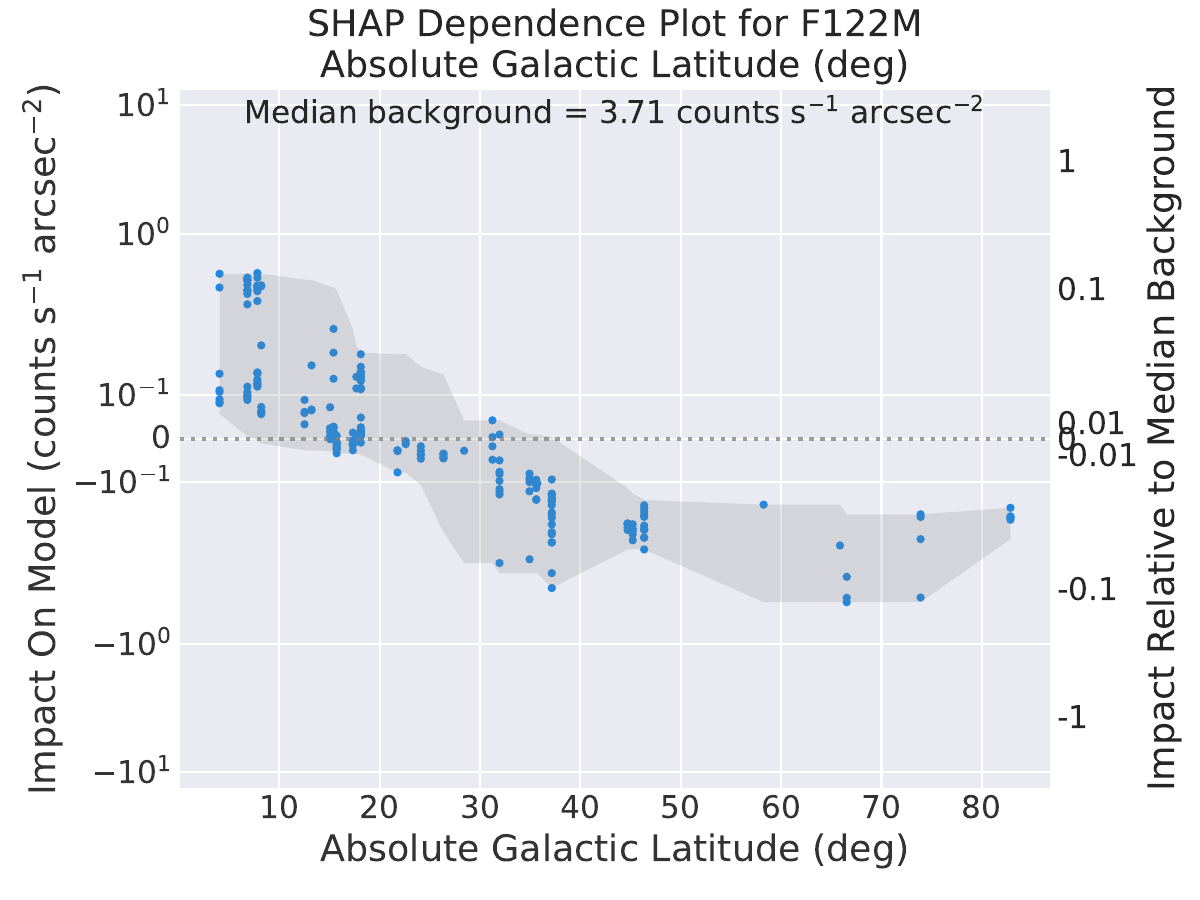}
\includegraphics[width=0.24\textwidth]{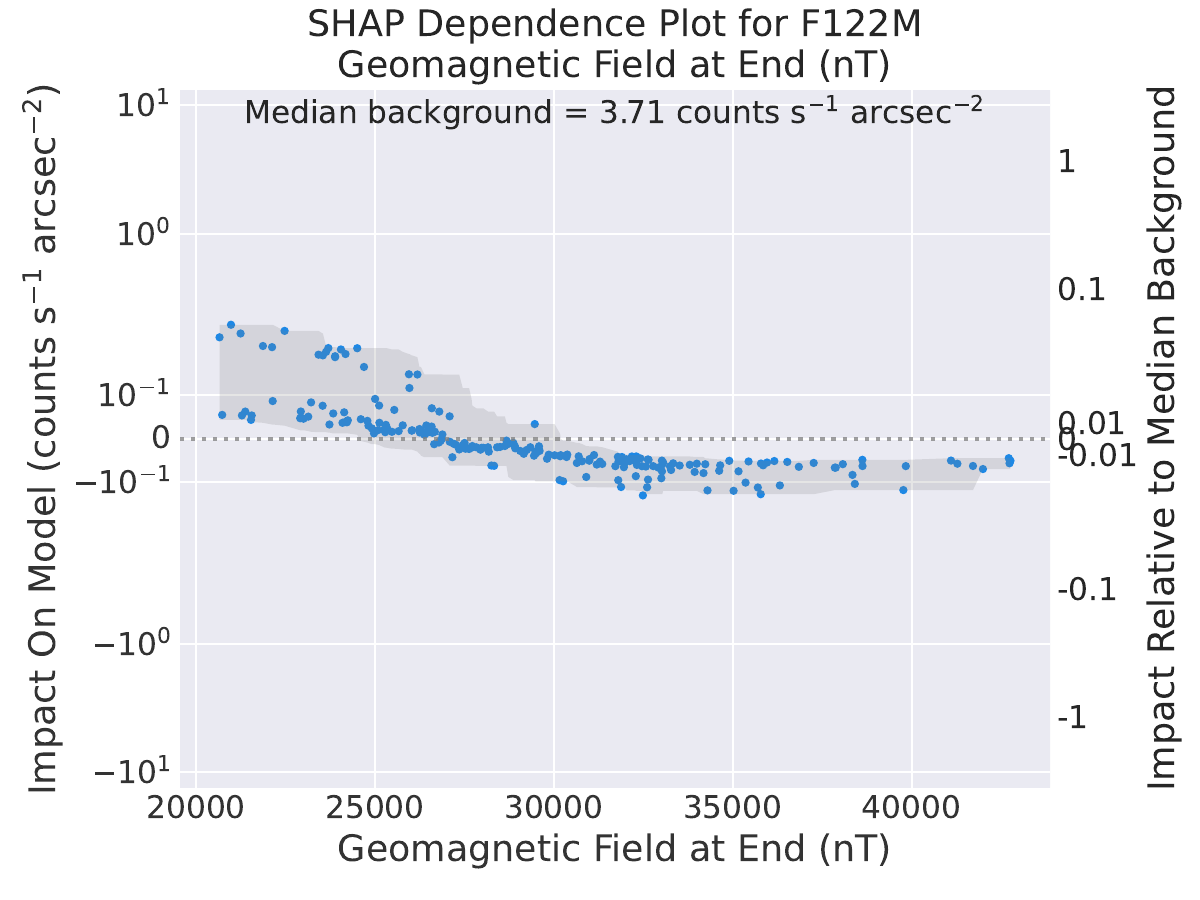}
\includegraphics[width=0.24\textwidth]{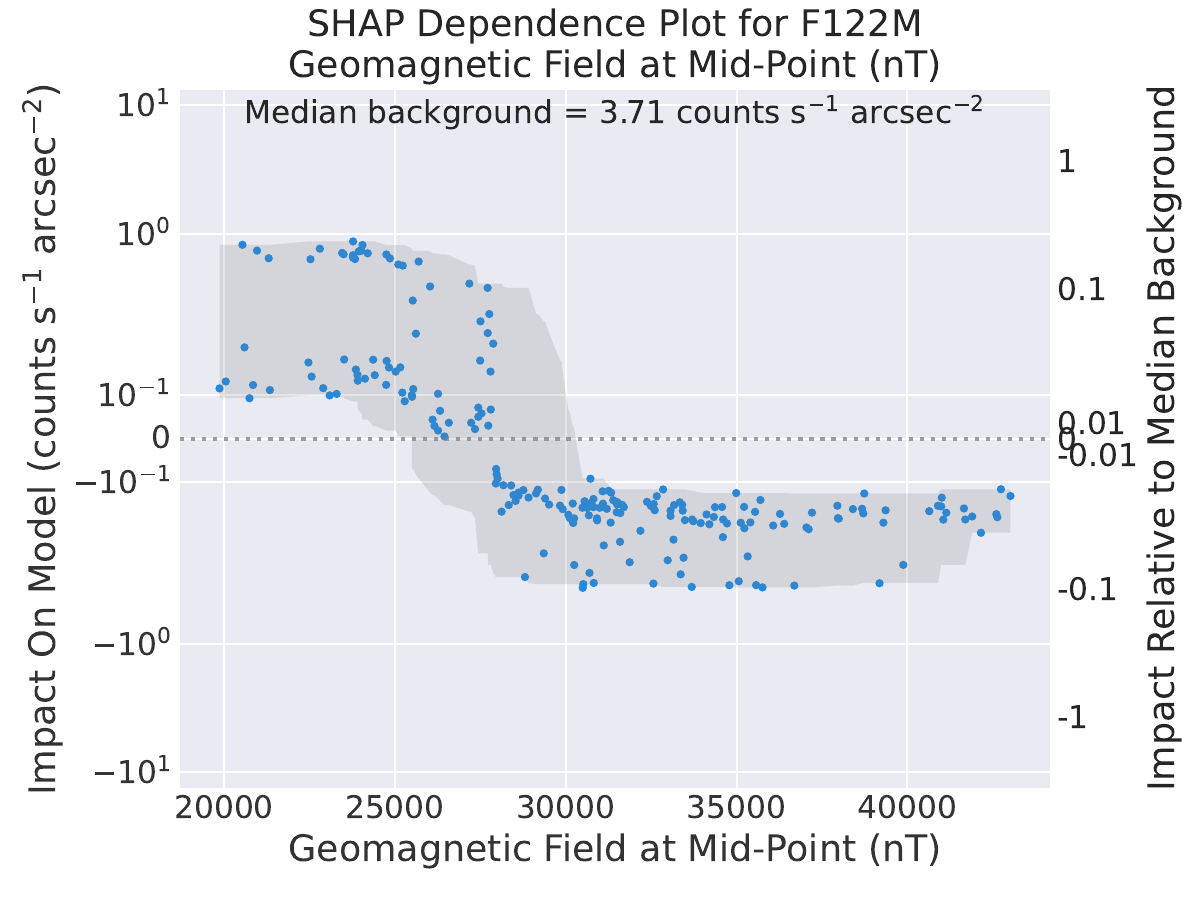}
\includegraphics[width=0.24\textwidth]{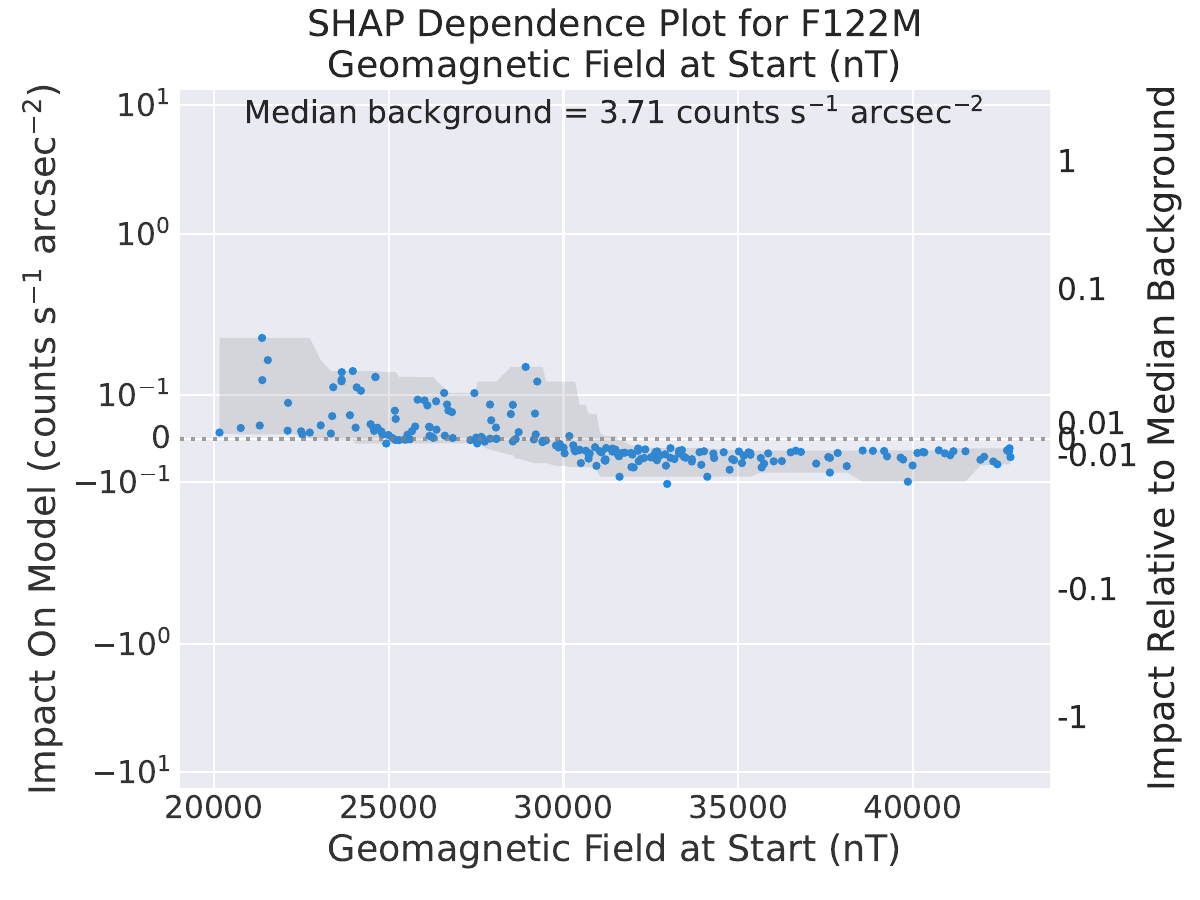}
\includegraphics[width=0.24\textwidth]{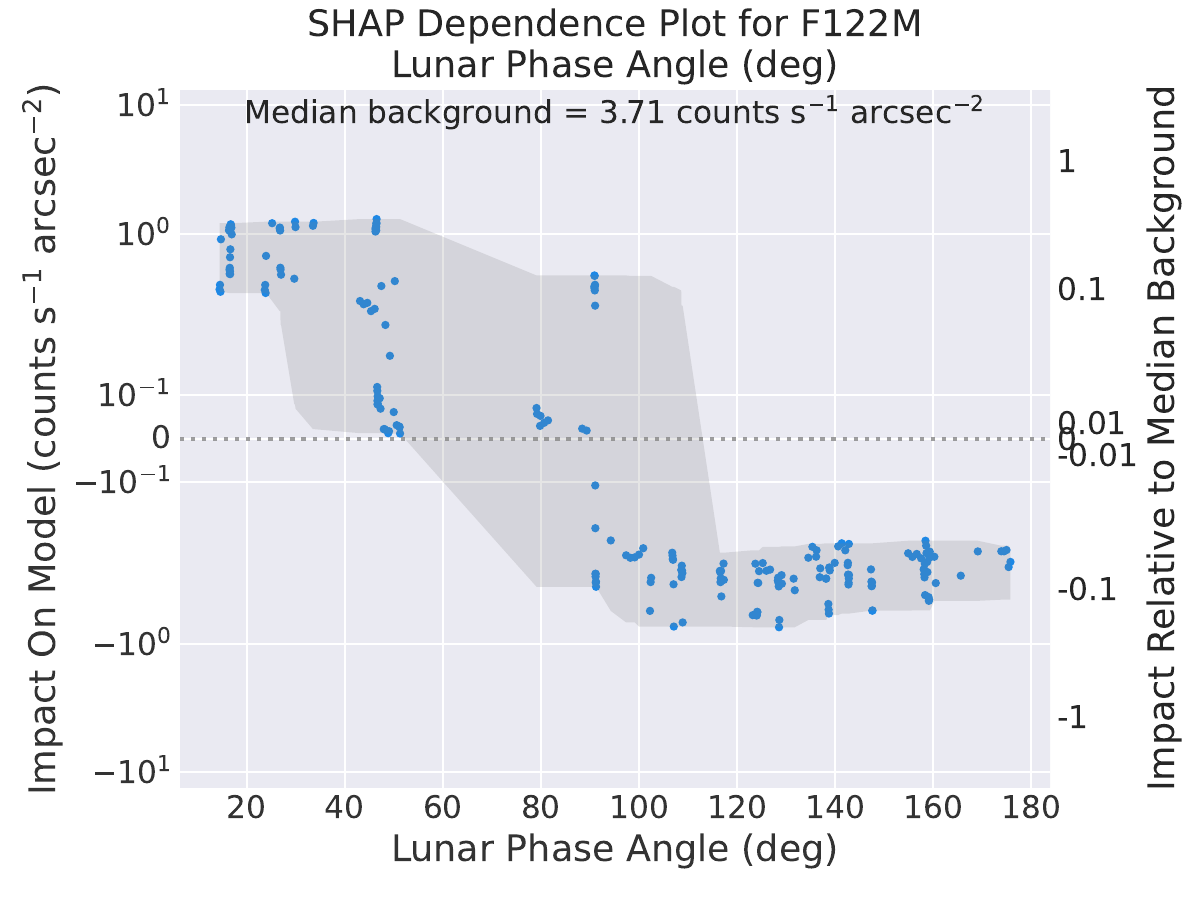}
\includegraphics[width=0.24\textwidth]{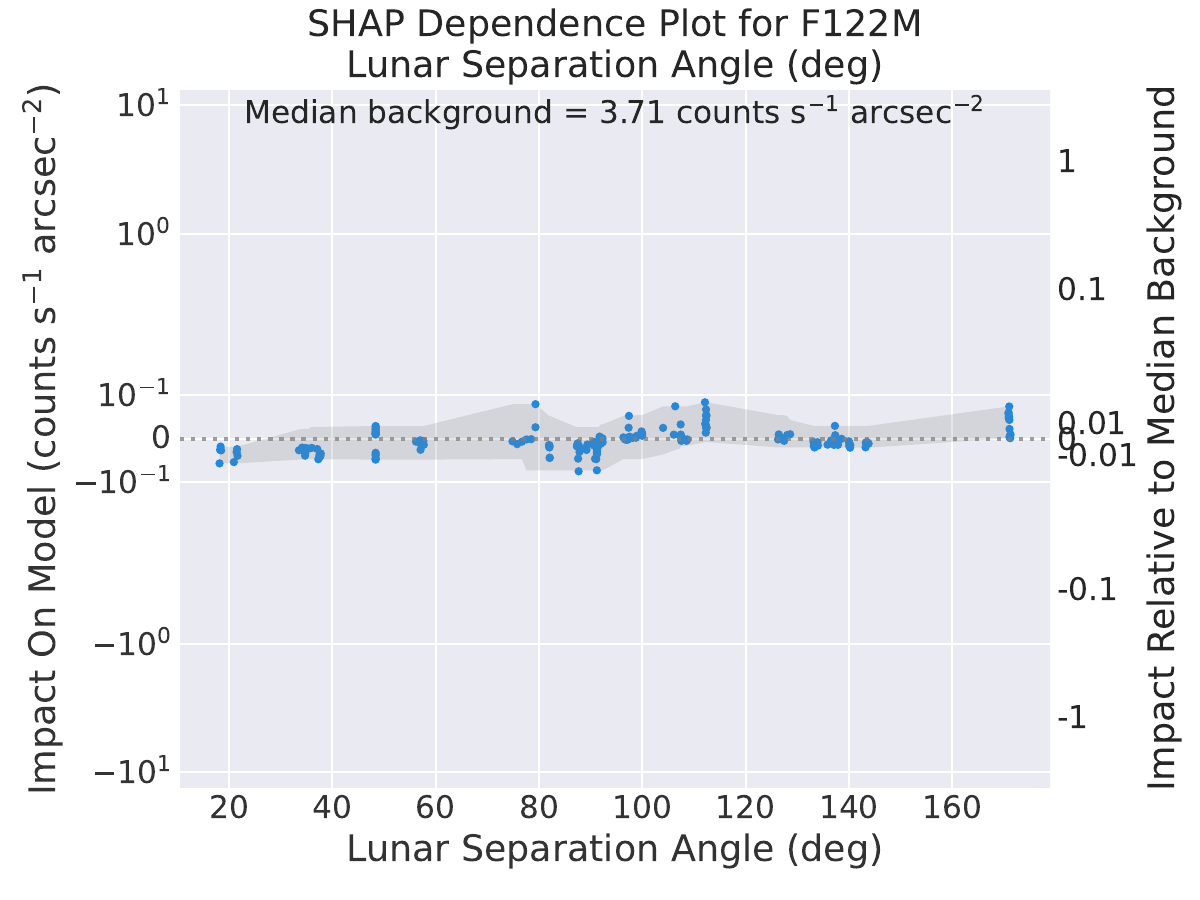}
\includegraphics[width=0.24\textwidth]{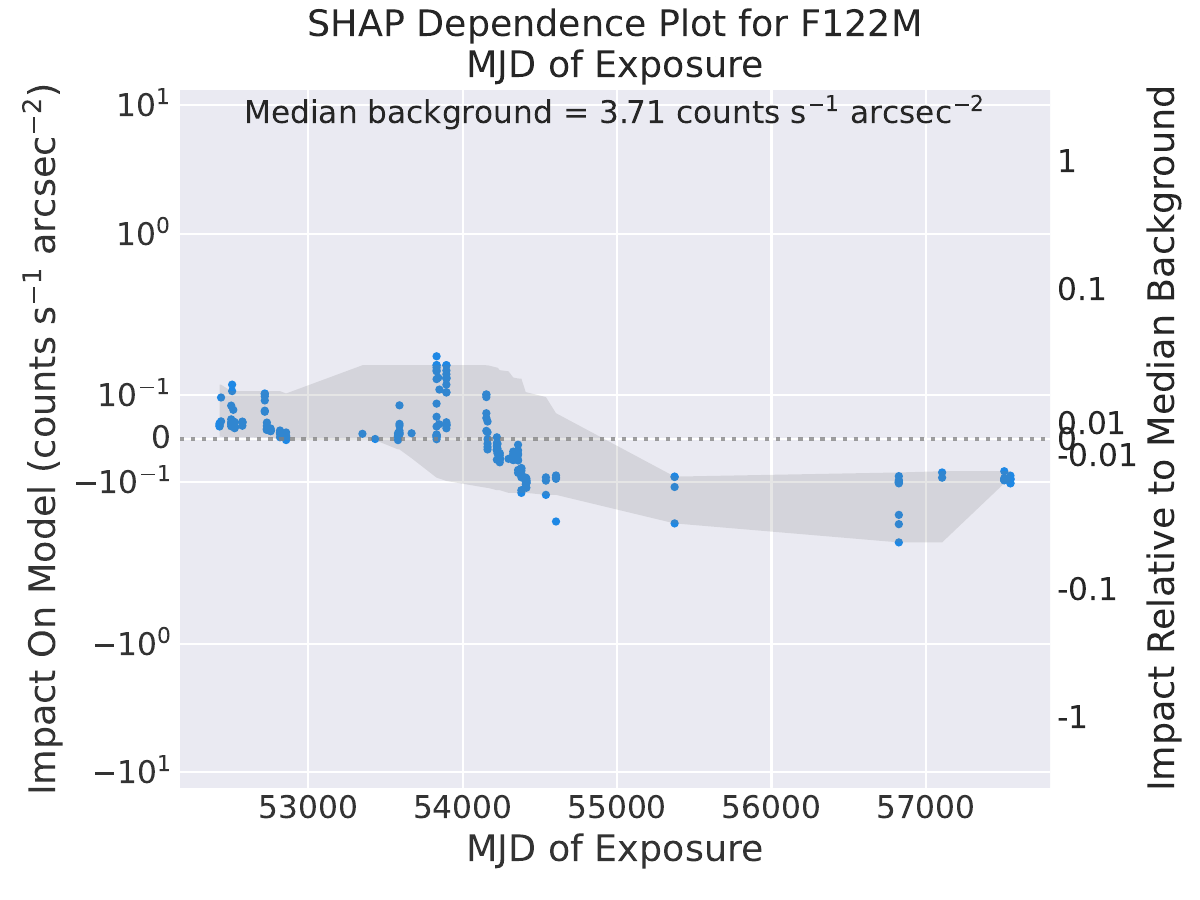}
\includegraphics[width=0.24\textwidth]{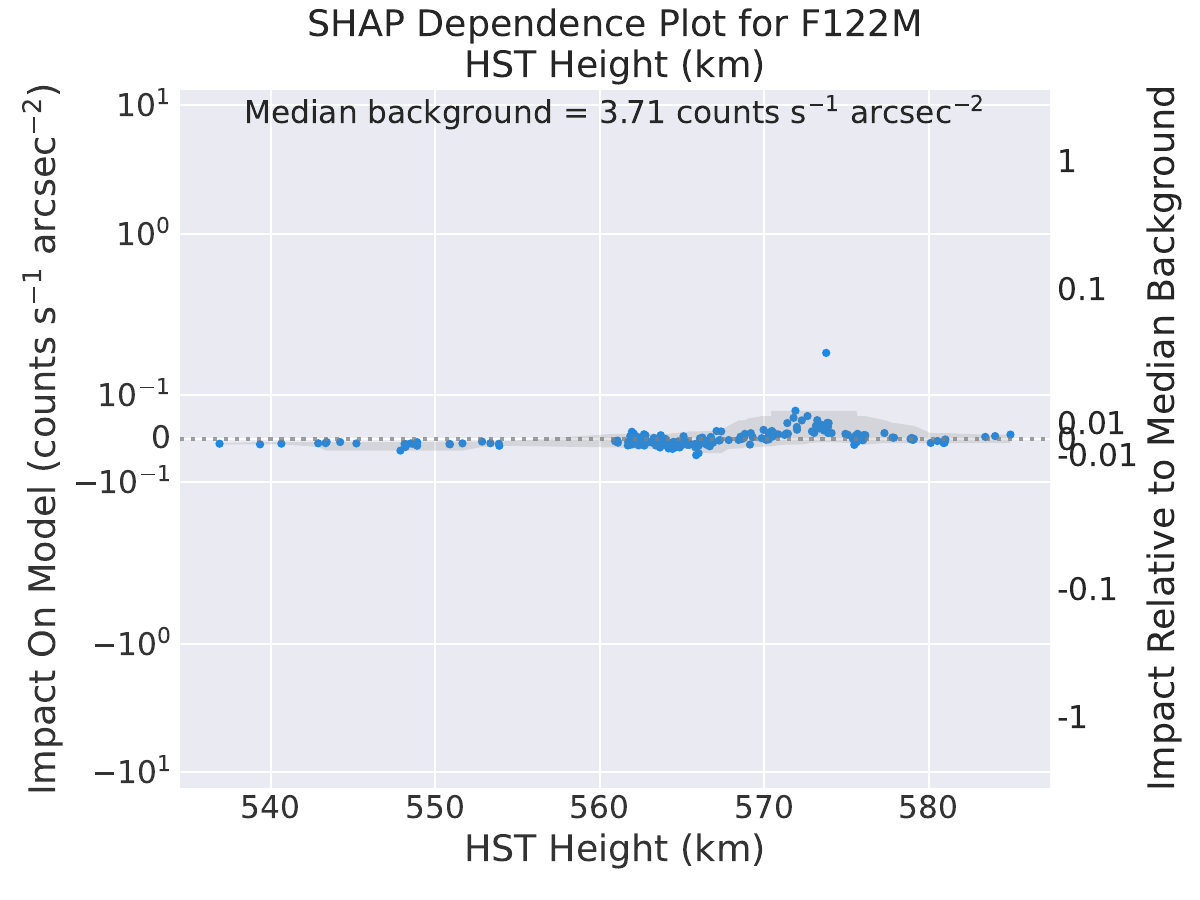}
\includegraphics[width=0.24\textwidth]{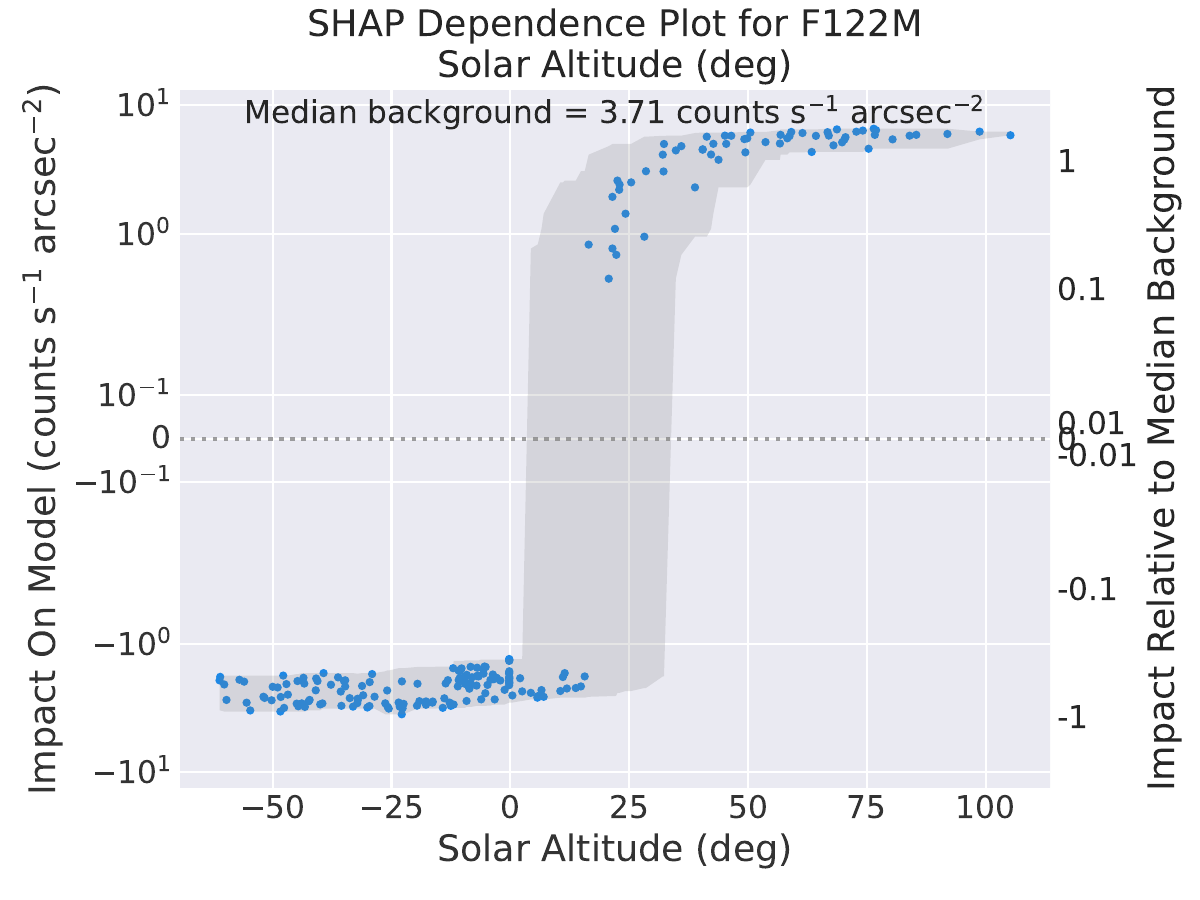}
\includegraphics[width=0.24\textwidth]{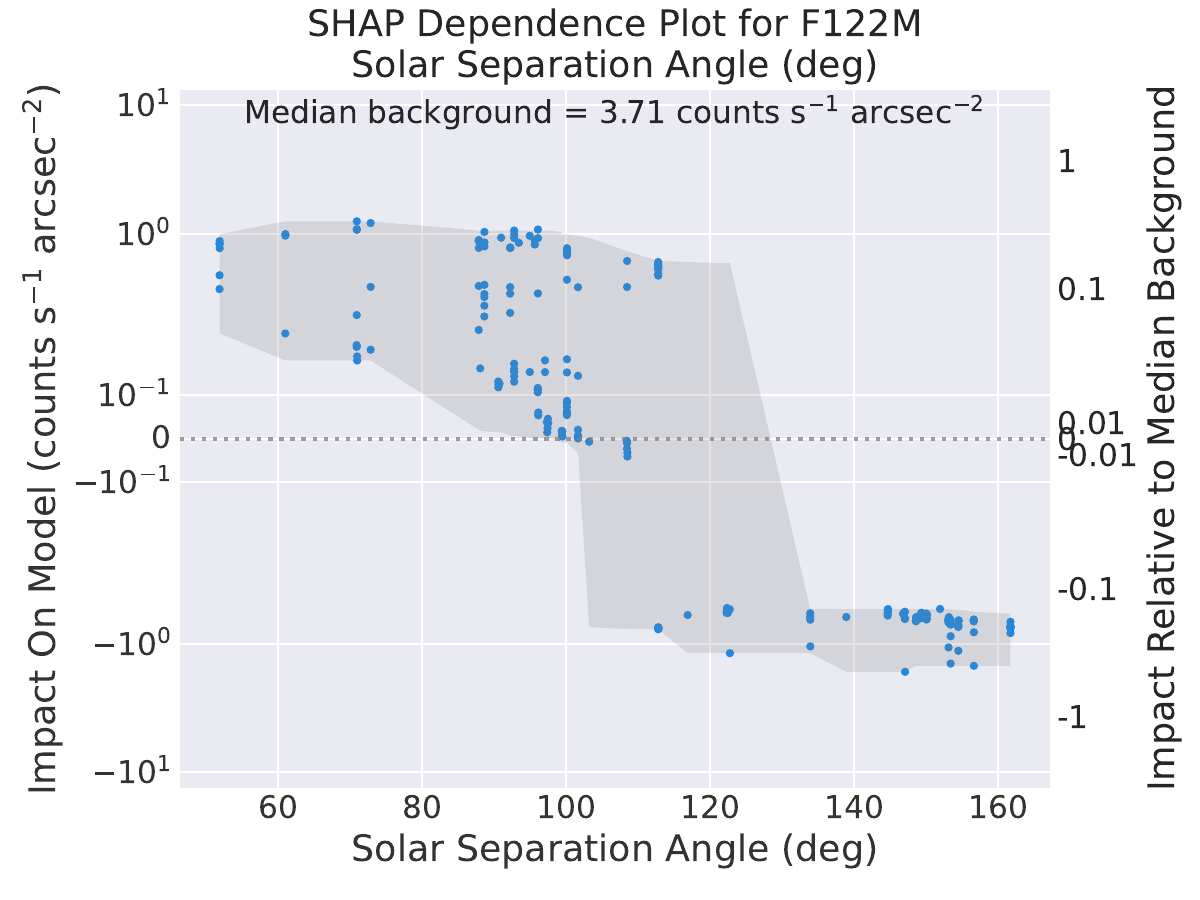}
\includegraphics[width=0.24\textwidth]{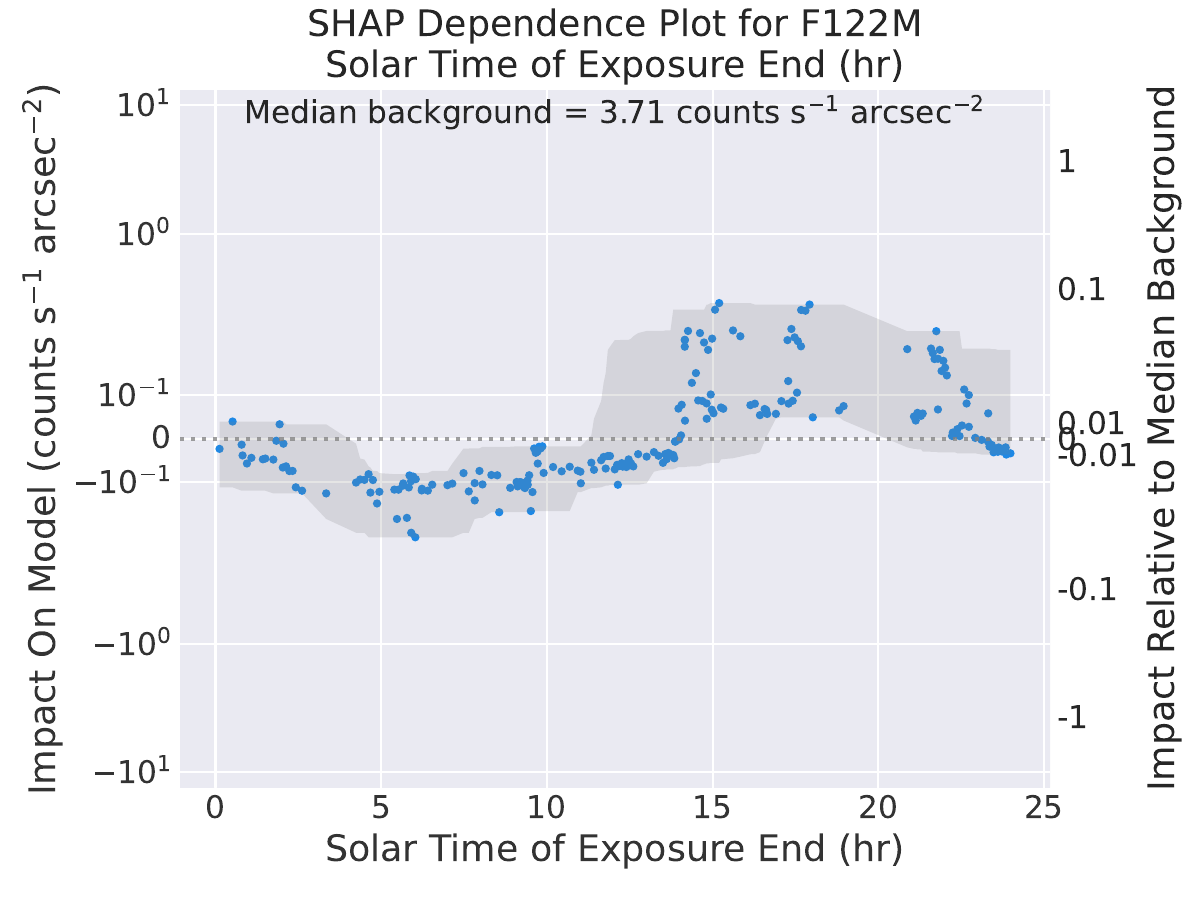}
\includegraphics[width=0.24\textwidth]{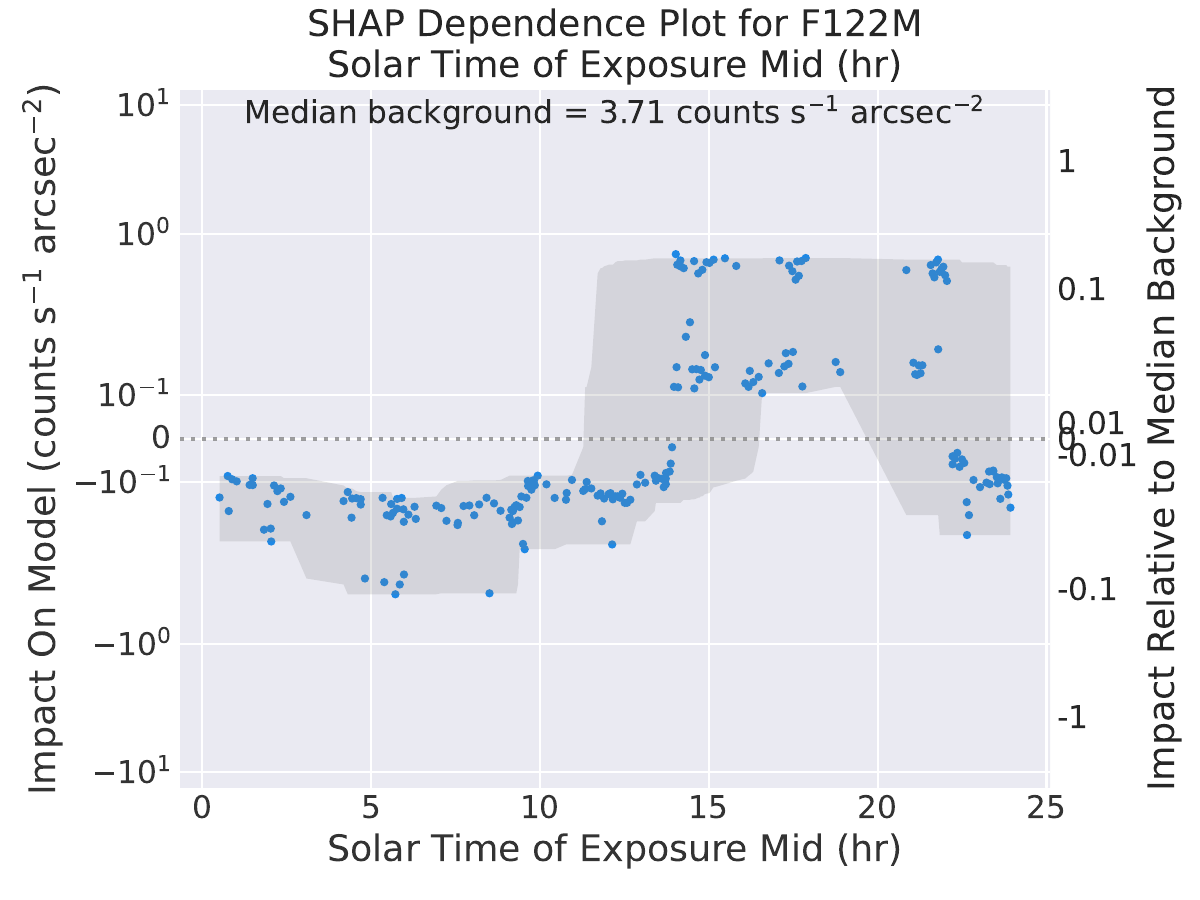}
\includegraphics[width=0.24\textwidth]{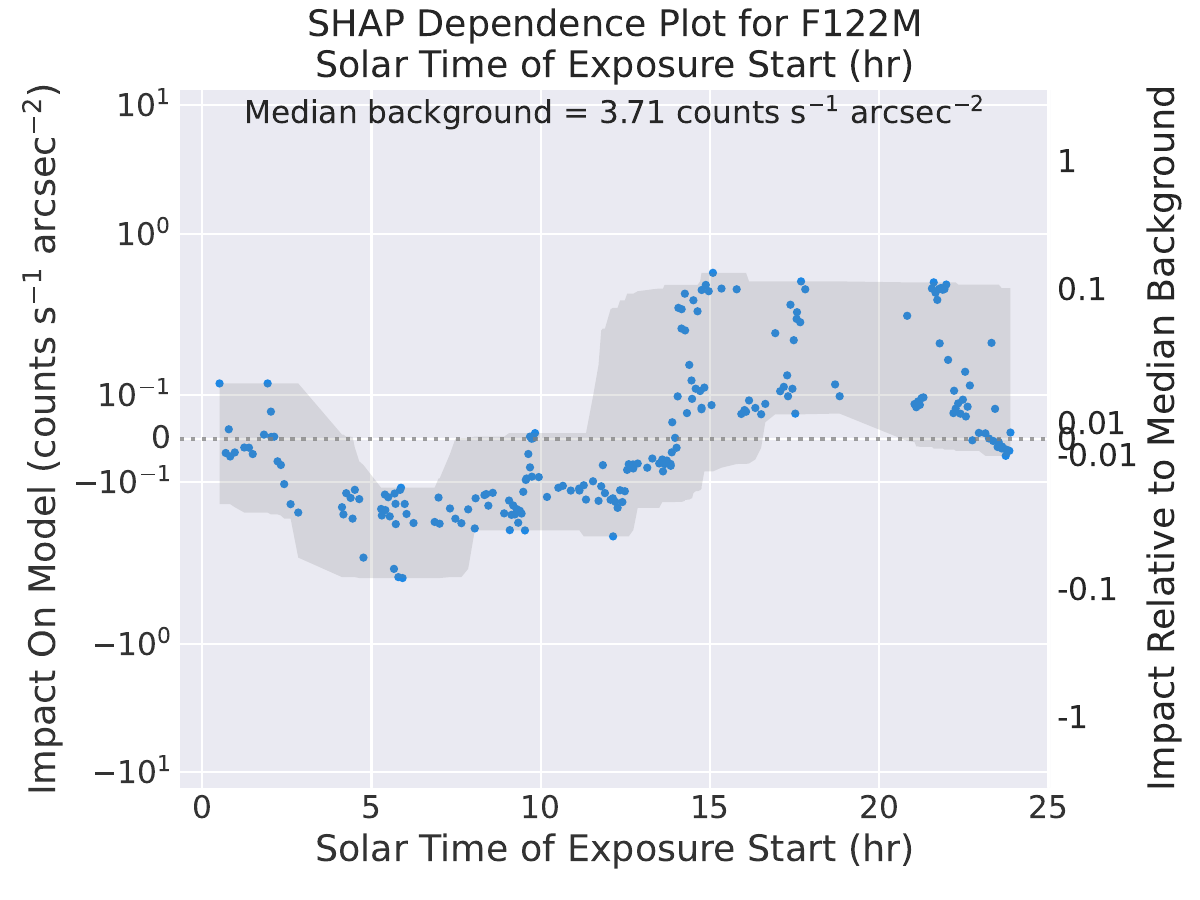}
\includegraphics[width=0.24\textwidth]{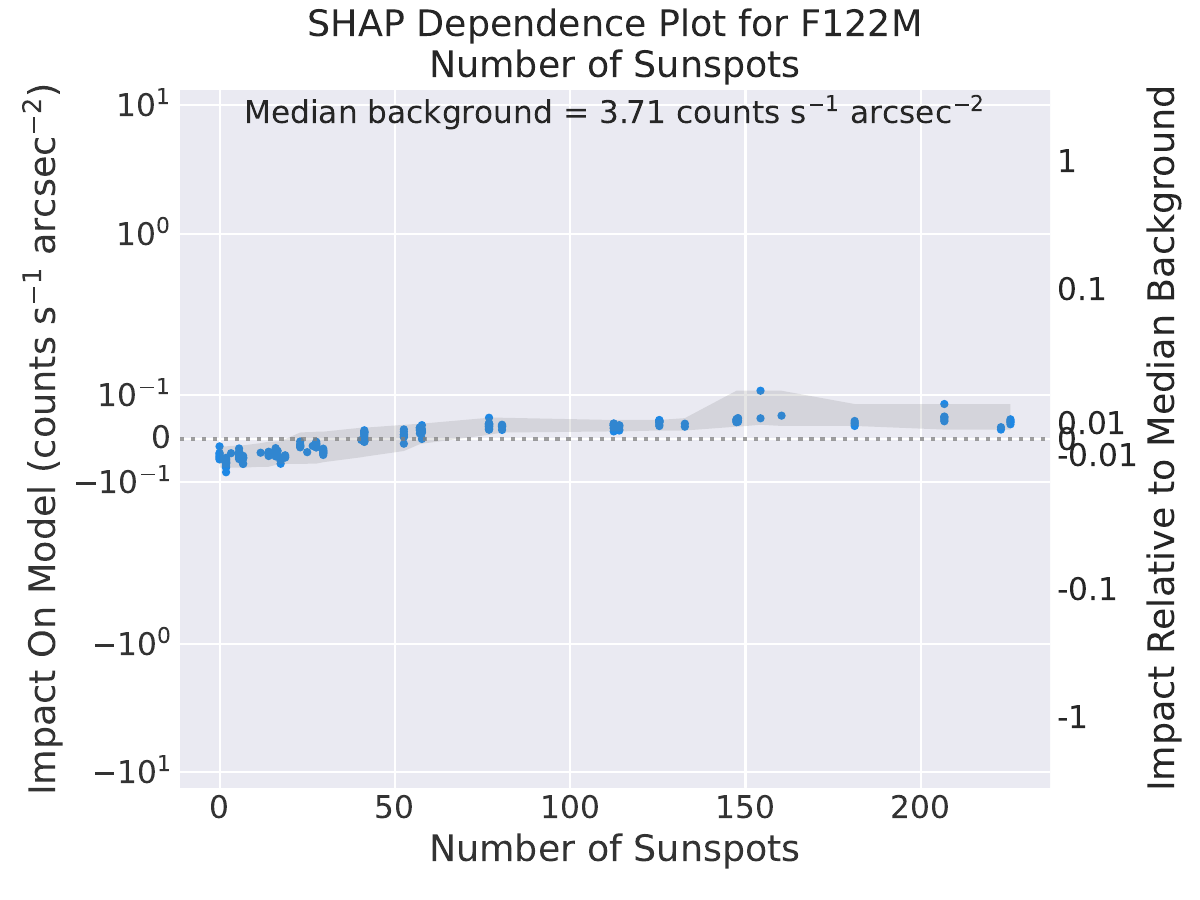}
\includegraphics[width=0.24\textwidth]{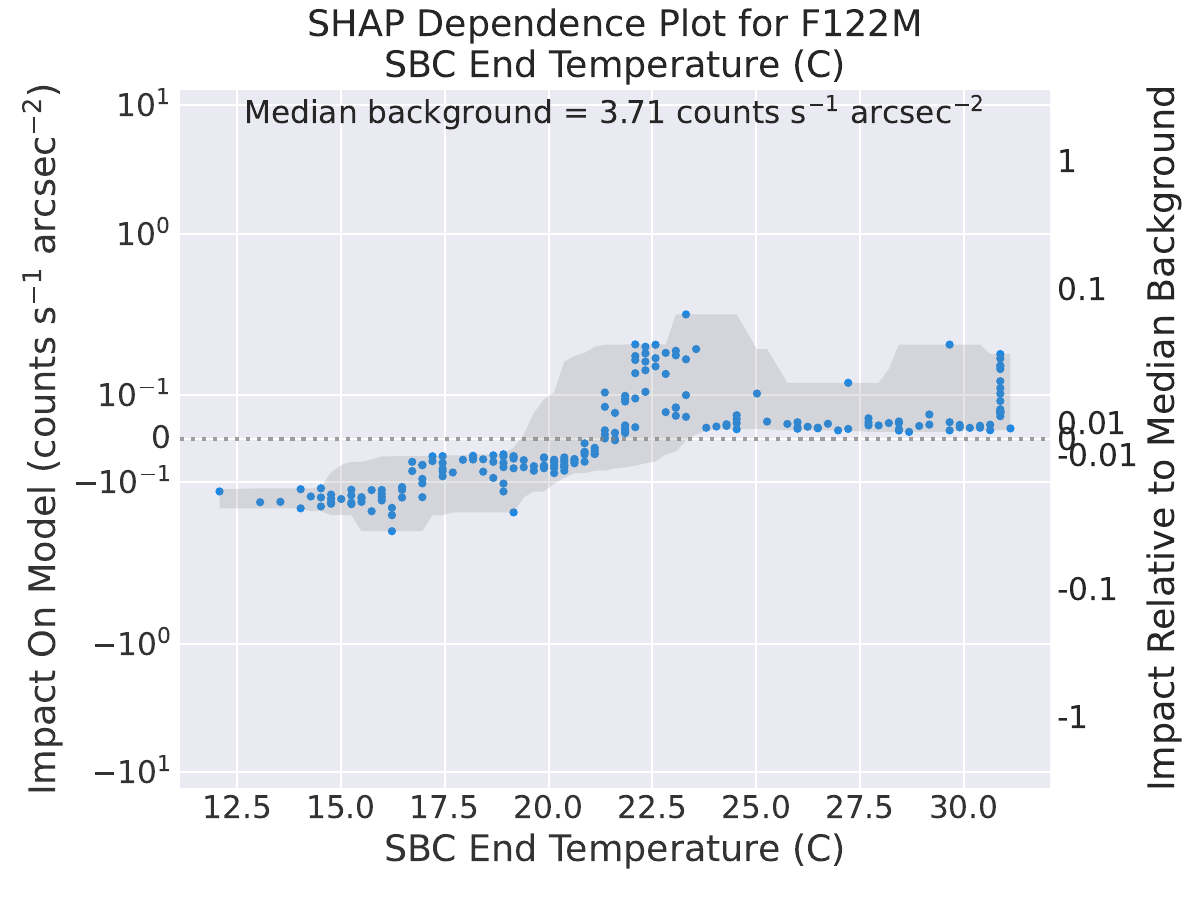}
\includegraphics[width=0.24\textwidth]{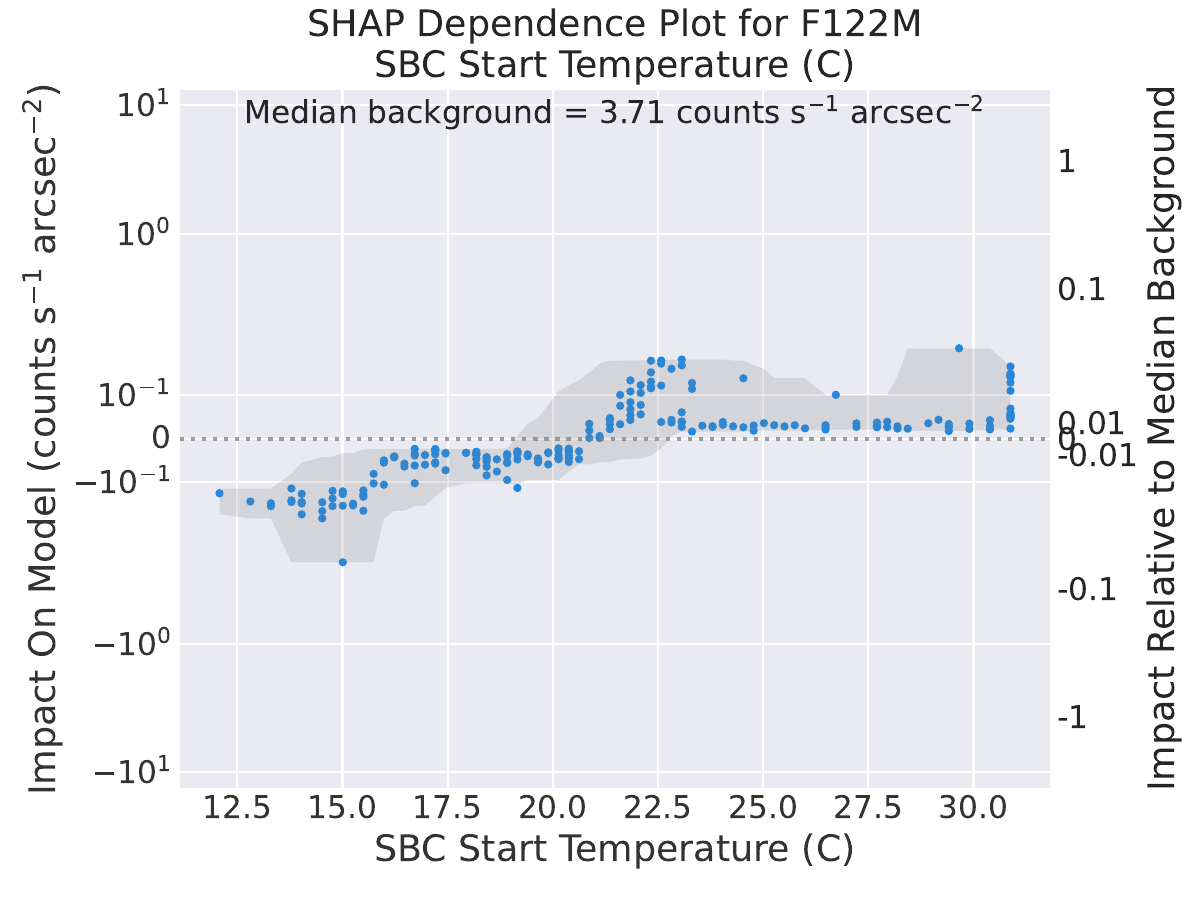}
\caption{SHAP dependence plots for QRF regression modeling of F122M. Otherwise as per Figure~\ref{Fig:SHAP_Dependence_F115LP}.}
\label{Fig:SHAP_Dependence_F122M}
\end{figure}

\begin{figure}
\centering
\includegraphics[width=0.24\textwidth]{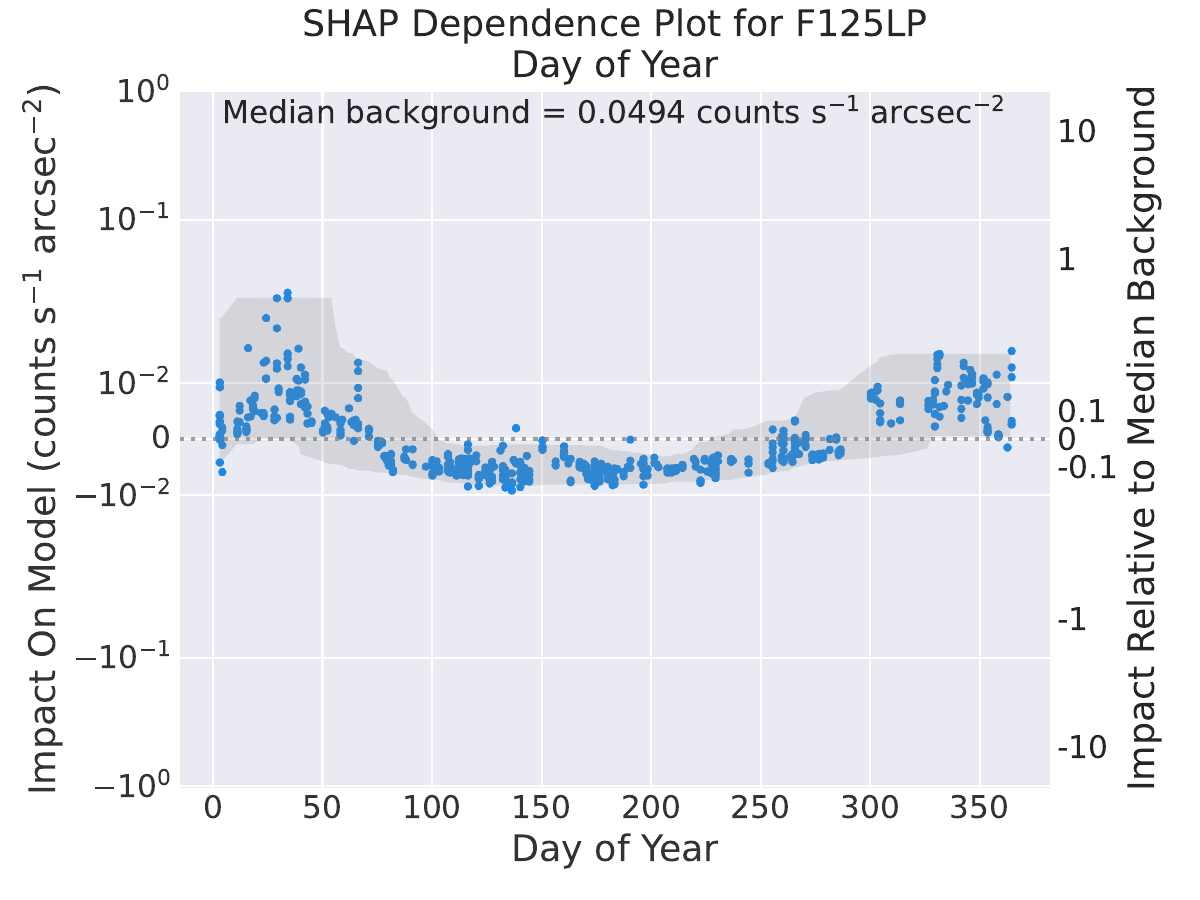}
\includegraphics[width=0.24\textwidth]{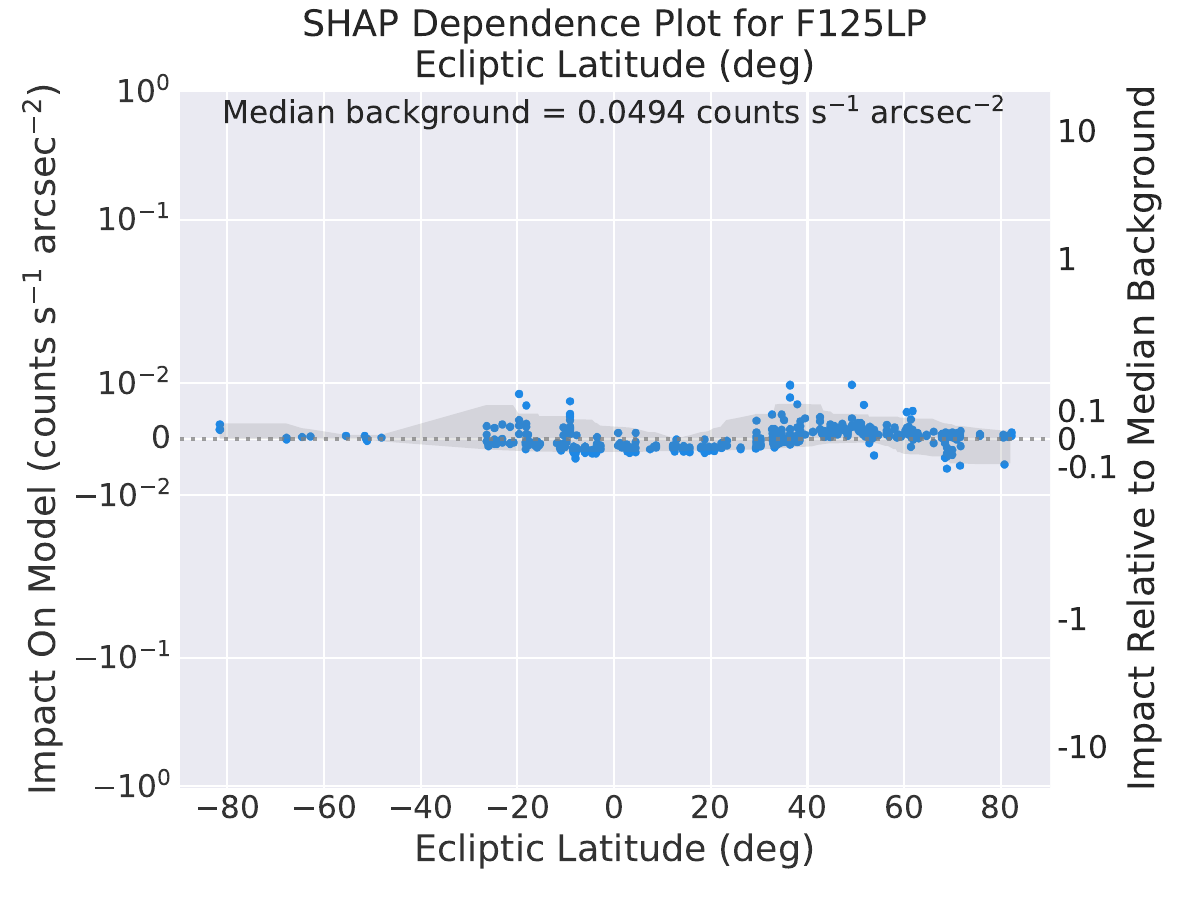}
\includegraphics[width=0.24\textwidth]{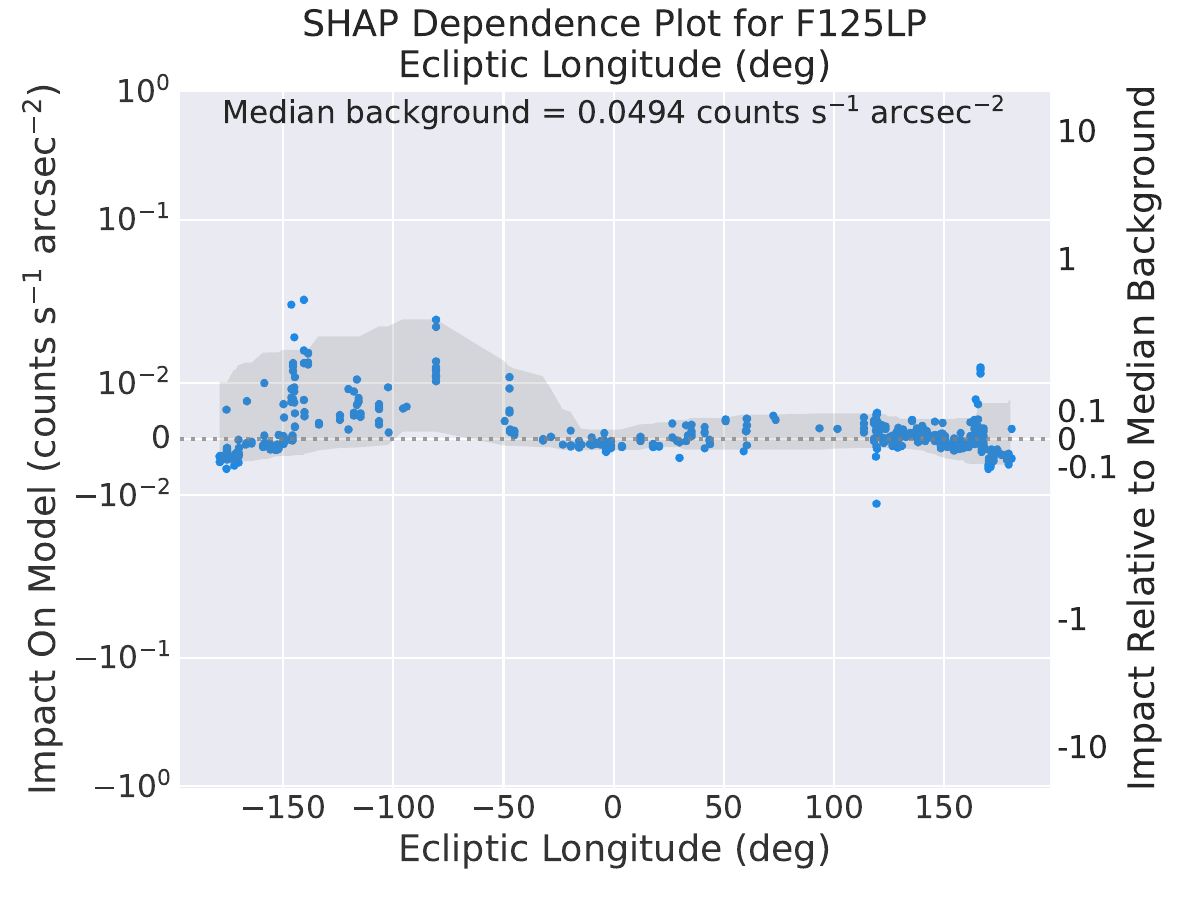}
\includegraphics[width=0.24\textwidth]{SHAP_Plots_QuantileForestRegr/F125LP_ela_avg_SHAP_Dependence.pdf}
\includegraphics[width=0.24\textwidth]{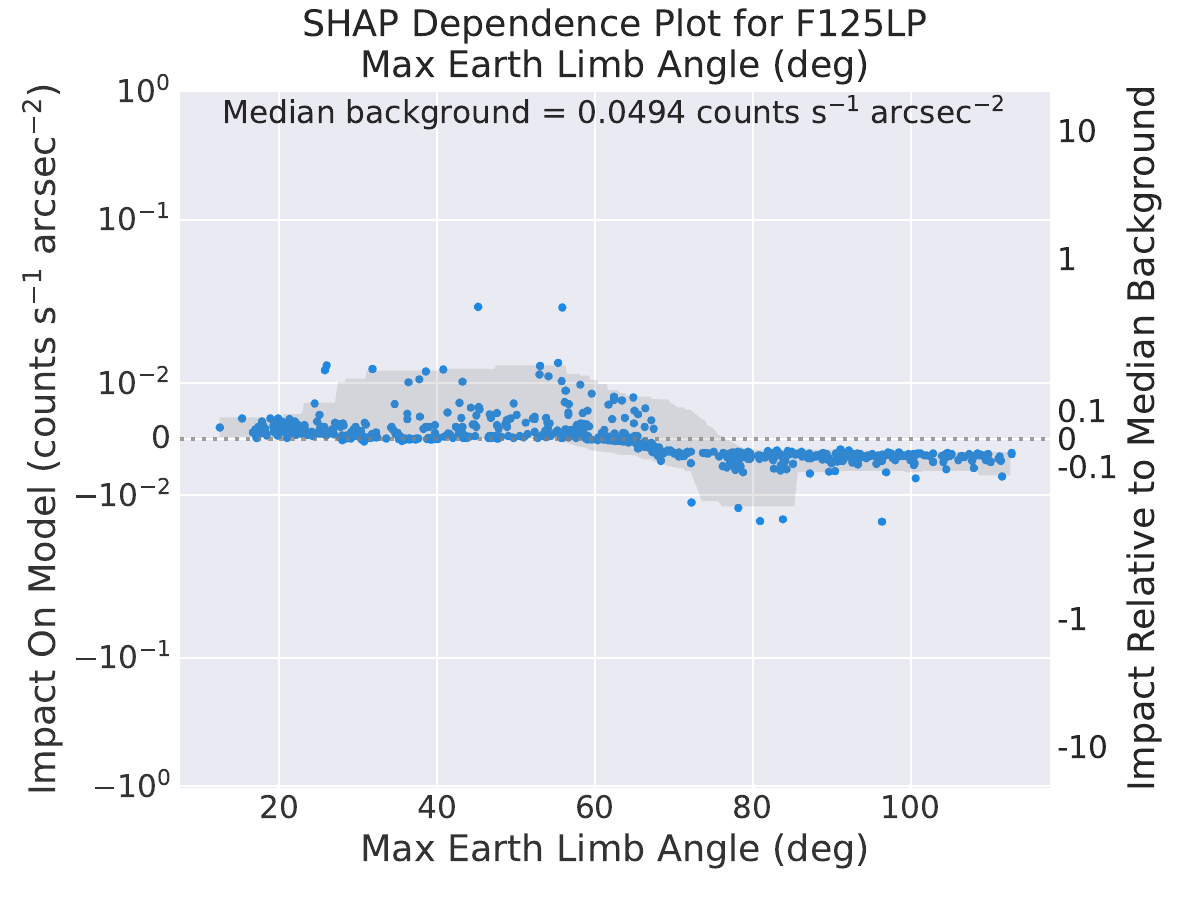}
\includegraphics[width=0.24\textwidth]{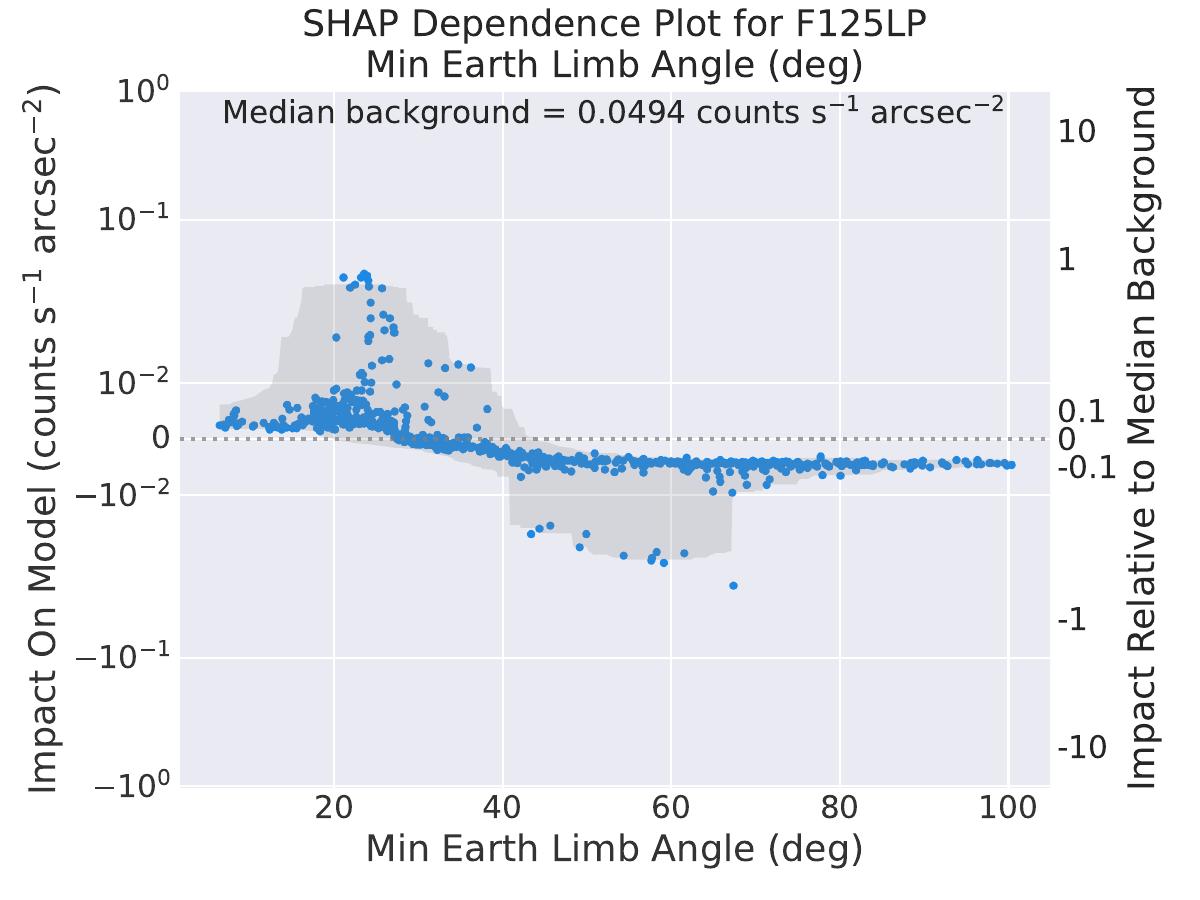}
\includegraphics[width=0.24\textwidth]{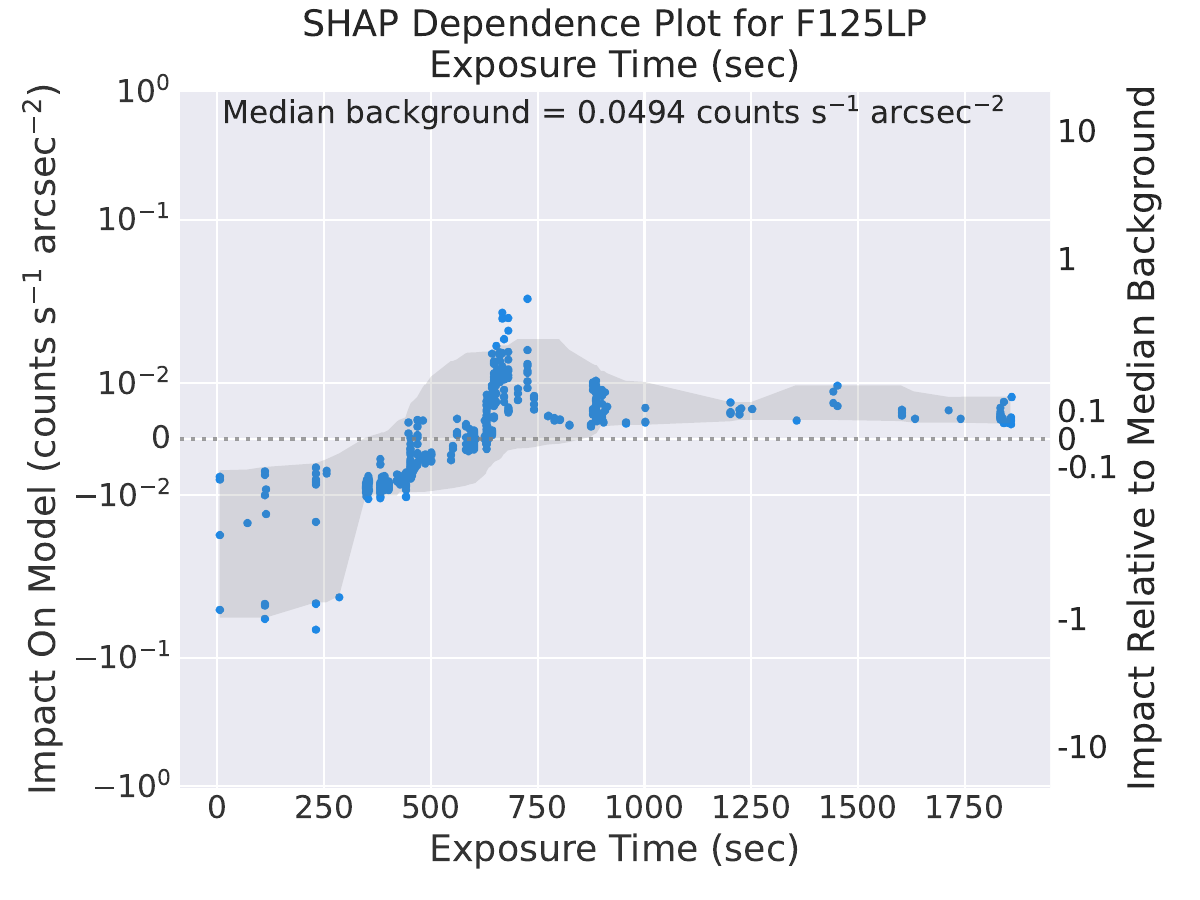}
\includegraphics[width=0.24\textwidth]{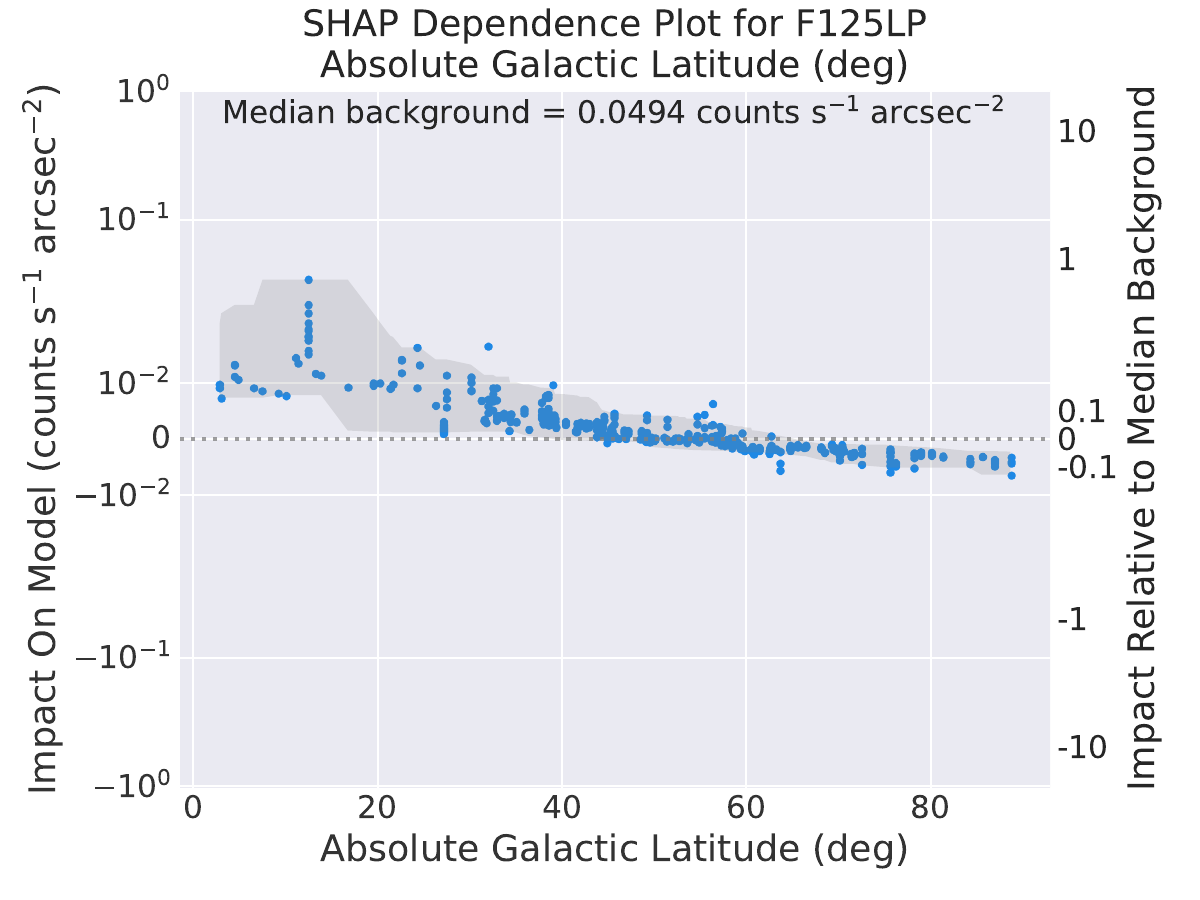}
\includegraphics[width=0.24\textwidth]{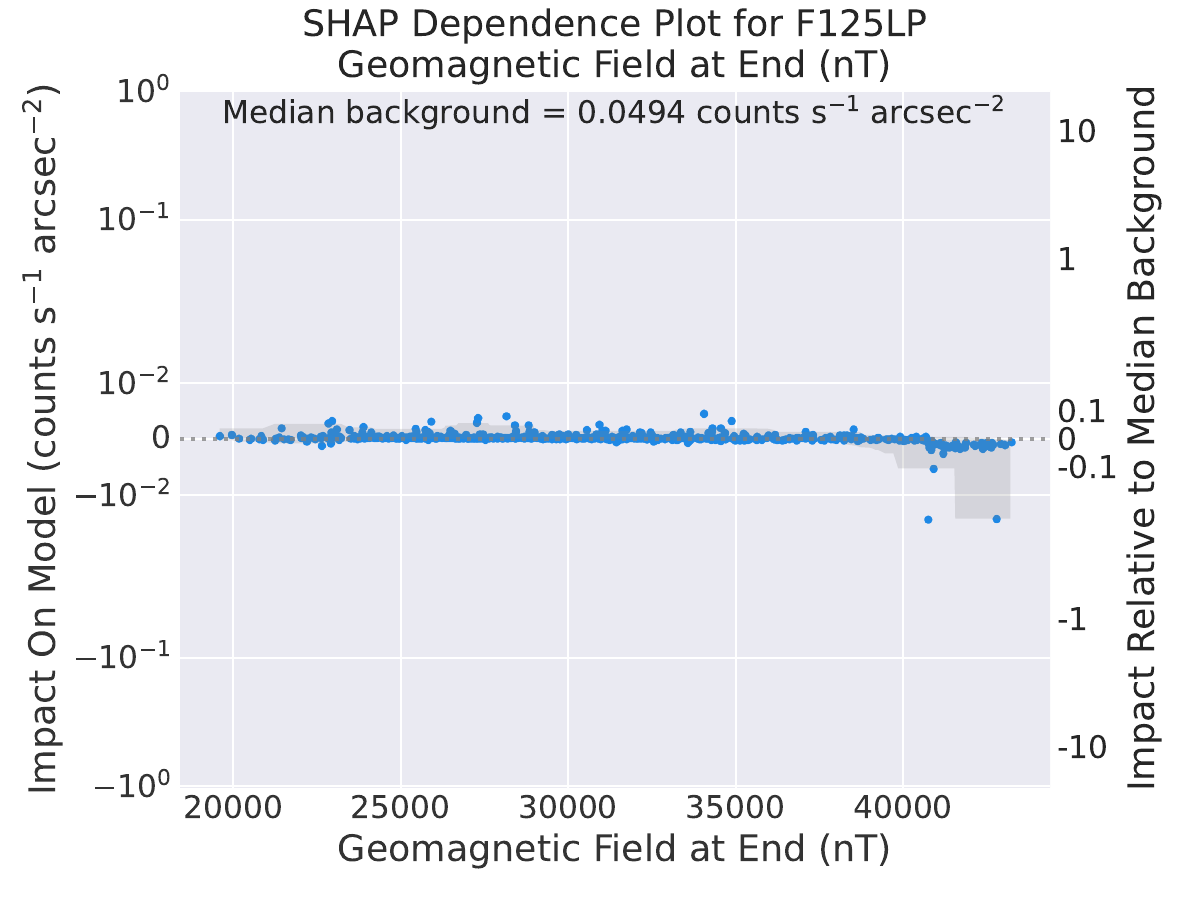}
\includegraphics[width=0.24\textwidth]{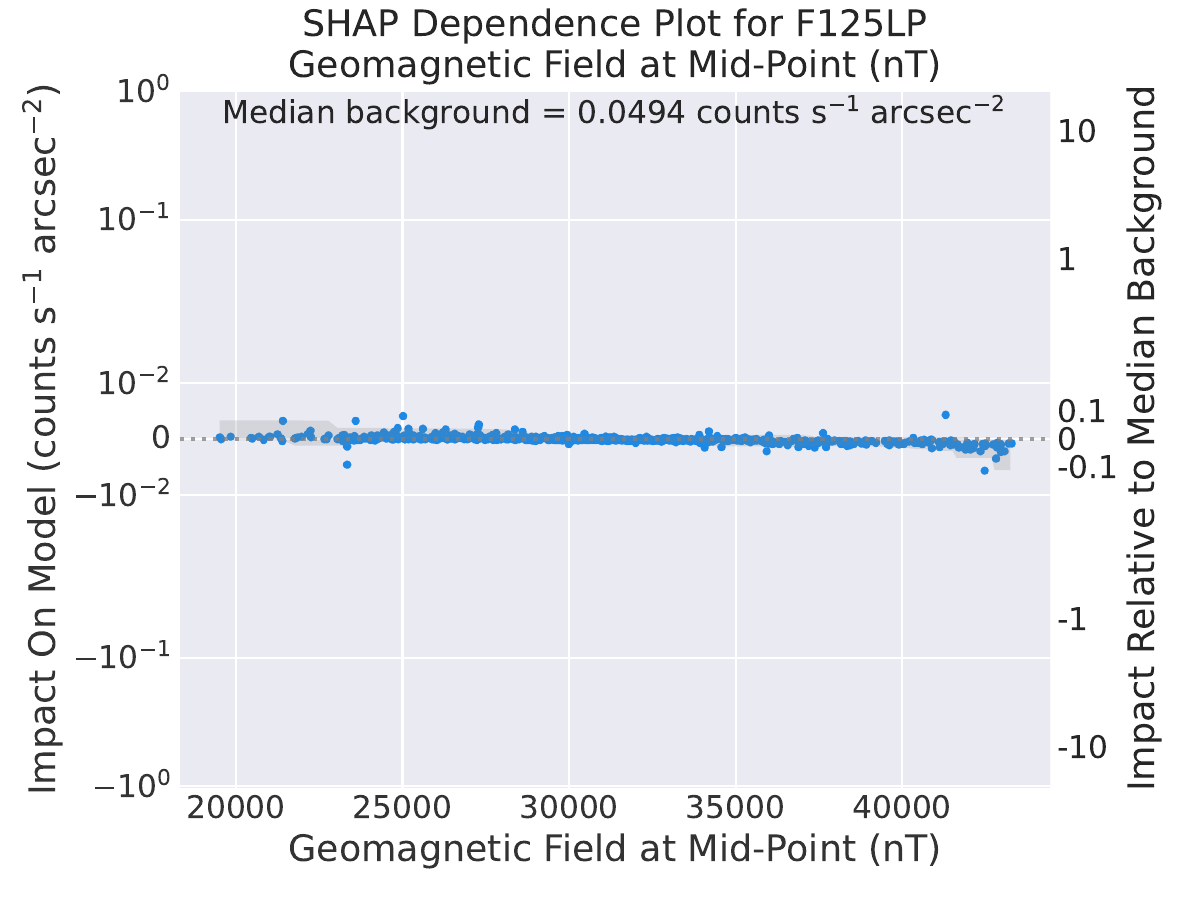}
\includegraphics[width=0.24\textwidth]{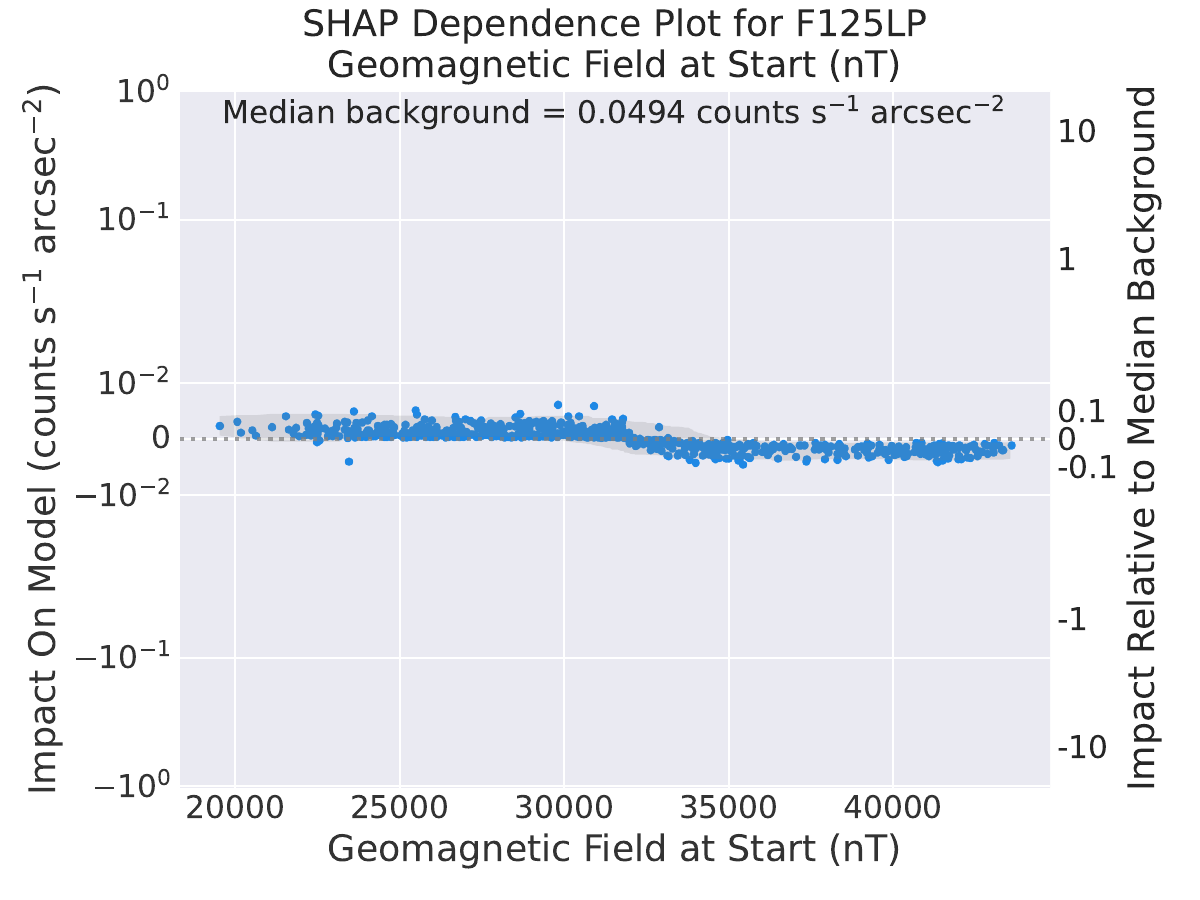}
\includegraphics[width=0.24\textwidth]{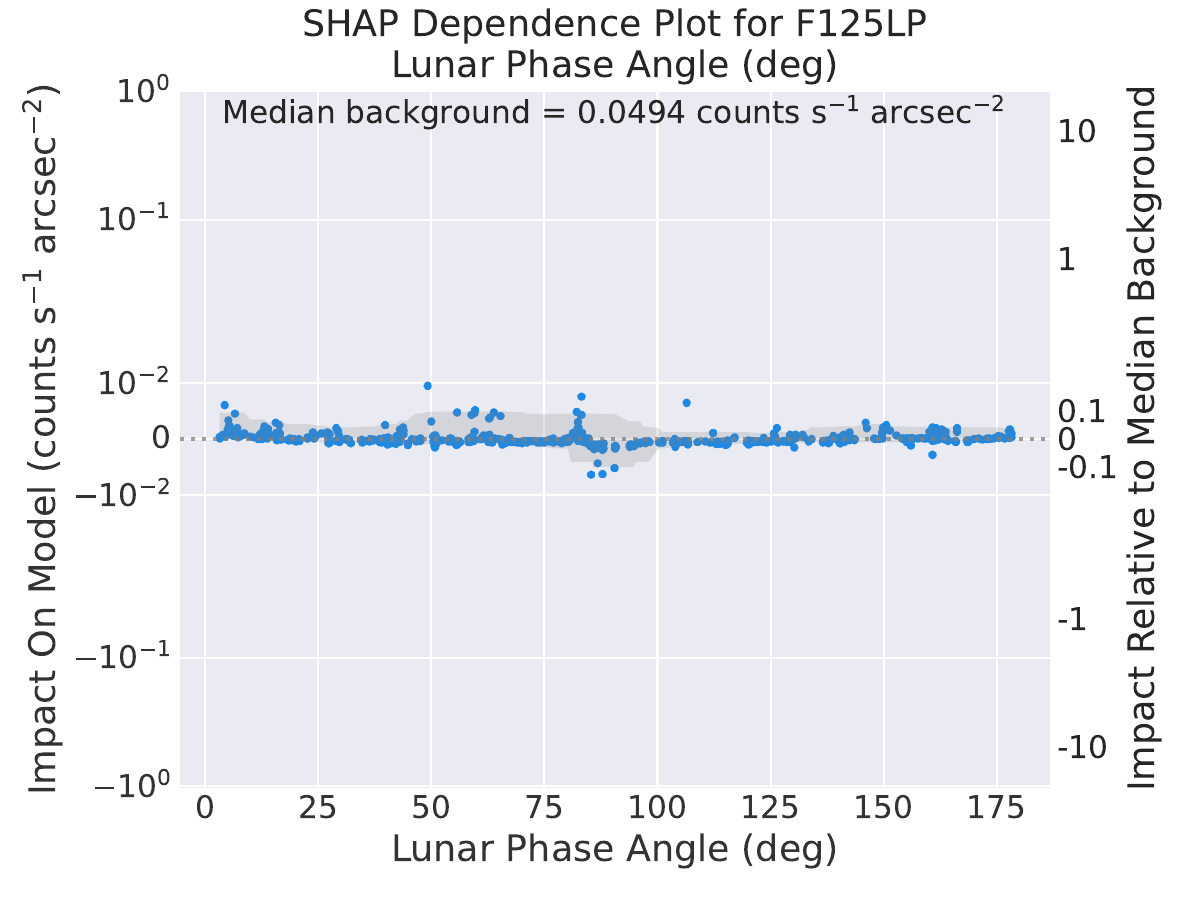}
\includegraphics[width=0.24\textwidth]{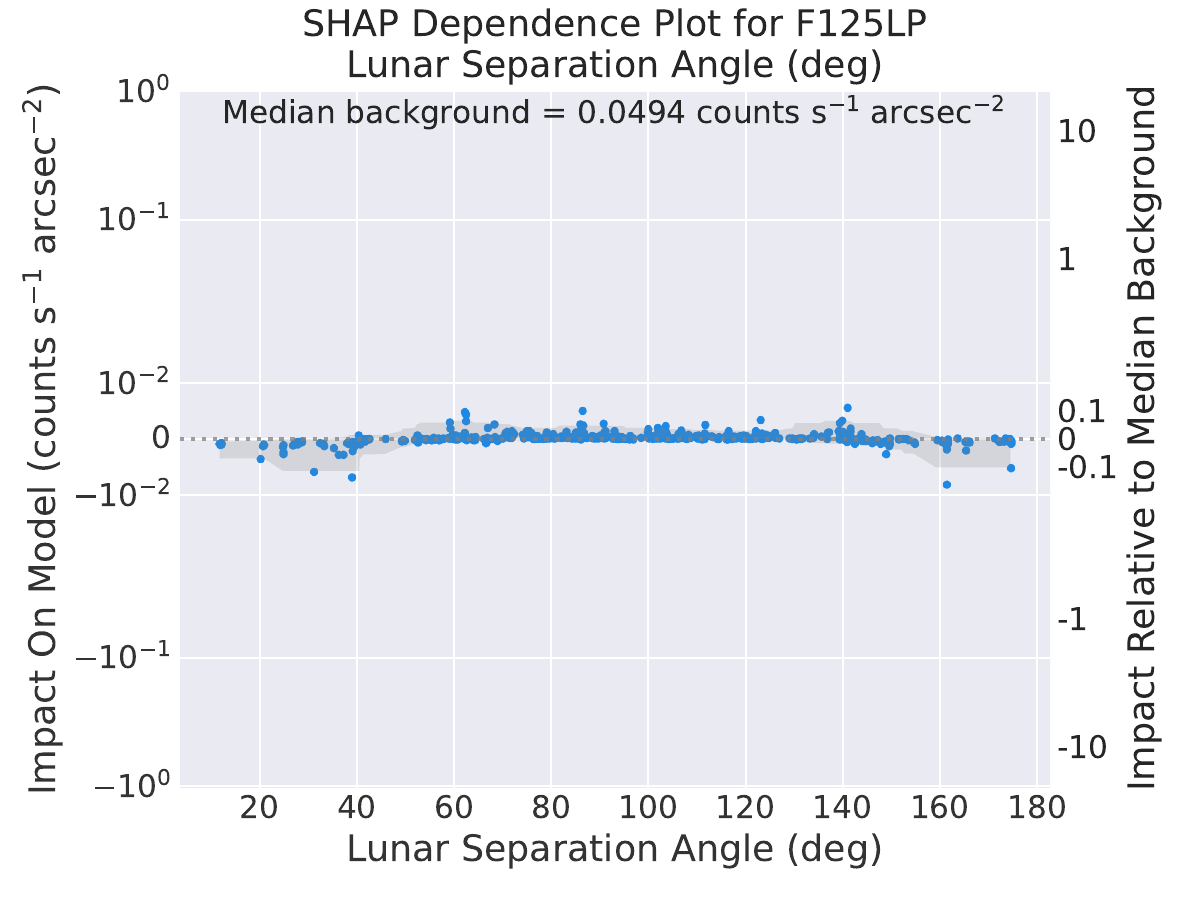}
\includegraphics[width=0.24\textwidth]{SHAP_Plots_QuantileForestRegr/F125LP_mjd_obs_mid_SHAP_Dependence.pdf}
\includegraphics[width=0.24\textwidth]{SHAP_Plots_QuantileForestRegr/F125LP_orbit_alt_SHAP_Dependence.pdf}
\includegraphics[width=0.24\textwidth]{SHAP_Plots_QuantileForestRegr/F125LP_solar_alt_SHAP_Dependence.pdf}
\includegraphics[width=0.24\textwidth]{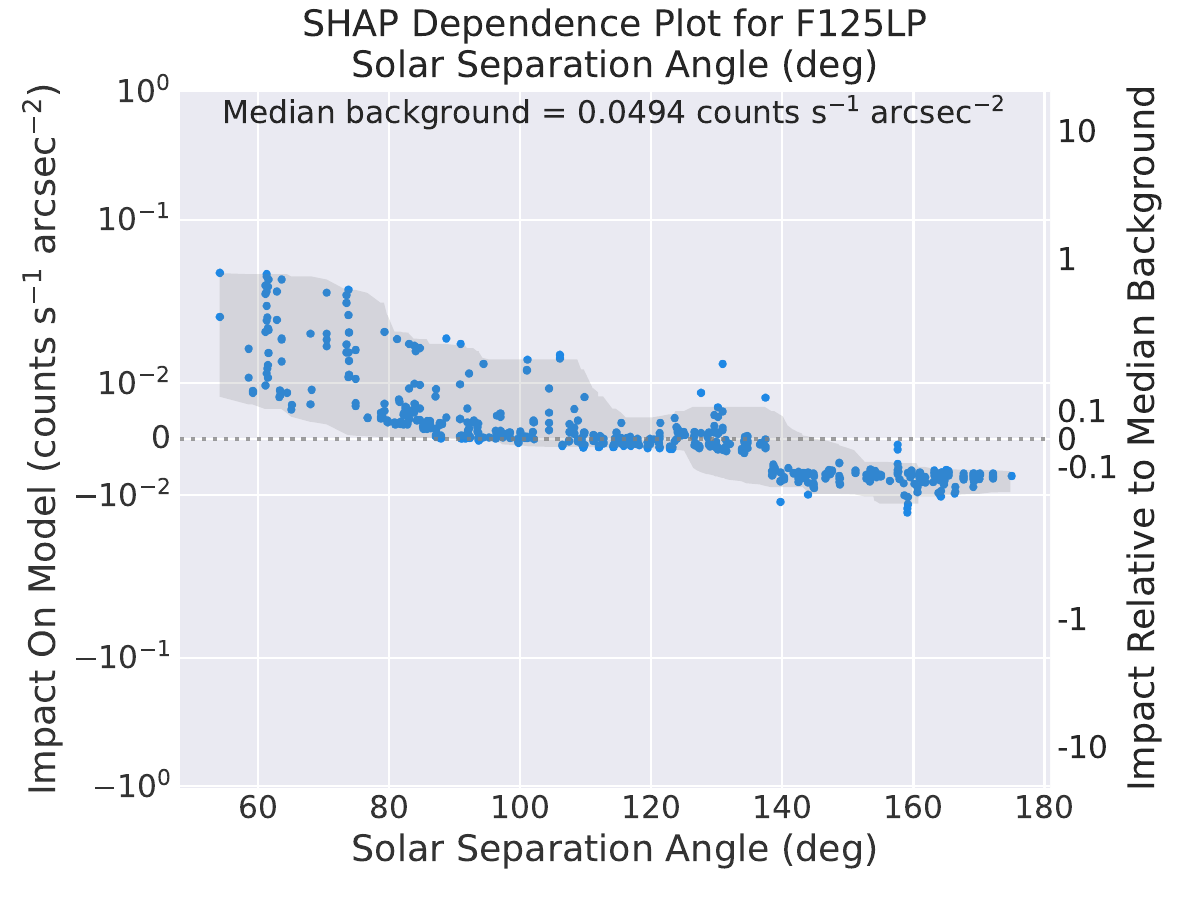}
\includegraphics[width=0.24\textwidth]{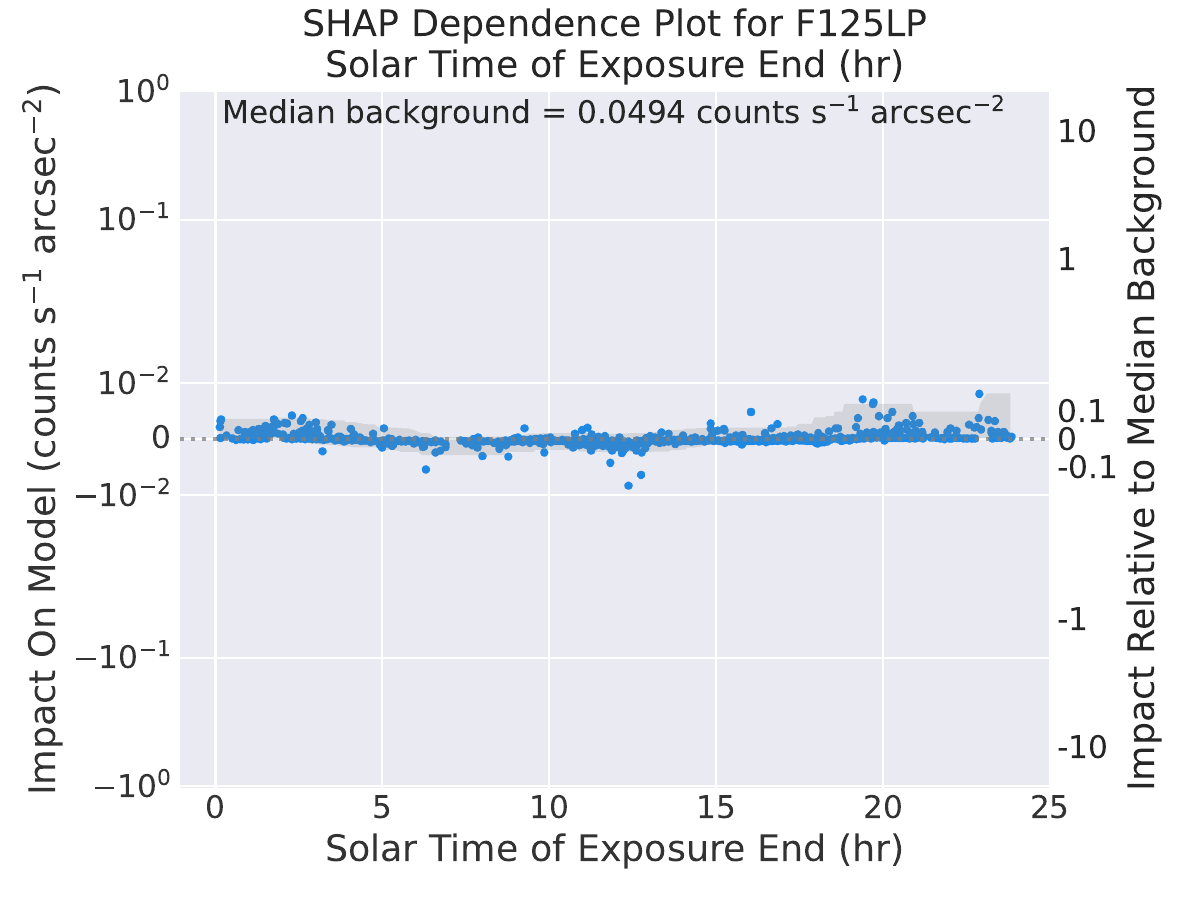}
\includegraphics[width=0.24\textwidth]{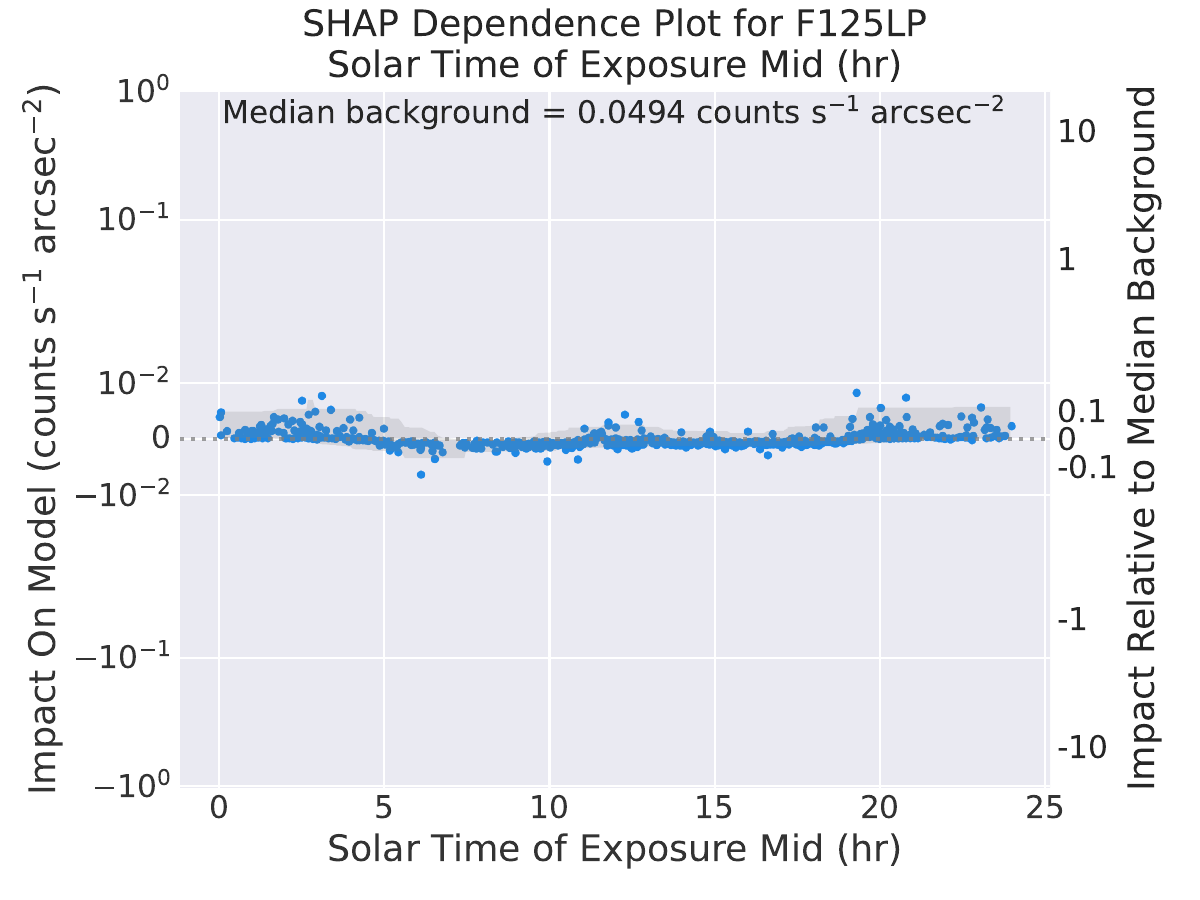}
\includegraphics[width=0.24\textwidth]{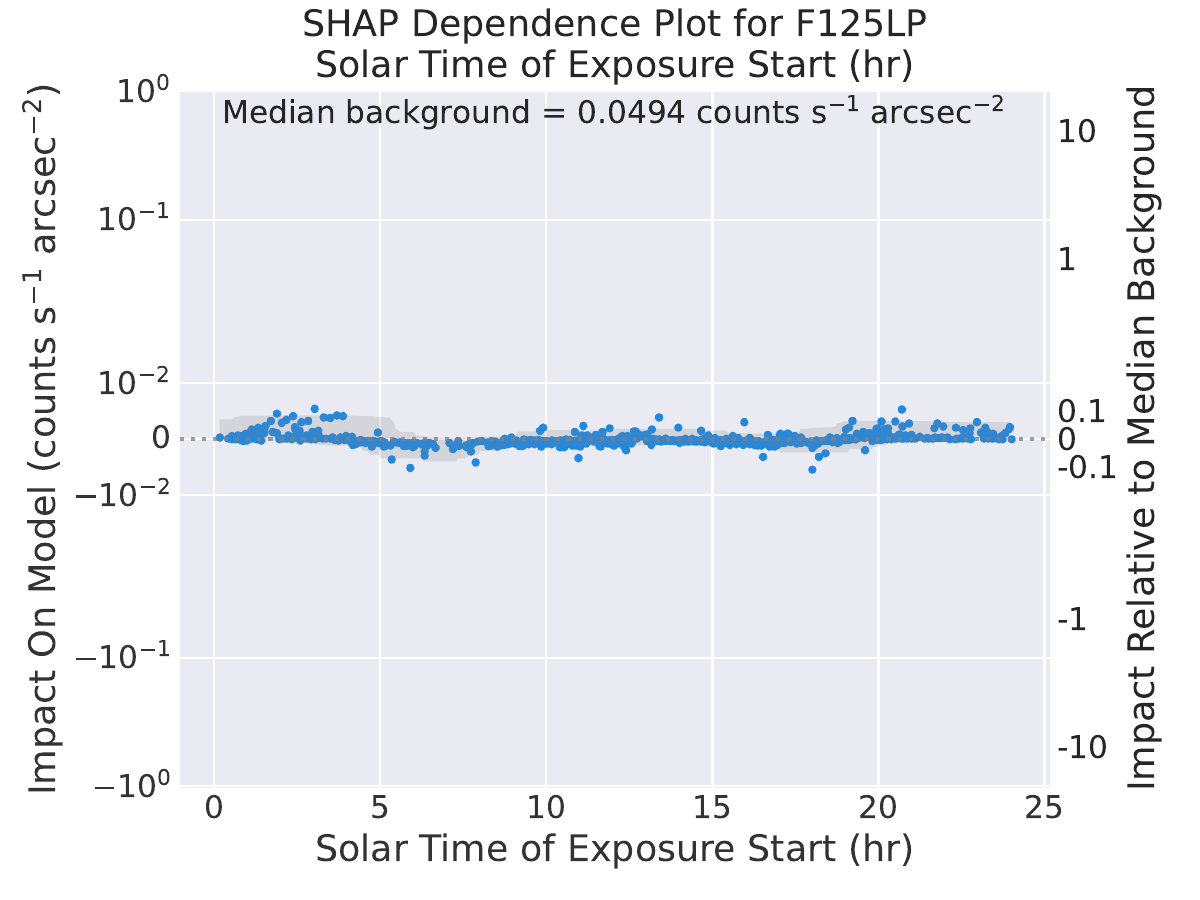}
\includegraphics[width=0.24\textwidth]{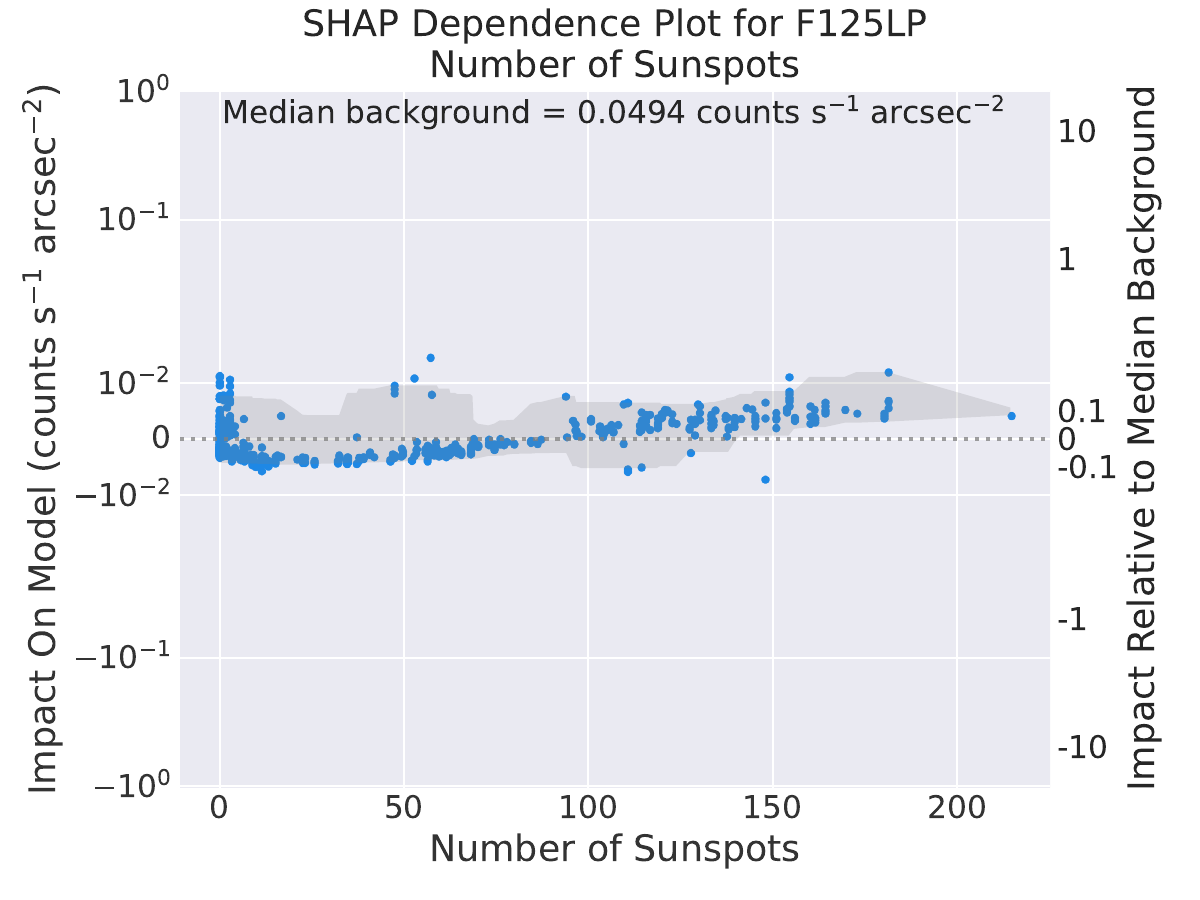}
\includegraphics[width=0.24\textwidth]{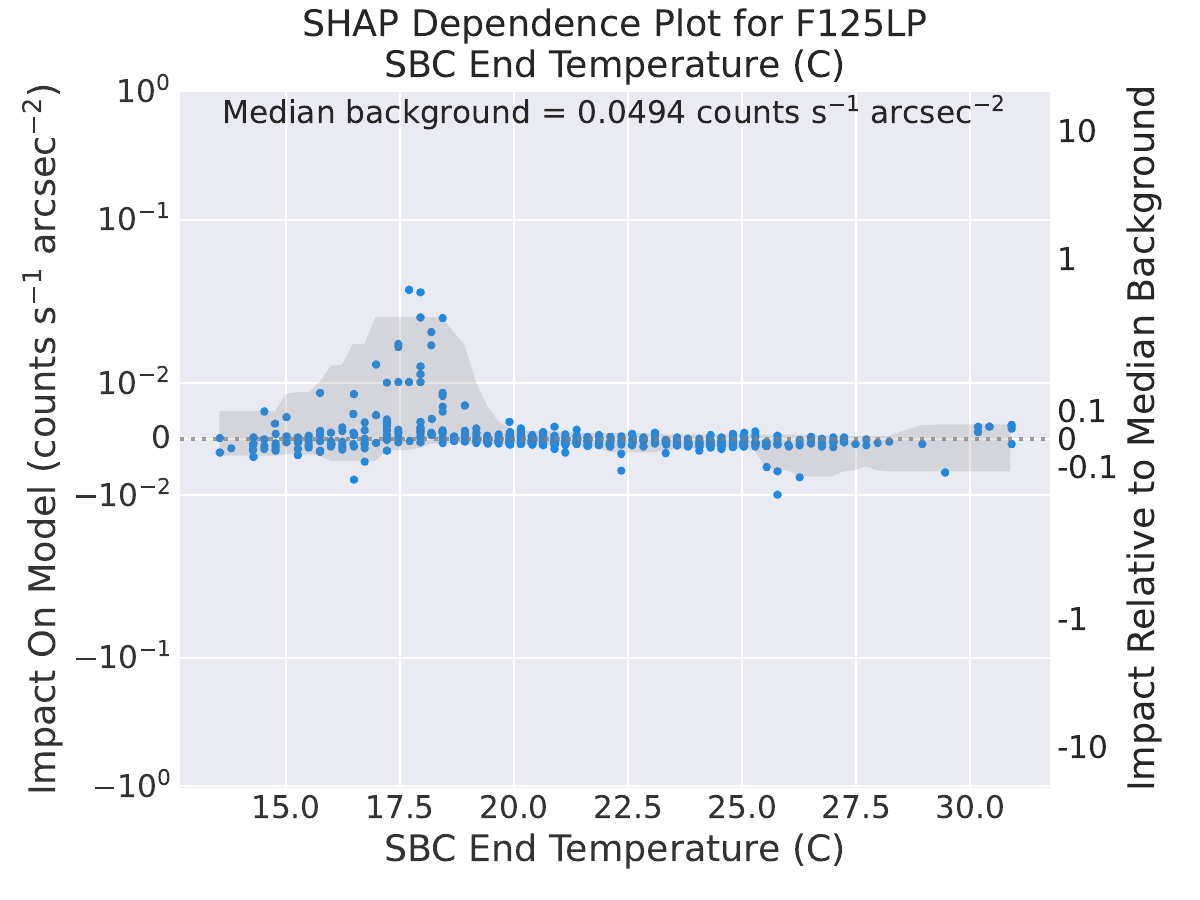}
\includegraphics[width=0.24\textwidth]{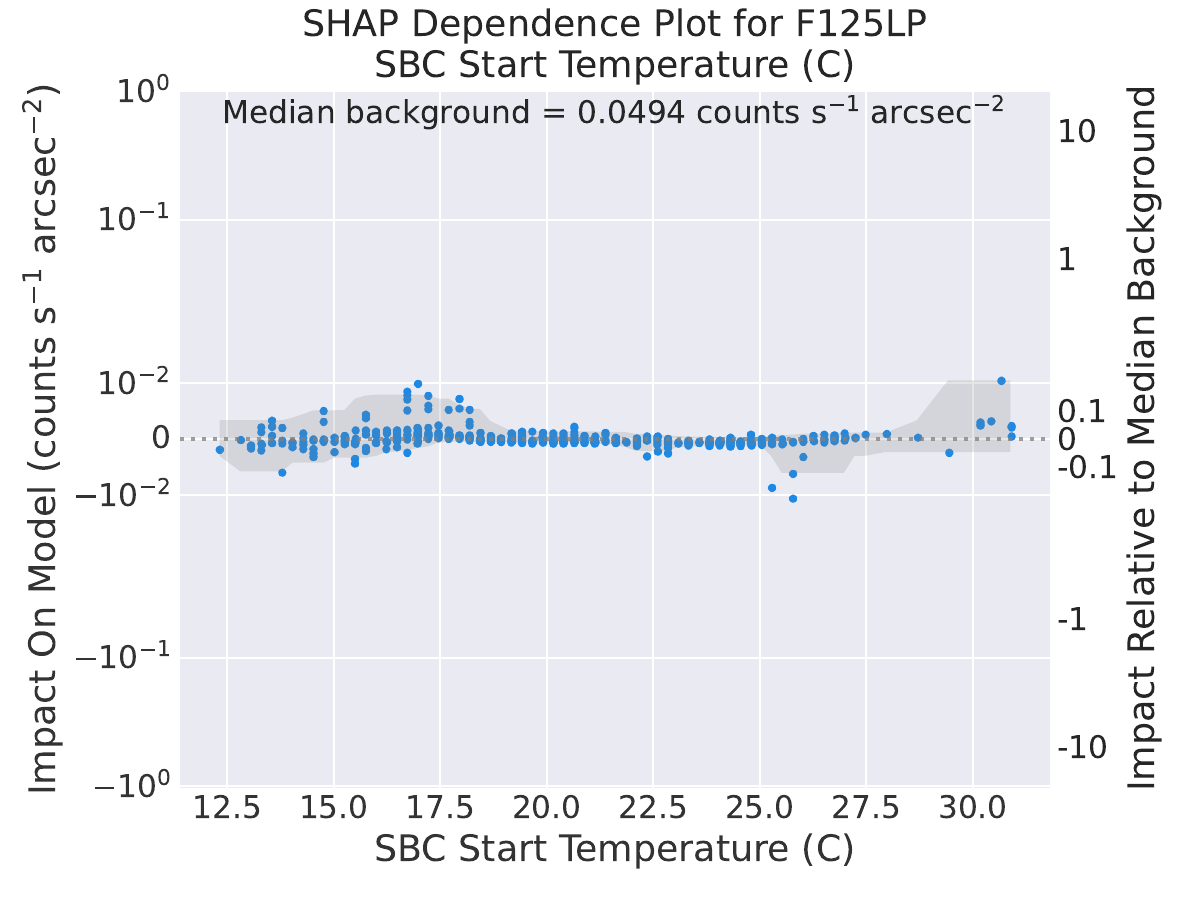}
\caption{SHAP dependence plots for QRF regression modeling of F125LP. Otherwise as per Figure~\ref{Fig:SHAP_Dependence_F115LP}.}
\label{Fig:SHAP_Dependence_F125LP}
\end{figure}

\begin{figure}
\centering
\includegraphics[width=0.24\textwidth]{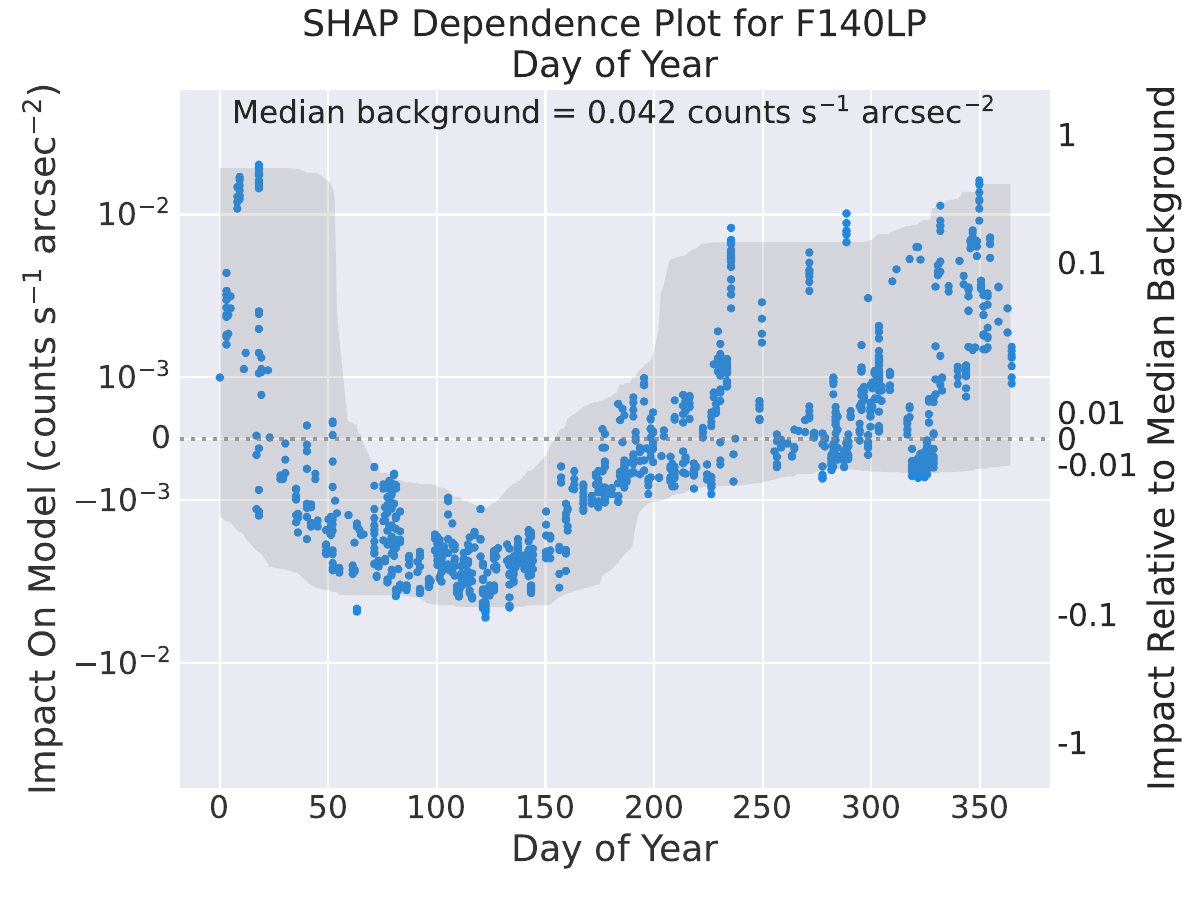}
\includegraphics[width=0.24\textwidth]{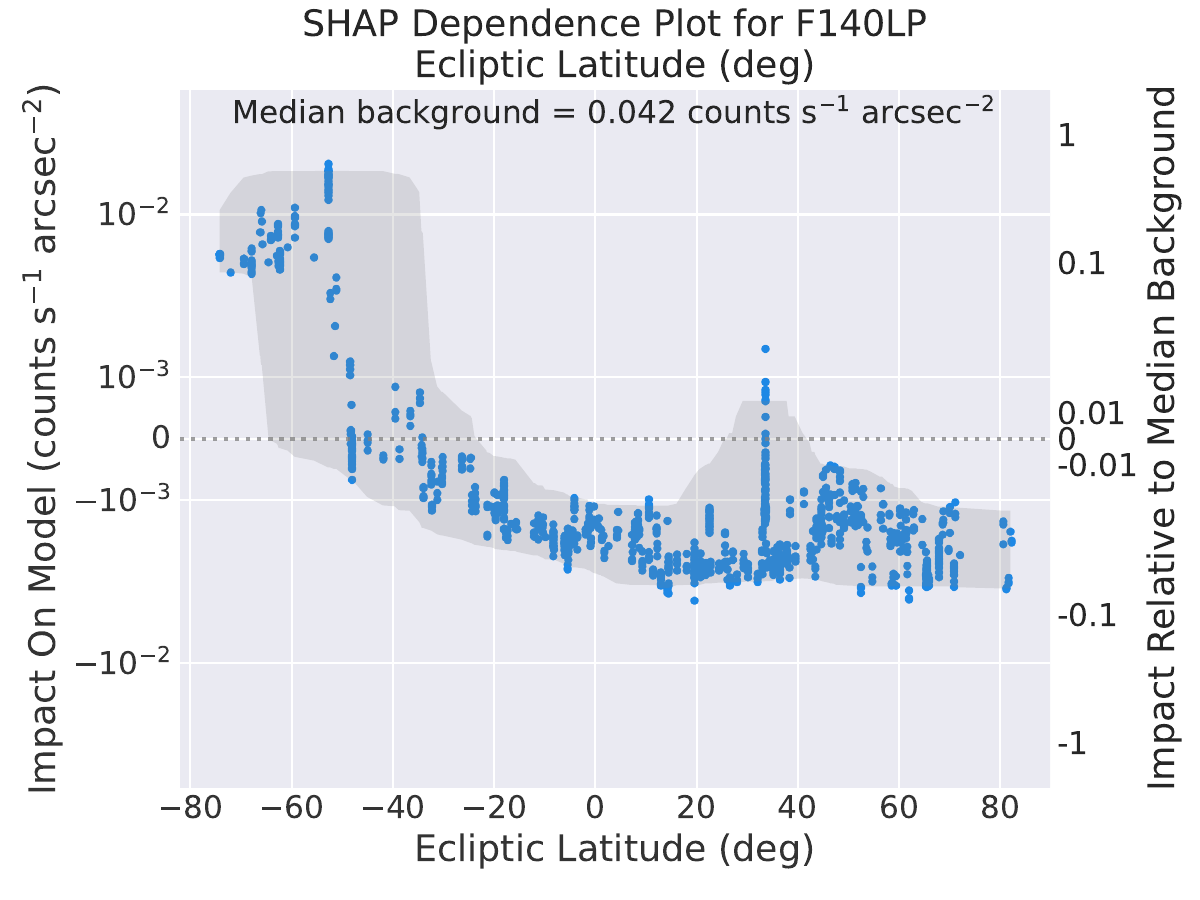}
\includegraphics[width=0.24\textwidth]{SHAP_Plots_QuantileForestRegr/F140LP_ecliptic_longitude_SHAP_Dependence.pdf}
\includegraphics[width=0.24\textwidth]{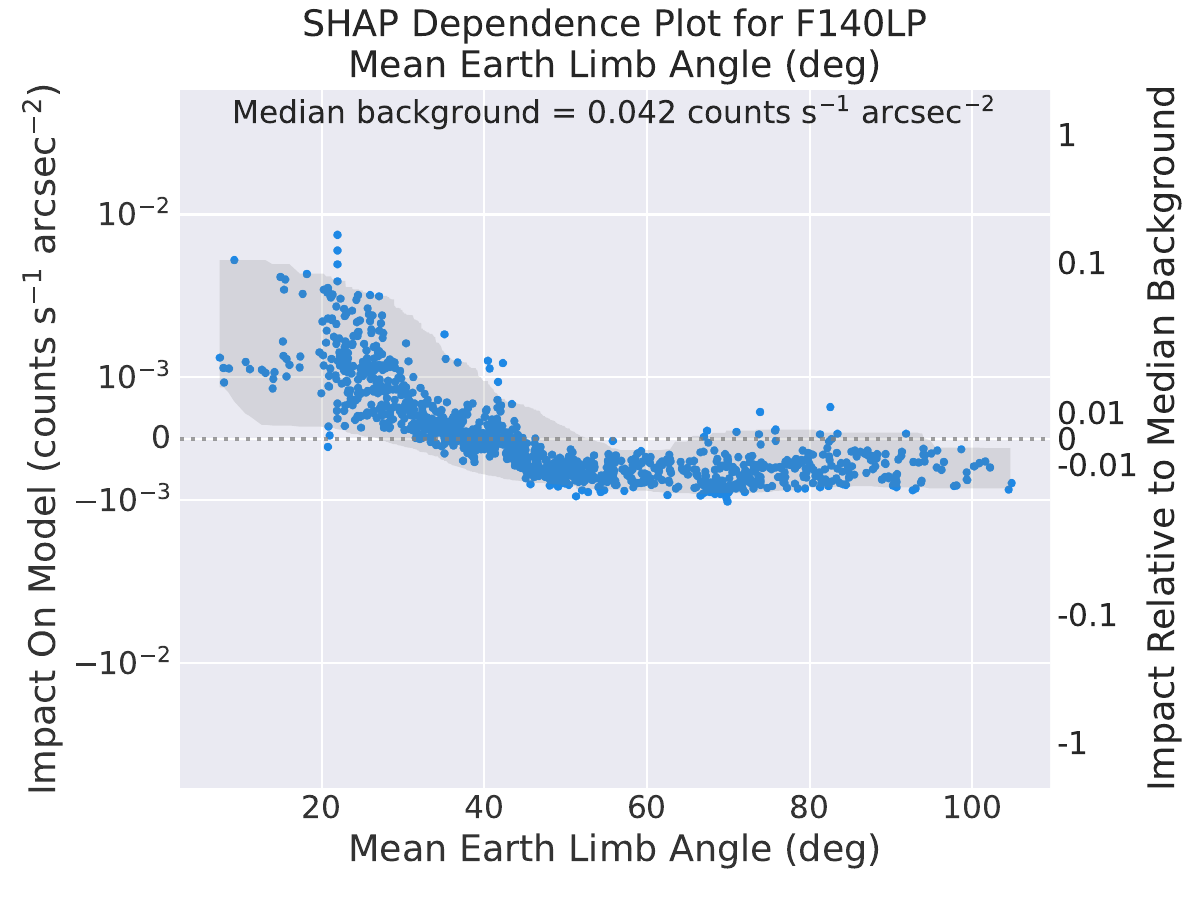}
\includegraphics[width=0.24\textwidth]{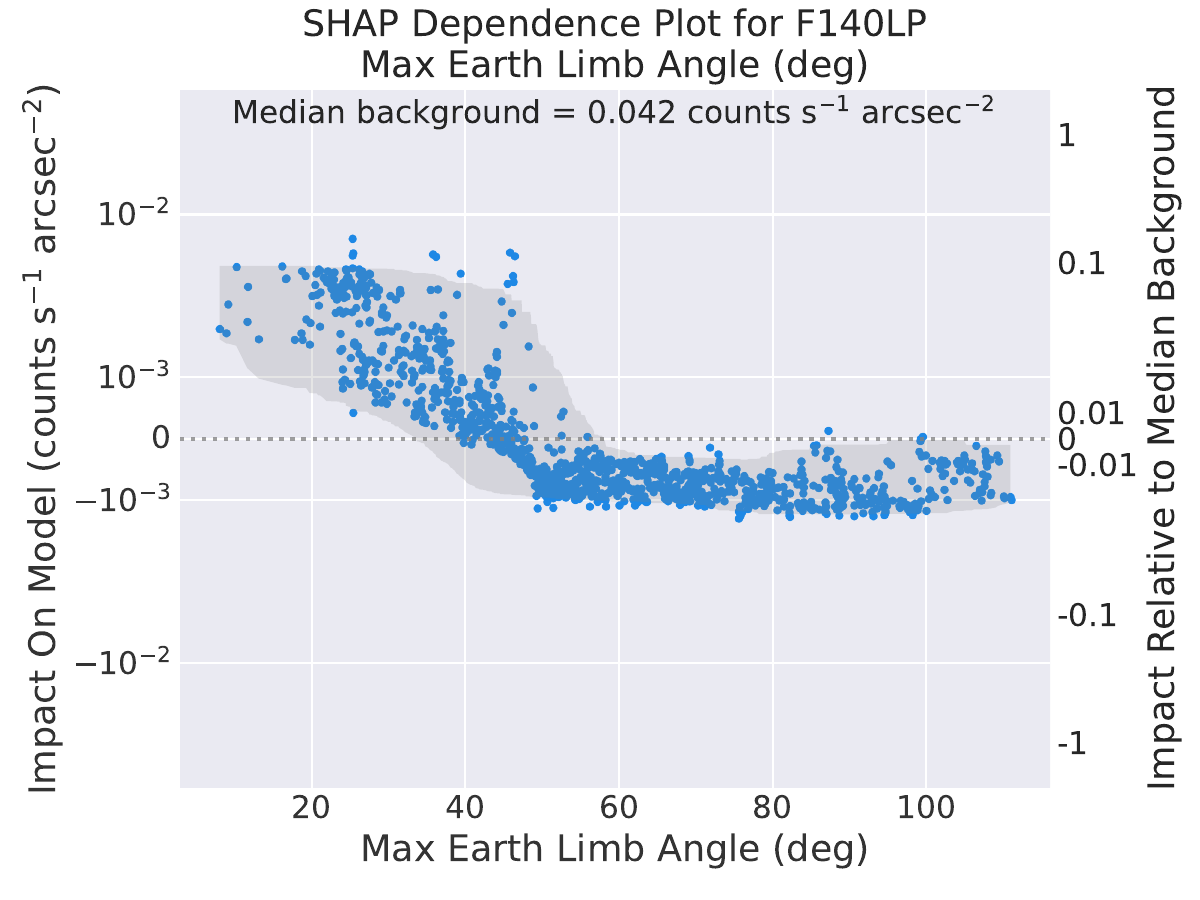}
\includegraphics[width=0.24\textwidth]{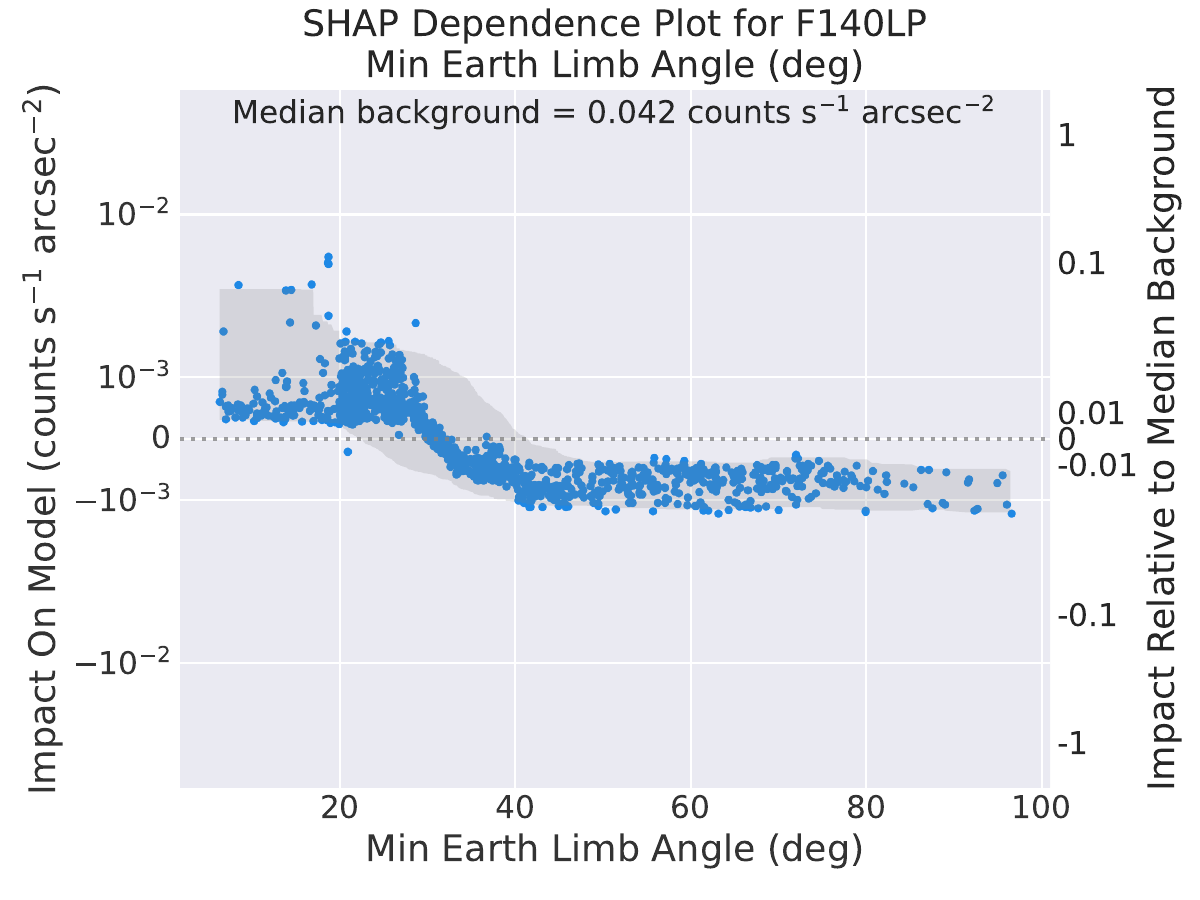}
\includegraphics[width=0.24\textwidth]{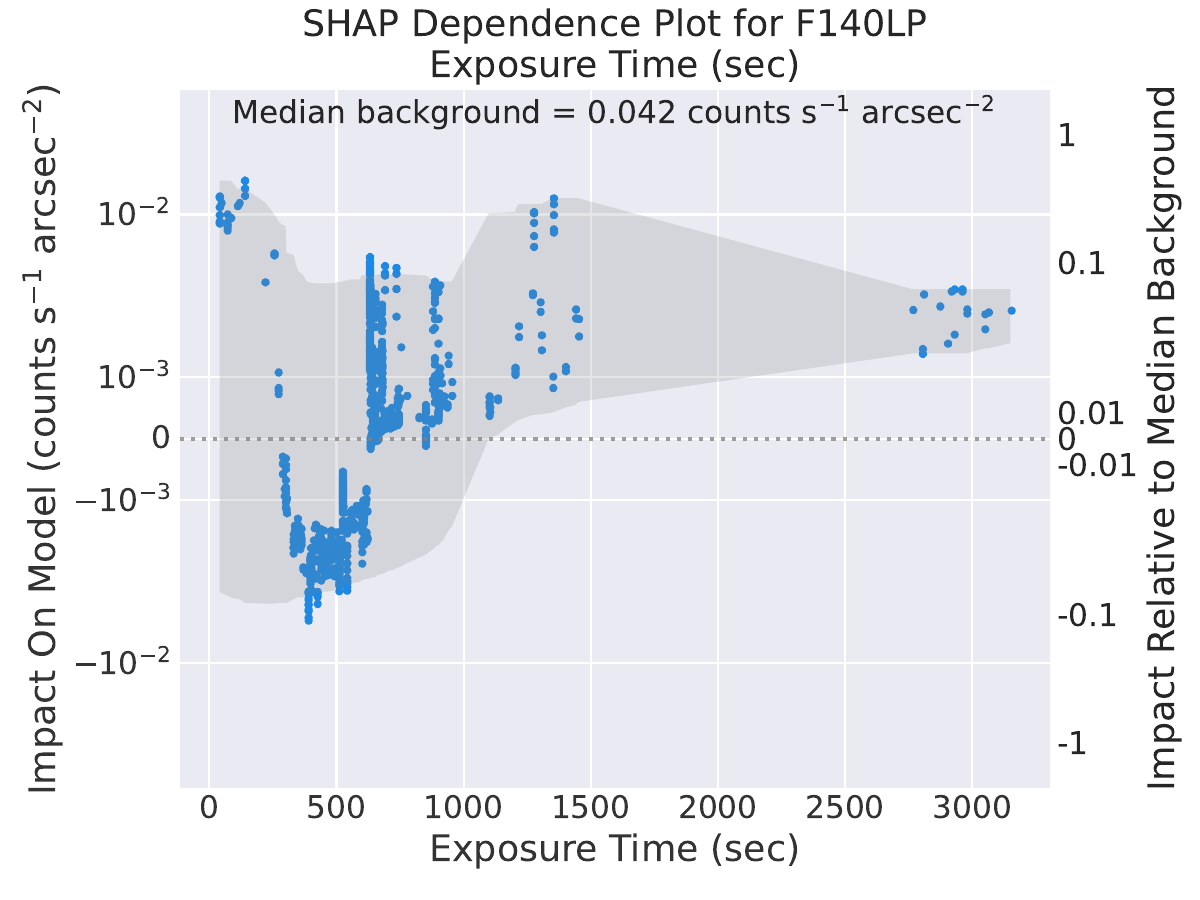}
\includegraphics[width=0.24\textwidth]{SHAP_Plots_QuantileForestRegr/F140LP_gal_lat_abs_SHAP_Dependence.pdf}
\includegraphics[width=0.24\textwidth]{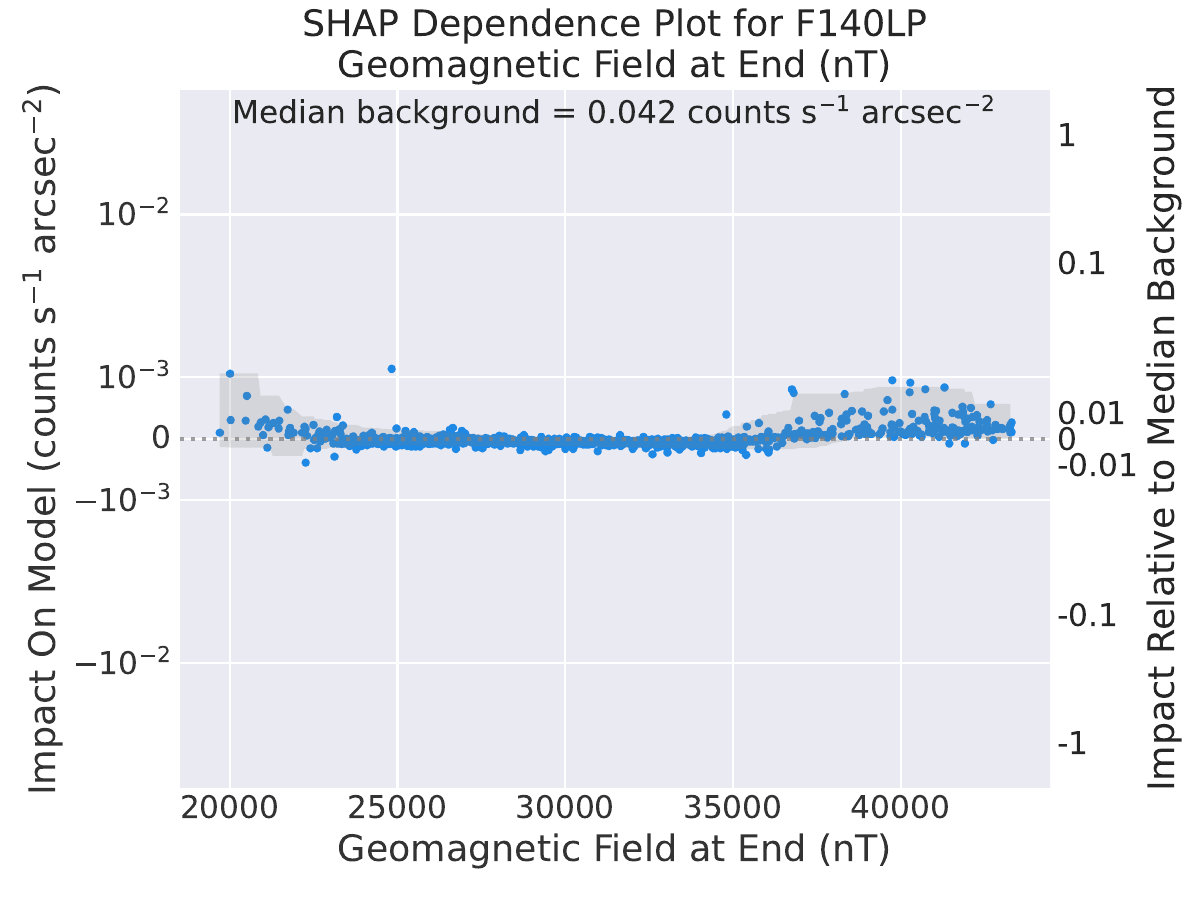}
\includegraphics[width=0.24\textwidth]{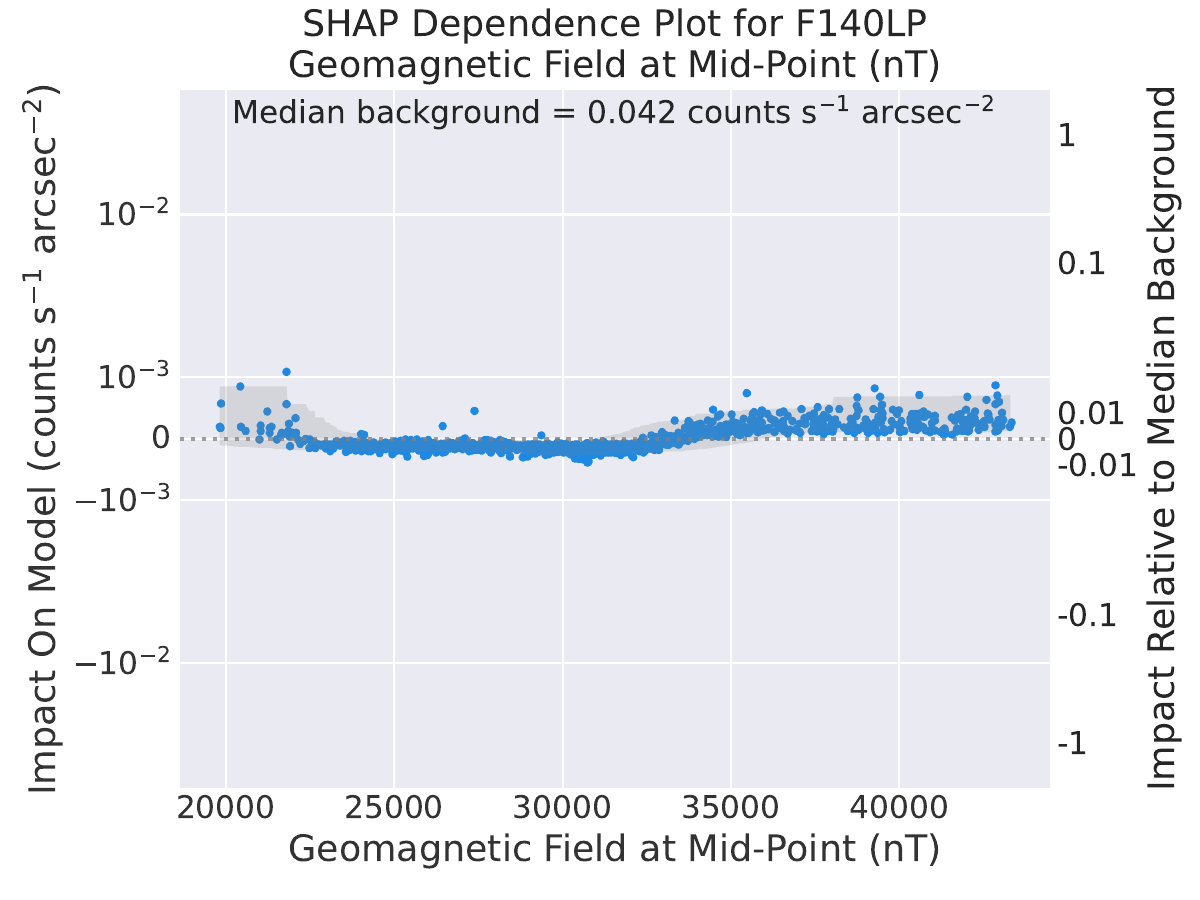}
\includegraphics[width=0.24\textwidth]{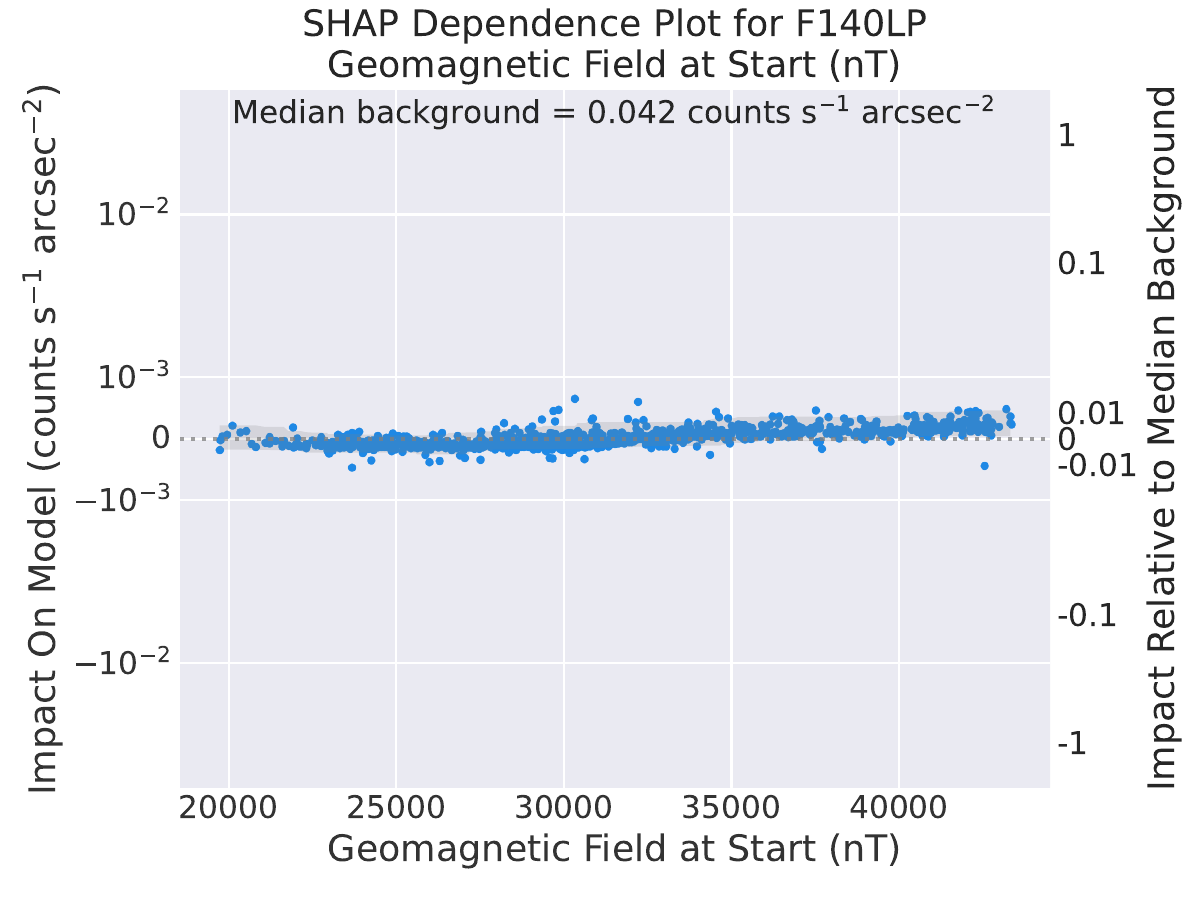}
\includegraphics[width=0.24\textwidth]{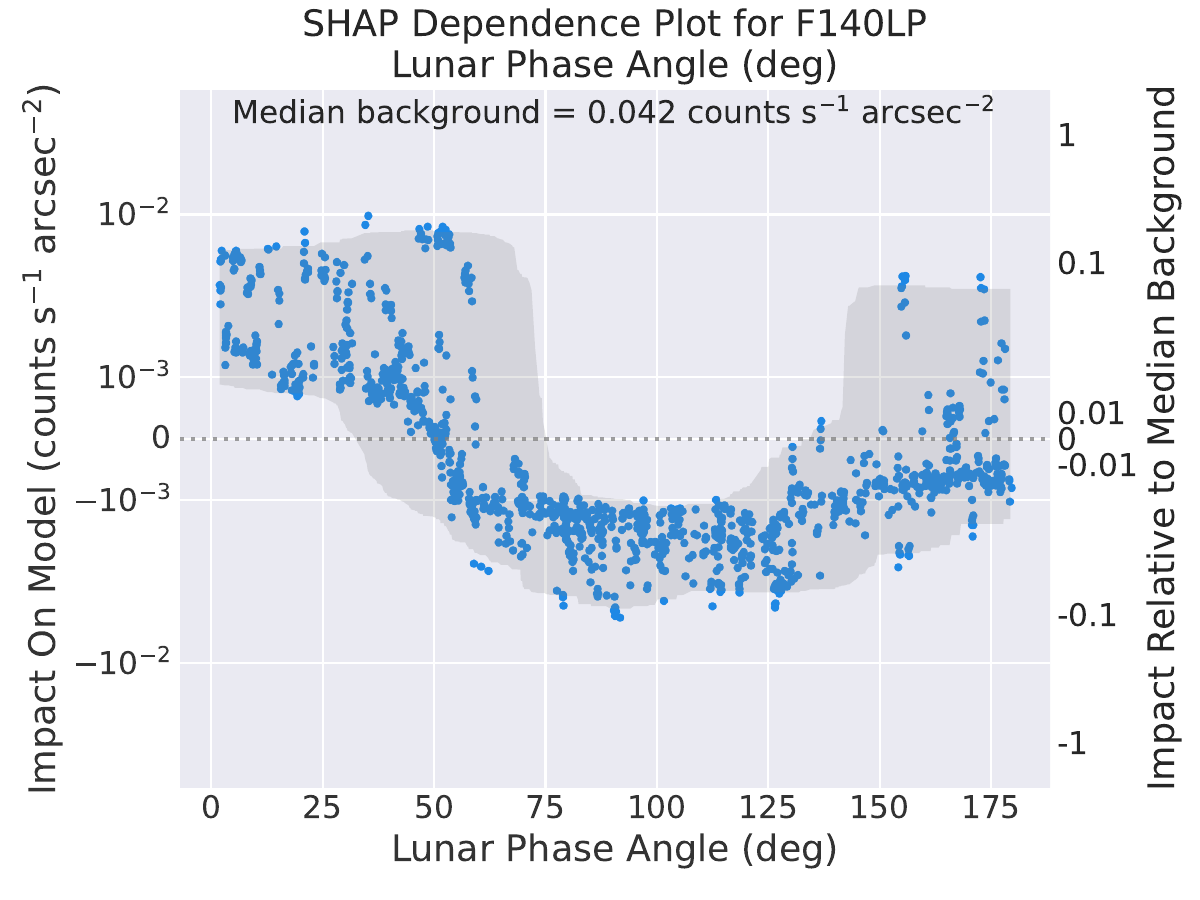}
\includegraphics[width=0.24\textwidth]{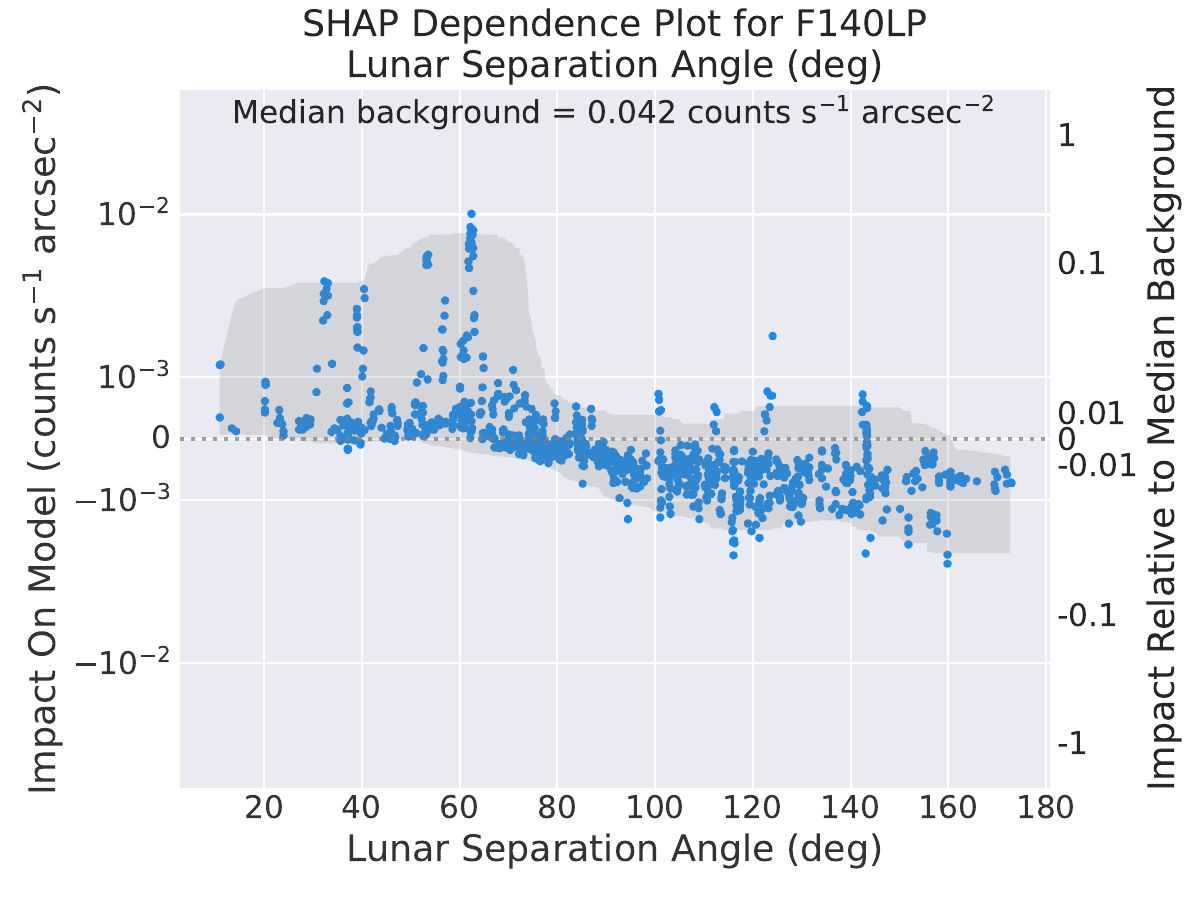}
\includegraphics[width=0.24\textwidth]{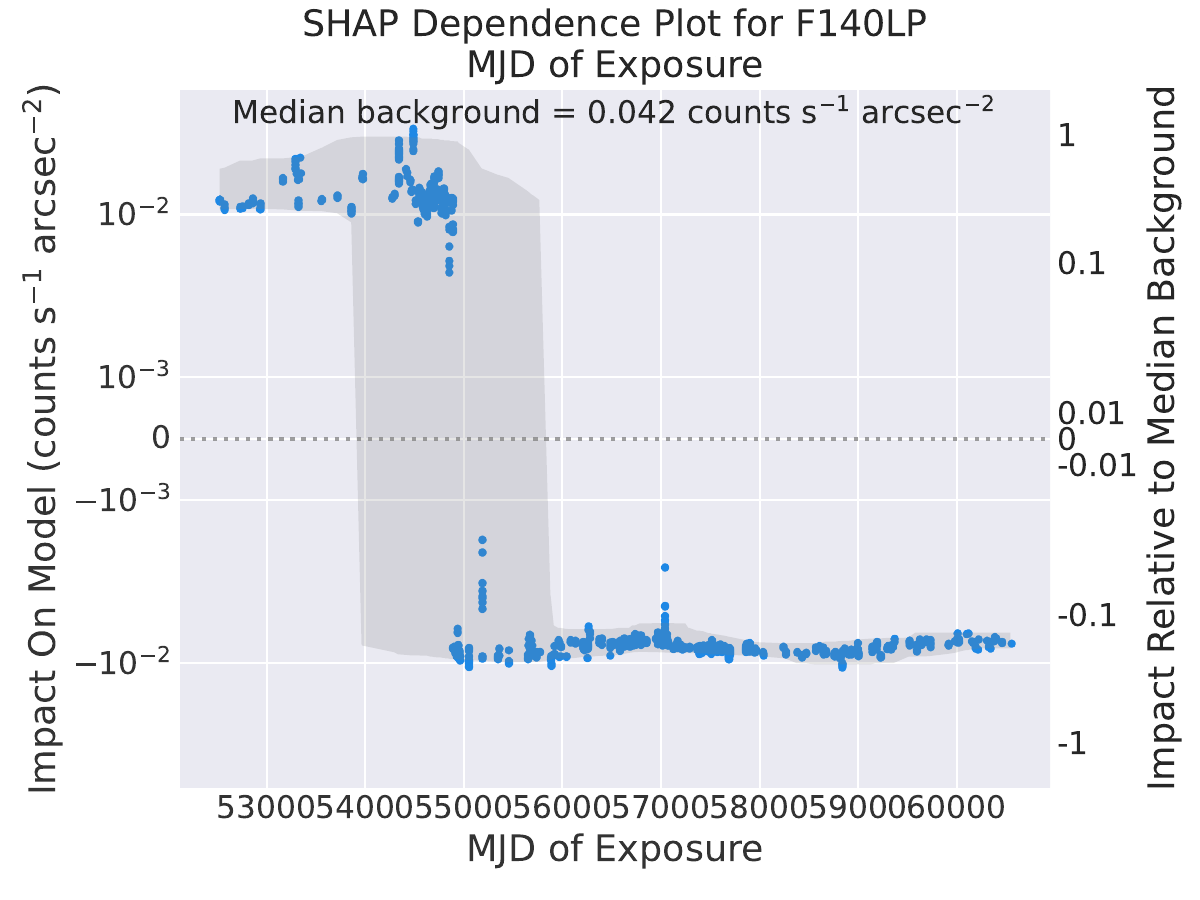}
\includegraphics[width=0.24\textwidth]{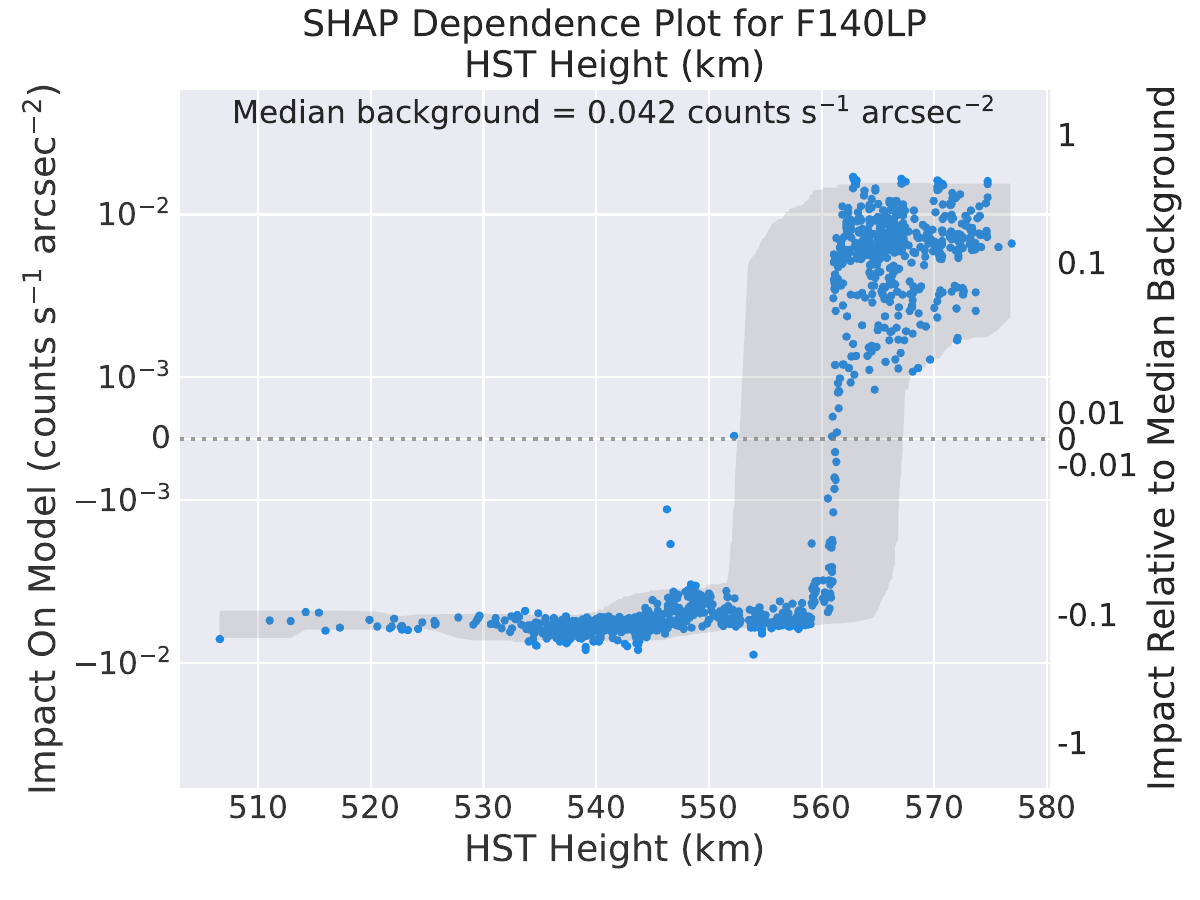}
\includegraphics[width=0.24\textwidth]{SHAP_Plots_QuantileForestRegr/F140LP_solar_alt_SHAP_Dependence.pdf}
\includegraphics[width=0.24\textwidth]{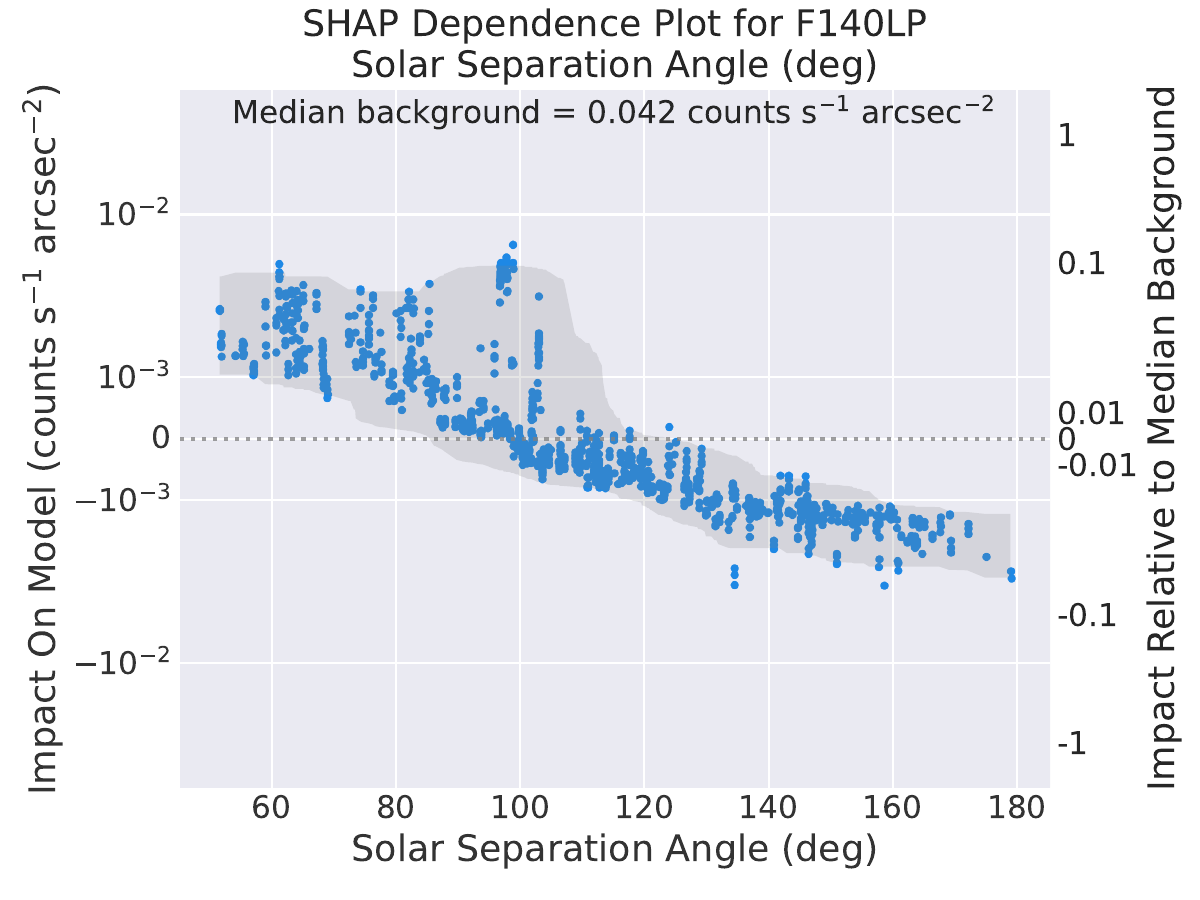}
\includegraphics[width=0.24\textwidth]{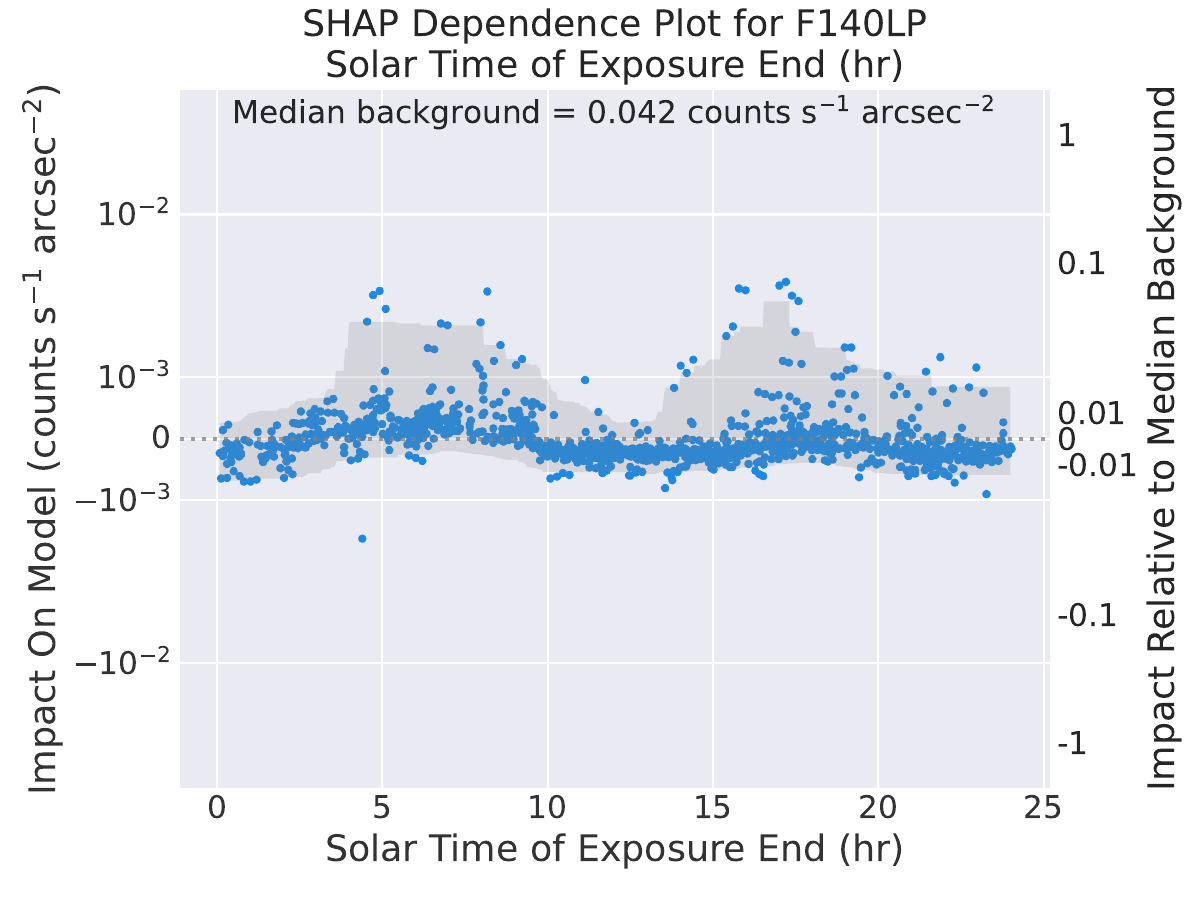}
\includegraphics[width=0.24\textwidth]{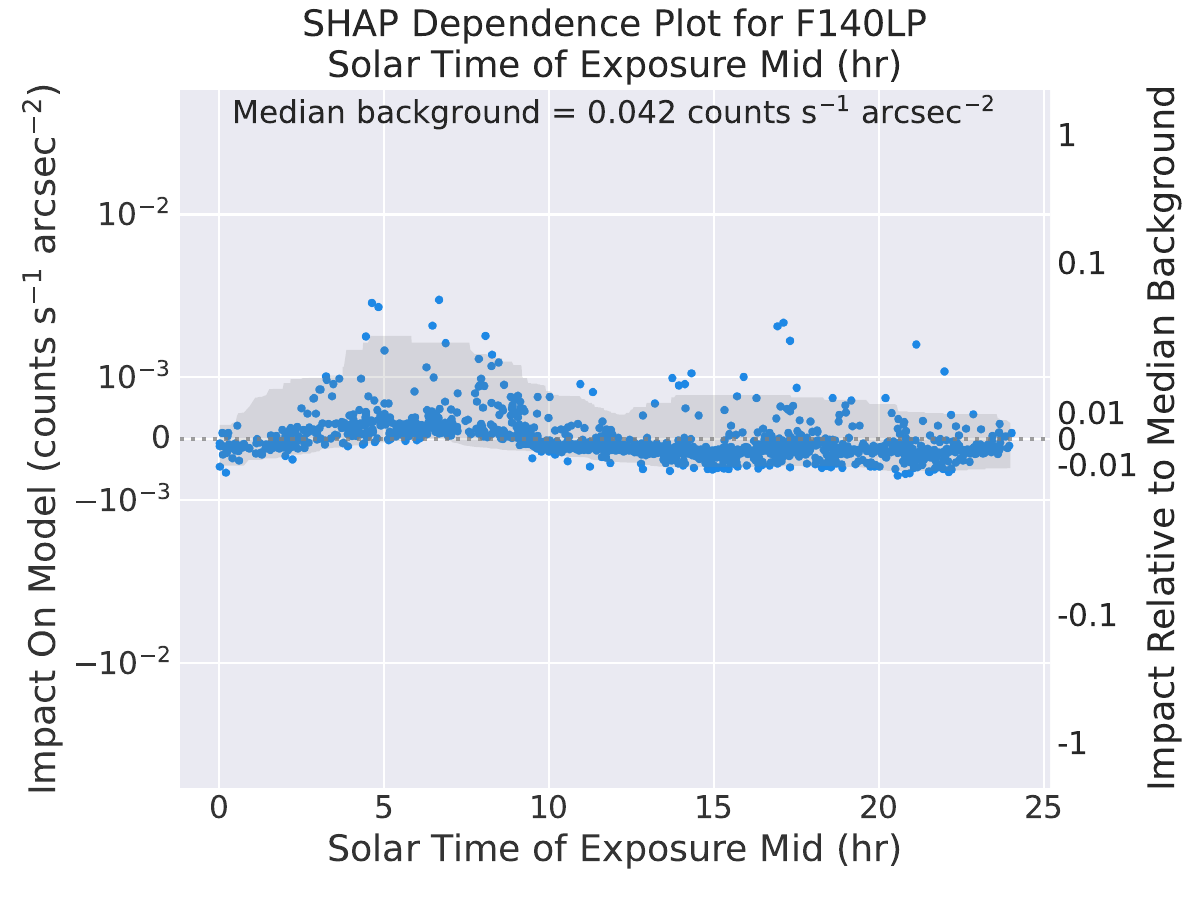}
\includegraphics[width=0.24\textwidth]{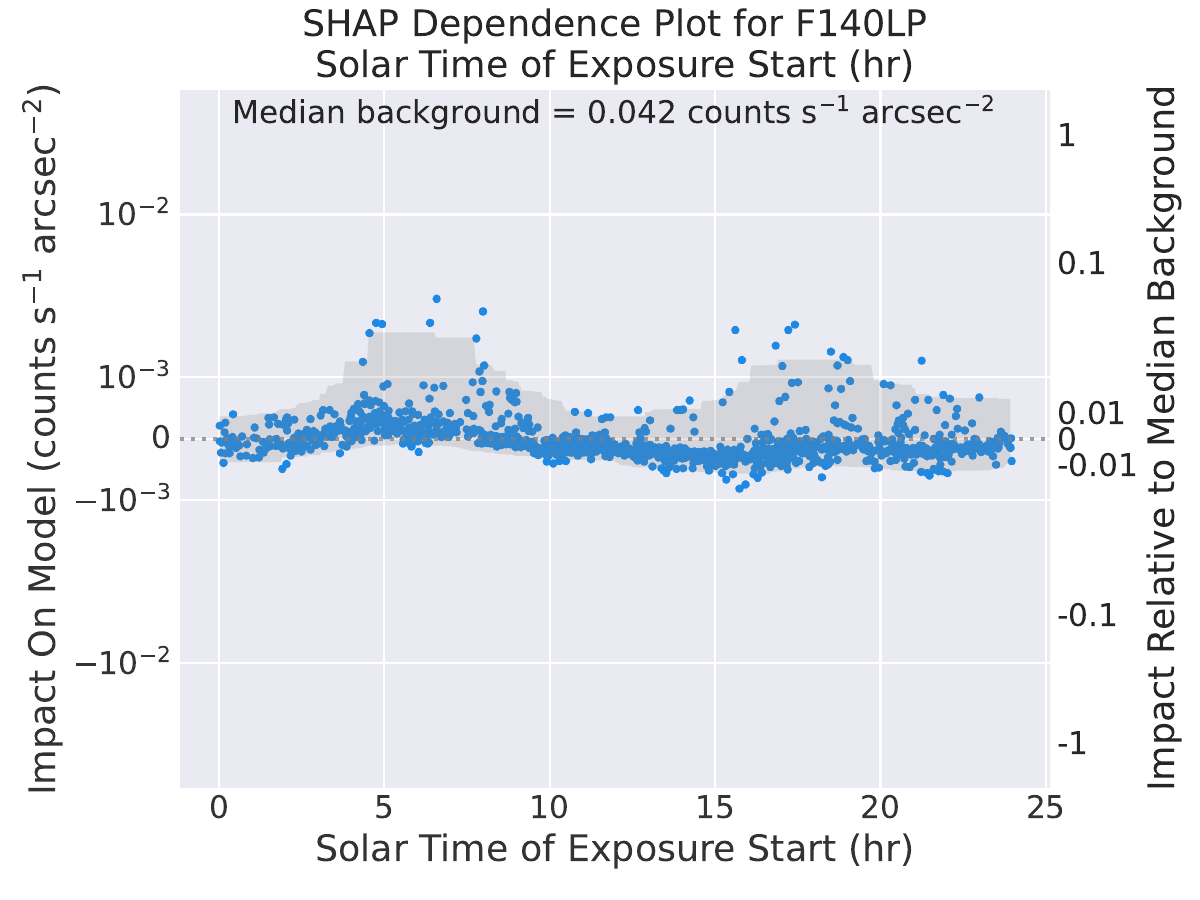}
\includegraphics[width=0.24\textwidth]{SHAP_Plots_QuantileForestRegr/F140LP_sunspots_SHAP_Dependence.pdf}
\includegraphics[width=0.24\textwidth]{SHAP_Plots_QuantileForestRegr/F140LP_temp_end_SHAP_Dependence.pdf}
\includegraphics[width=0.24\textwidth]{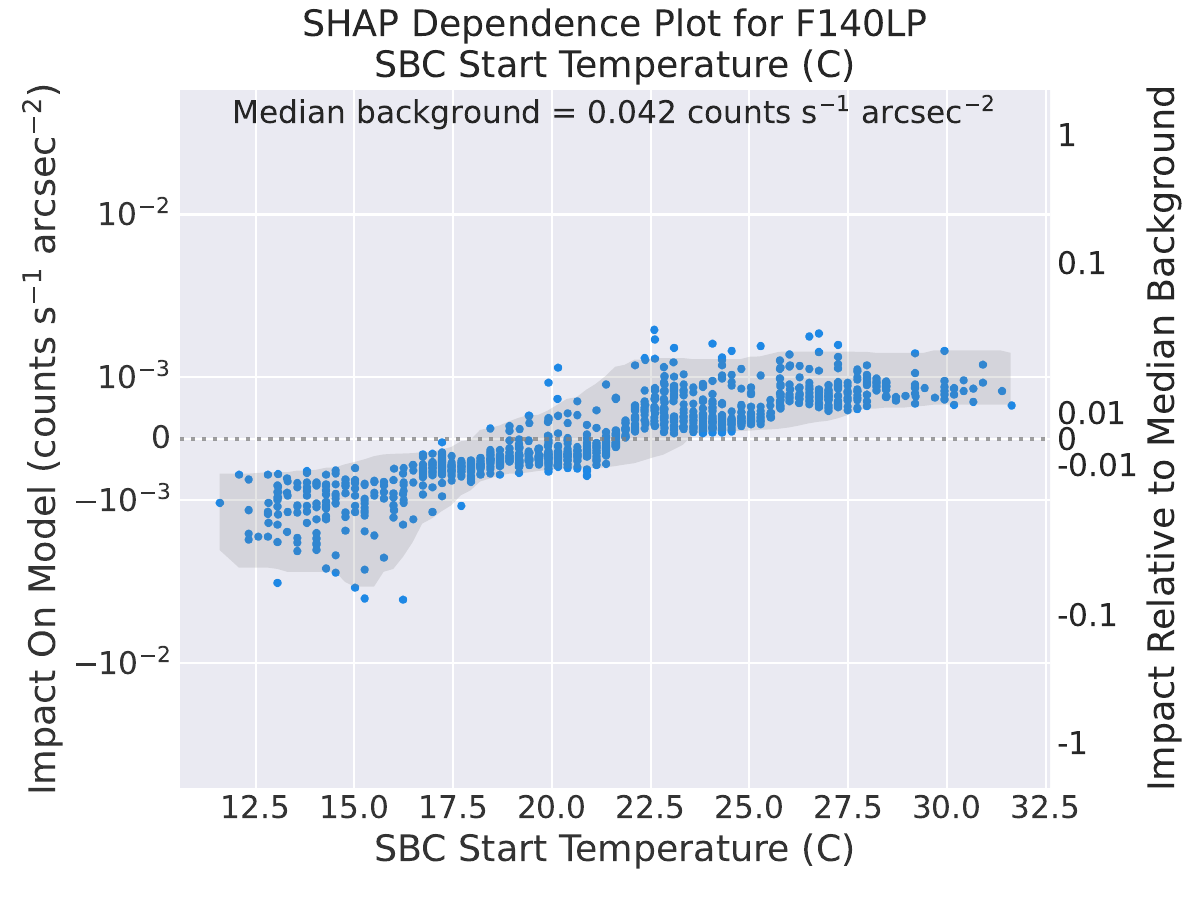}
\caption{SHAP dependence plots for QRF regression modeling of F140LP. Otherwise as per Figure~\ref{Fig:SHAP_Dependence_F115LP}.}
\label{Fig:SHAP_Dependence_F140LP}
\end{figure}

\begin{figure}
\centering
\includegraphics[width=0.24\textwidth]{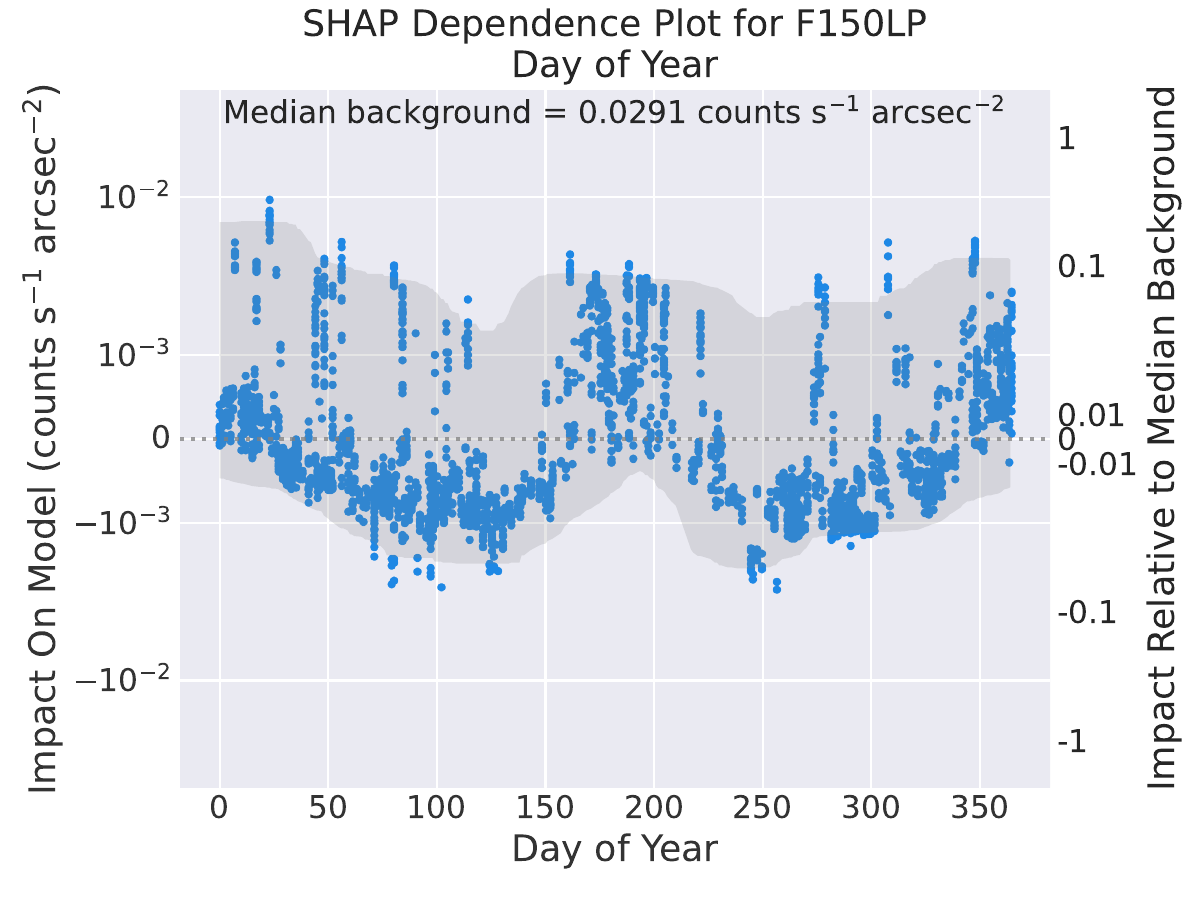}
\includegraphics[width=0.24\textwidth]{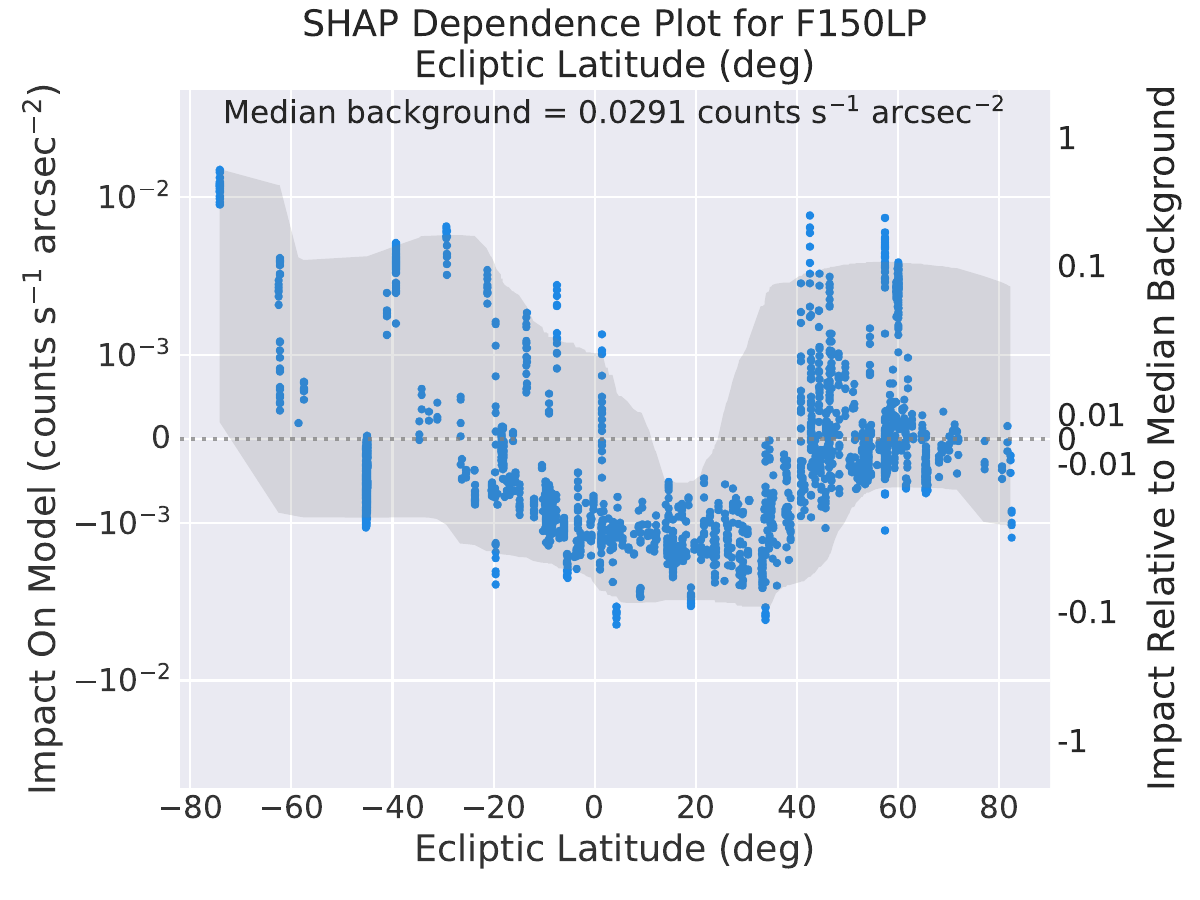}
\includegraphics[width=0.24\textwidth]{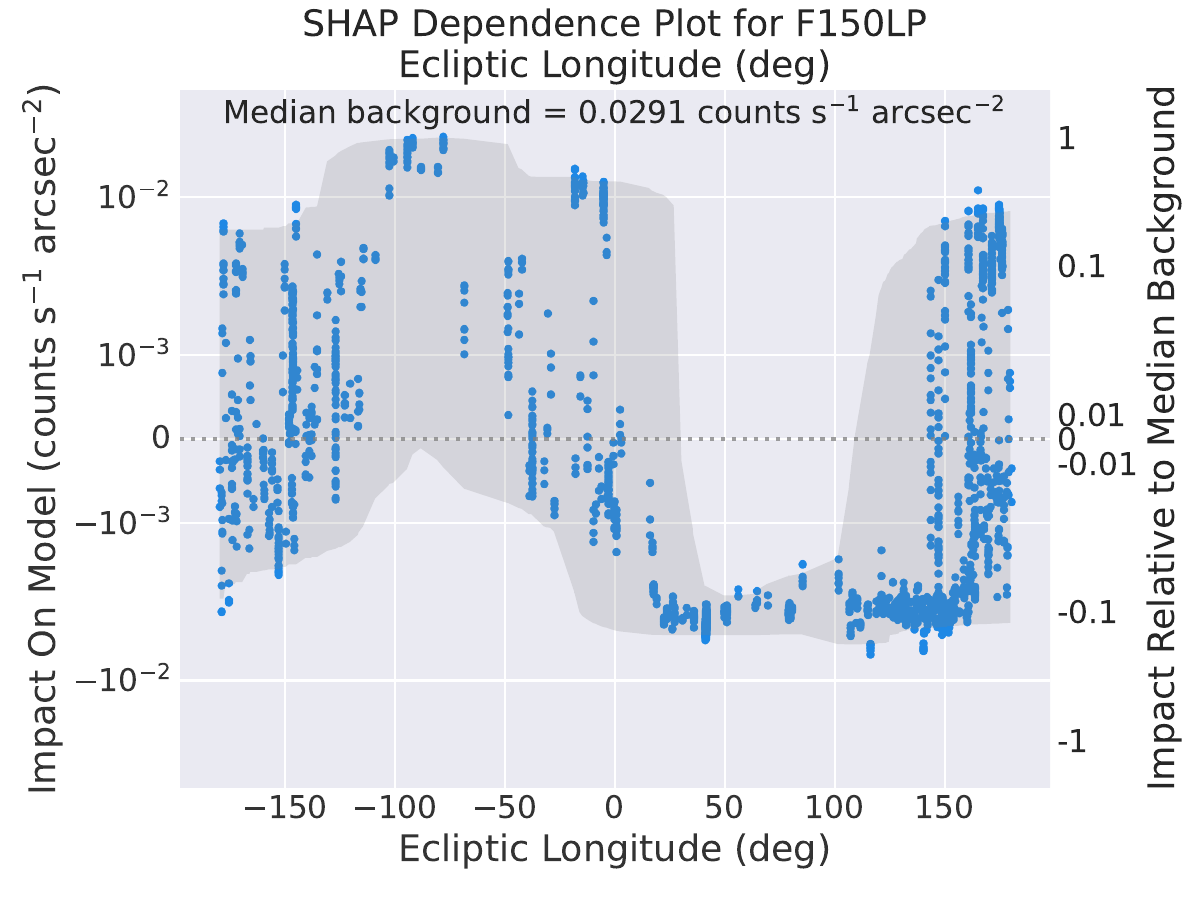}
\includegraphics[width=0.24\textwidth]{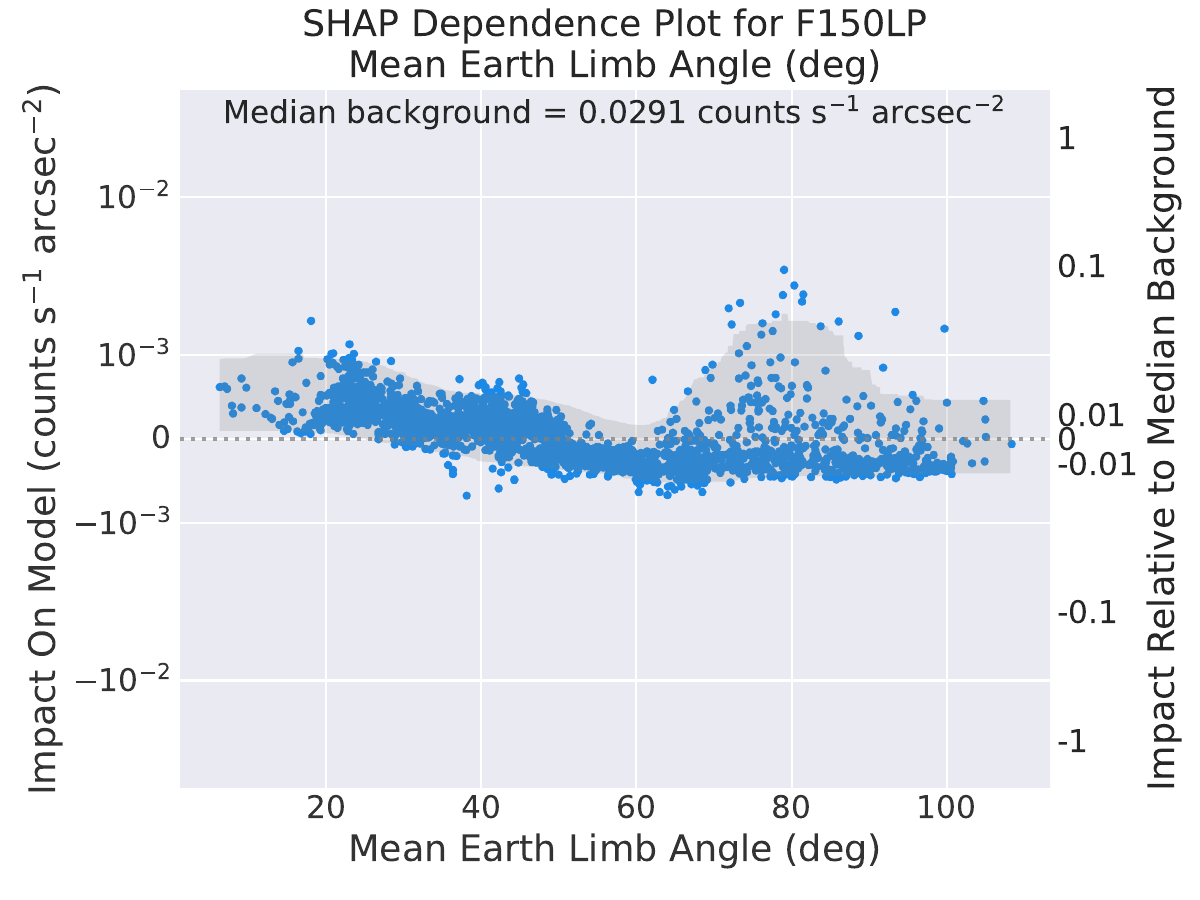}
\includegraphics[width=0.24\textwidth]{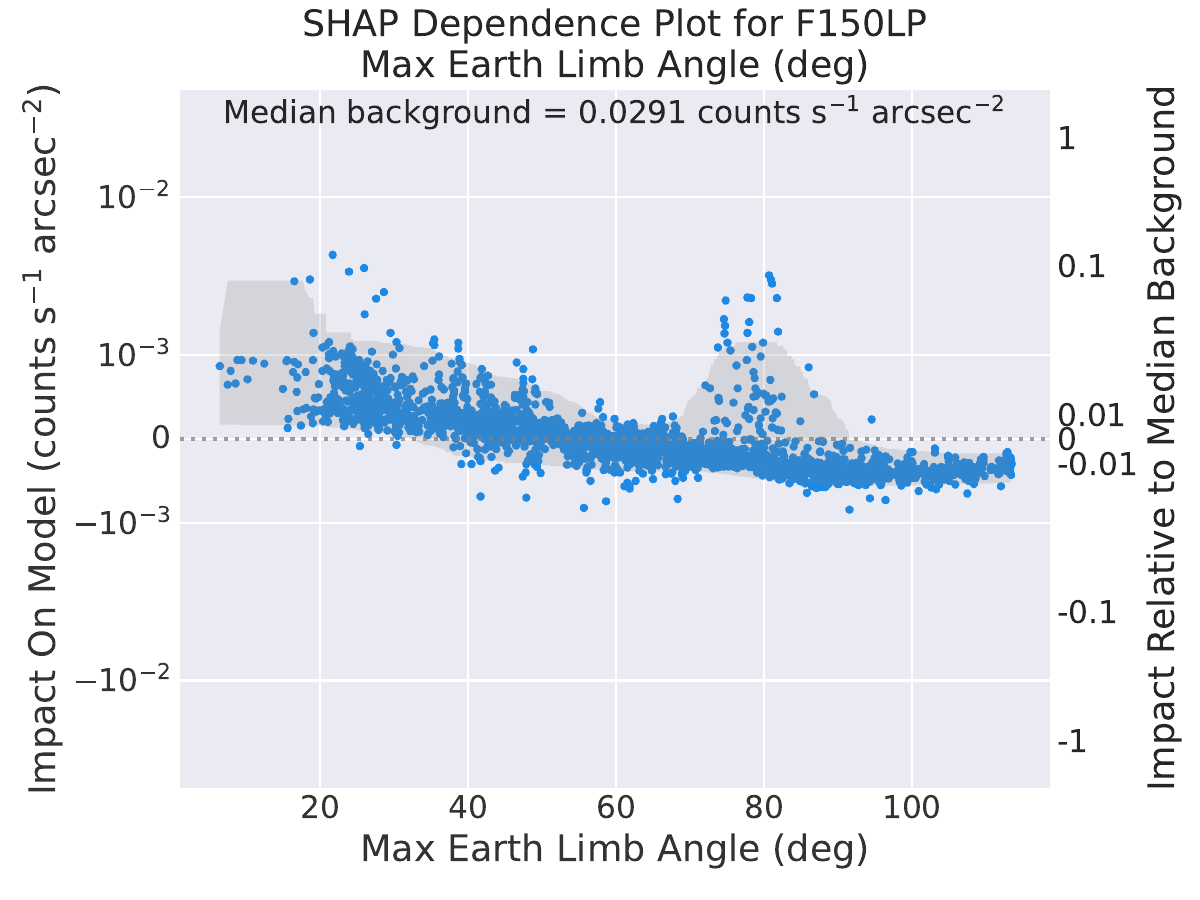}
\includegraphics[width=0.24\textwidth]{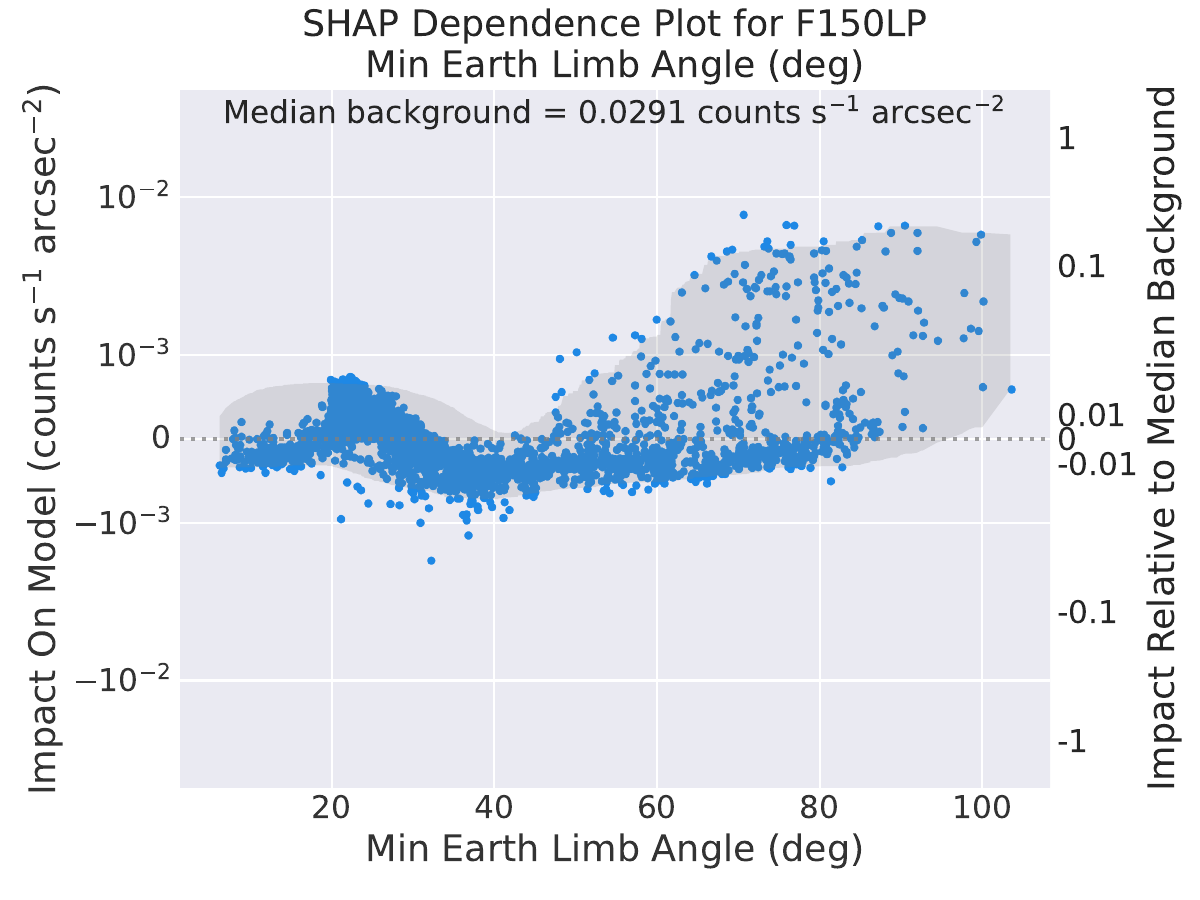}
\includegraphics[width=0.24\textwidth]{SHAP_Plots_QuantileForestRegr/F150LP_exp_time_SHAP_Dependence.pdf}
\includegraphics[width=0.24\textwidth]{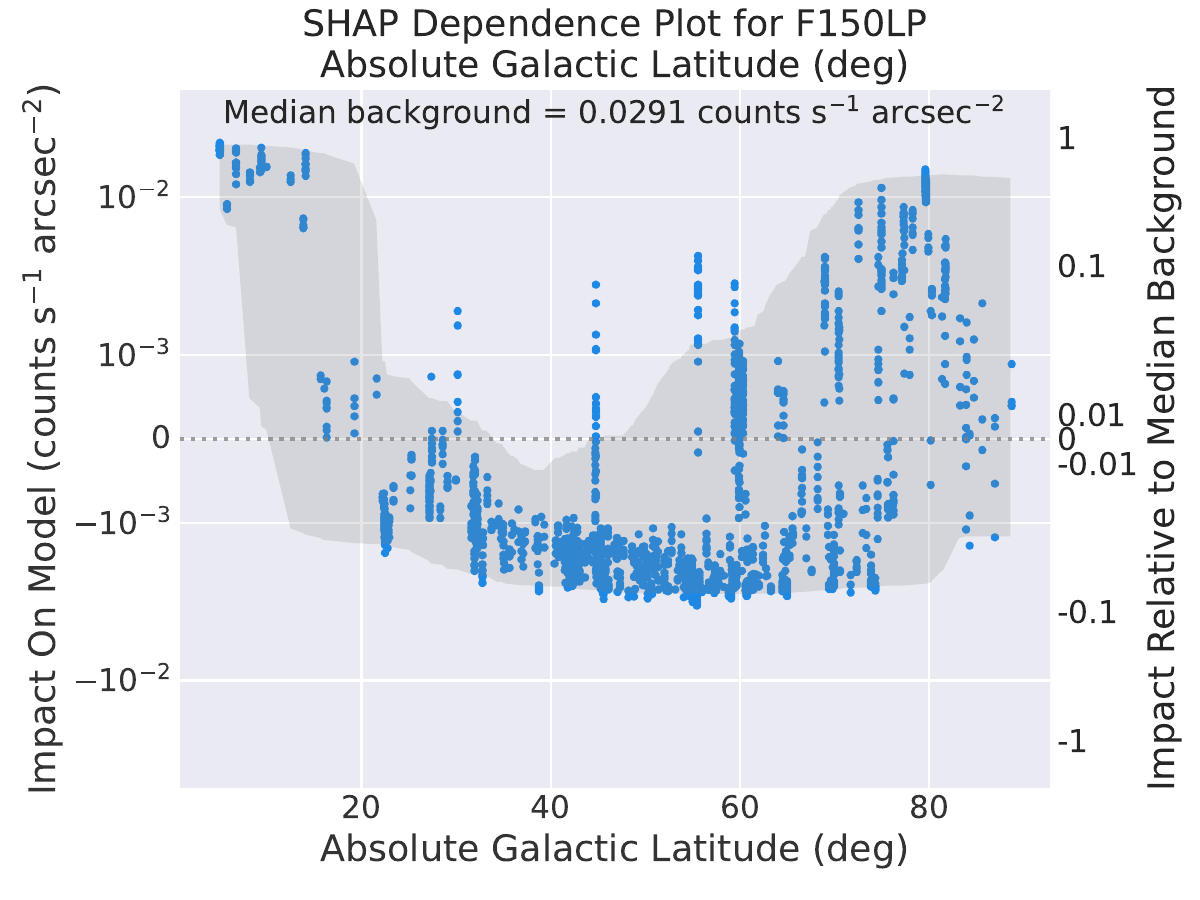}
\includegraphics[width=0.24\textwidth]{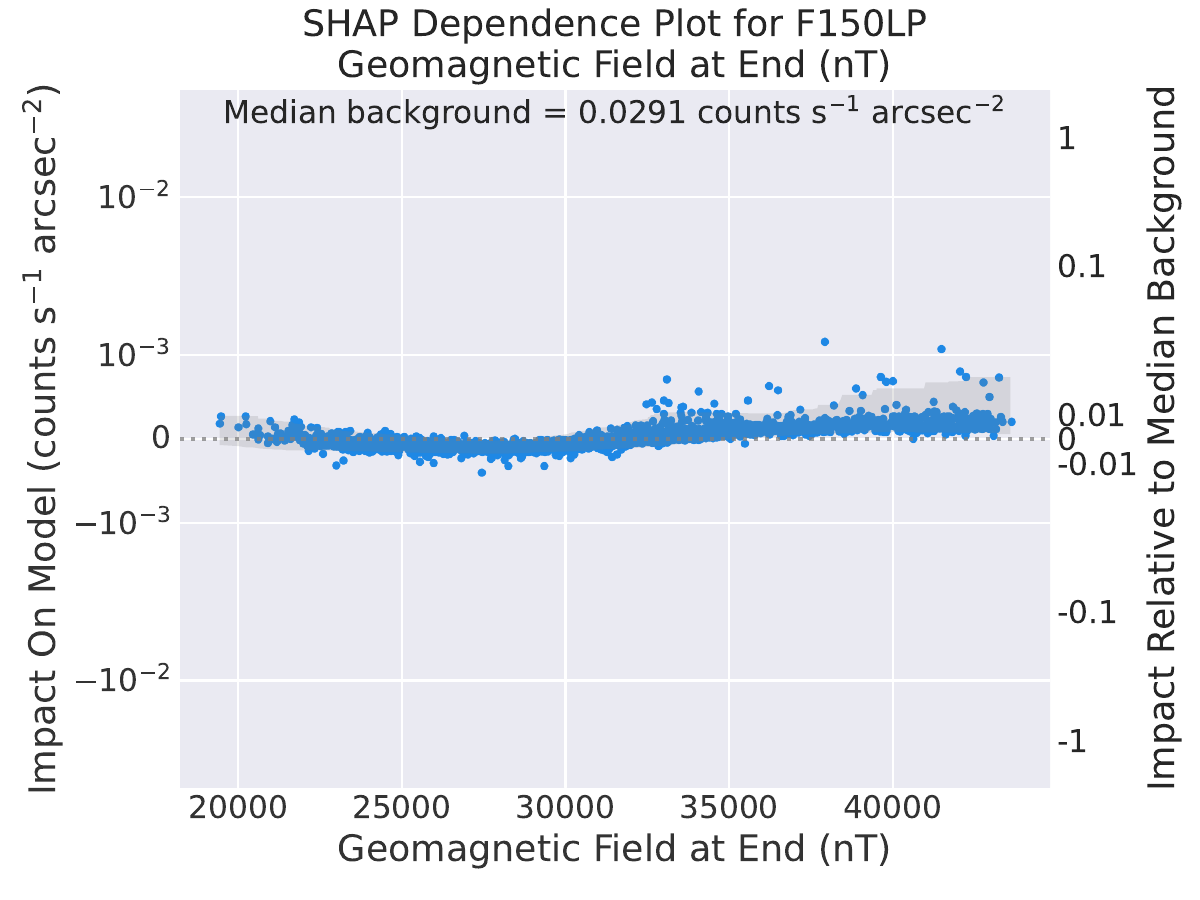}
\includegraphics[width=0.24\textwidth]{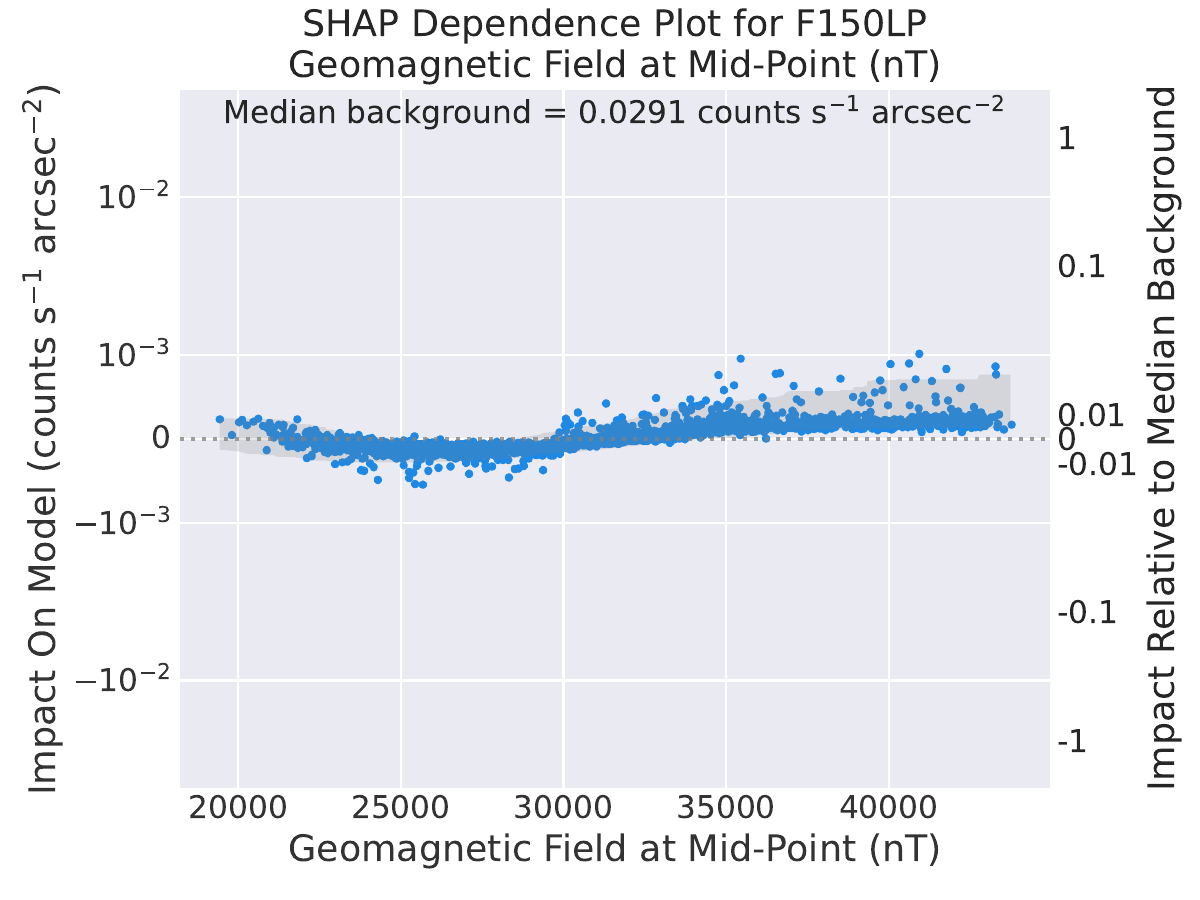}
\includegraphics[width=0.24\textwidth]{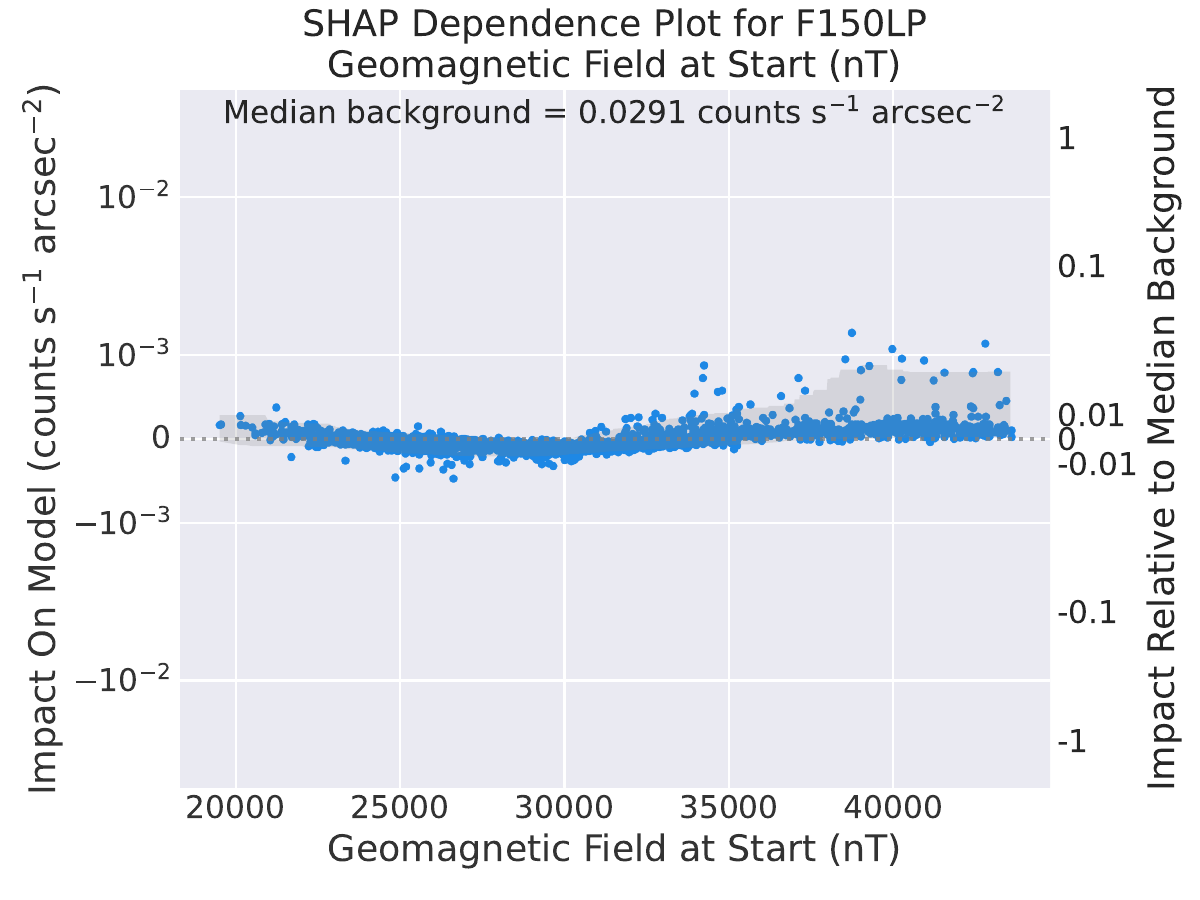}
\includegraphics[width=0.24\textwidth]{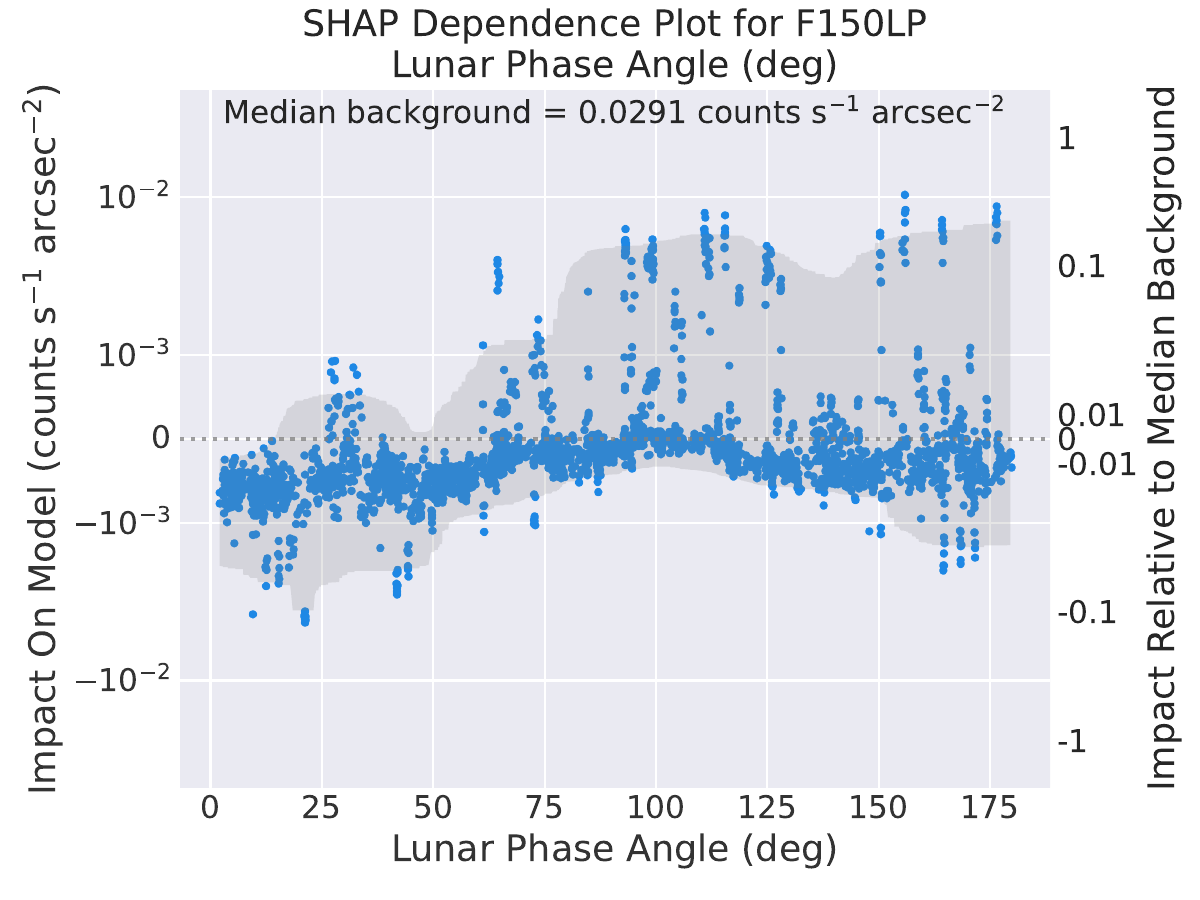}
\includegraphics[width=0.24\textwidth]{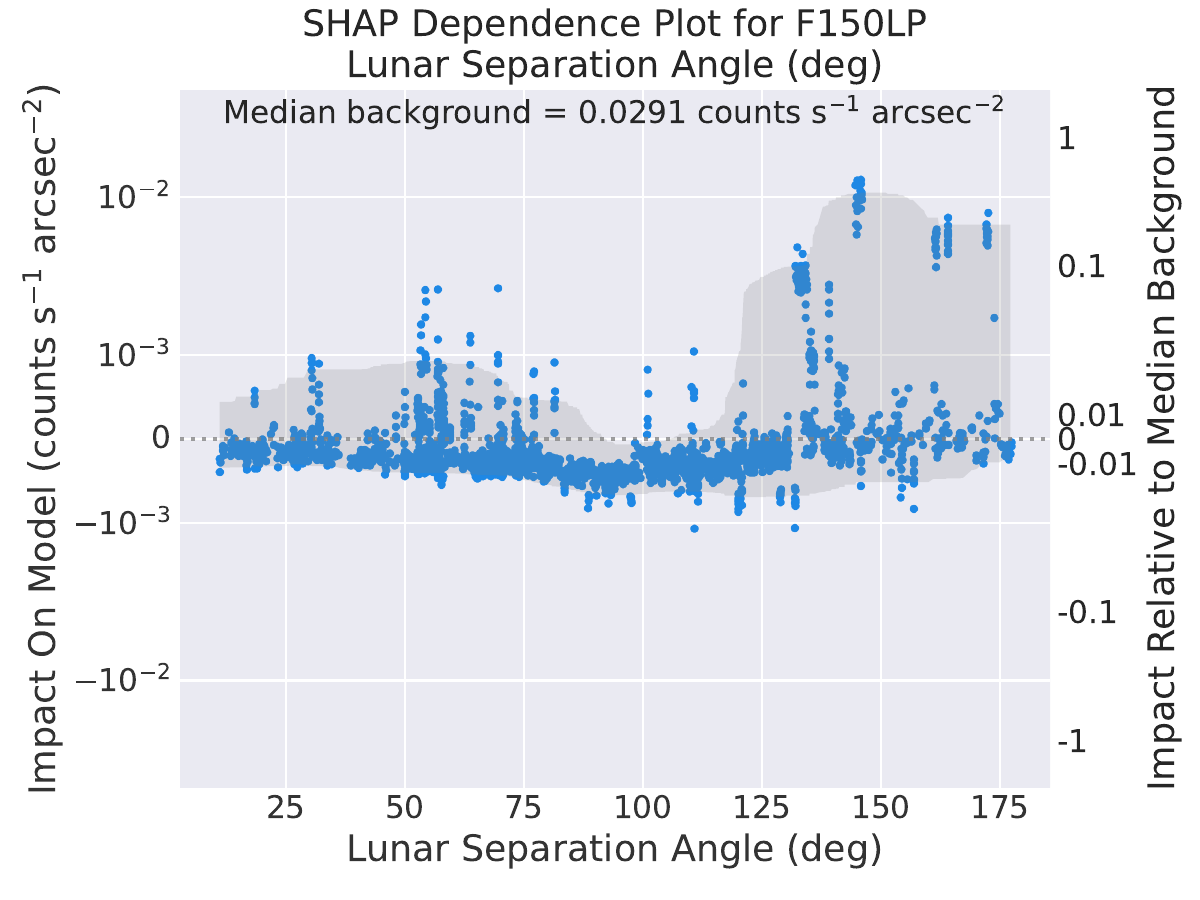}
\includegraphics[width=0.24\textwidth]{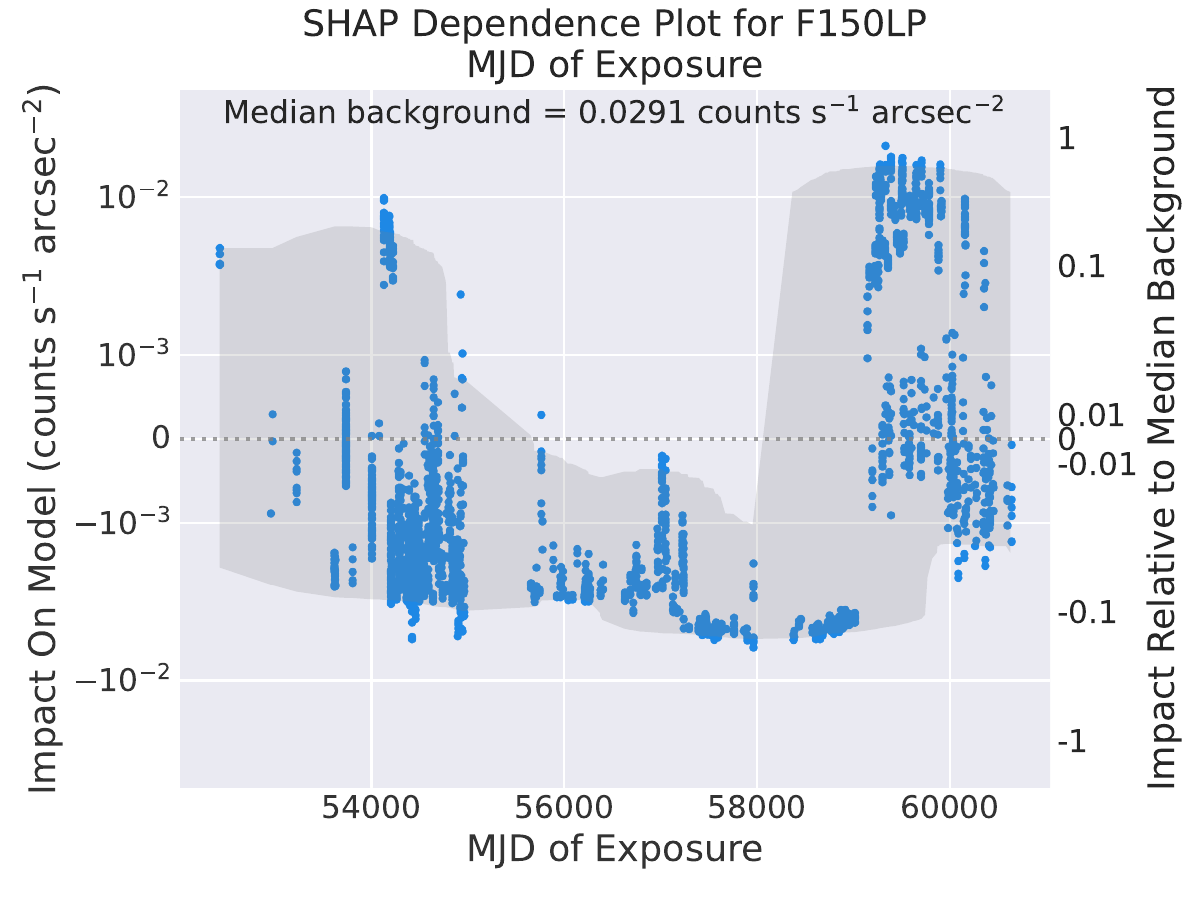}
\includegraphics[width=0.24\textwidth]{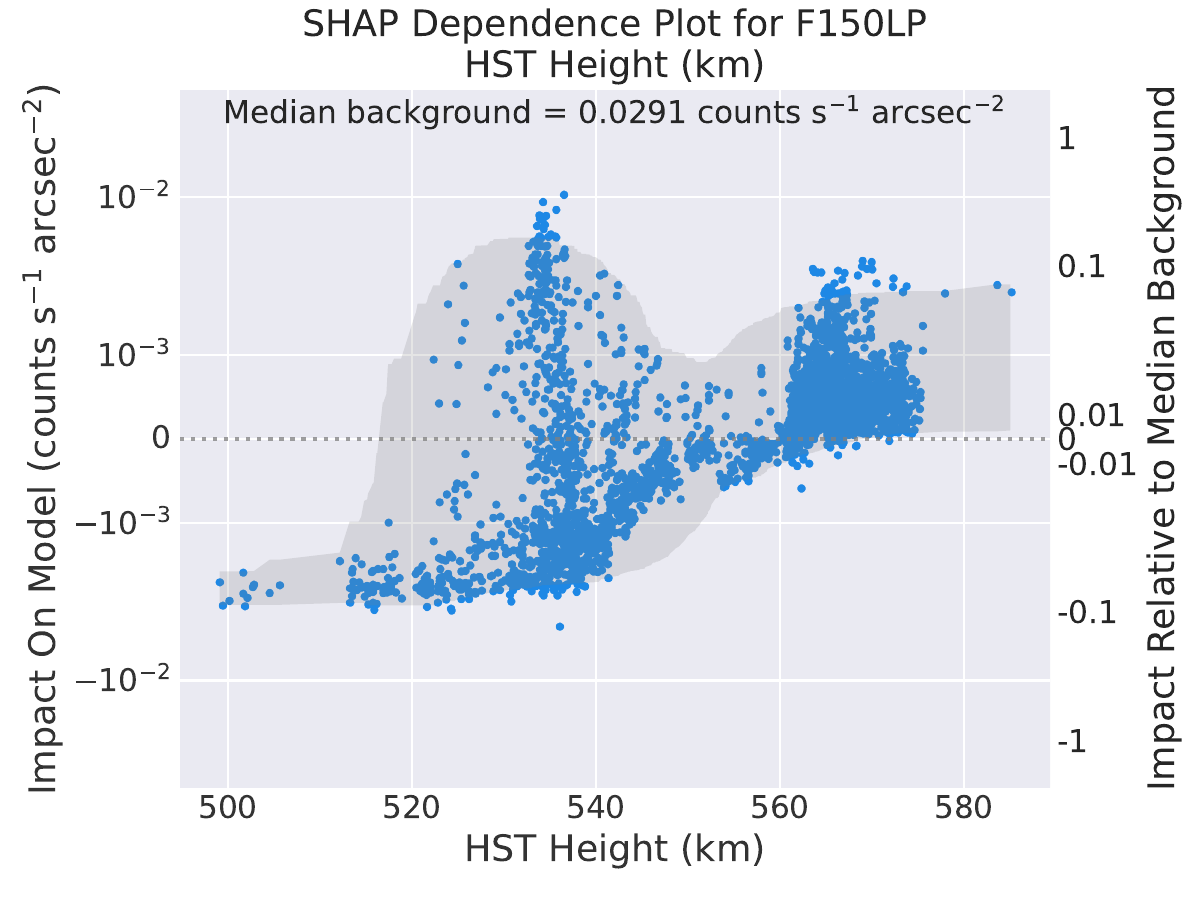}
\includegraphics[width=0.24\textwidth]{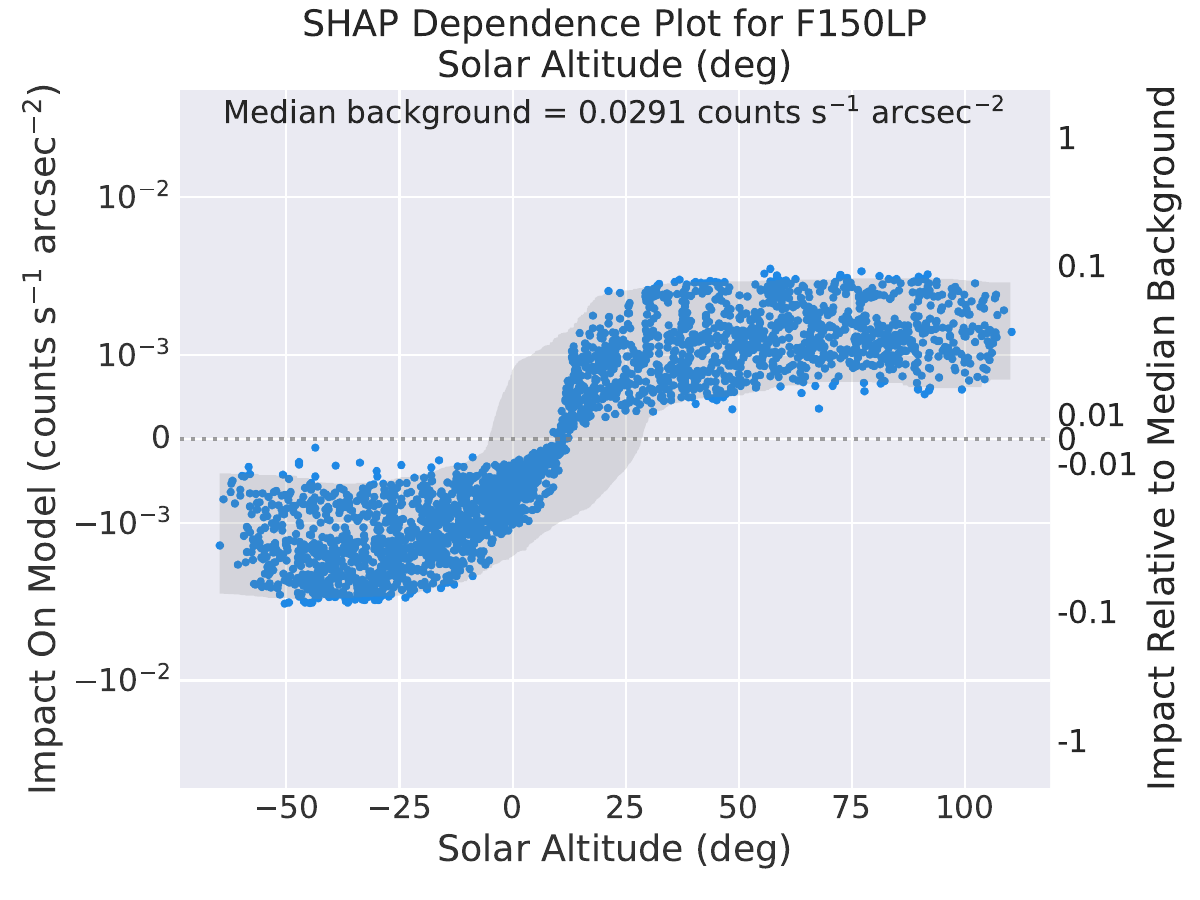}
\includegraphics[width=0.24\textwidth]{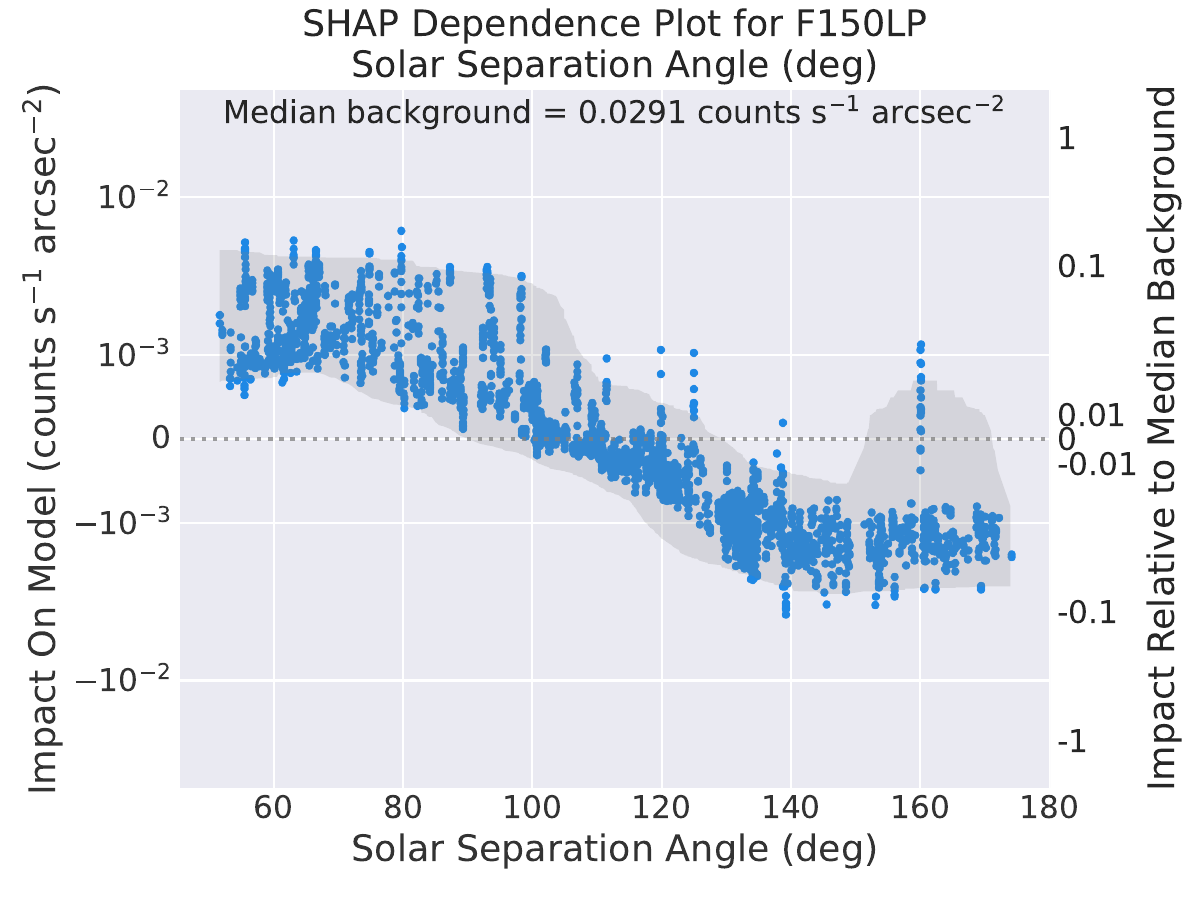}
\includegraphics[width=0.24\textwidth]{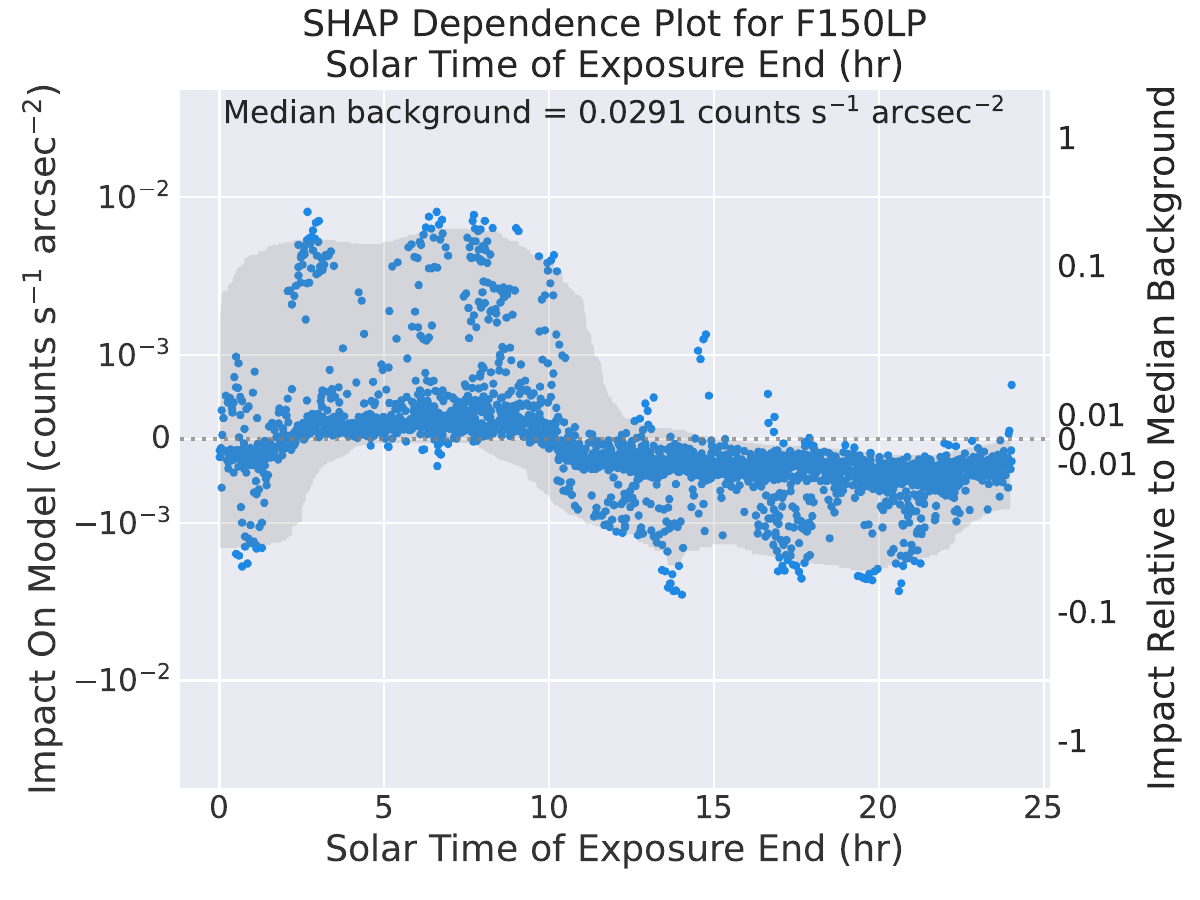}
\includegraphics[width=0.24\textwidth]{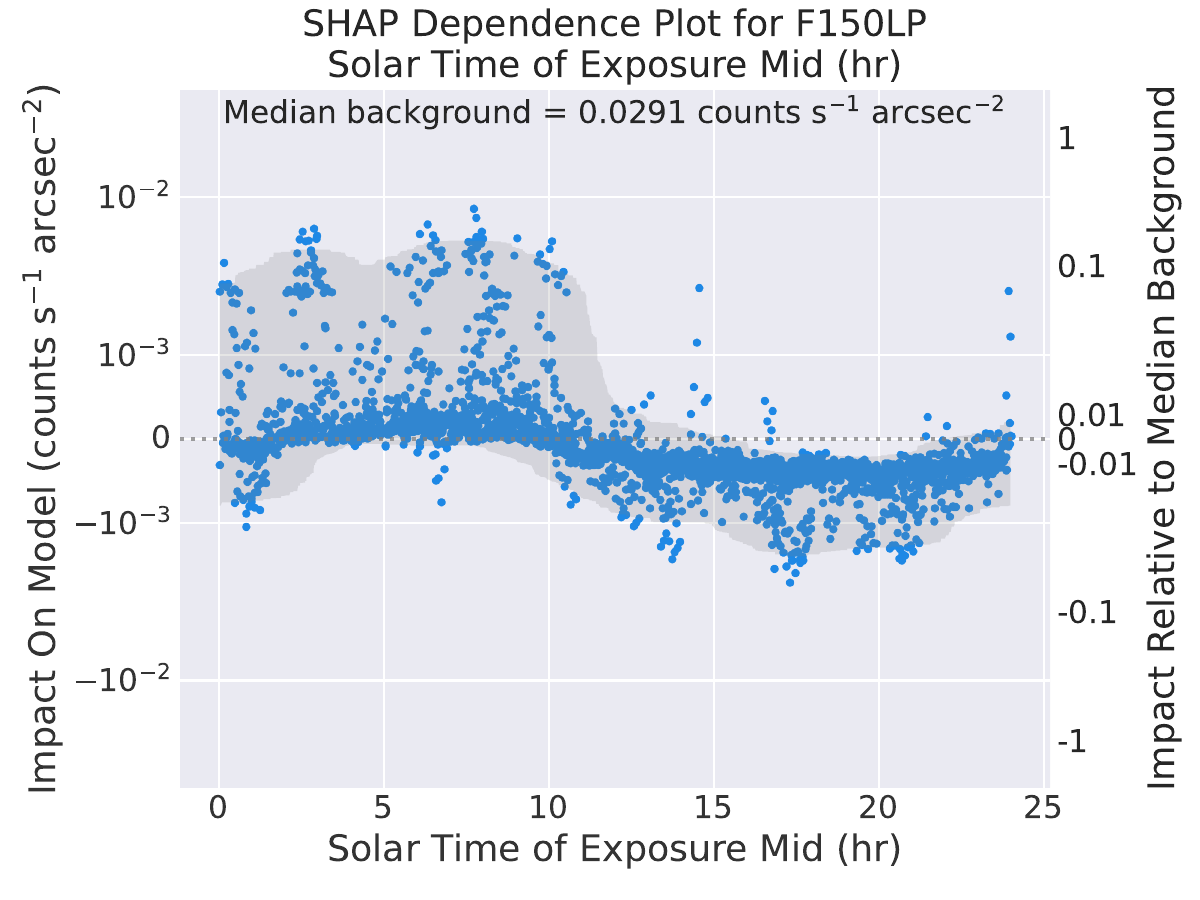}
\includegraphics[width=0.24\textwidth]{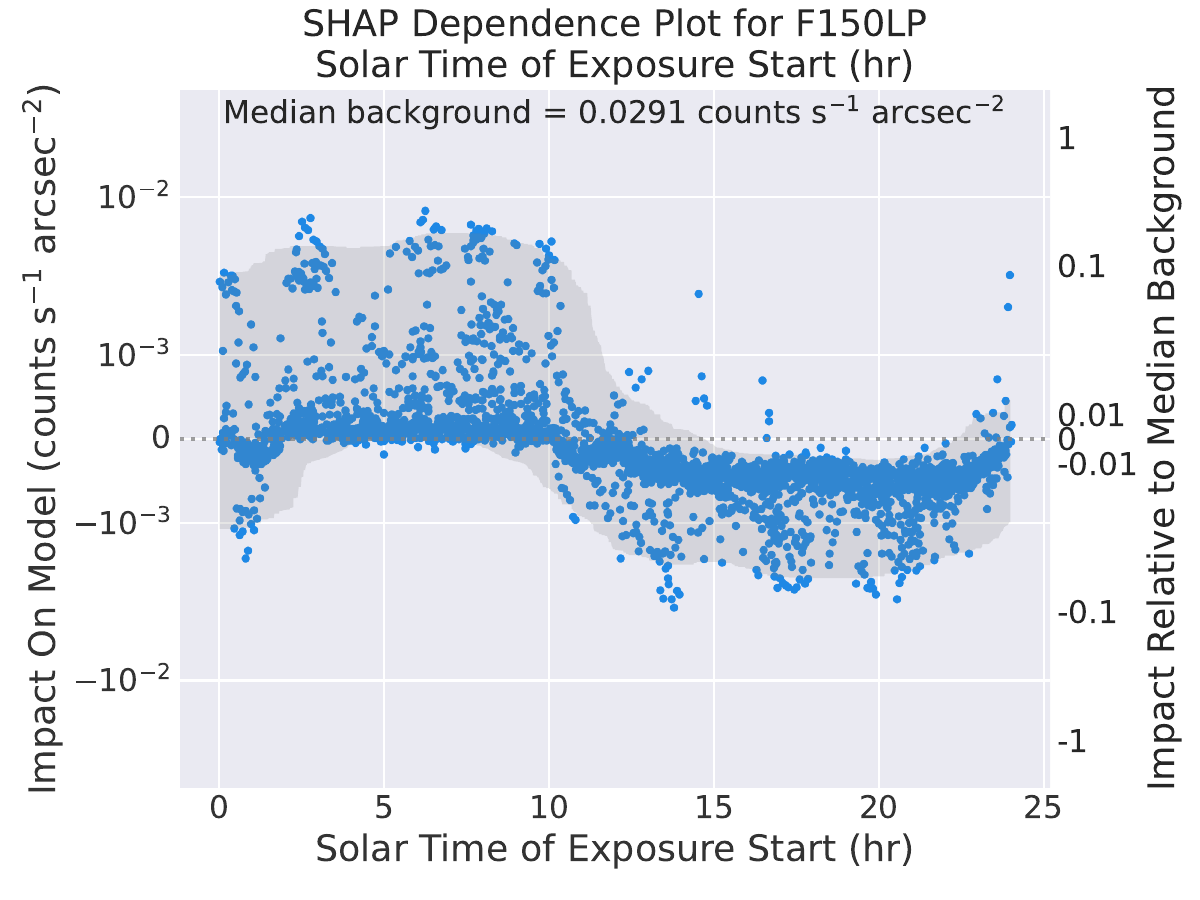}
\includegraphics[width=0.24\textwidth]{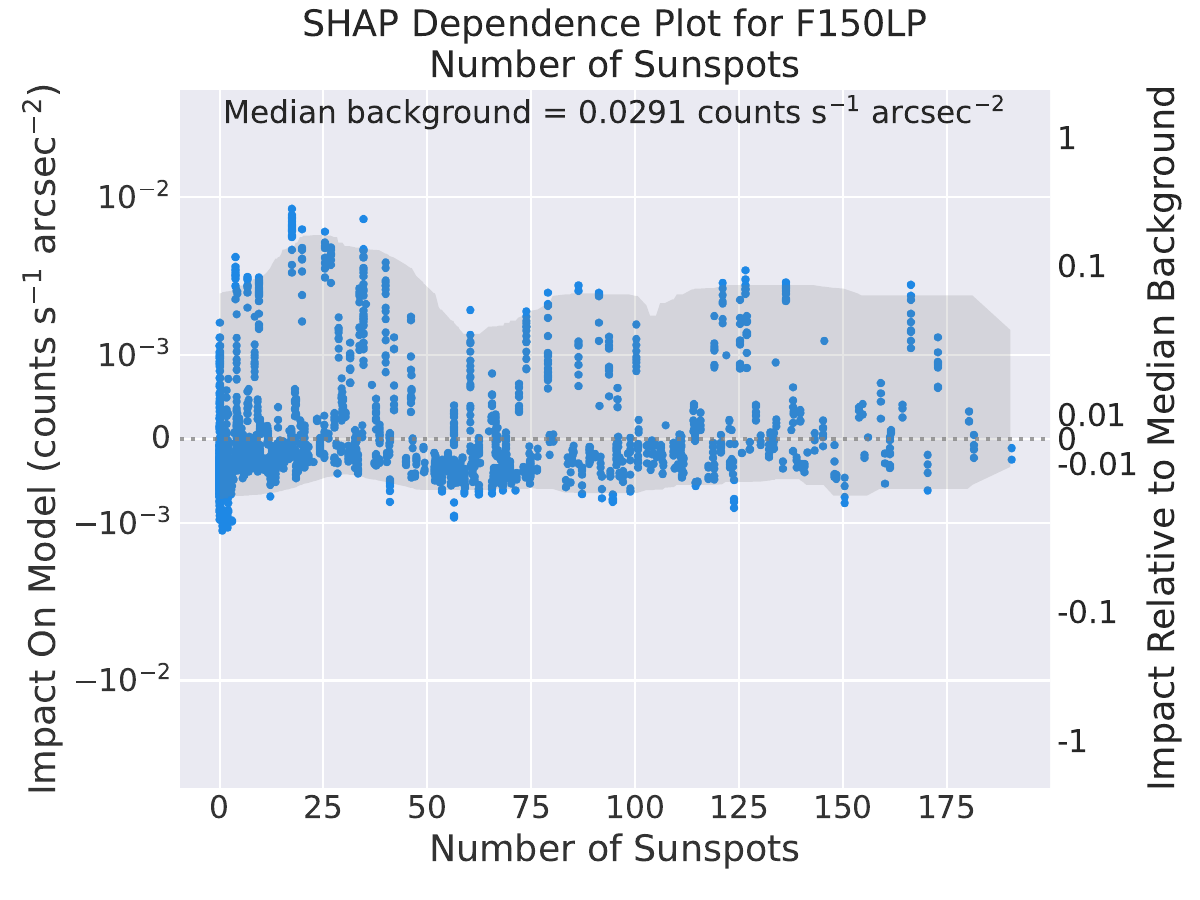}
\includegraphics[width=0.24\textwidth]{SHAP_Plots_QuantileForestRegr/F150LP_temp_end_SHAP_Dependence.pdf}
\includegraphics[width=0.24\textwidth]{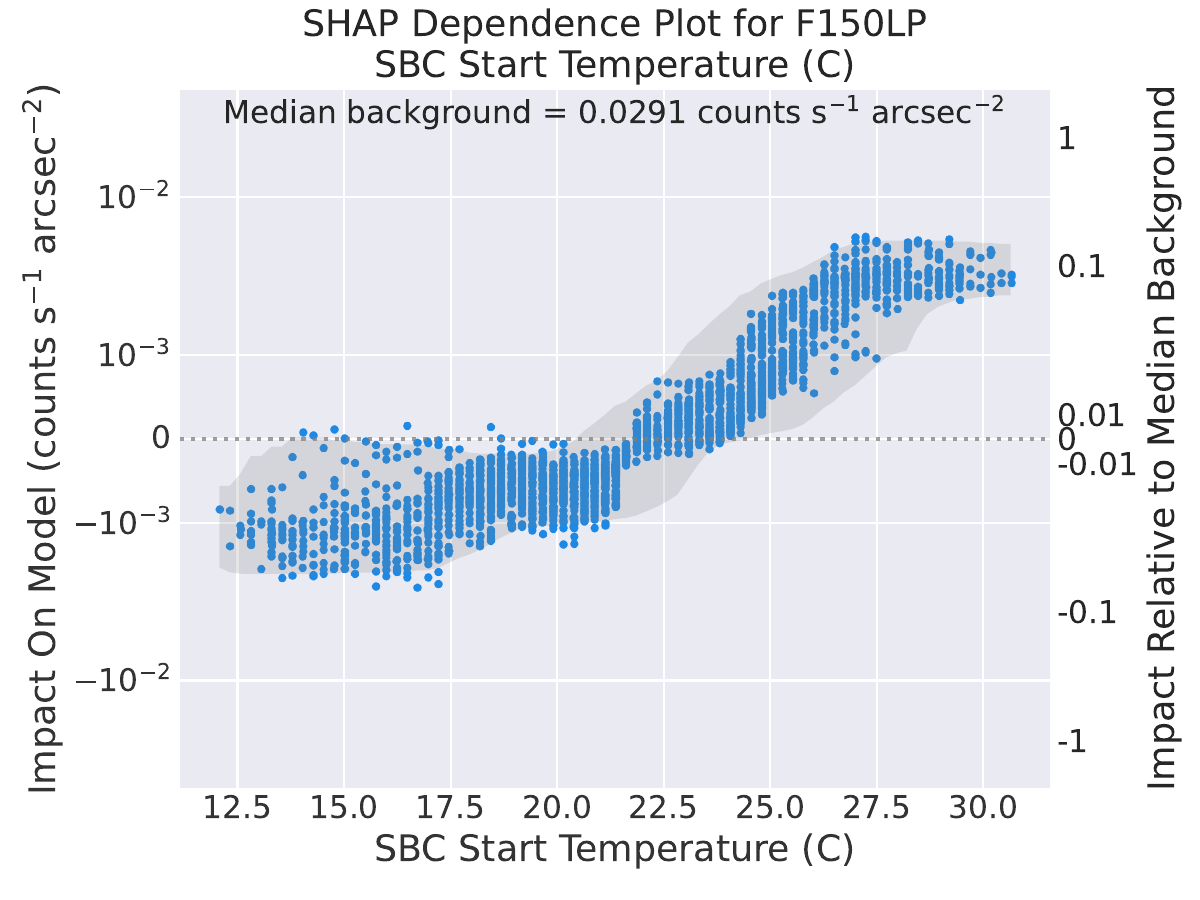}
\caption{SHAP dependence plots for QRF regression modeling of F150LP. Otherwise as per Figure~\ref{Fig:SHAP_Dependence_F115LP}.}
\label{Fig:SHAP_Dependence_F150LP}
\end{figure}

\begin{figure}
\centering
\includegraphics[width=0.24\textwidth]{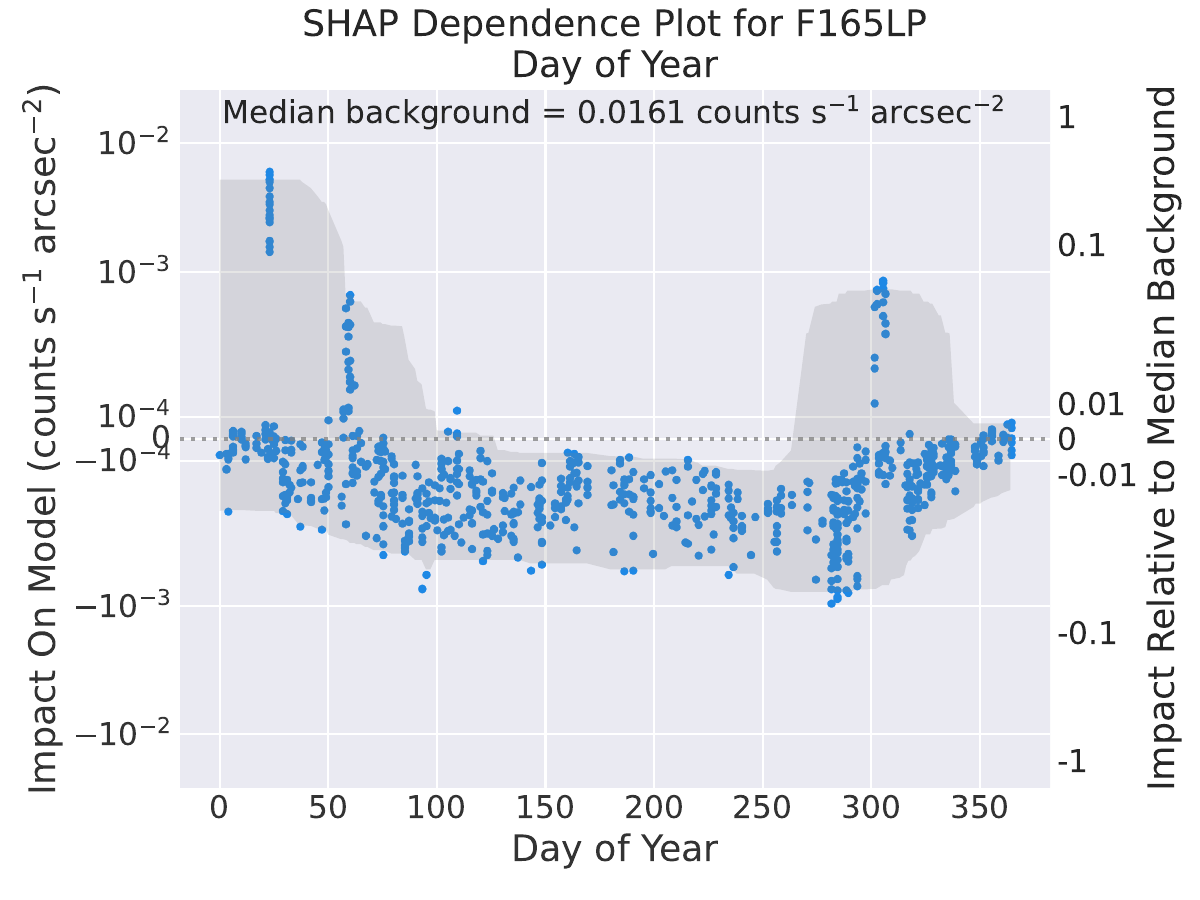}
\includegraphics[width=0.24\textwidth]{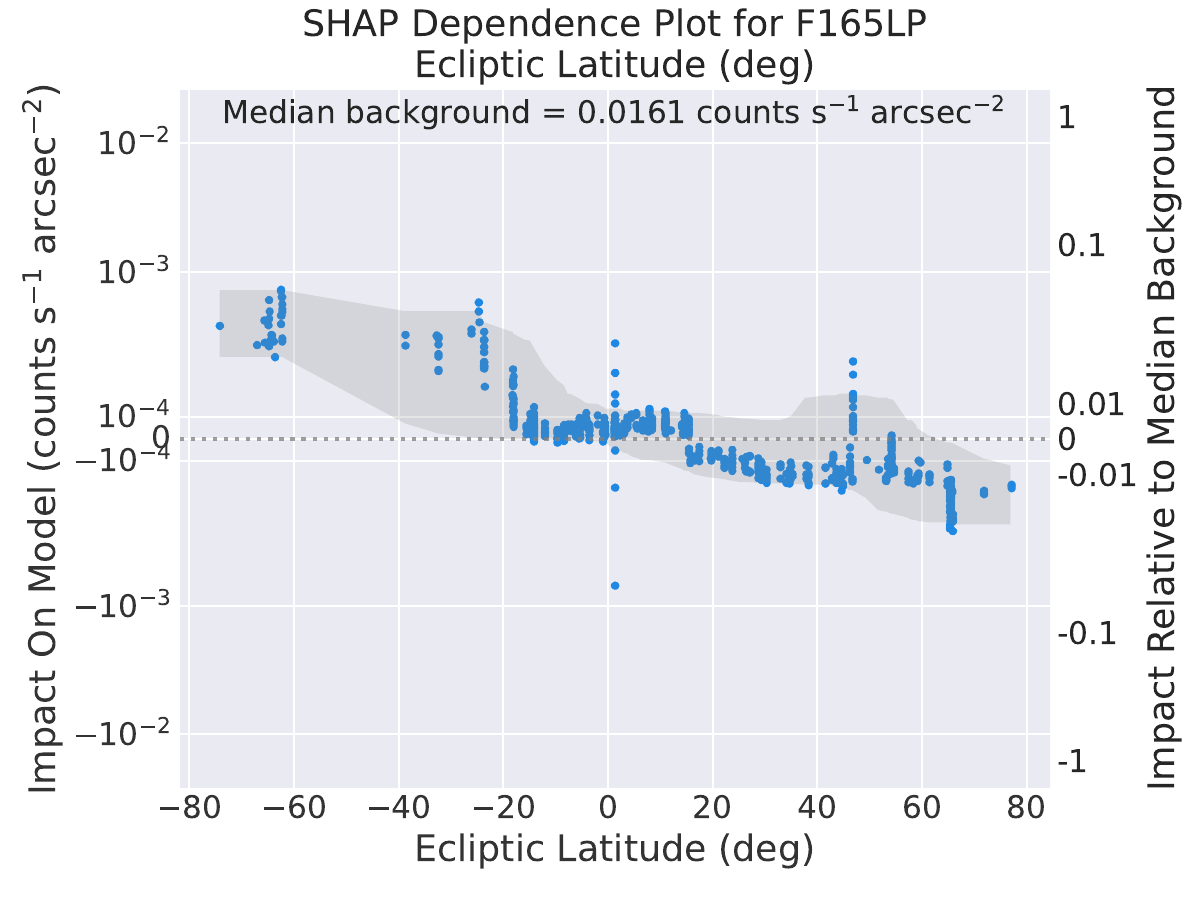}
\includegraphics[width=0.24\textwidth]{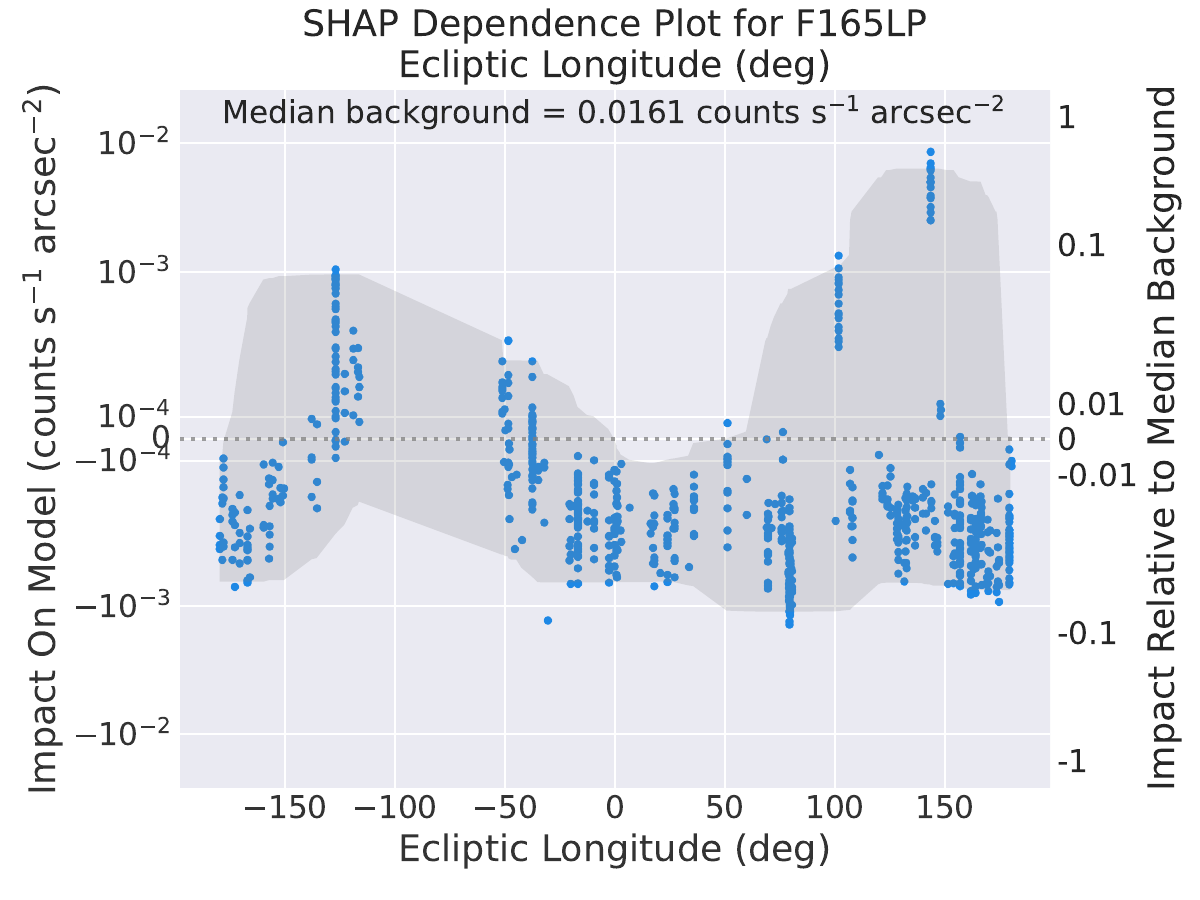}
\includegraphics[width=0.24\textwidth]{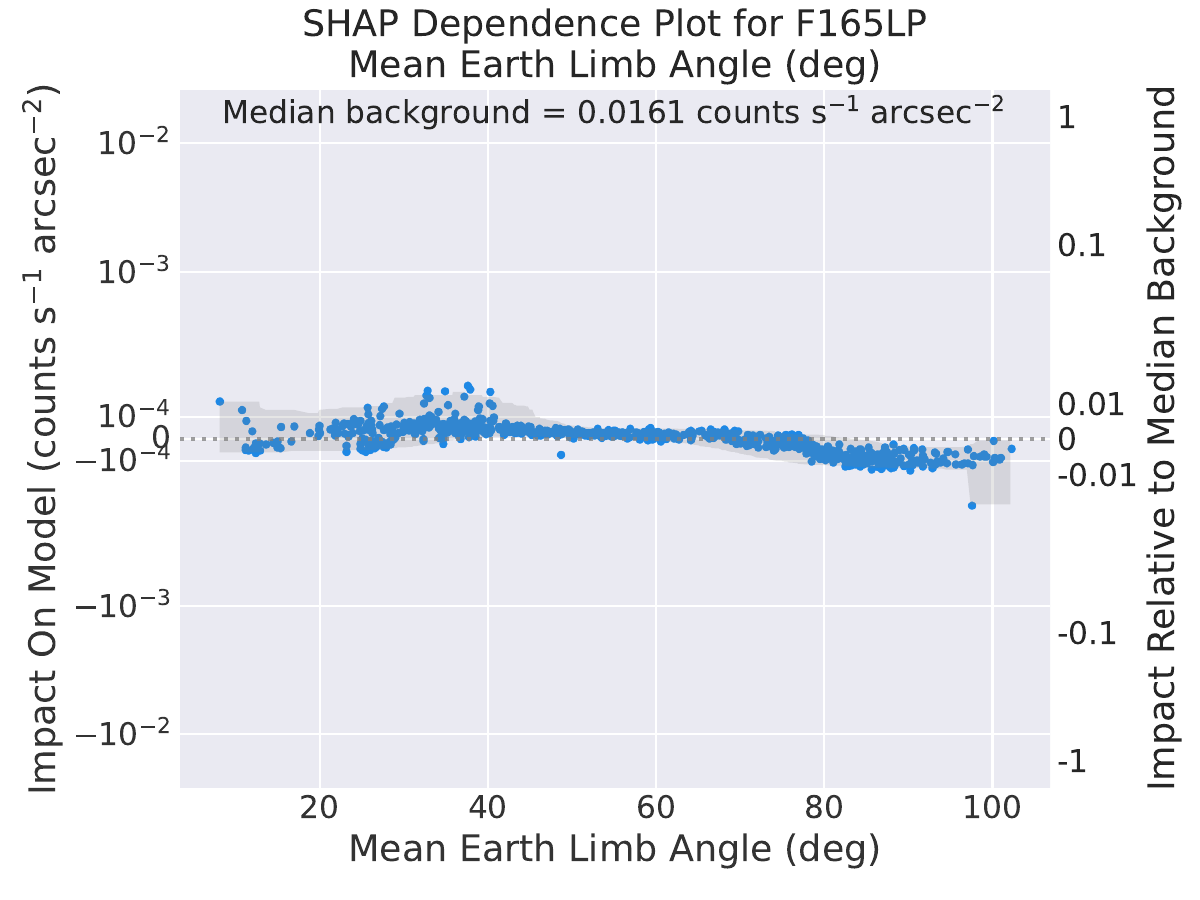}
\includegraphics[width=0.24\textwidth]{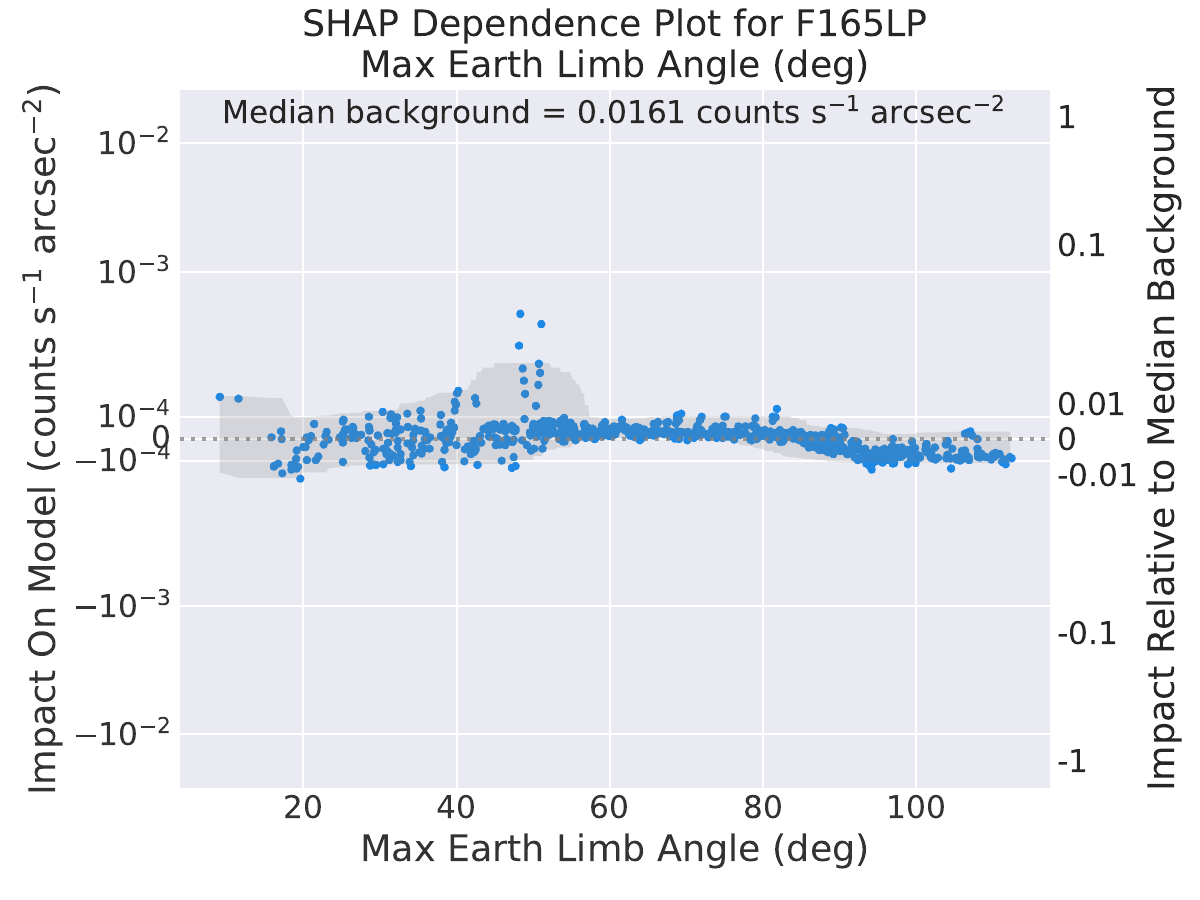}
\includegraphics[width=0.24\textwidth]{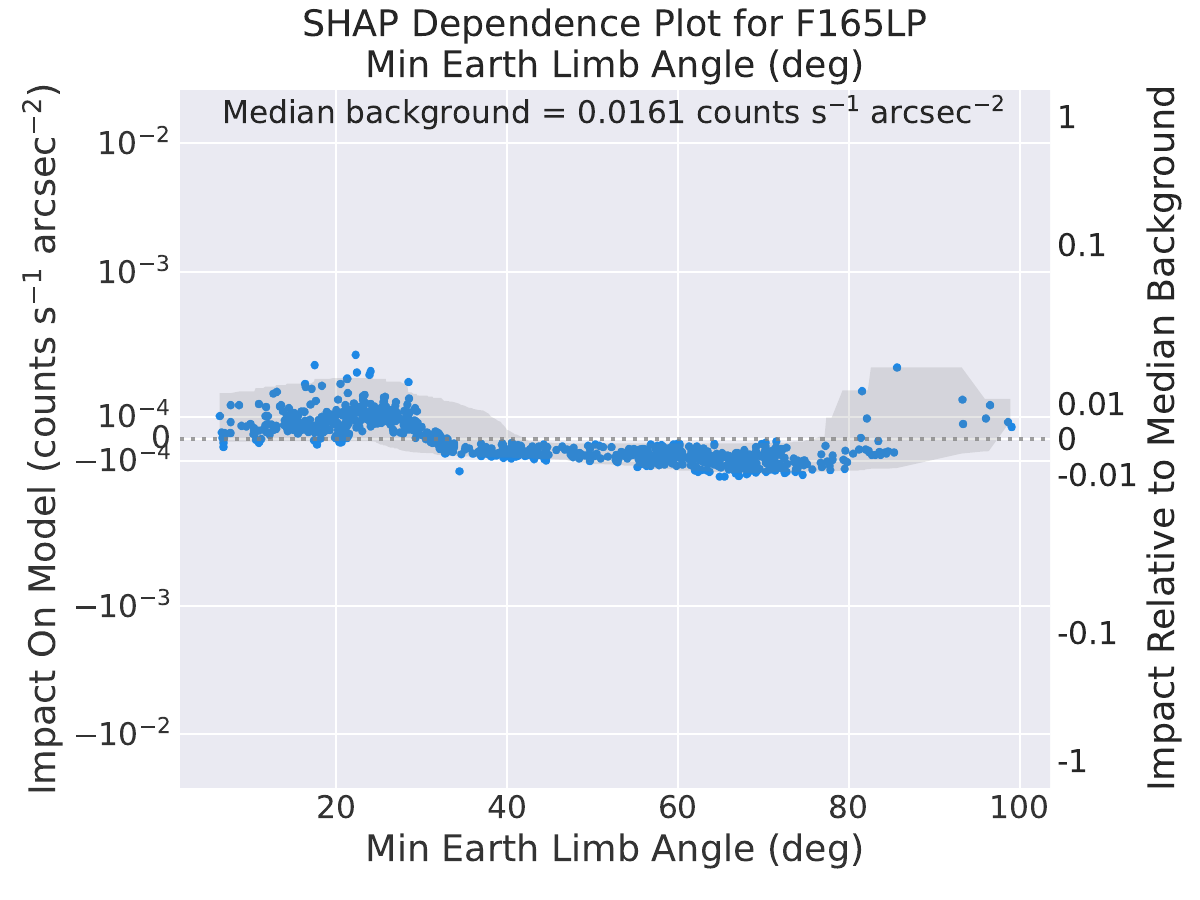}
\includegraphics[width=0.24\textwidth]{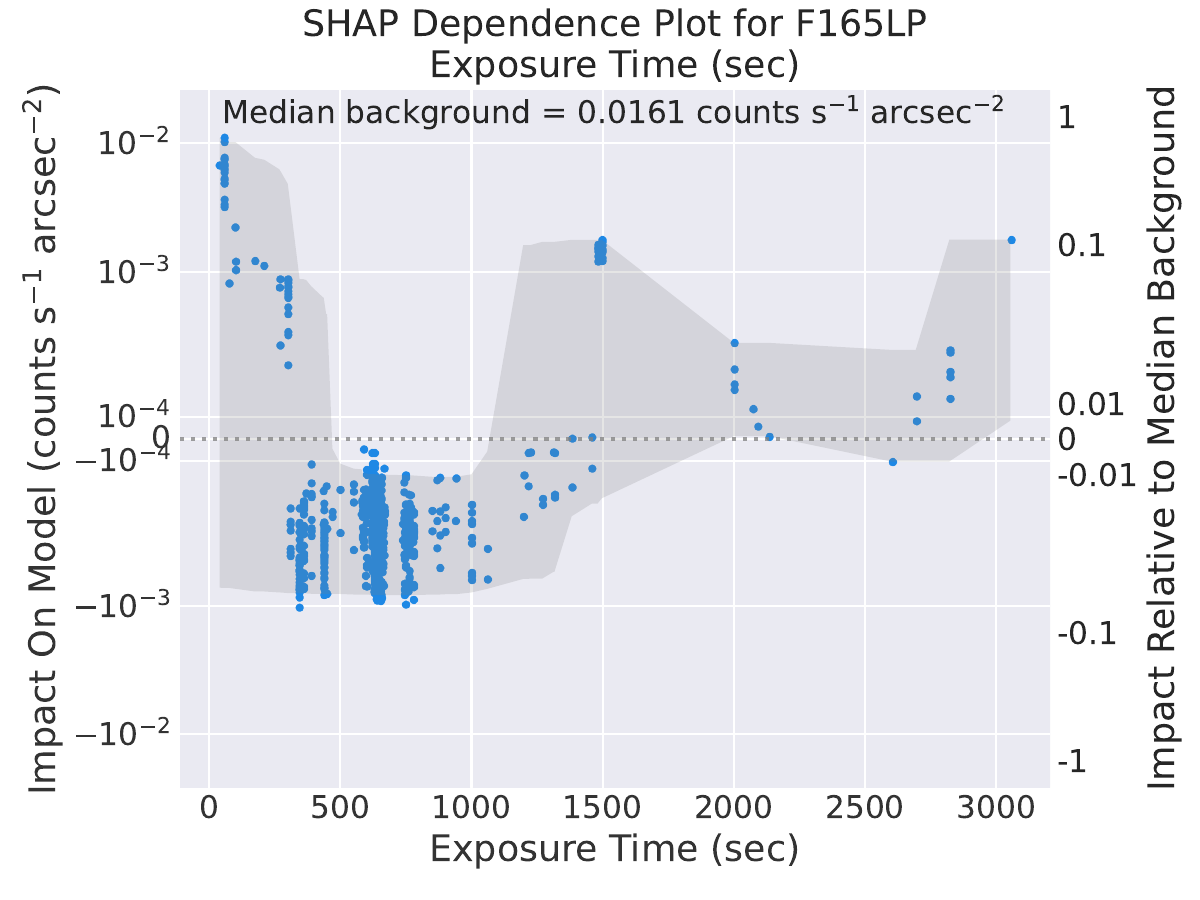}
\includegraphics[width=0.24\textwidth]{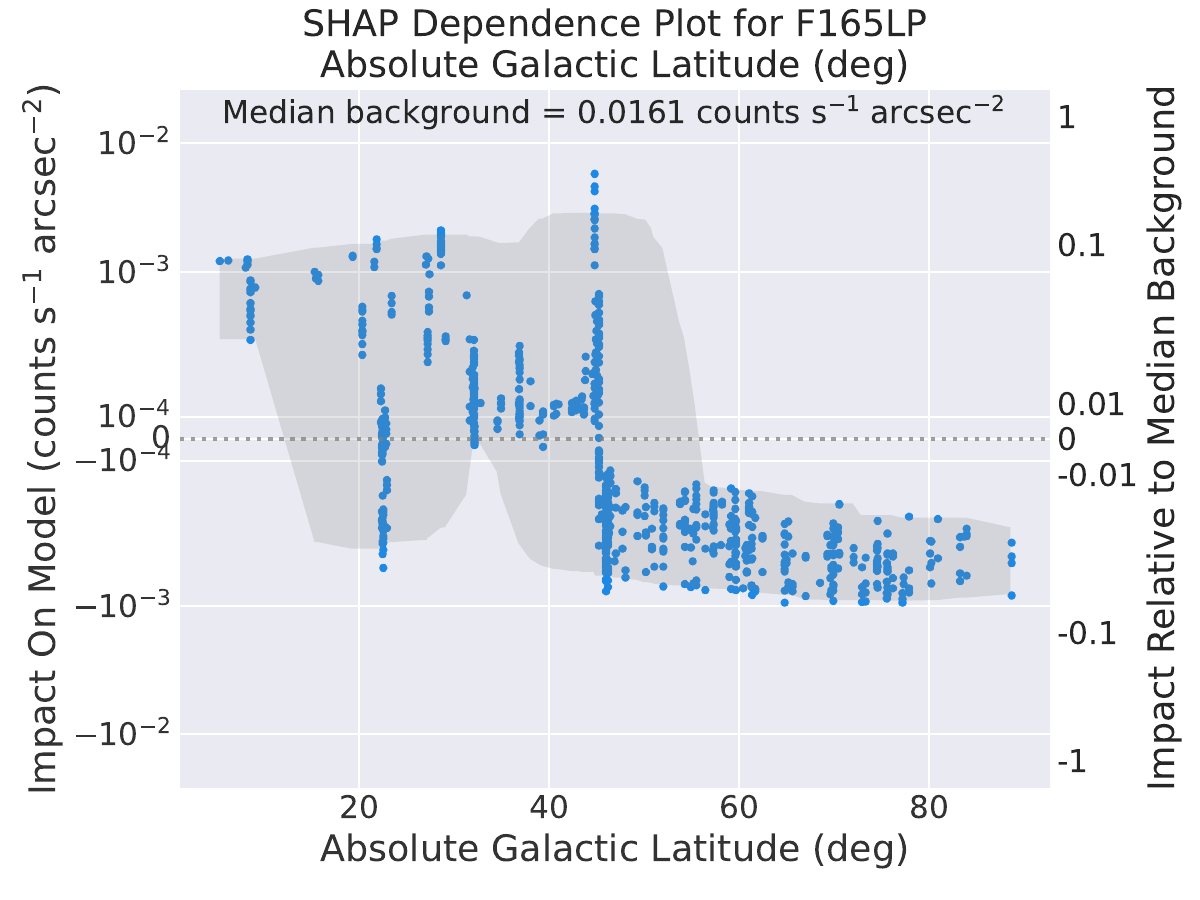}
\includegraphics[width=0.24\textwidth]{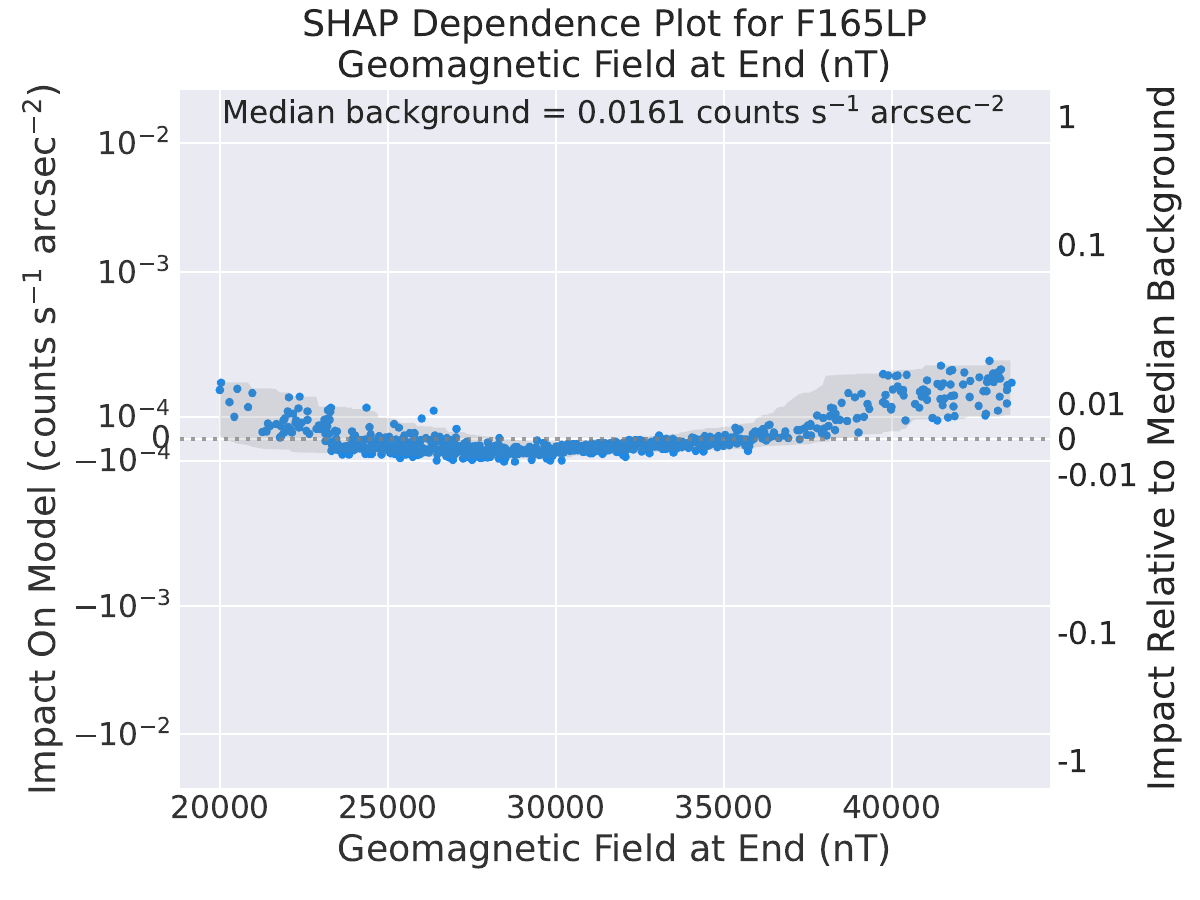}
\includegraphics[width=0.24\textwidth]{SHAP_Plots_QuantileForestRegr/F165LP_geomag_mid_SHAP_Dependence.pdf}
\includegraphics[width=0.24\textwidth]{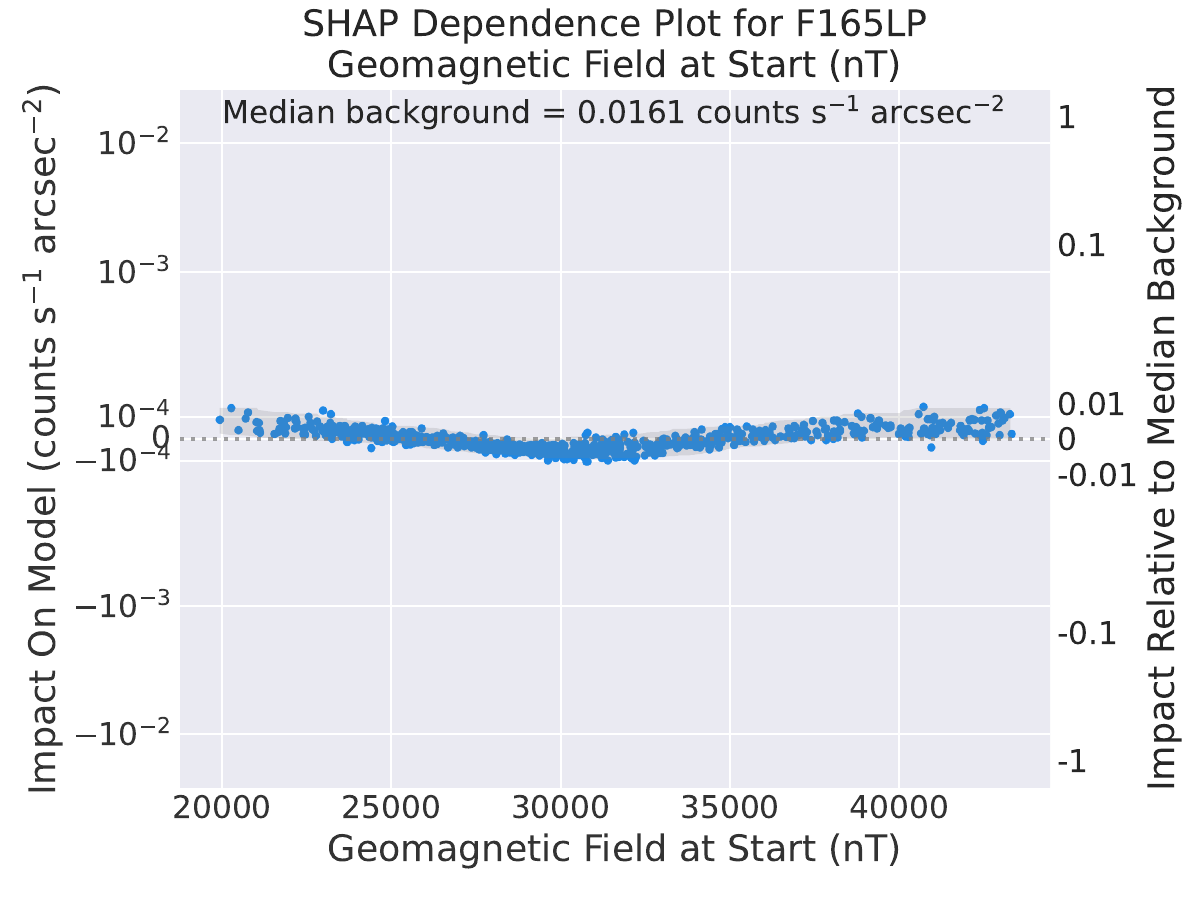}
\includegraphics[width=0.24\textwidth]{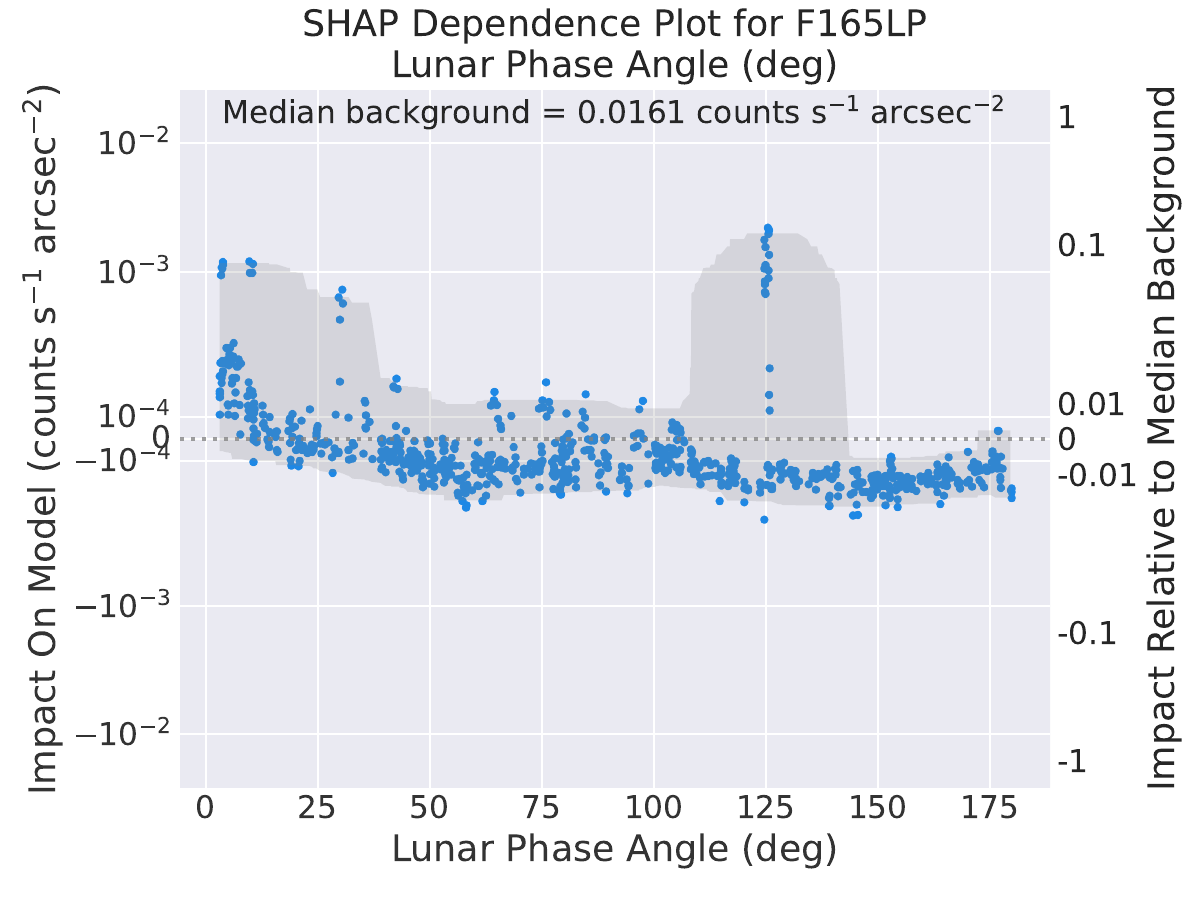}
\includegraphics[width=0.24\textwidth]{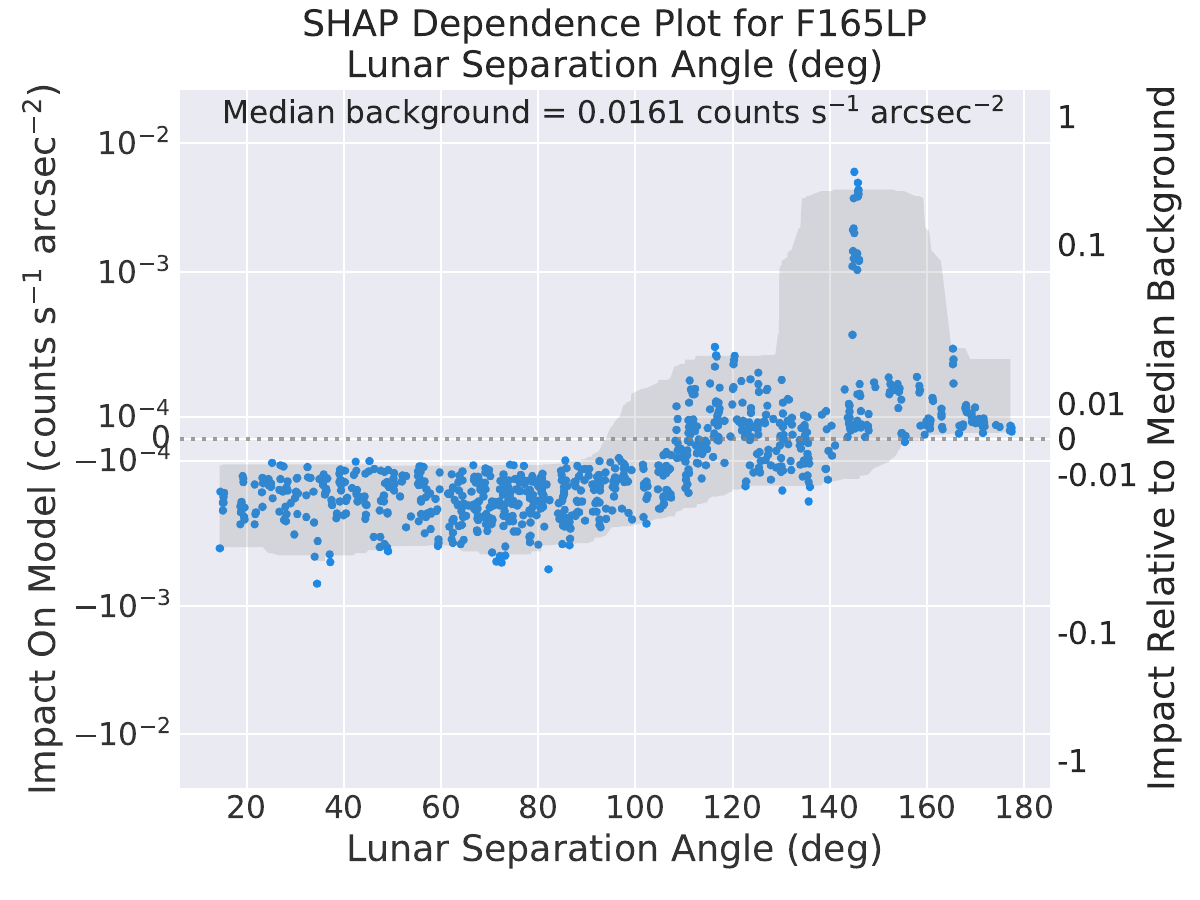}
\includegraphics[width=0.24\textwidth]{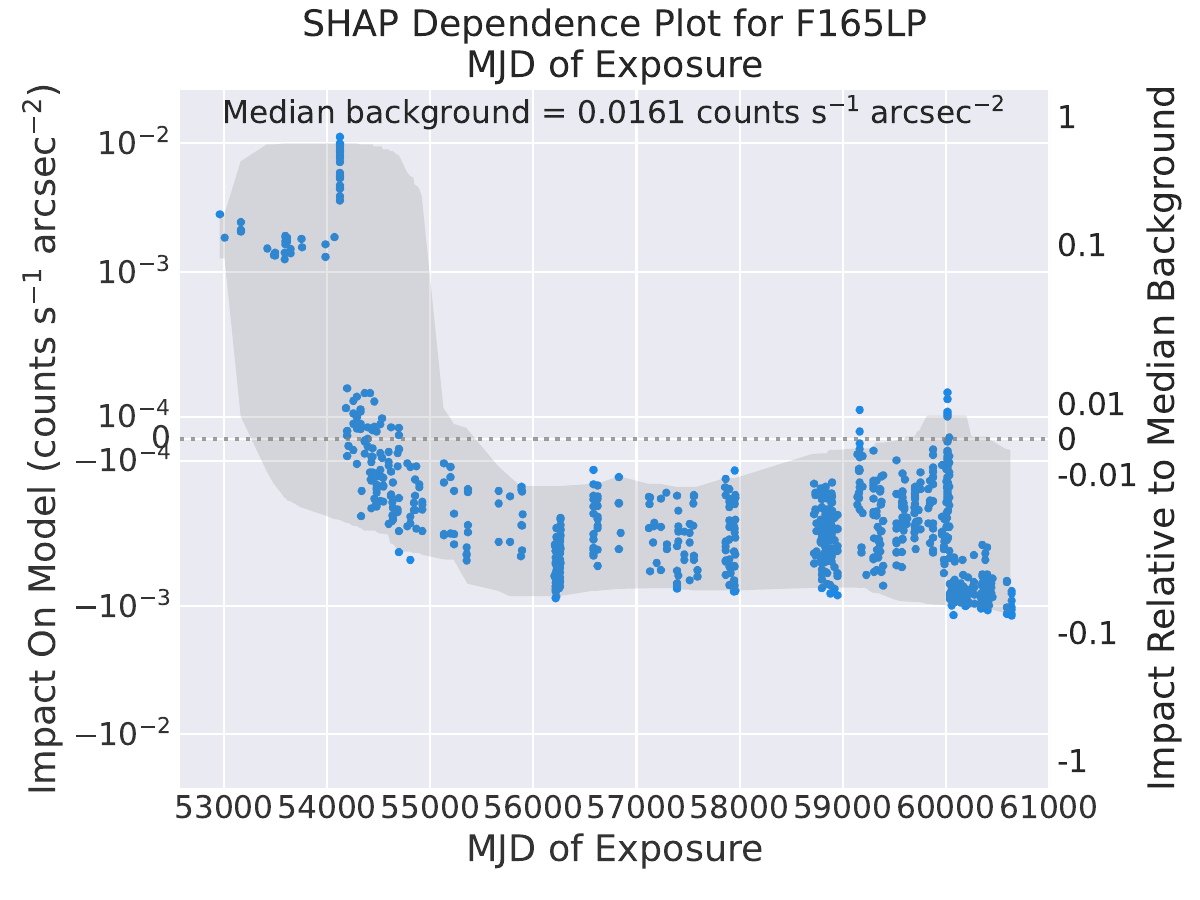}
\includegraphics[width=0.24\textwidth]{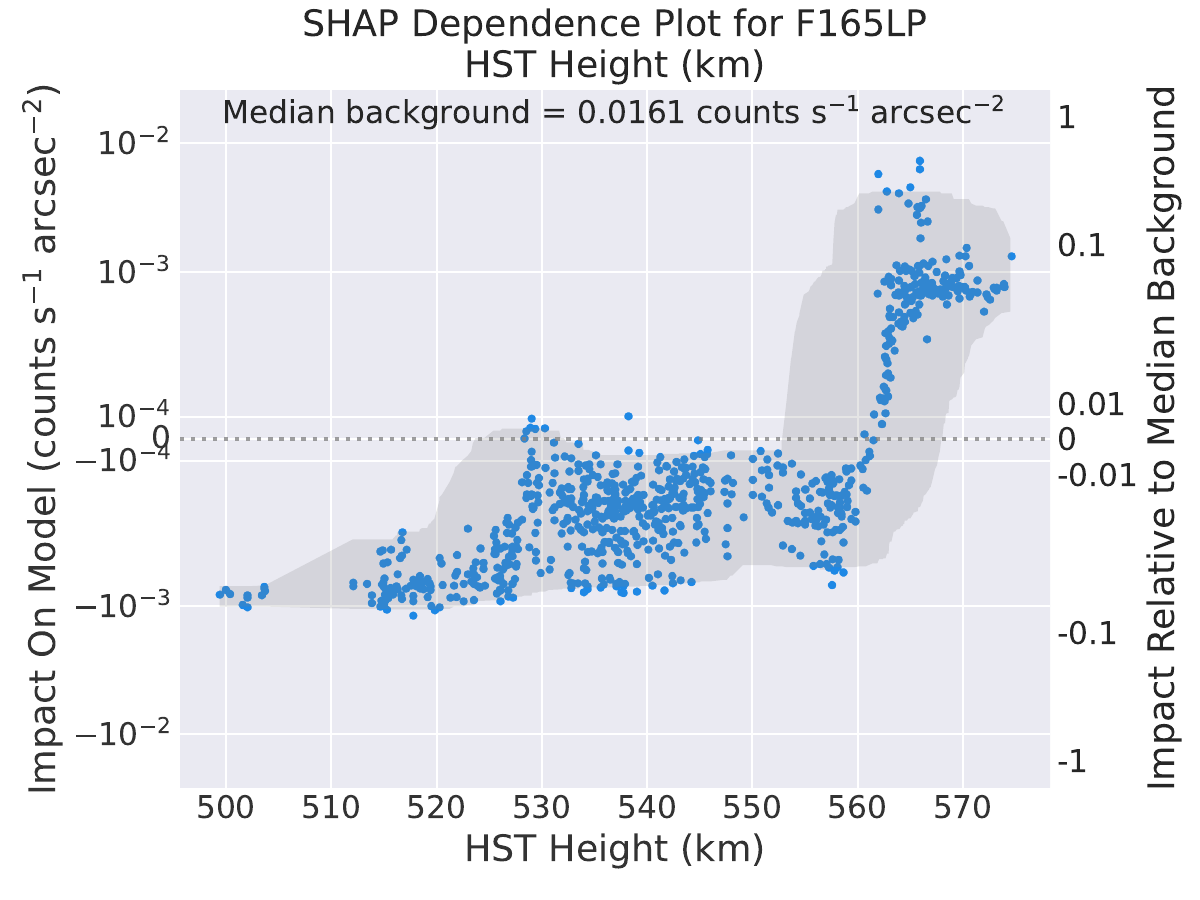}
\includegraphics[width=0.24\textwidth]{SHAP_Plots_QuantileForestRegr/F165LP_solar_alt_SHAP_Dependence.pdf}
\includegraphics[width=0.24\textwidth]{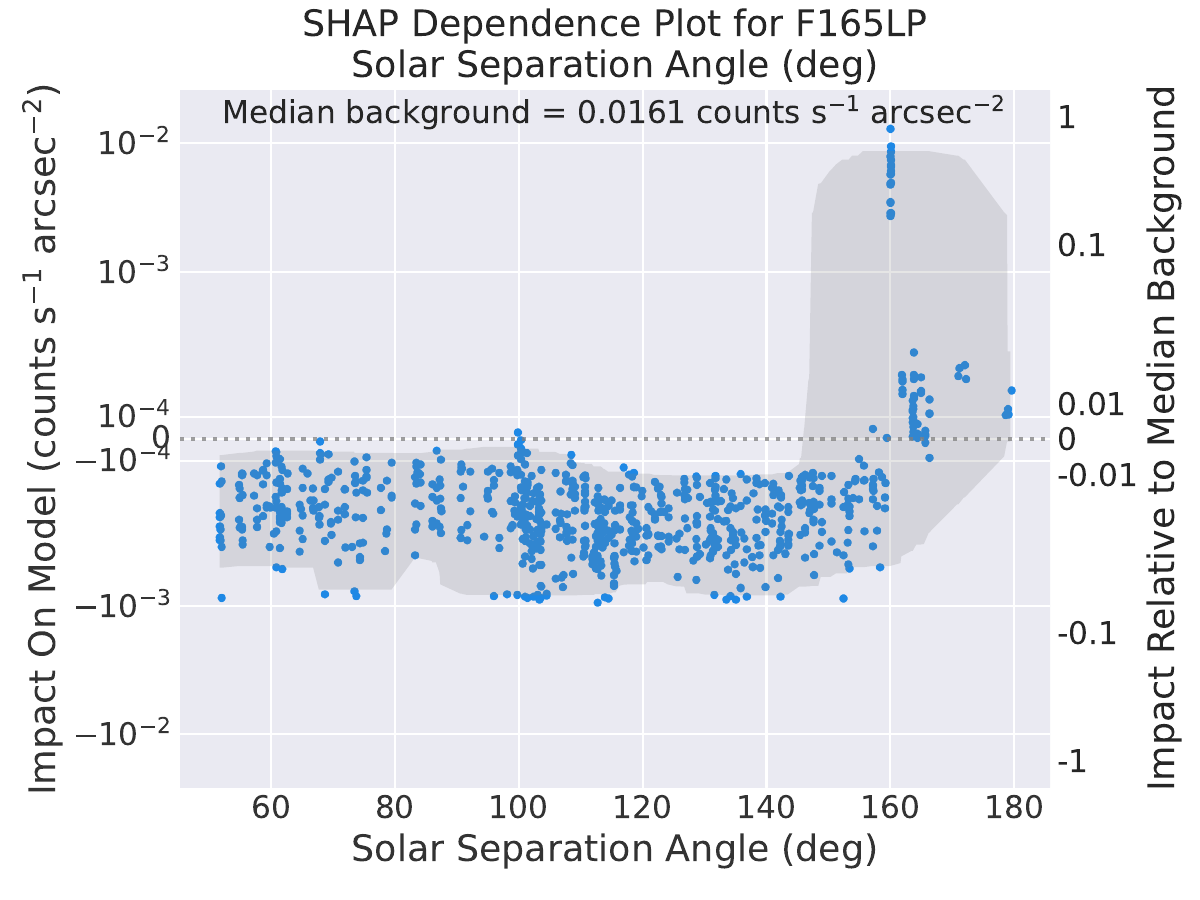}
\includegraphics[width=0.24\textwidth]{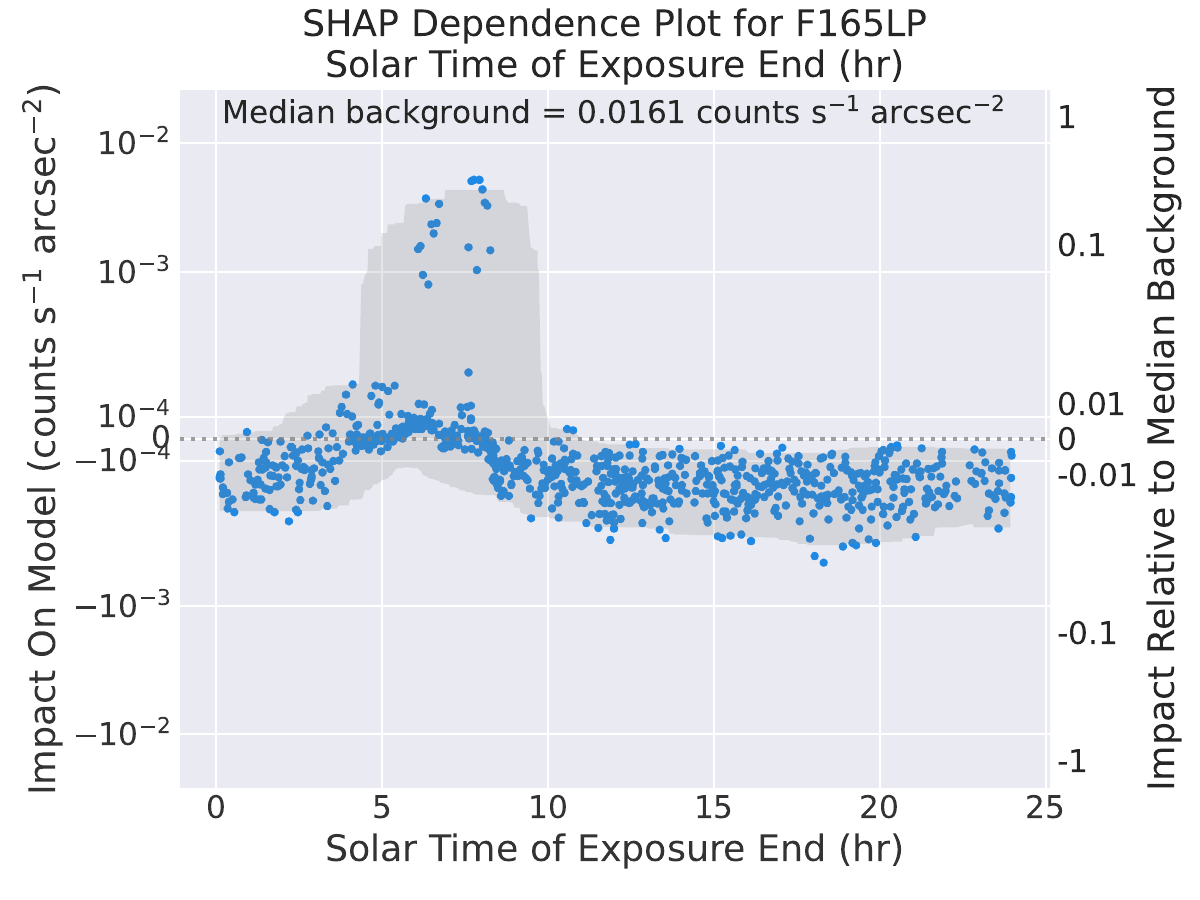}
\includegraphics[width=0.24\textwidth]{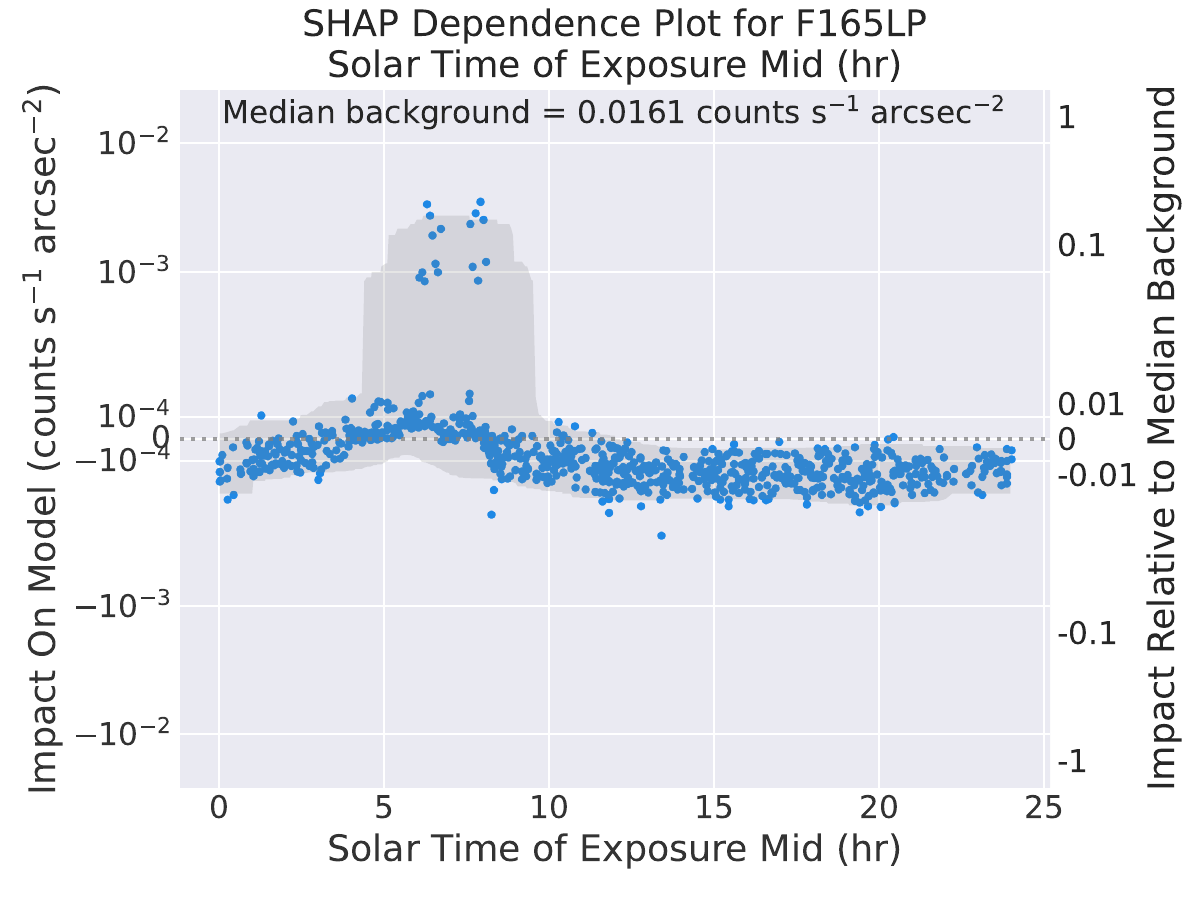}
\includegraphics[width=0.24\textwidth]{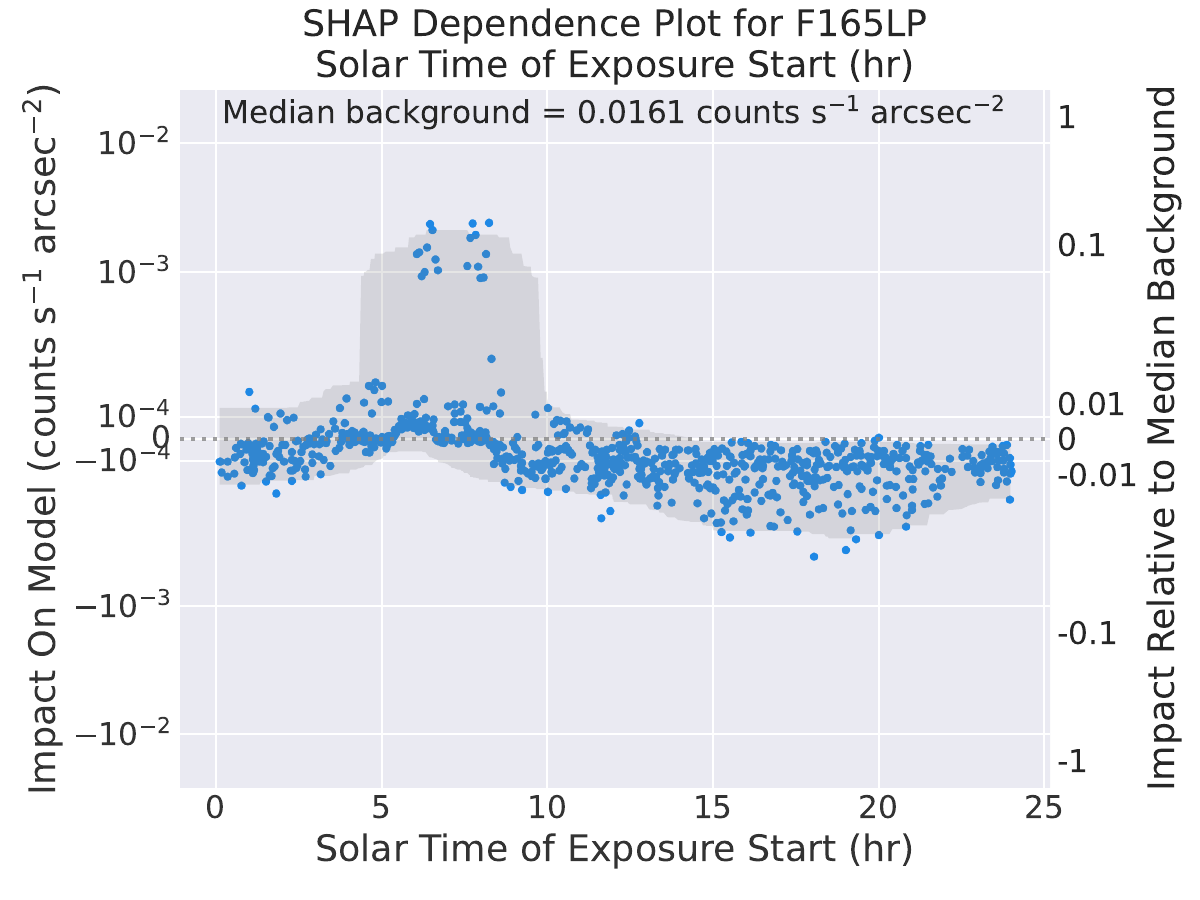}
\includegraphics[width=0.24\textwidth]{SHAP_Plots_QuantileForestRegr/F165LP_sunspots_SHAP_Dependence.pdf}
\includegraphics[width=0.24\textwidth]{SHAP_Plots_QuantileForestRegr/F165LP_temp_end_SHAP_Dependence.pdf}
\includegraphics[width=0.24\textwidth]{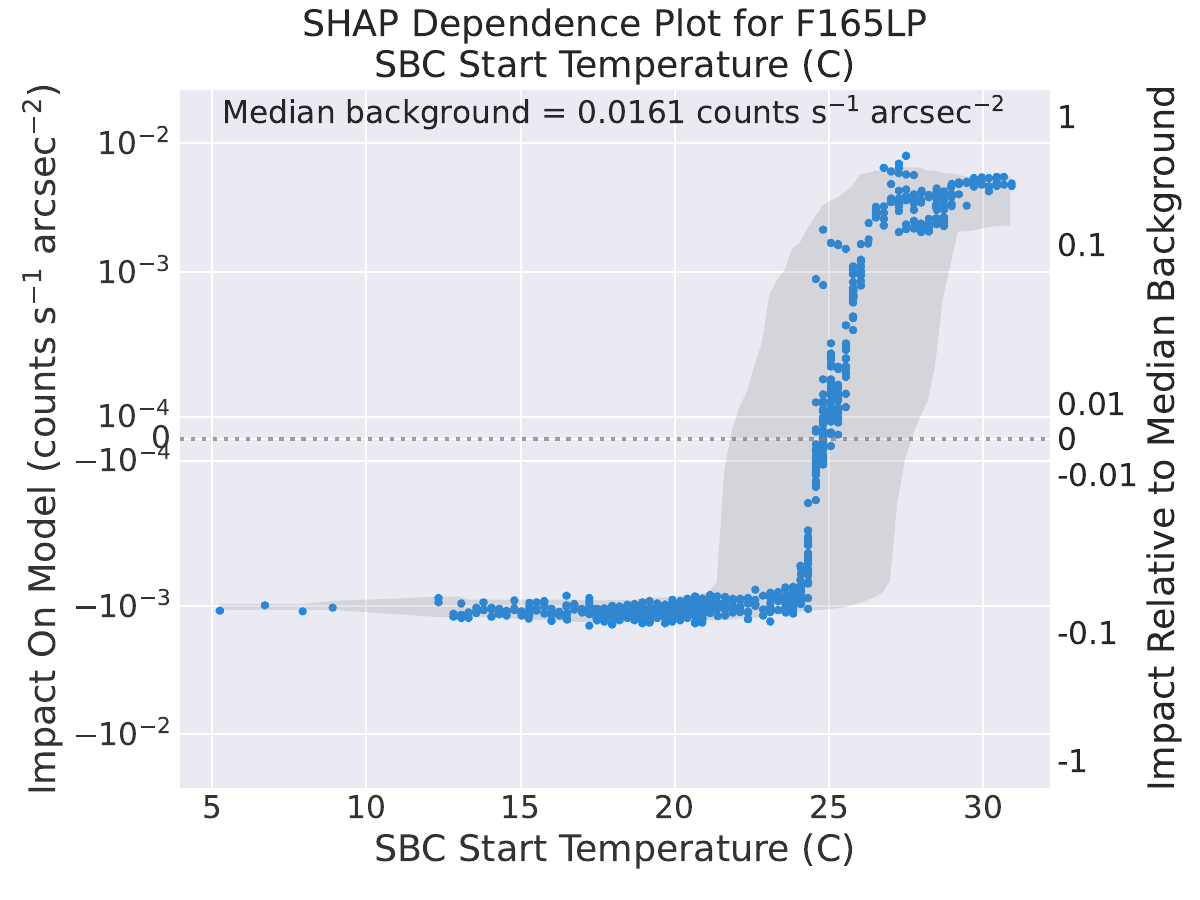}
\caption{SHAP dependence plotsfor QRF regression modeling of F165LP. Otherwise as per Figure~\ref{Fig:SHAP_Dependence_F115LP}.}
\label{Fig:SHAP_Dependence_F165LP}
\end{figure}

\begin{figure}
\centering
\includegraphics[width=0.24\textwidth]{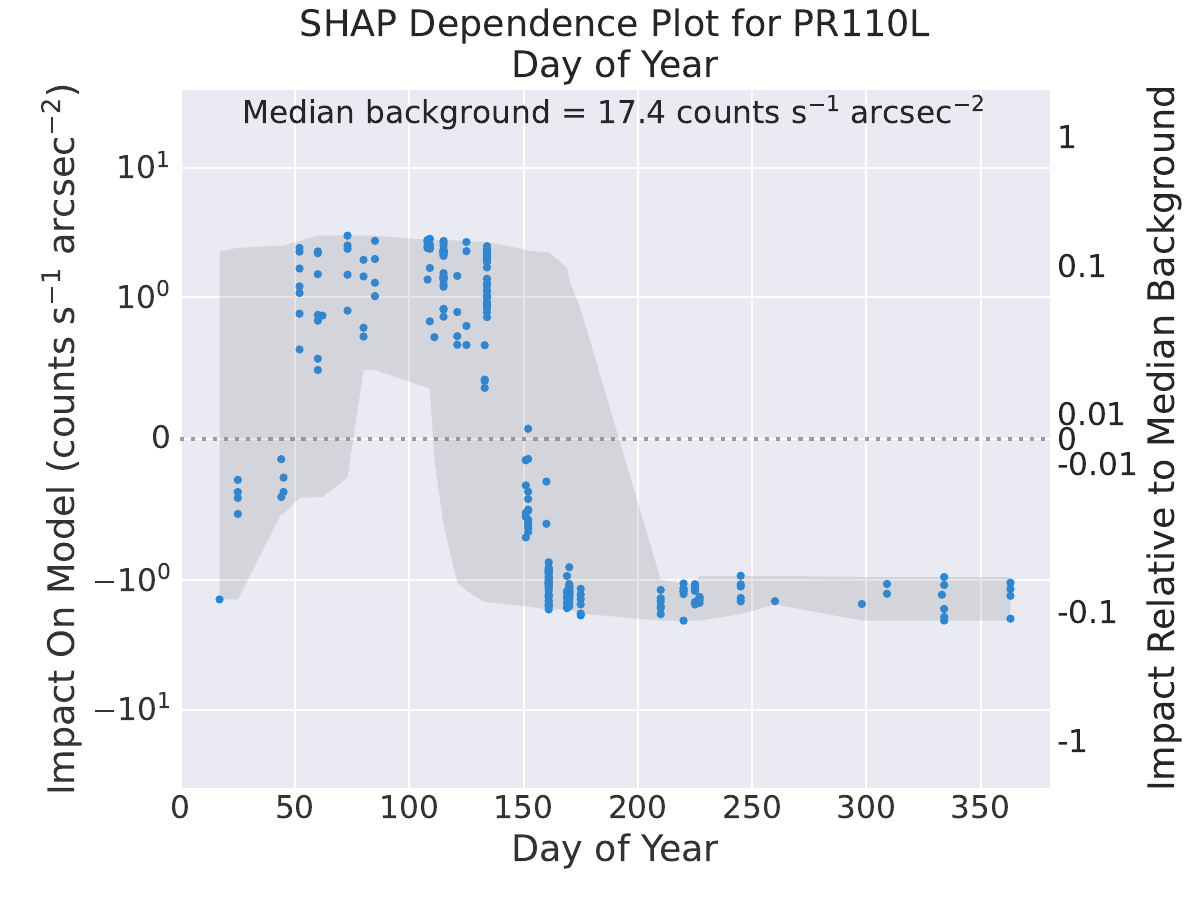}
\includegraphics[width=0.24\textwidth]{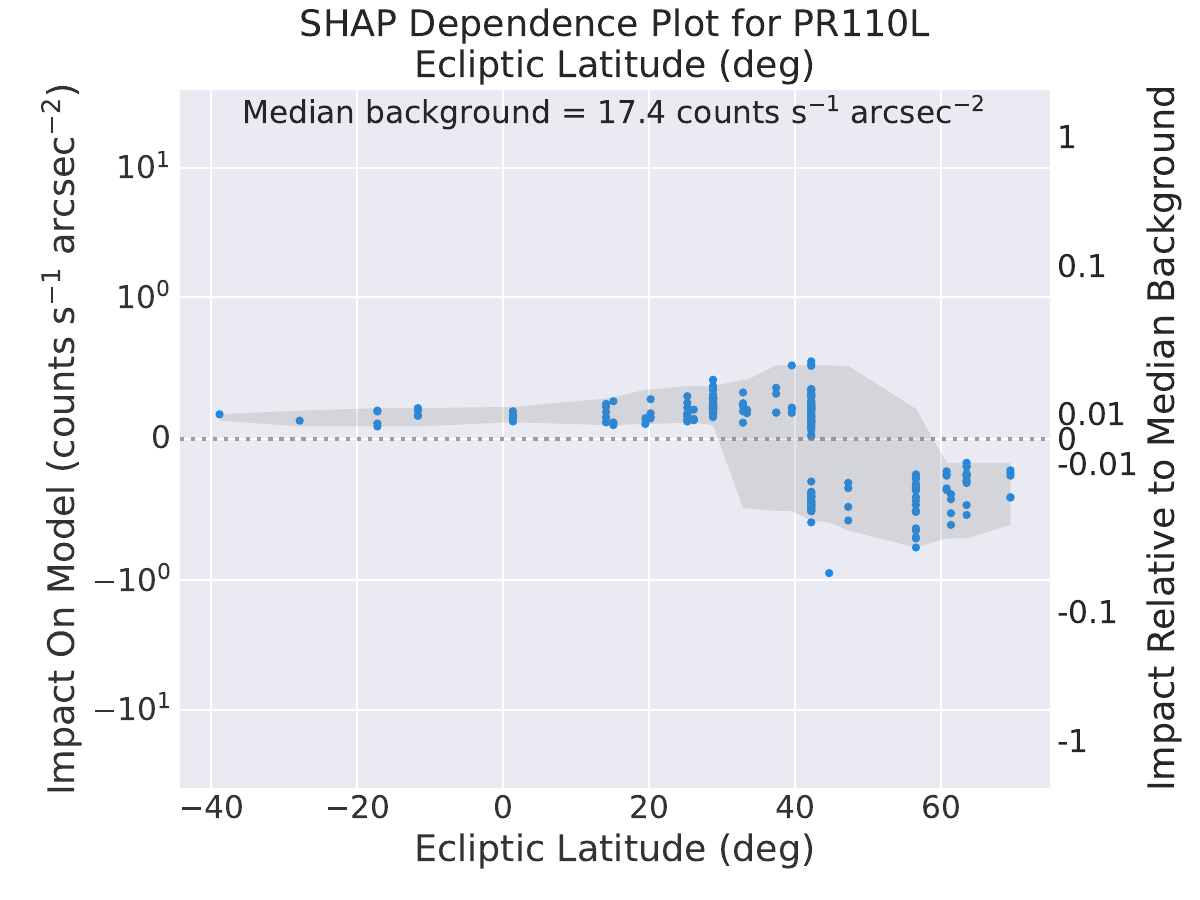}
\includegraphics[width=0.24\textwidth]{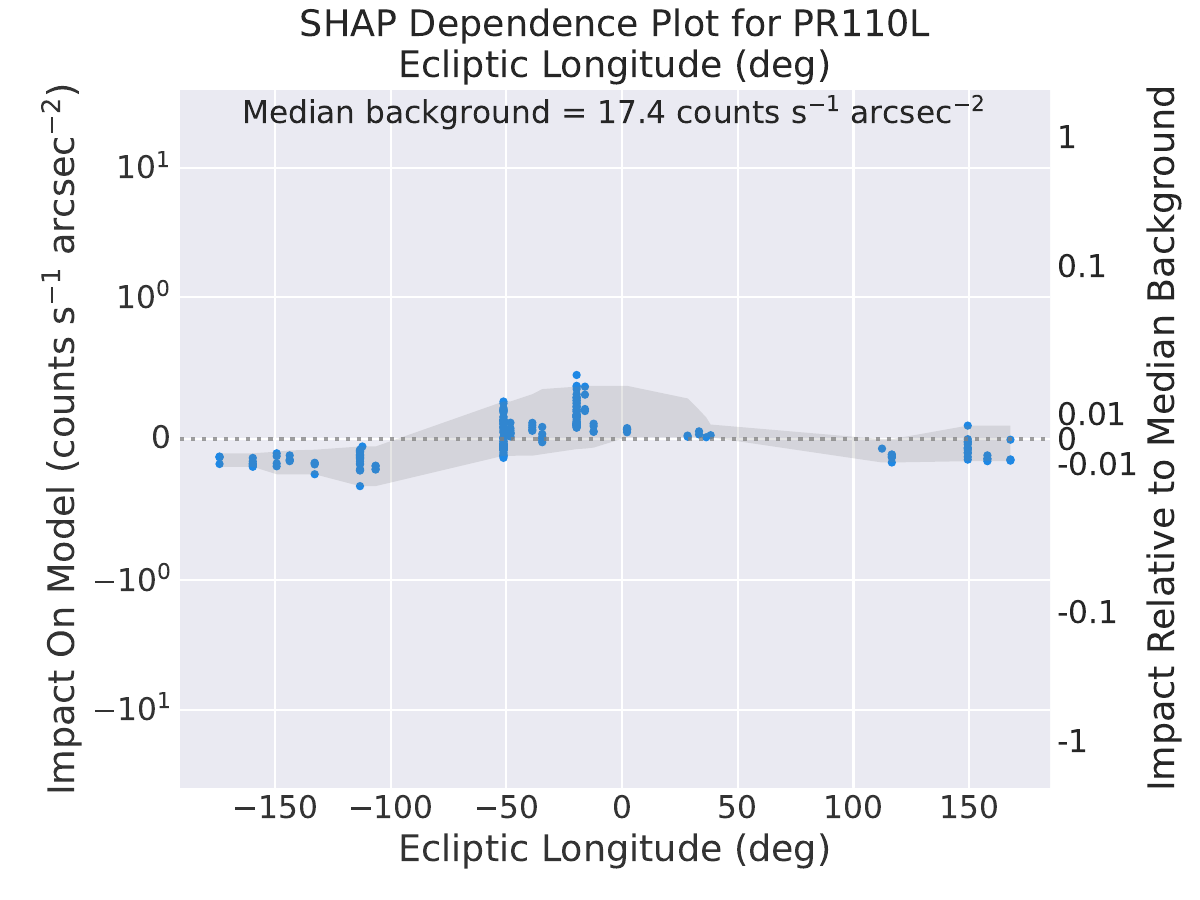}
\includegraphics[width=0.24\textwidth]{SHAP_Plots_QuantileForestRegr/PR110L_ela_avg_SHAP_Dependence.pdf}
\includegraphics[width=0.24\textwidth]{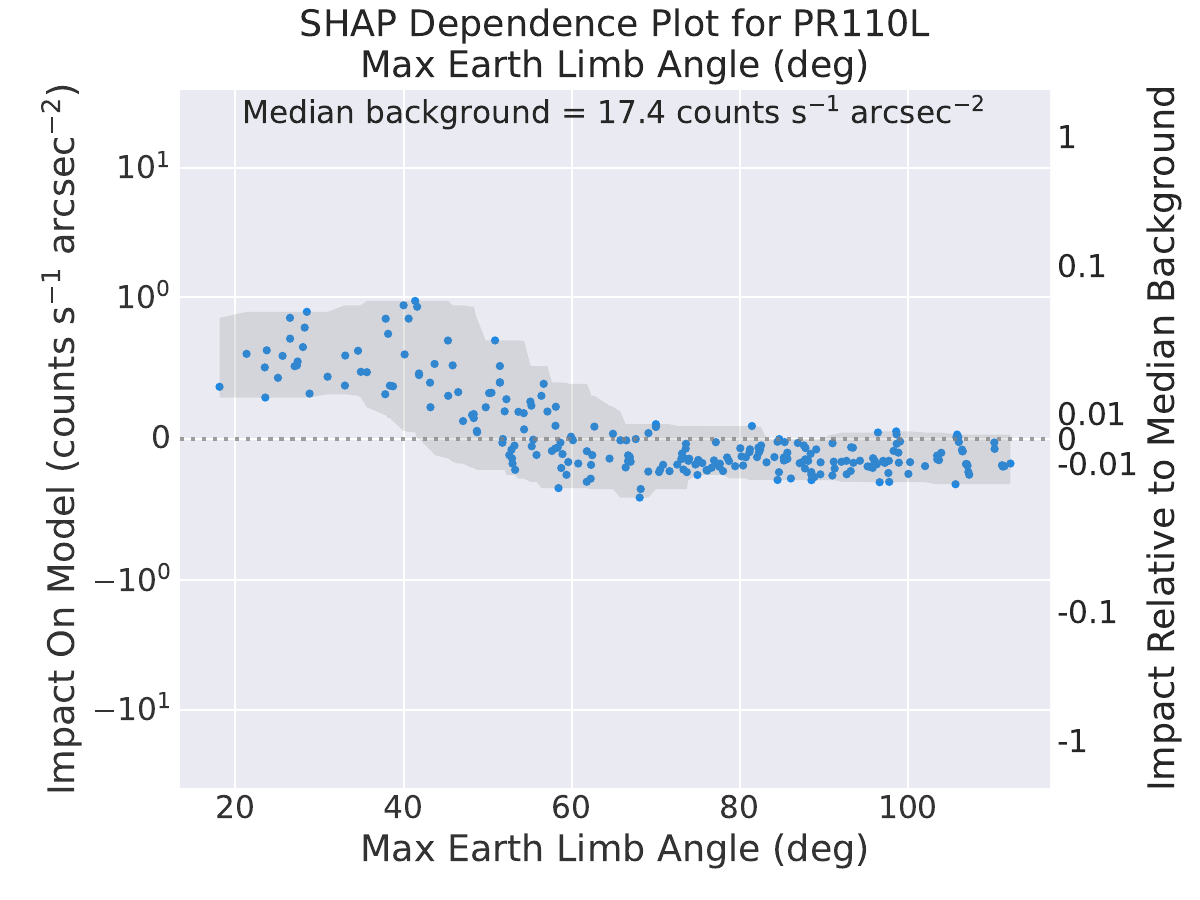}
\includegraphics[width=0.24\textwidth]{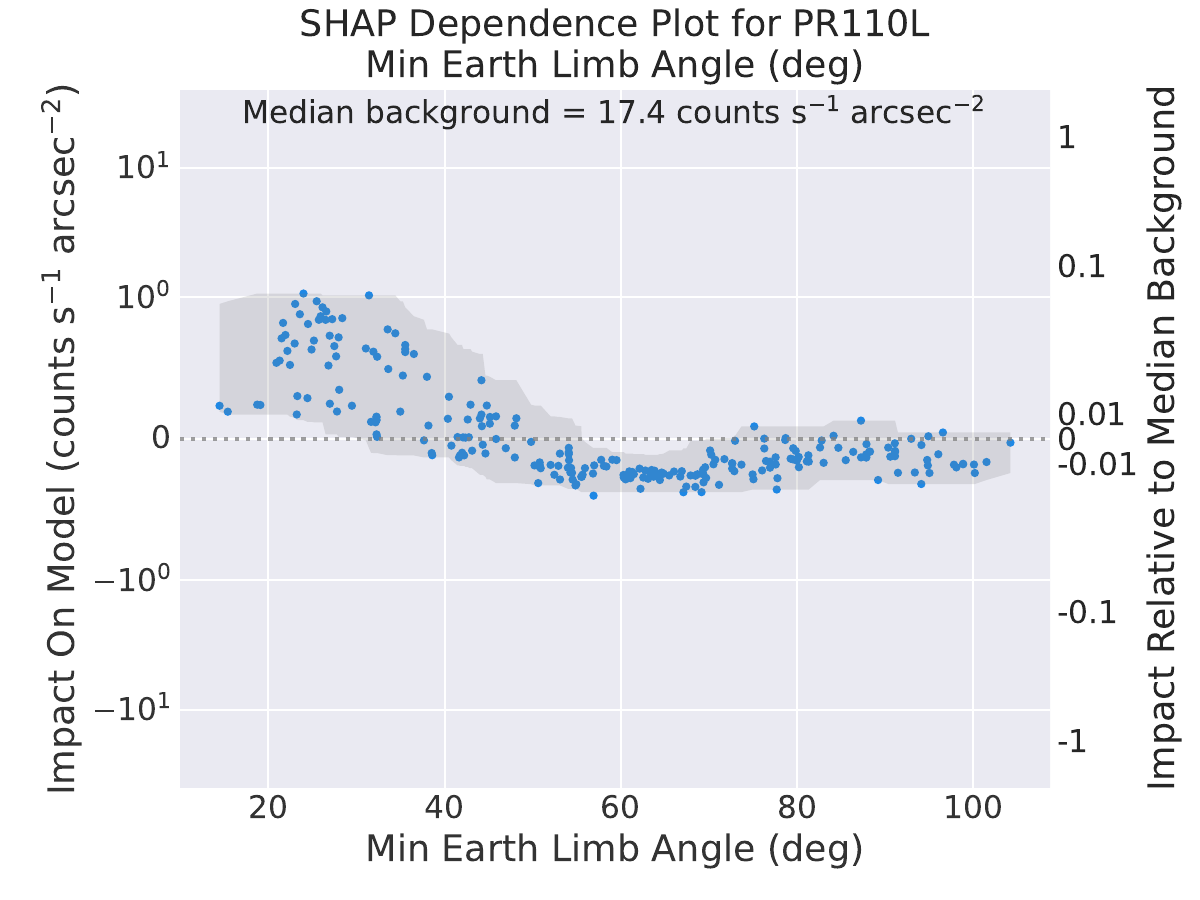}
\includegraphics[width=0.24\textwidth]{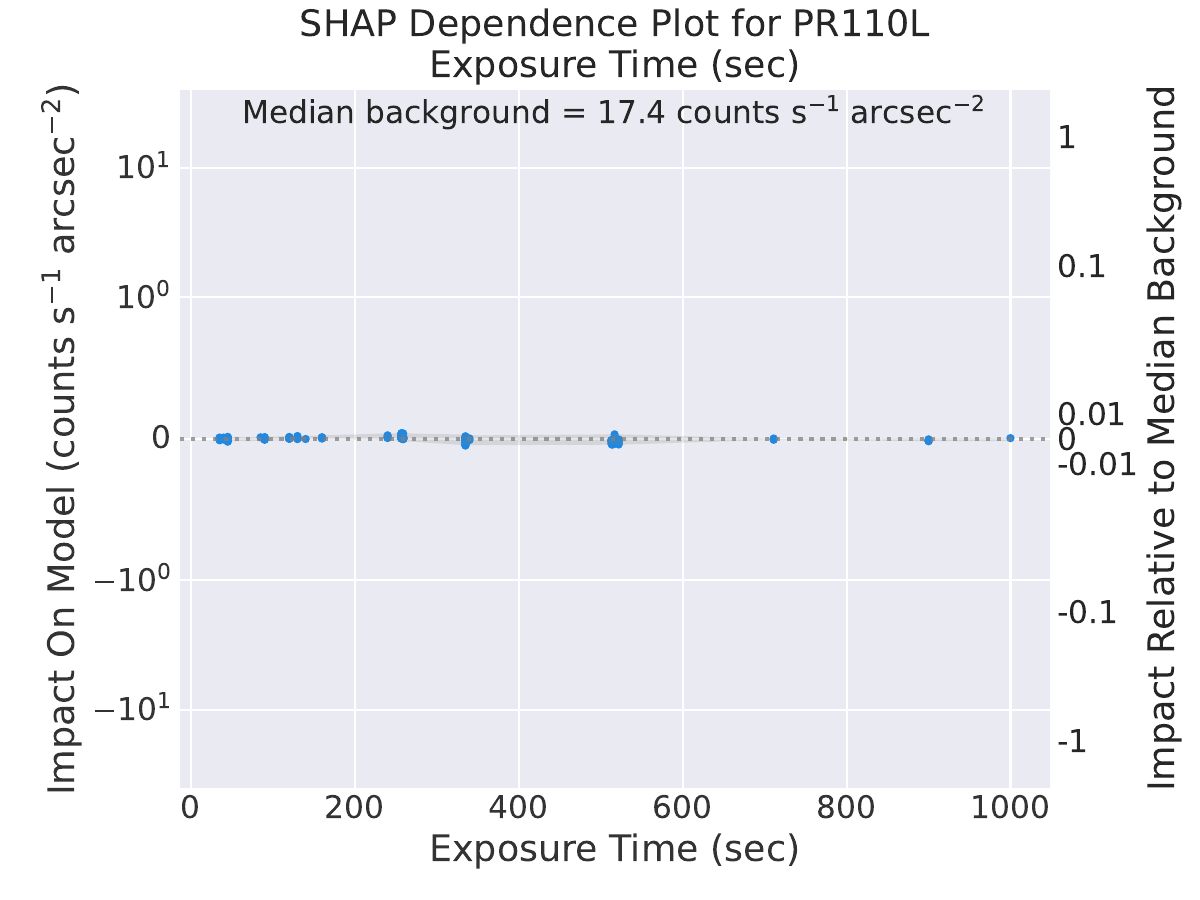}
\includegraphics[width=0.24\textwidth]{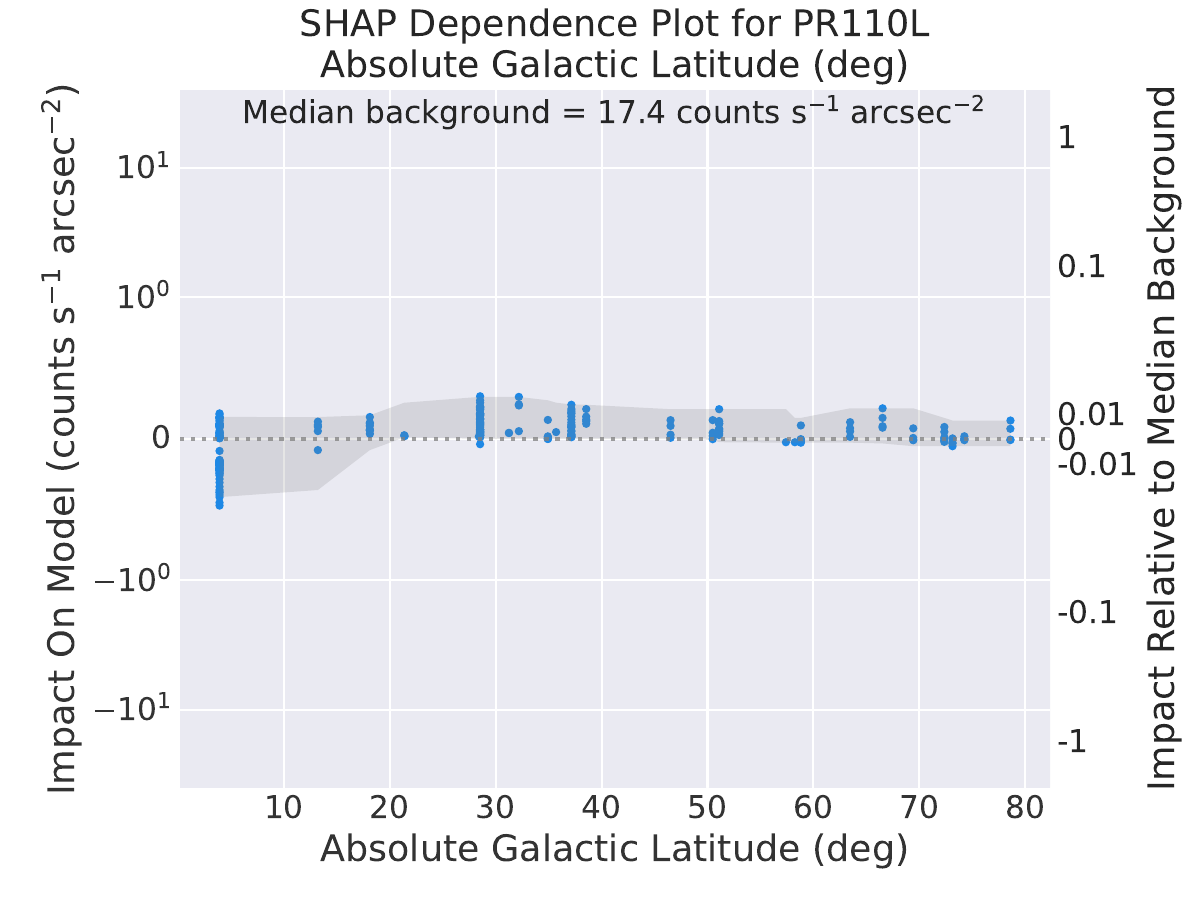}
\includegraphics[width=0.24\textwidth]{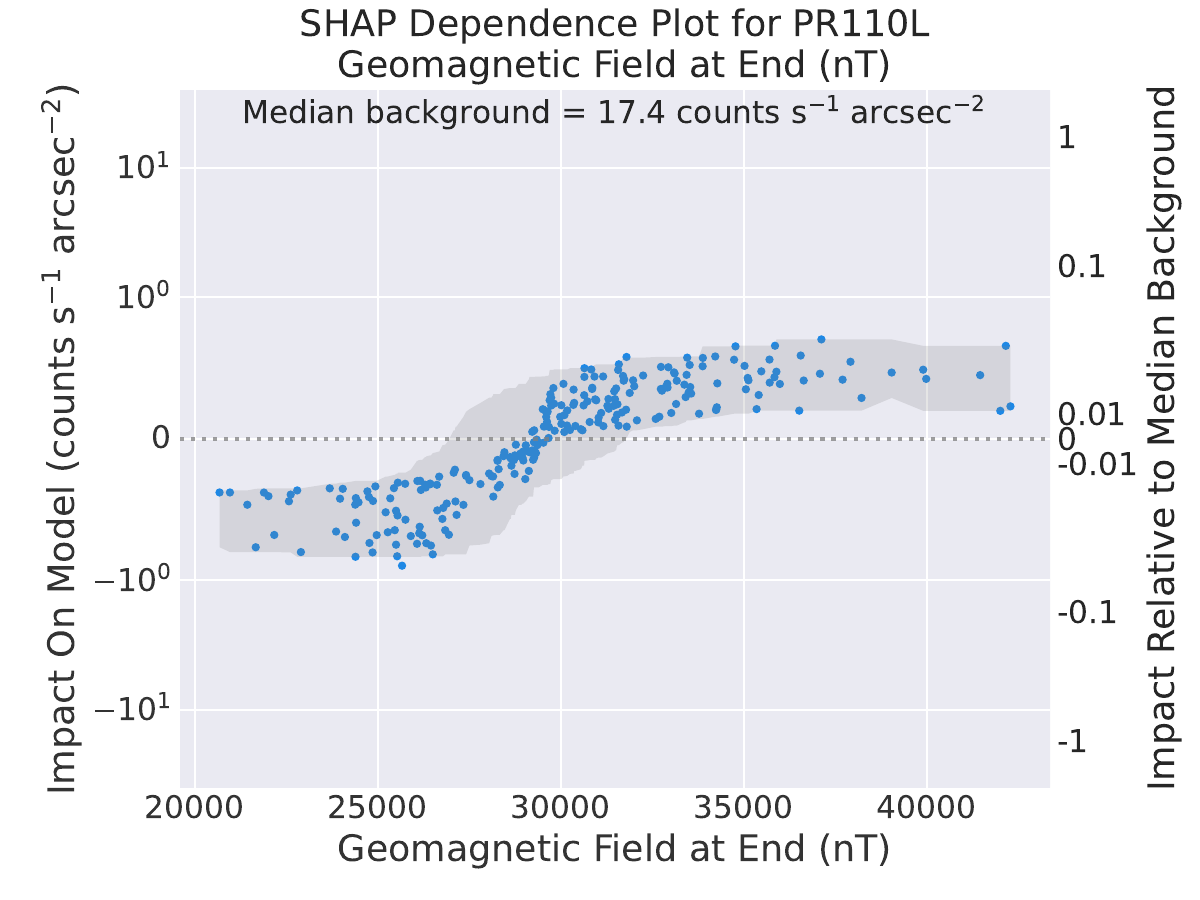}
\includegraphics[width=0.24\textwidth]{SHAP_Plots_QuantileForestRegr/PR110L_geomag_mid_SHAP_Dependence.pdf}
\includegraphics[width=0.24\textwidth]{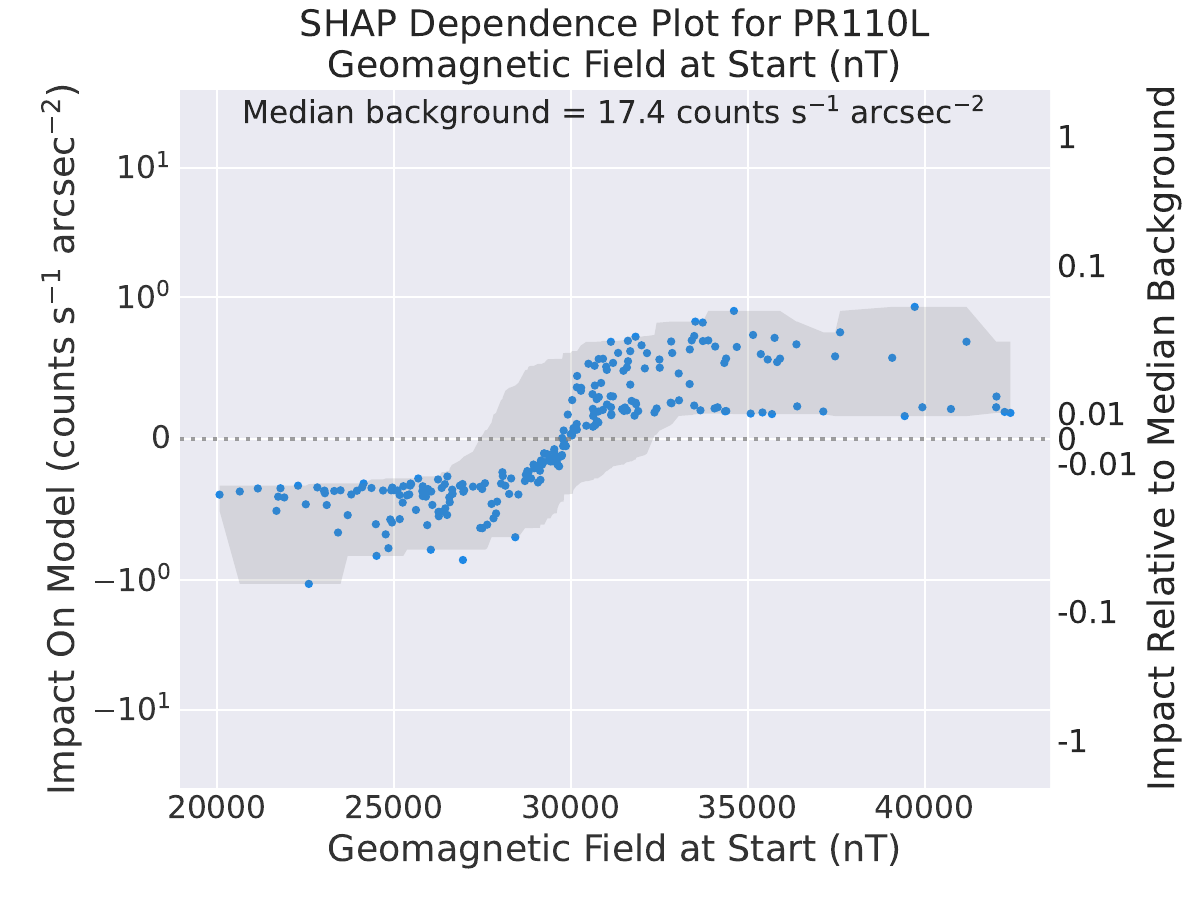}
\includegraphics[width=0.24\textwidth]{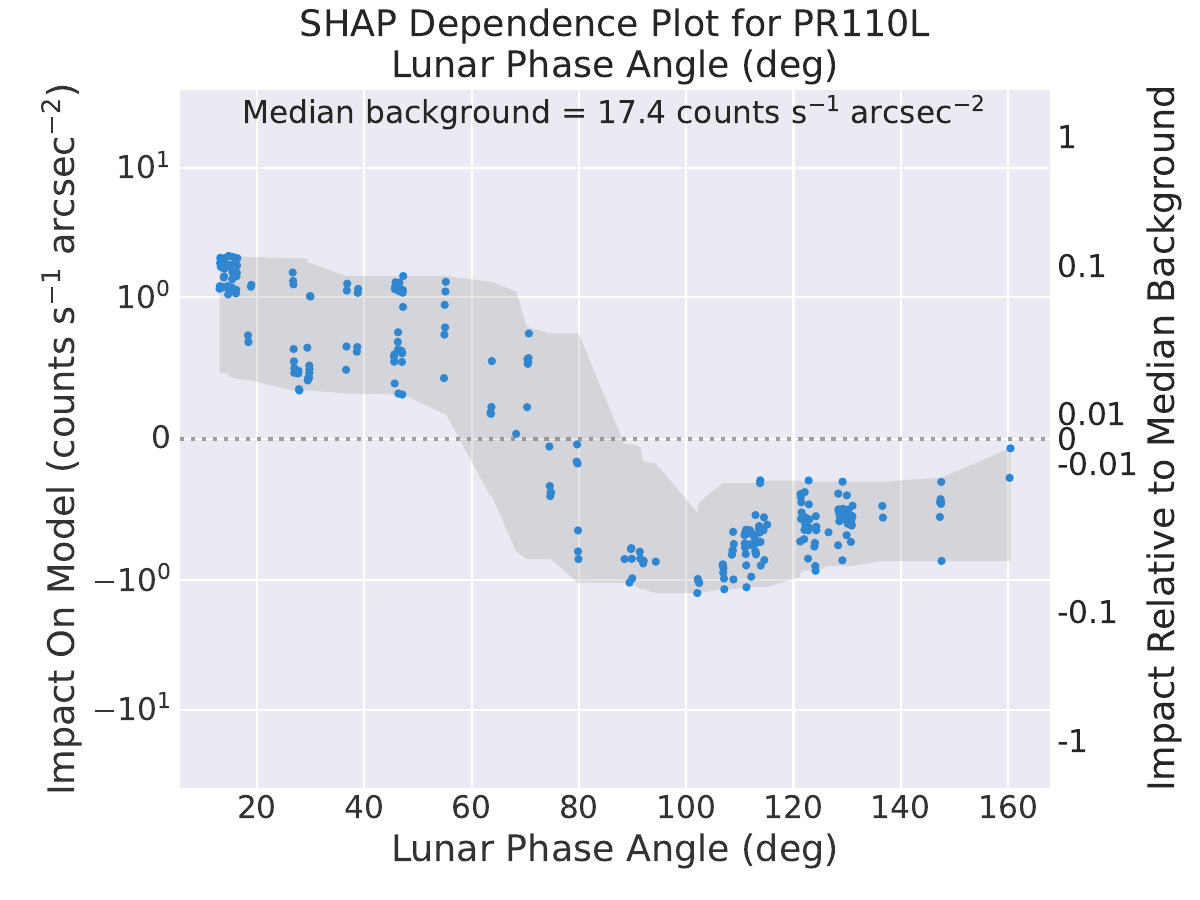}
\includegraphics[width=0.24\textwidth]{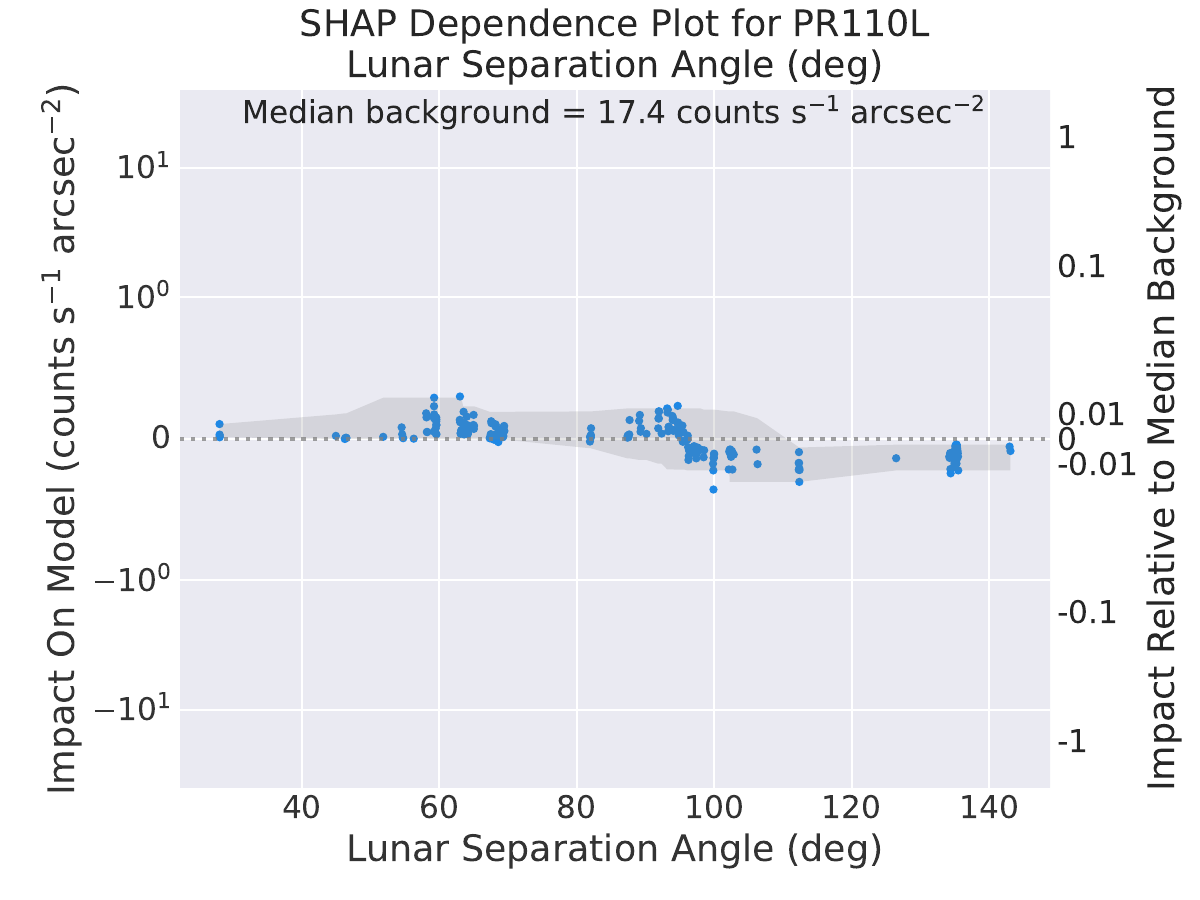}
\includegraphics[width=0.24\textwidth]{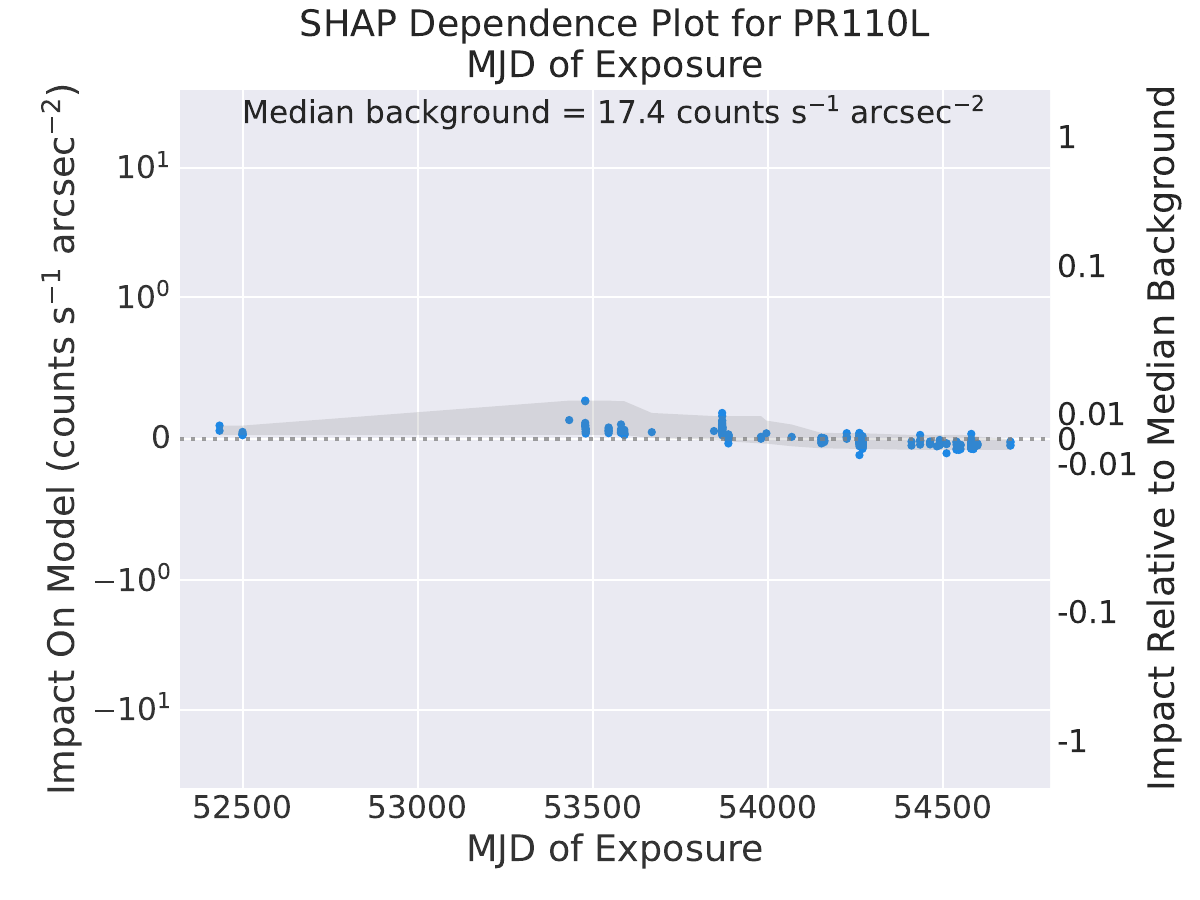}
\includegraphics[width=0.24\textwidth]{SHAP_Plots_QuantileForestRegr/PR110L_orbit_alt_SHAP_Dependence.pdf}
\includegraphics[width=0.24\textwidth]{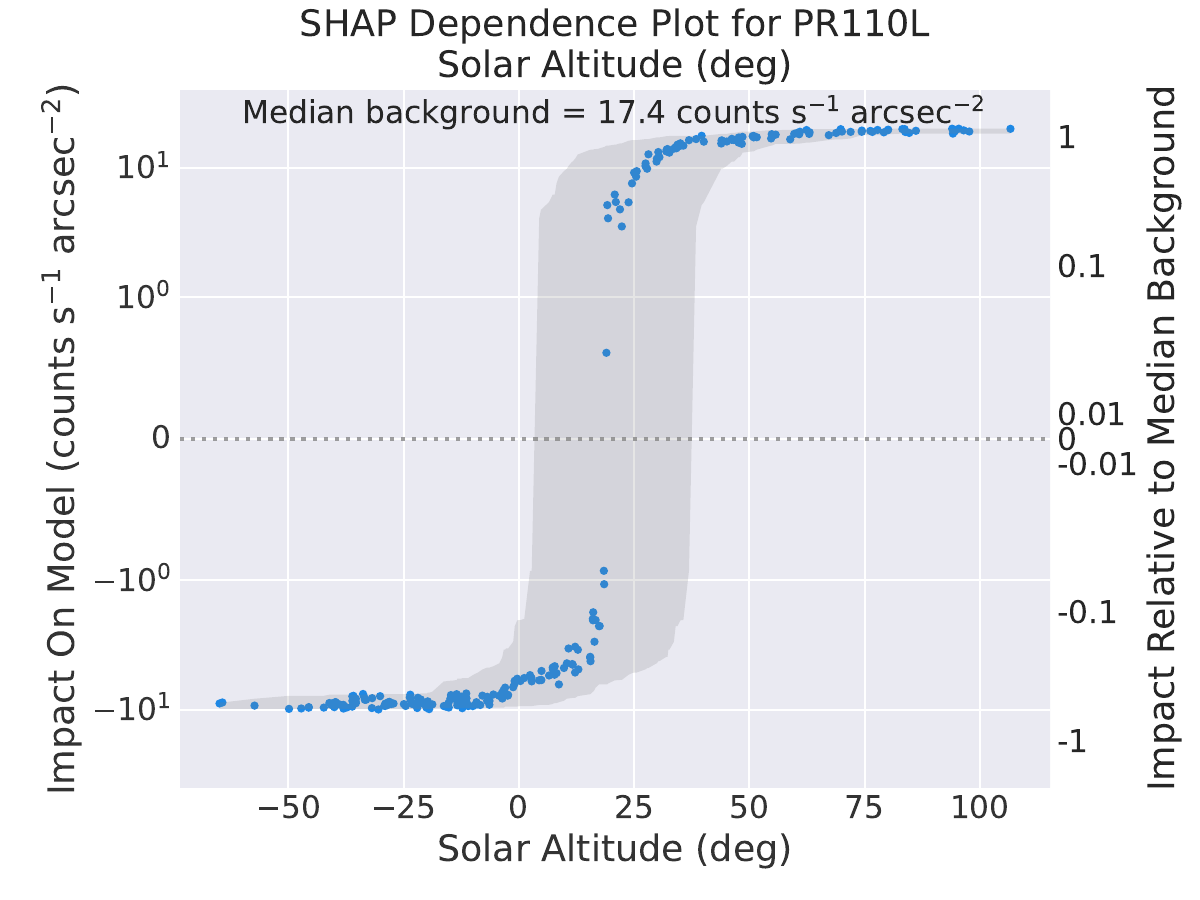}
\includegraphics[width=0.24\textwidth]{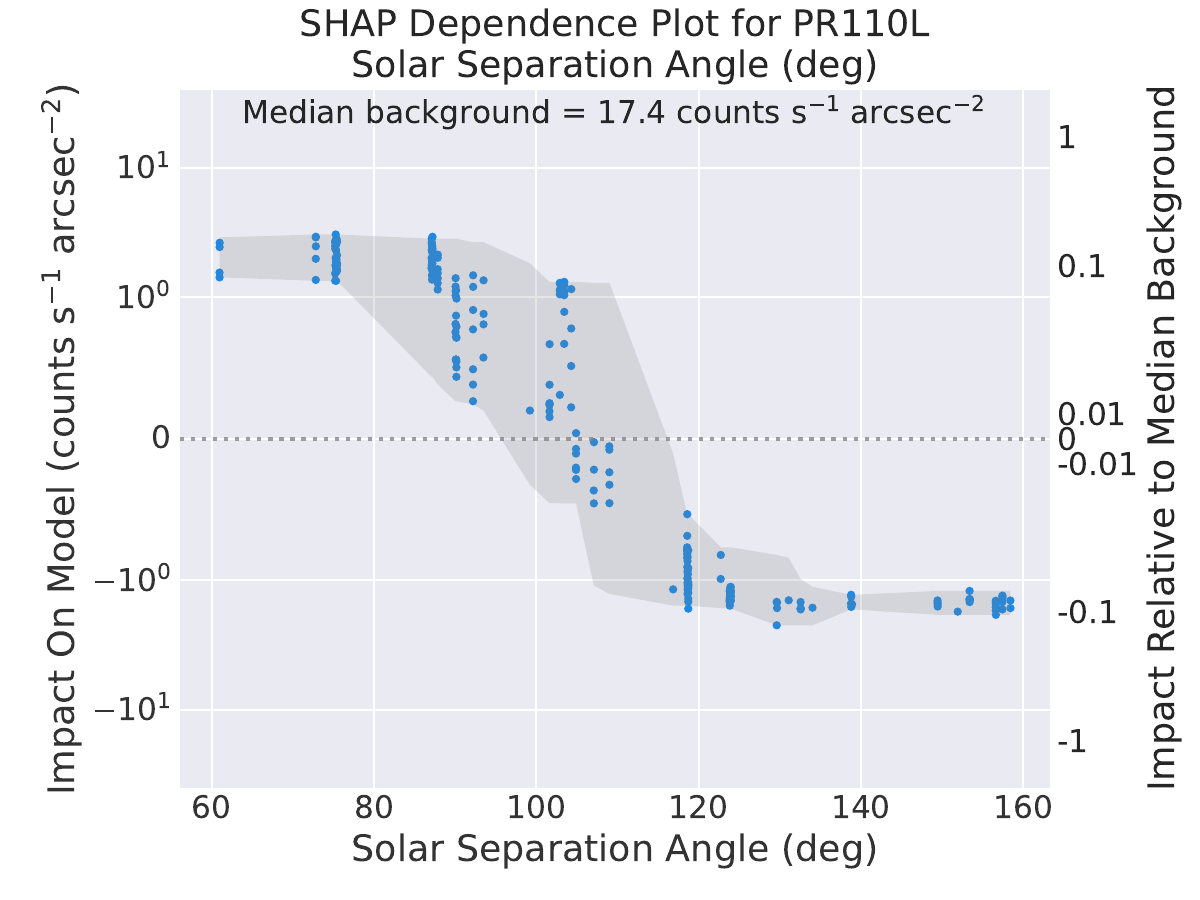}
\includegraphics[width=0.24\textwidth]{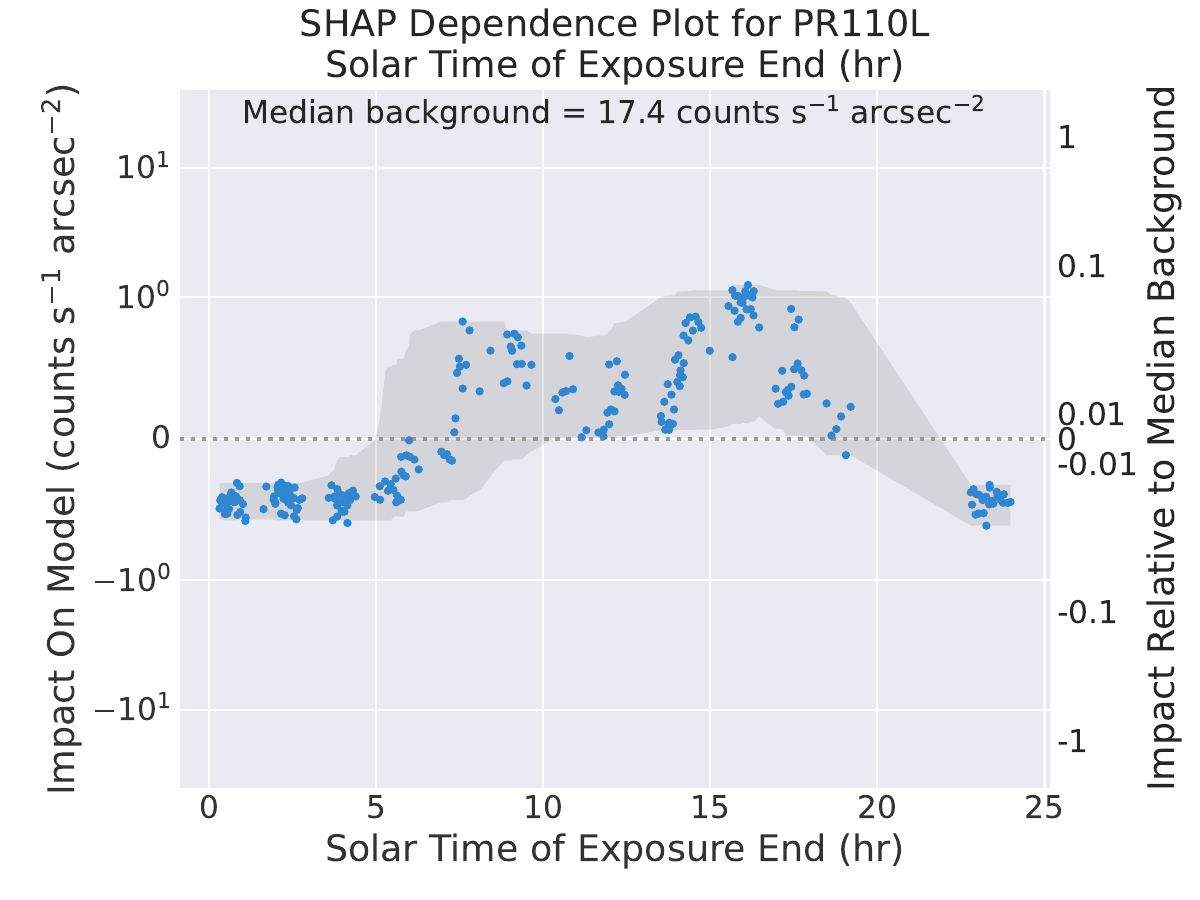}
\includegraphics[width=0.24\textwidth]{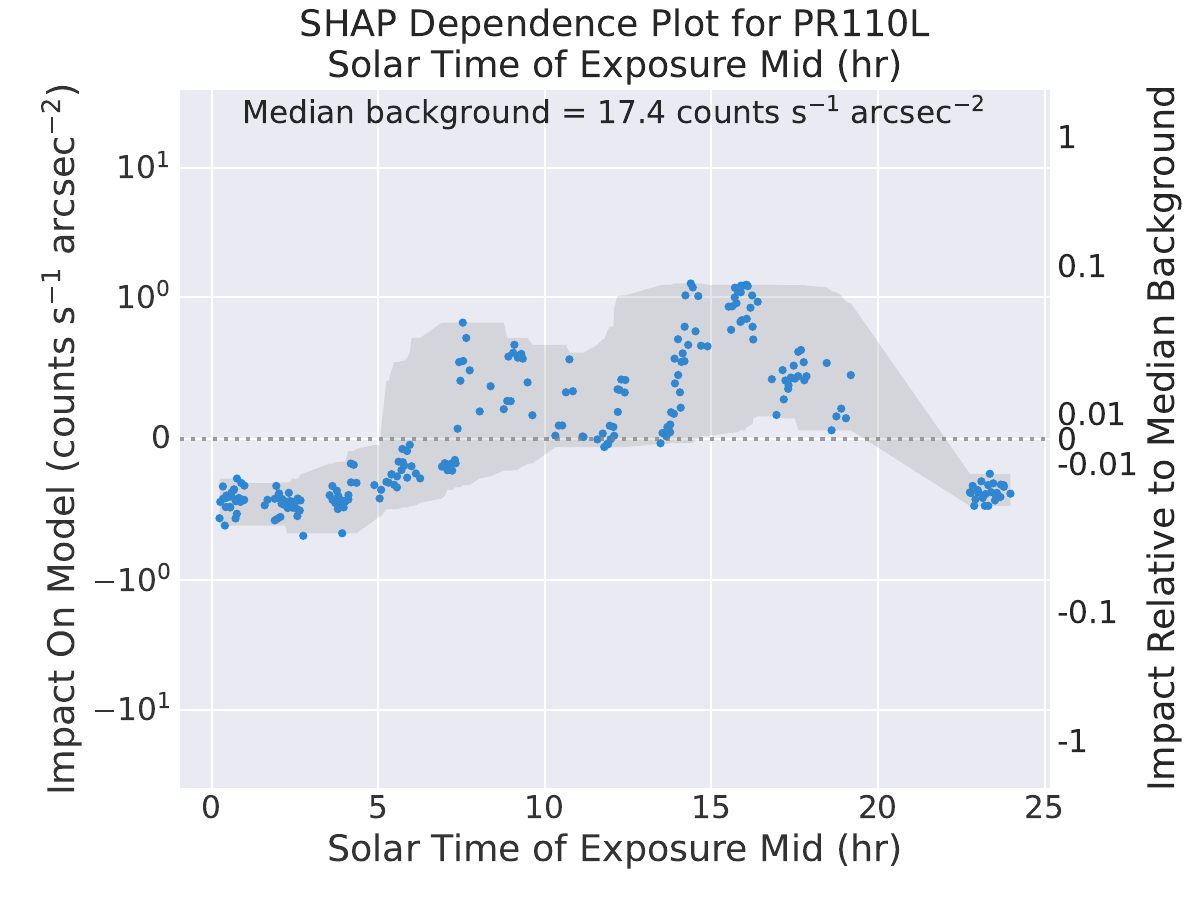}
\includegraphics[width=0.24\textwidth]{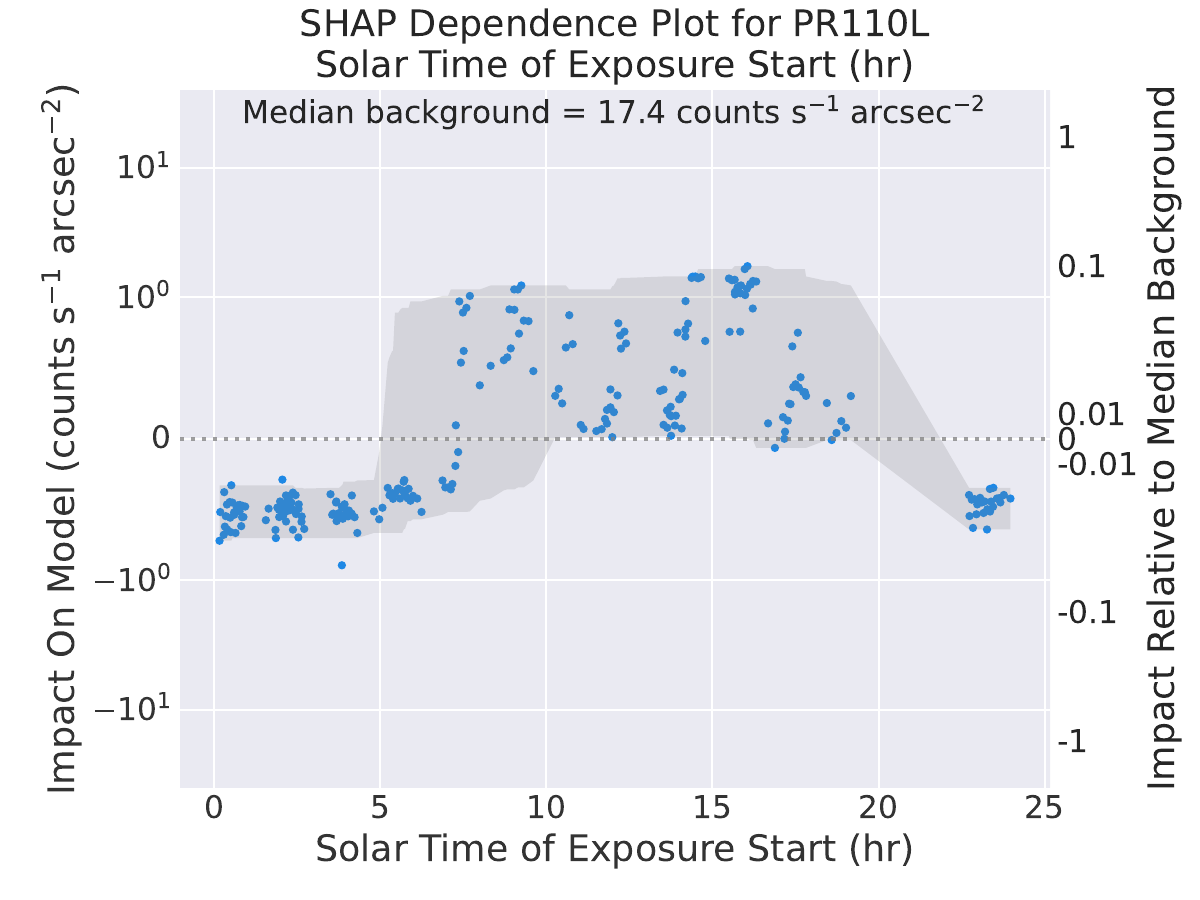}
\includegraphics[width=0.24\textwidth]{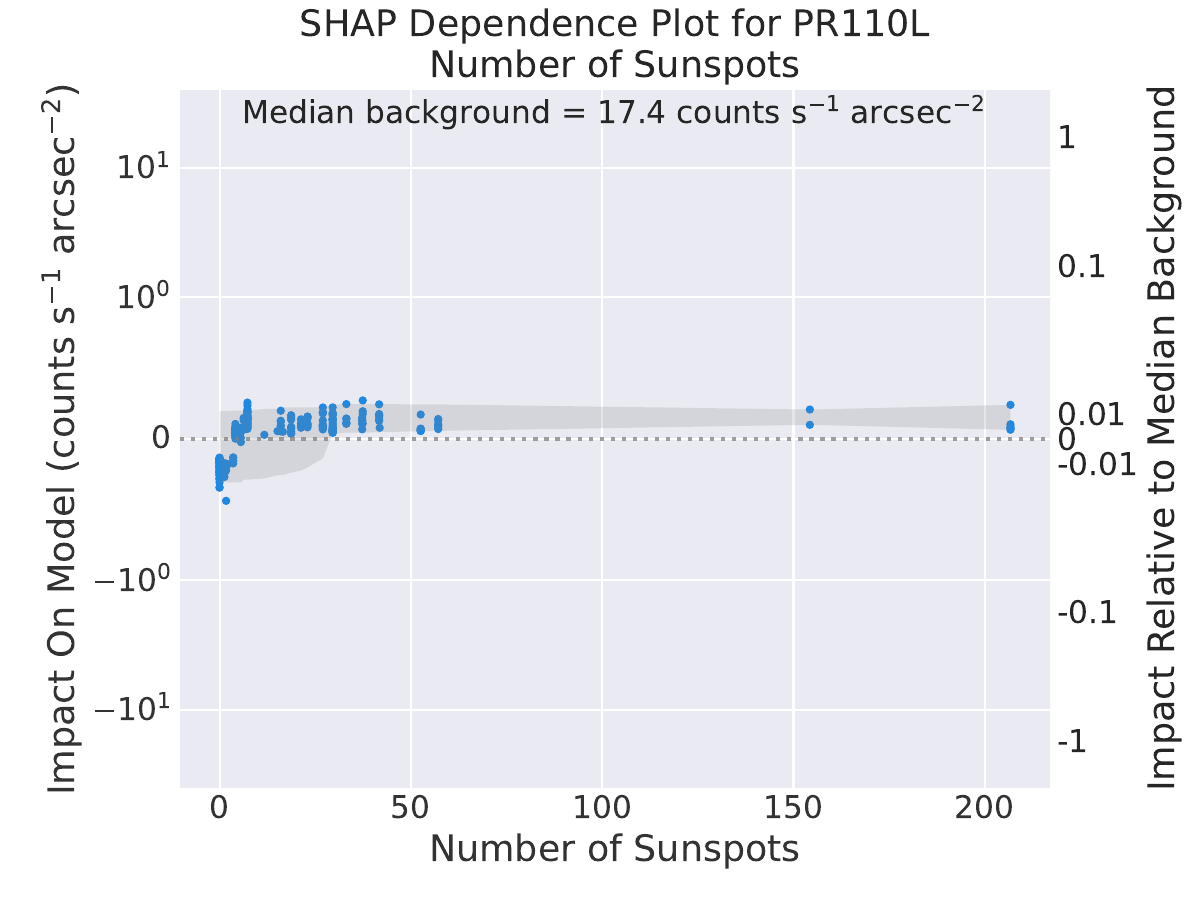}
\includegraphics[width=0.24\textwidth]{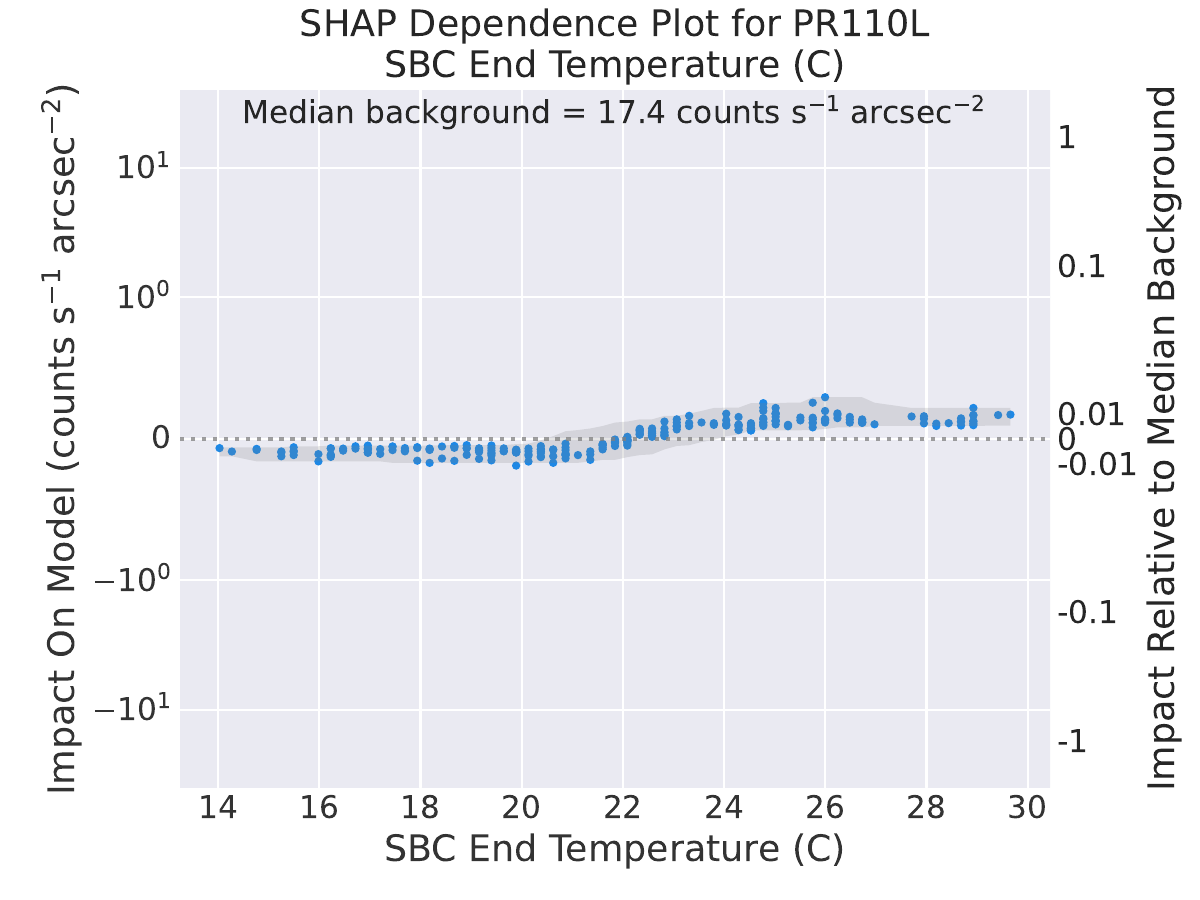}
\includegraphics[width=0.24\textwidth]{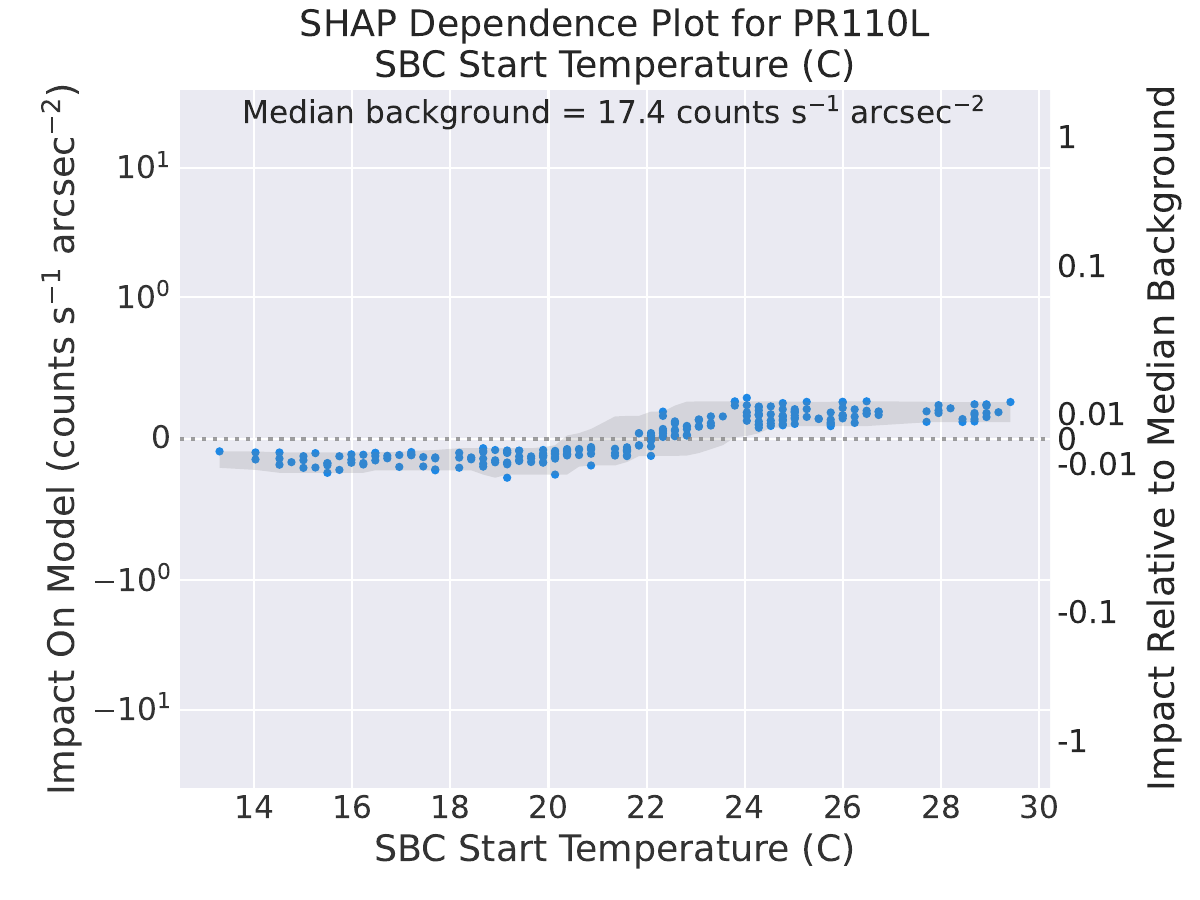}
\caption{SHAP dependence plots for QRF regression modeling of PR110L. Otherwise as per Figure~\ref{Fig:SHAP_Dependence_F115LP}.}
\label{Fig:SHAP_Dependence_PR110L}
\end{figure}

\begin{figure}
\centering
\includegraphics[width=0.24\textwidth]{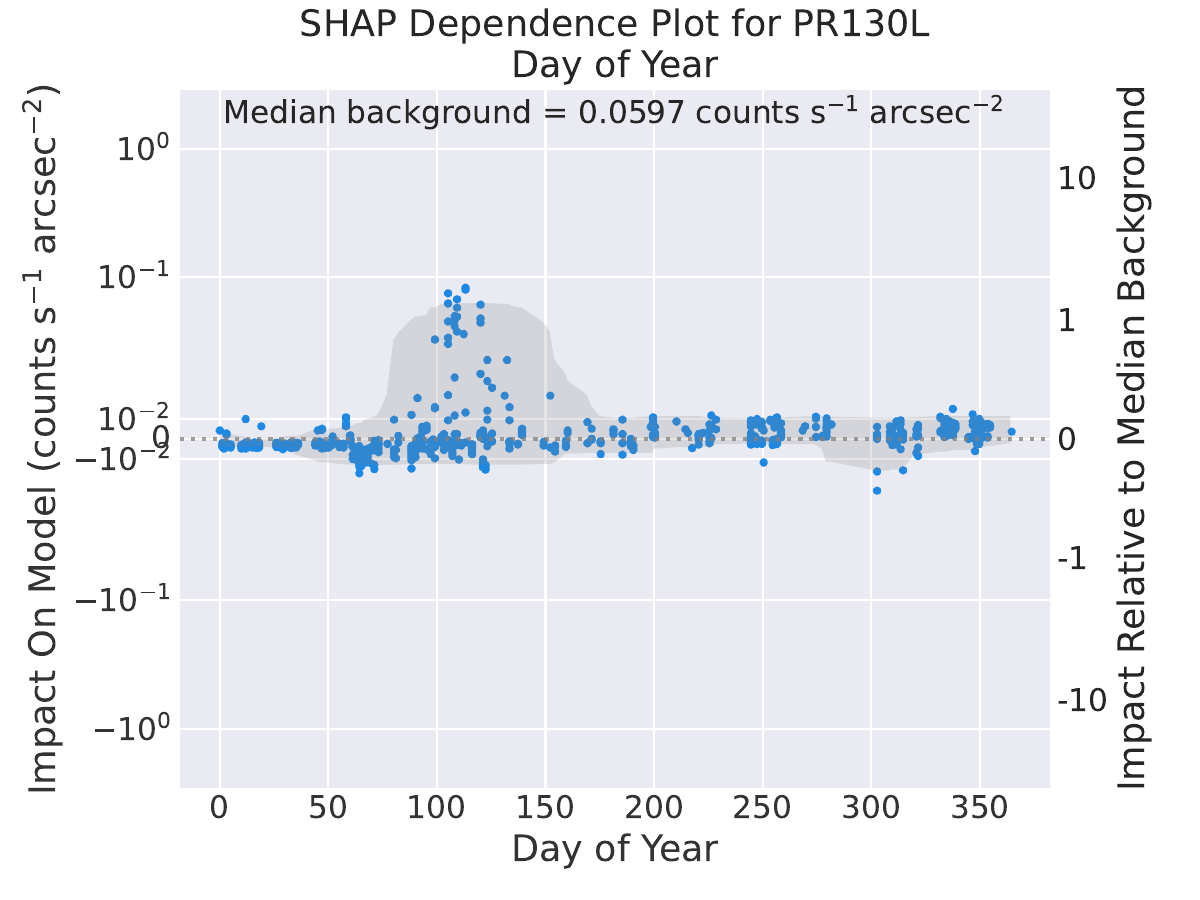}
\includegraphics[width=0.24\textwidth]{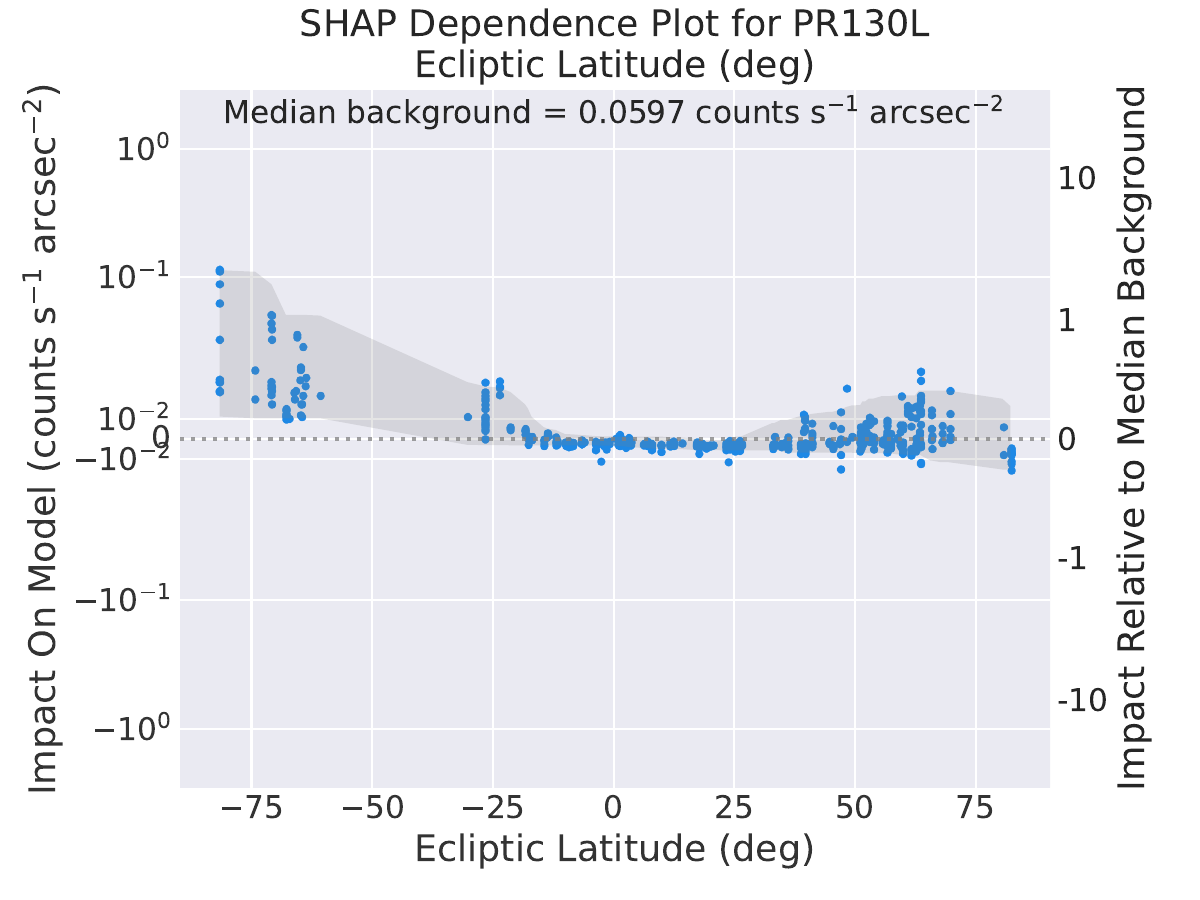}
\includegraphics[width=0.24\textwidth]{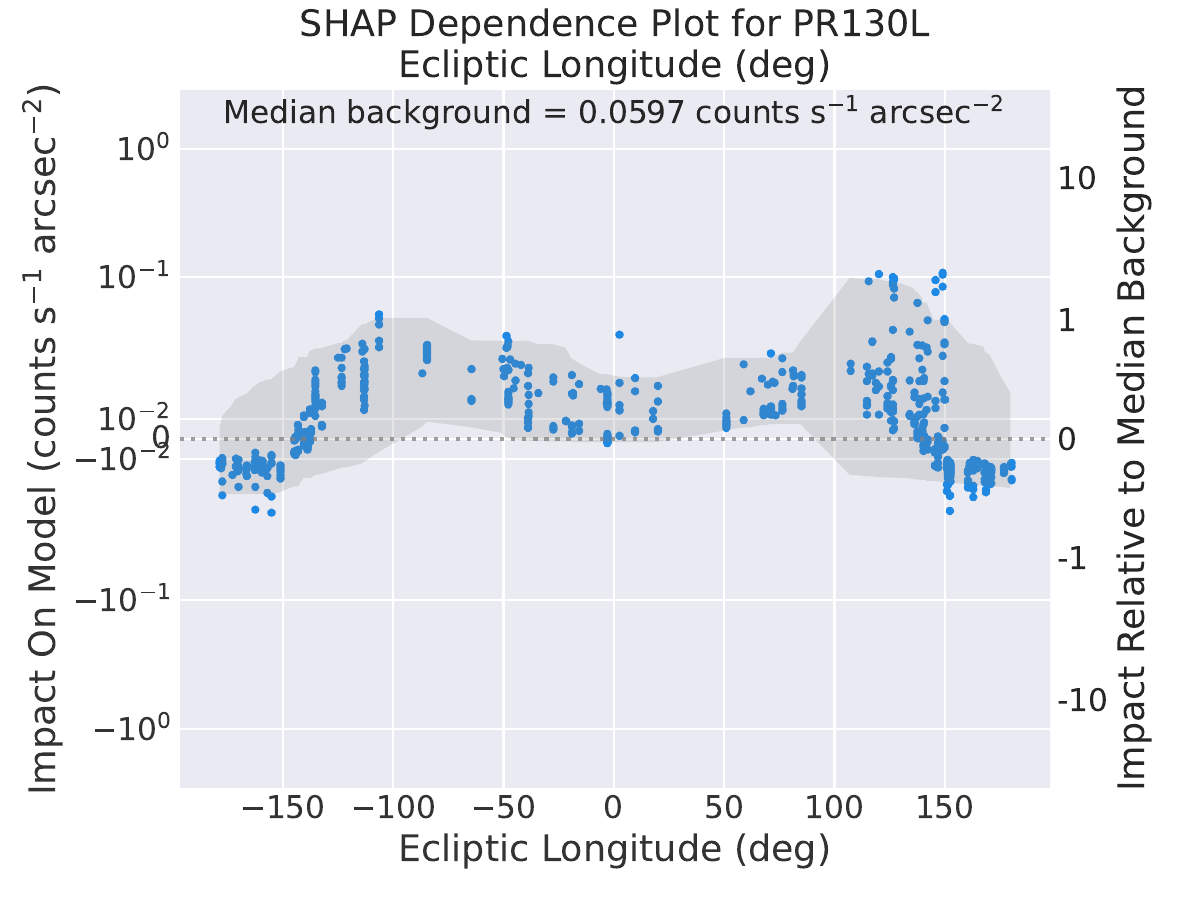}
\includegraphics[width=0.24\textwidth]{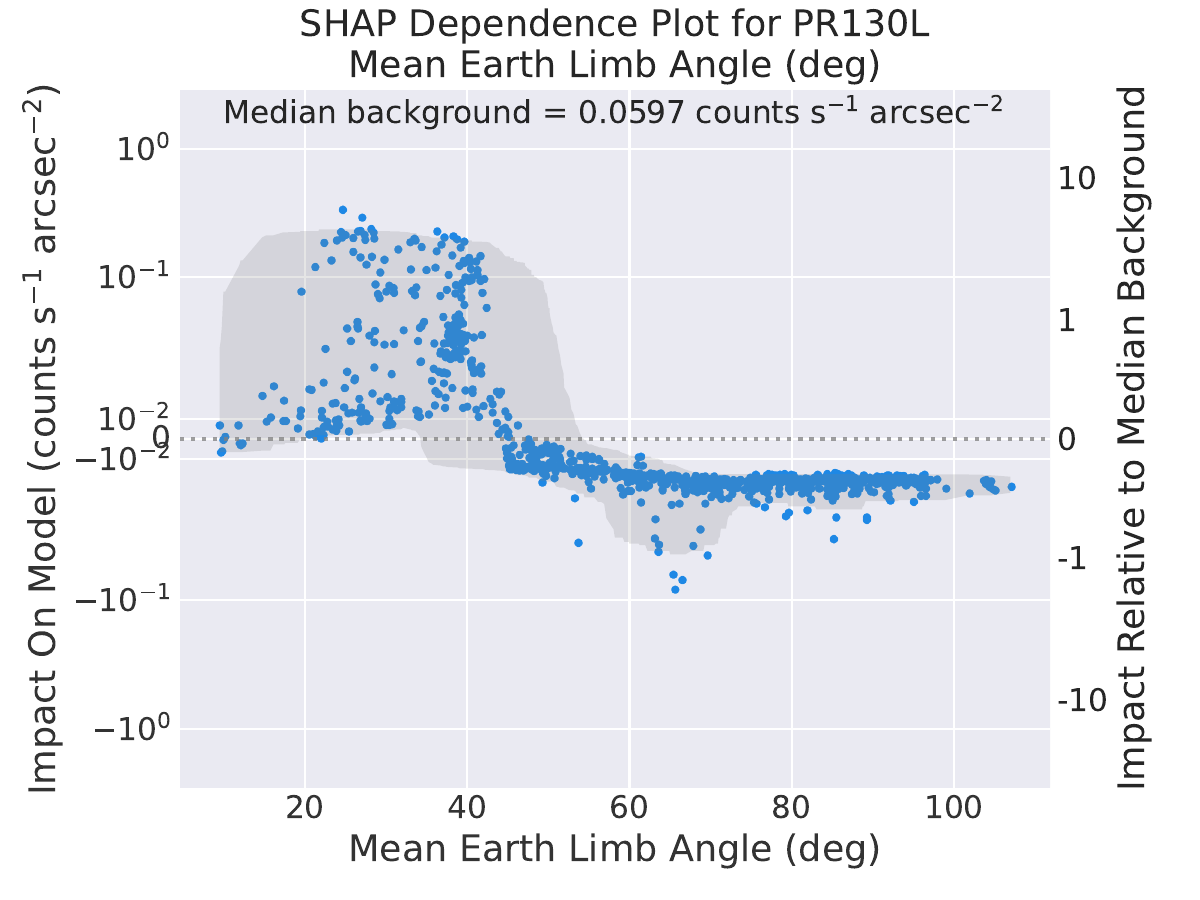}
\includegraphics[width=0.24\textwidth]{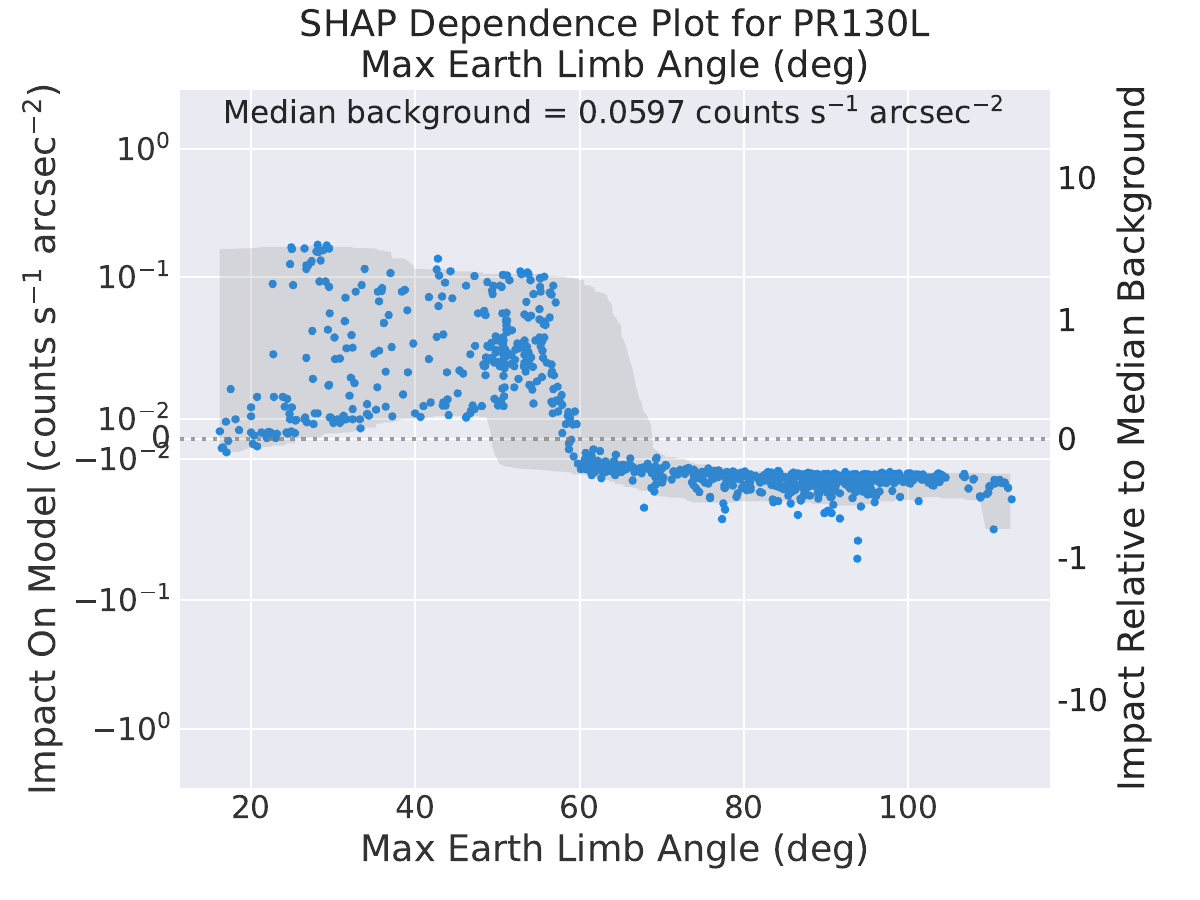}
\includegraphics[width=0.24\textwidth]{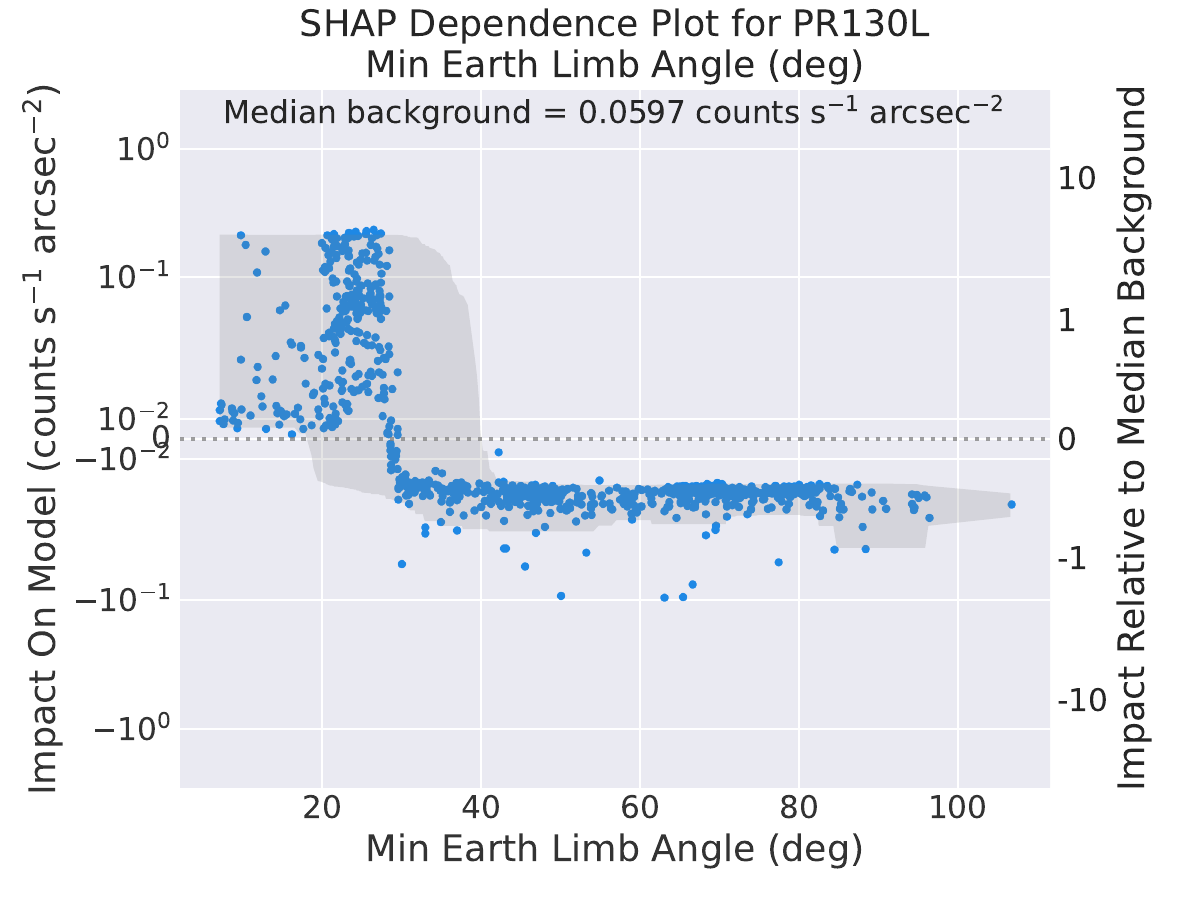}
\includegraphics[width=0.24\textwidth]{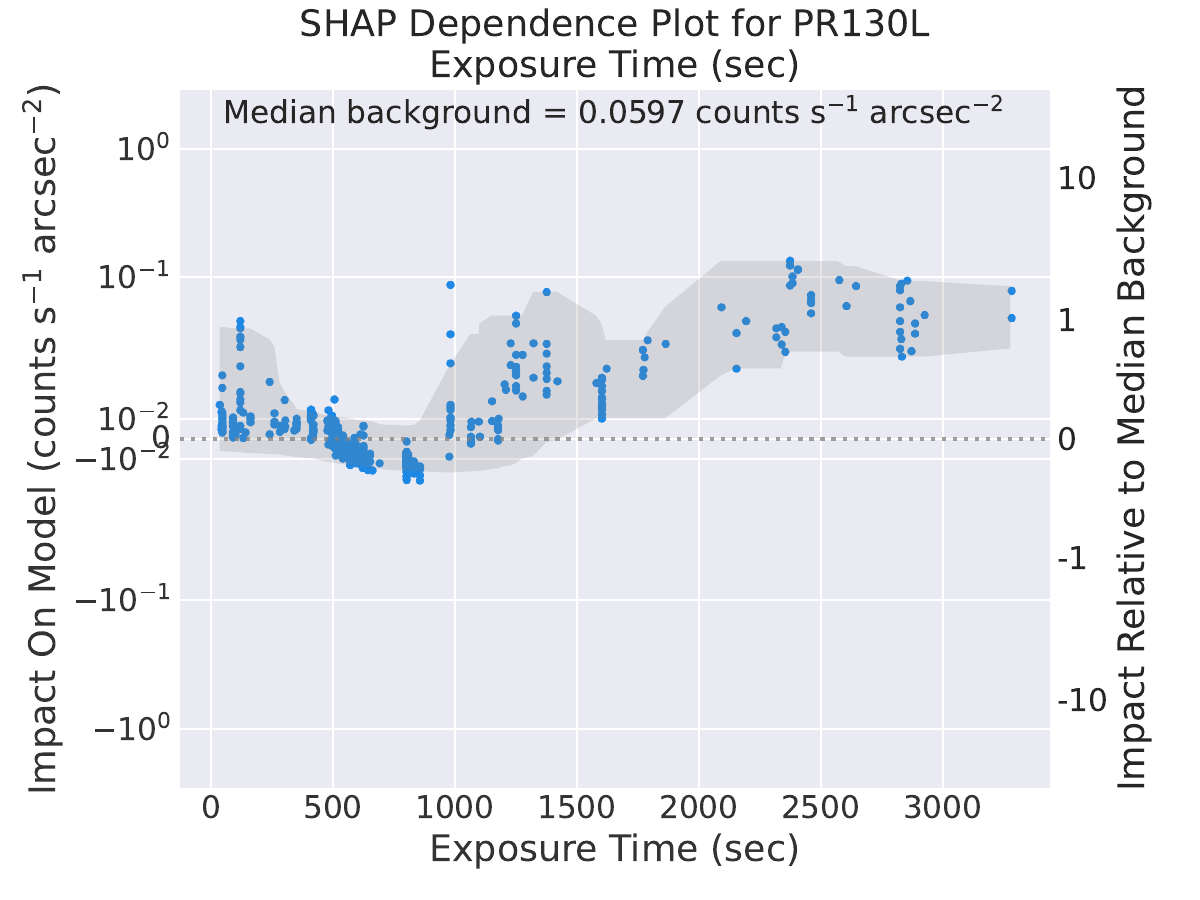}
\includegraphics[width=0.24\textwidth]{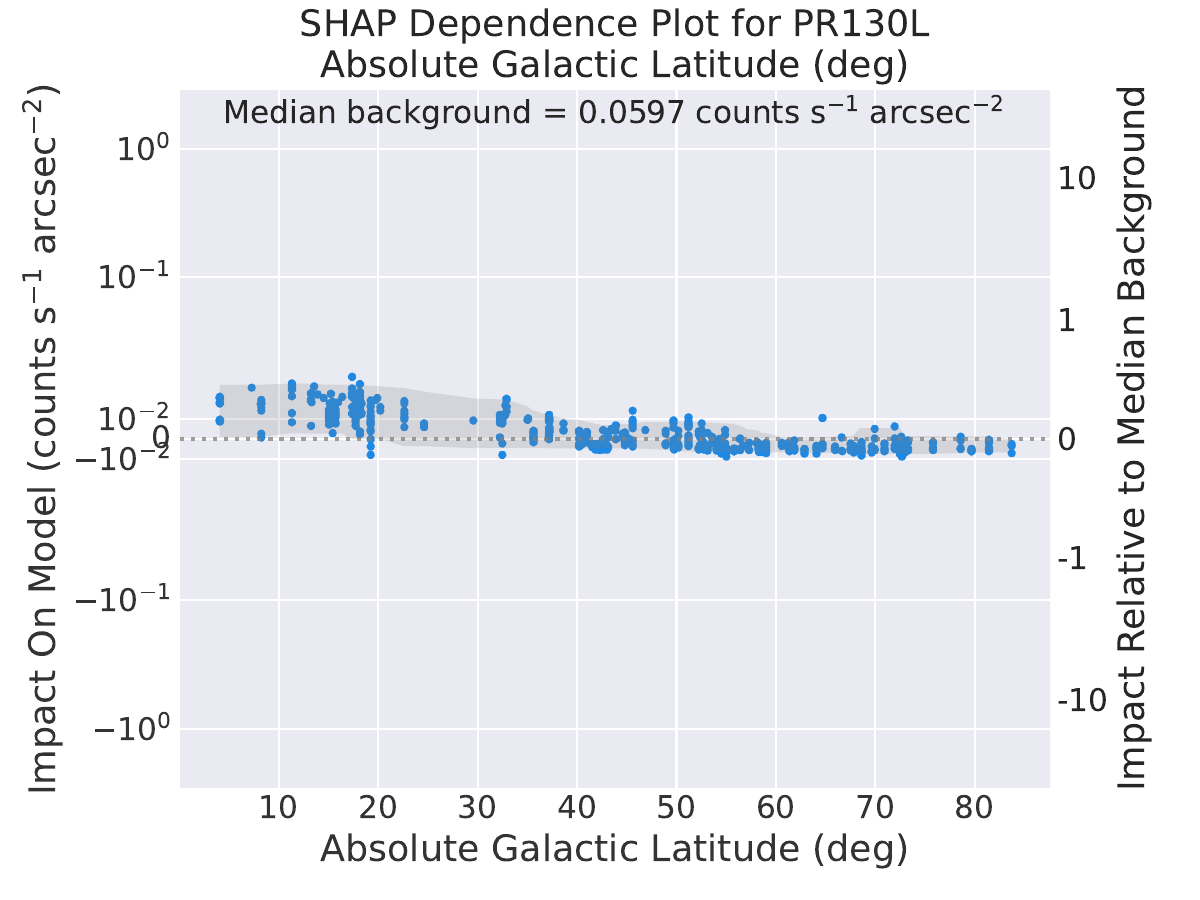}
\includegraphics[width=0.24\textwidth]{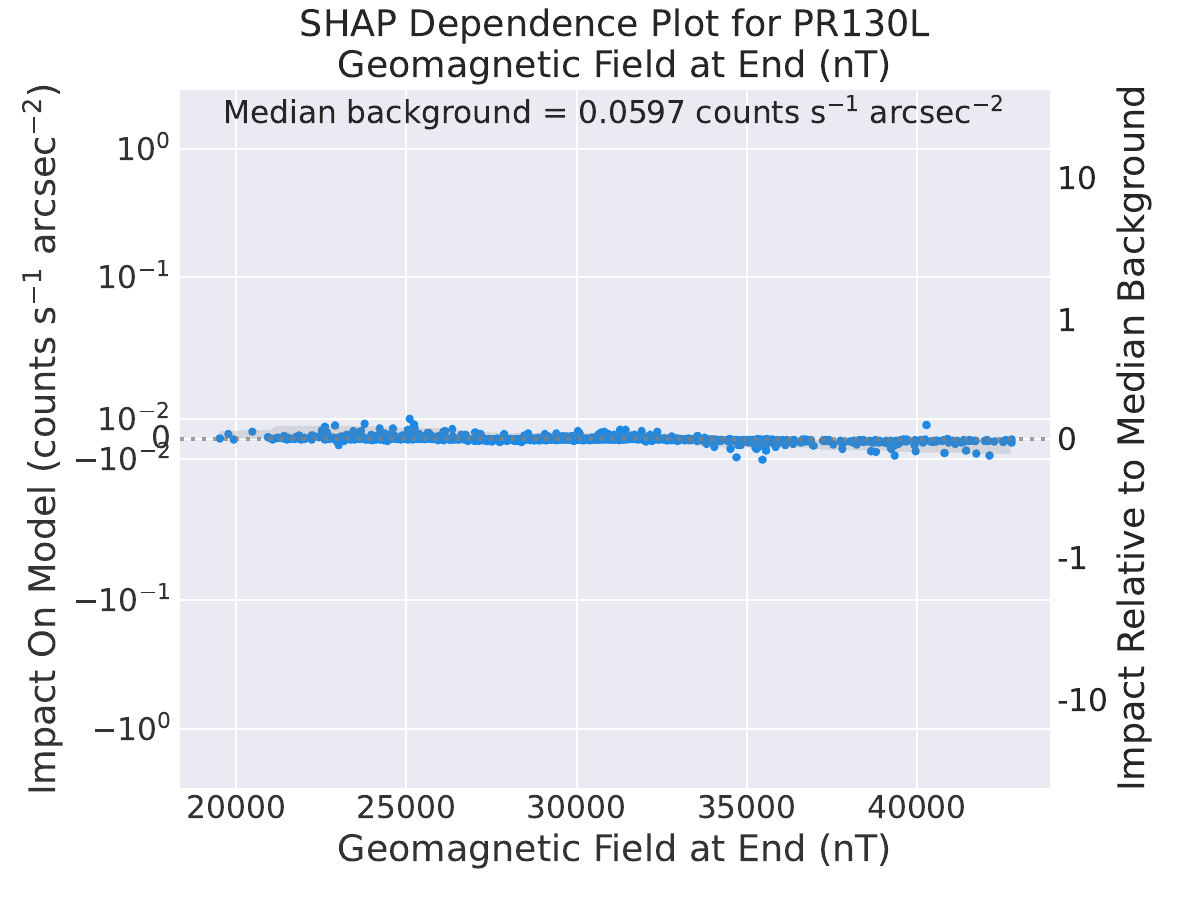}
\includegraphics[width=0.24\textwidth]{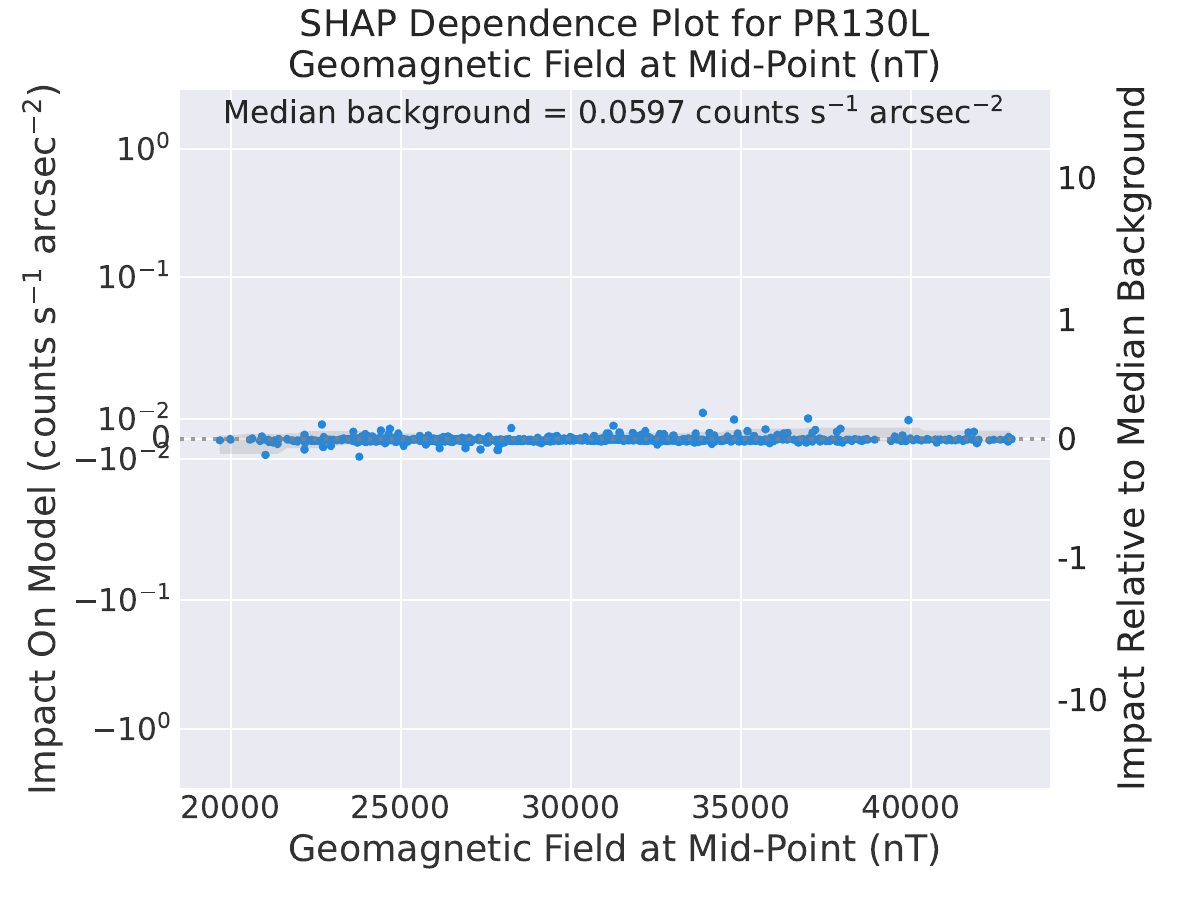}
\includegraphics[width=0.24\textwidth]{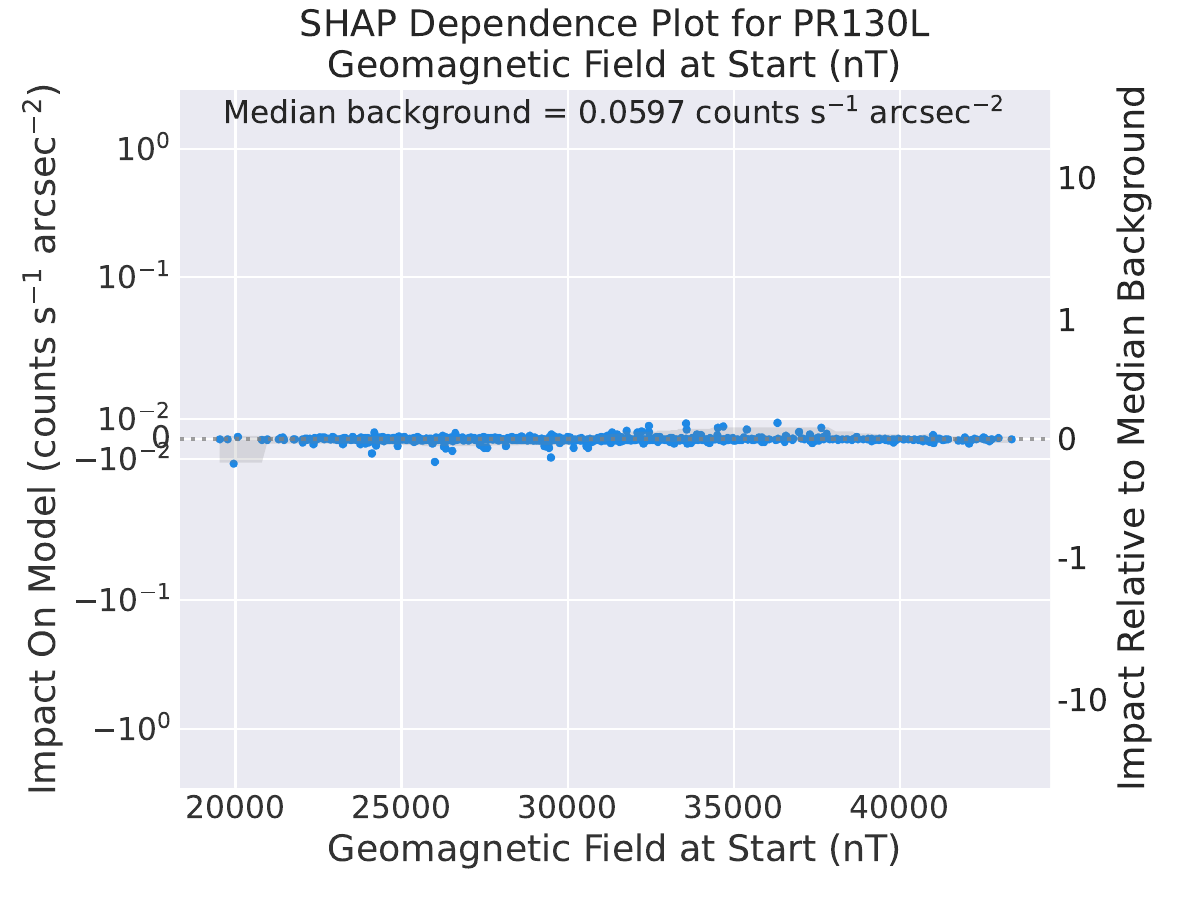}
\includegraphics[width=0.24\textwidth]{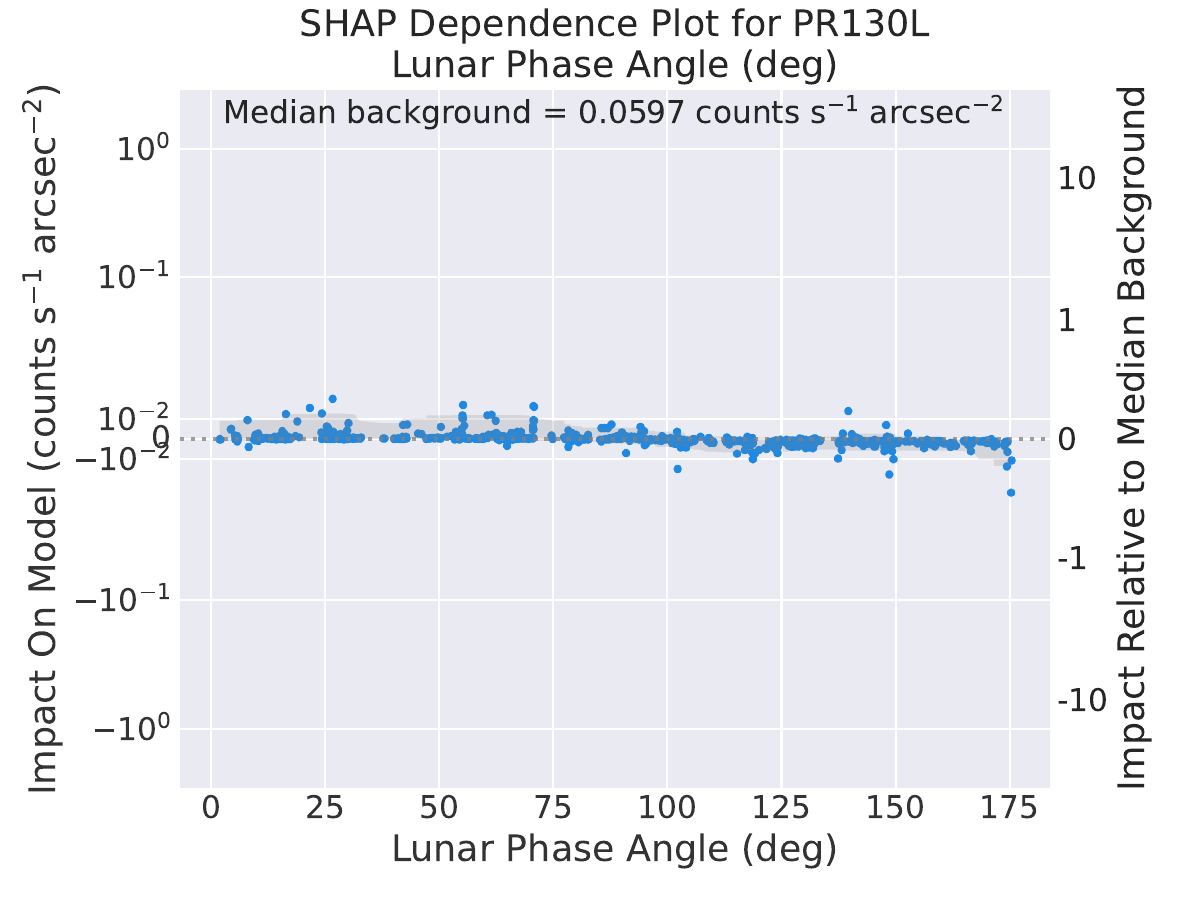}
\includegraphics[width=0.24\textwidth]{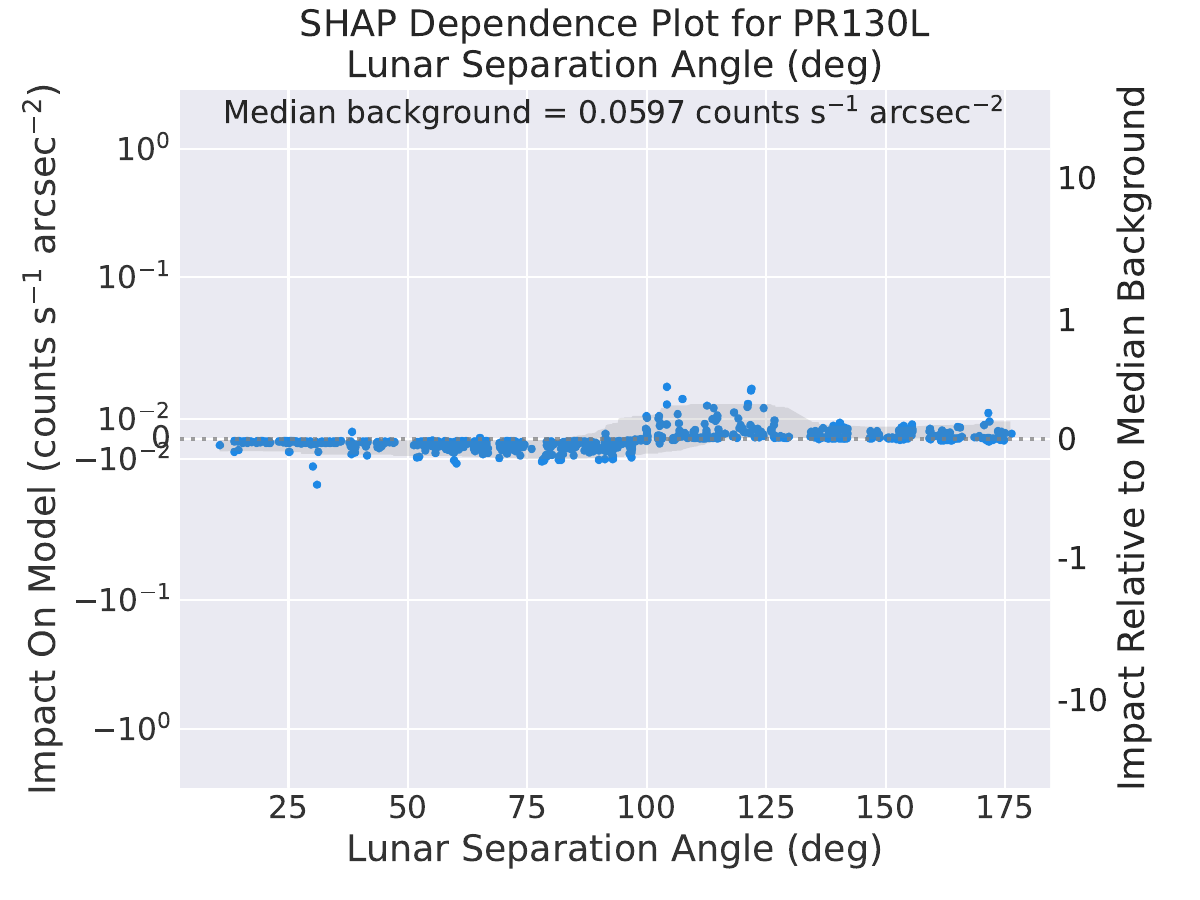}
\includegraphics[width=0.24\textwidth]{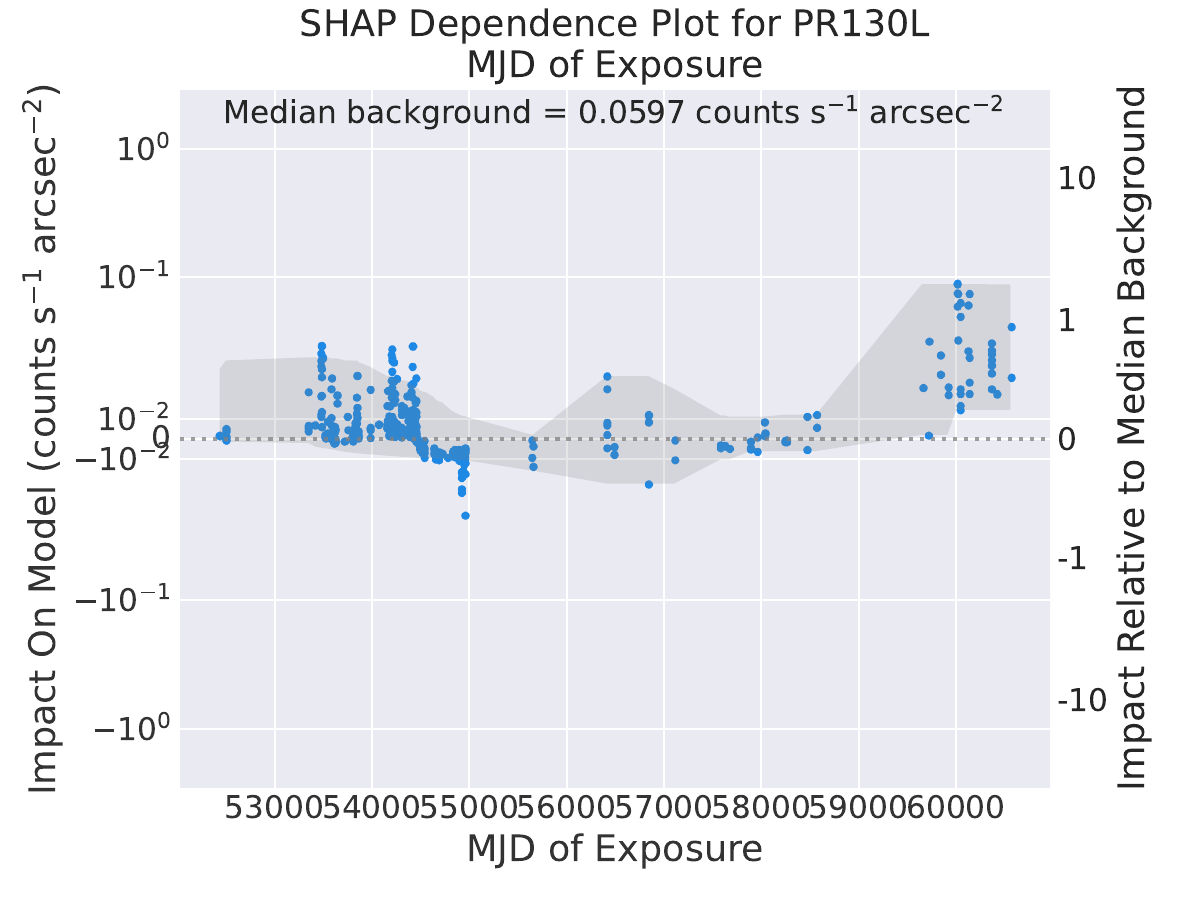}
\includegraphics[width=0.24\textwidth]{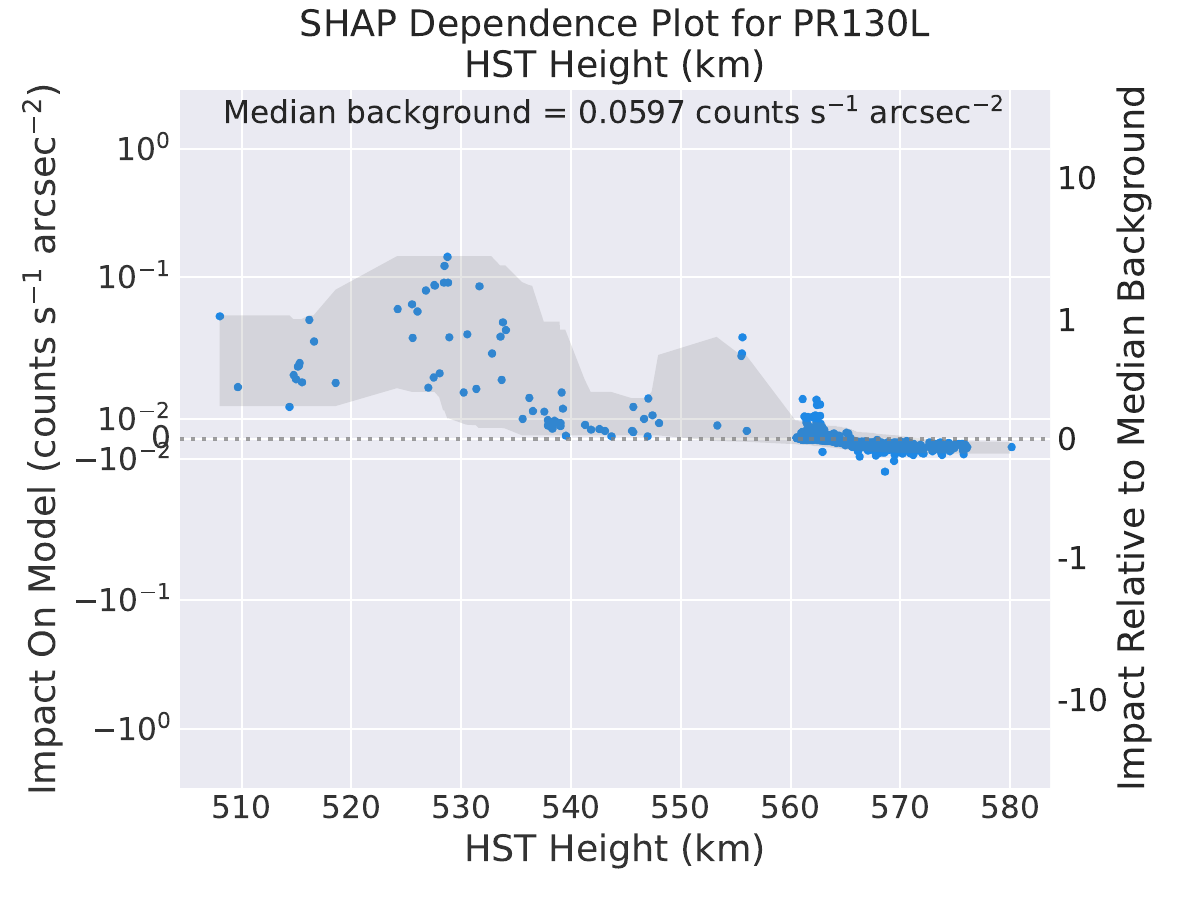}
\includegraphics[width=0.24\textwidth]{SHAP_Plots_QuantileForestRegr/PR130L_solar_alt_SHAP_Dependence.pdf}
\includegraphics[width=0.24\textwidth]{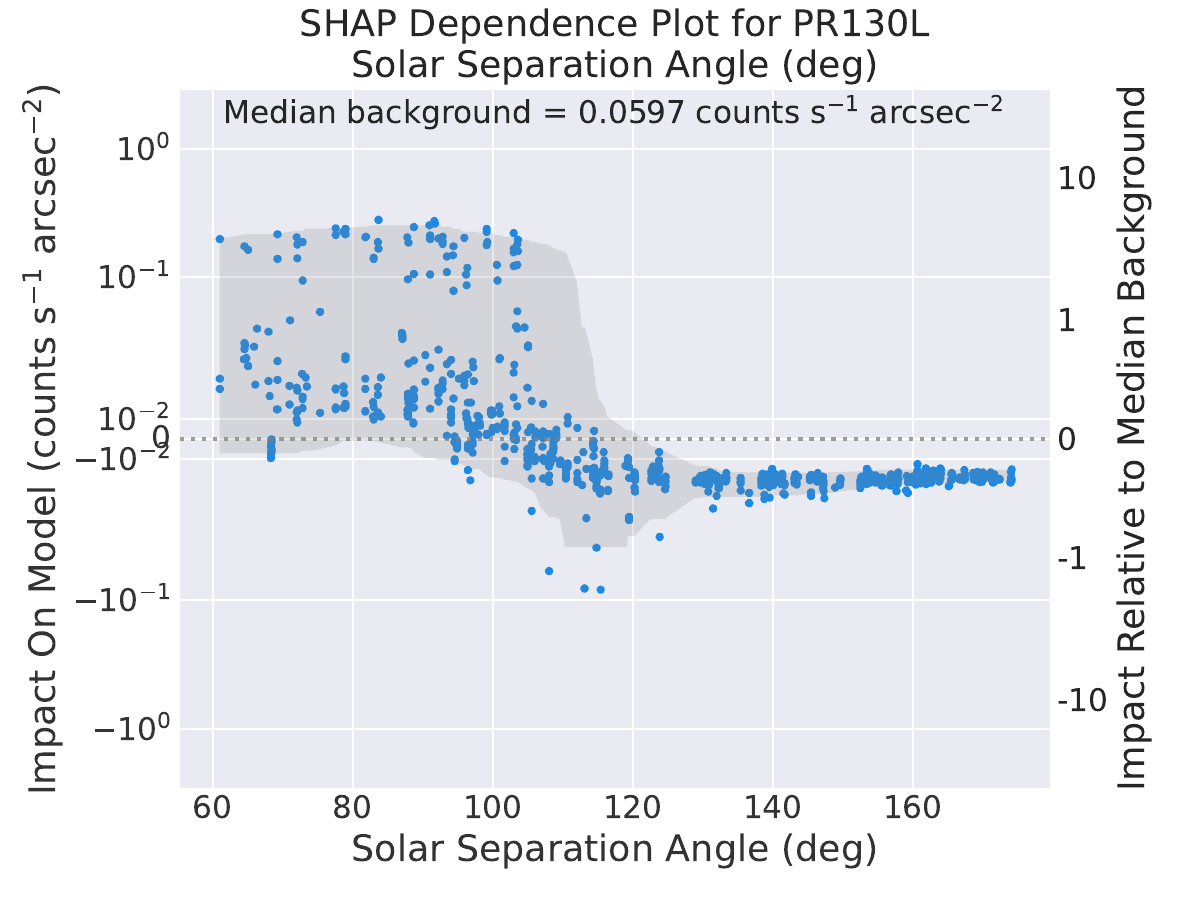}
\includegraphics[width=0.24\textwidth]{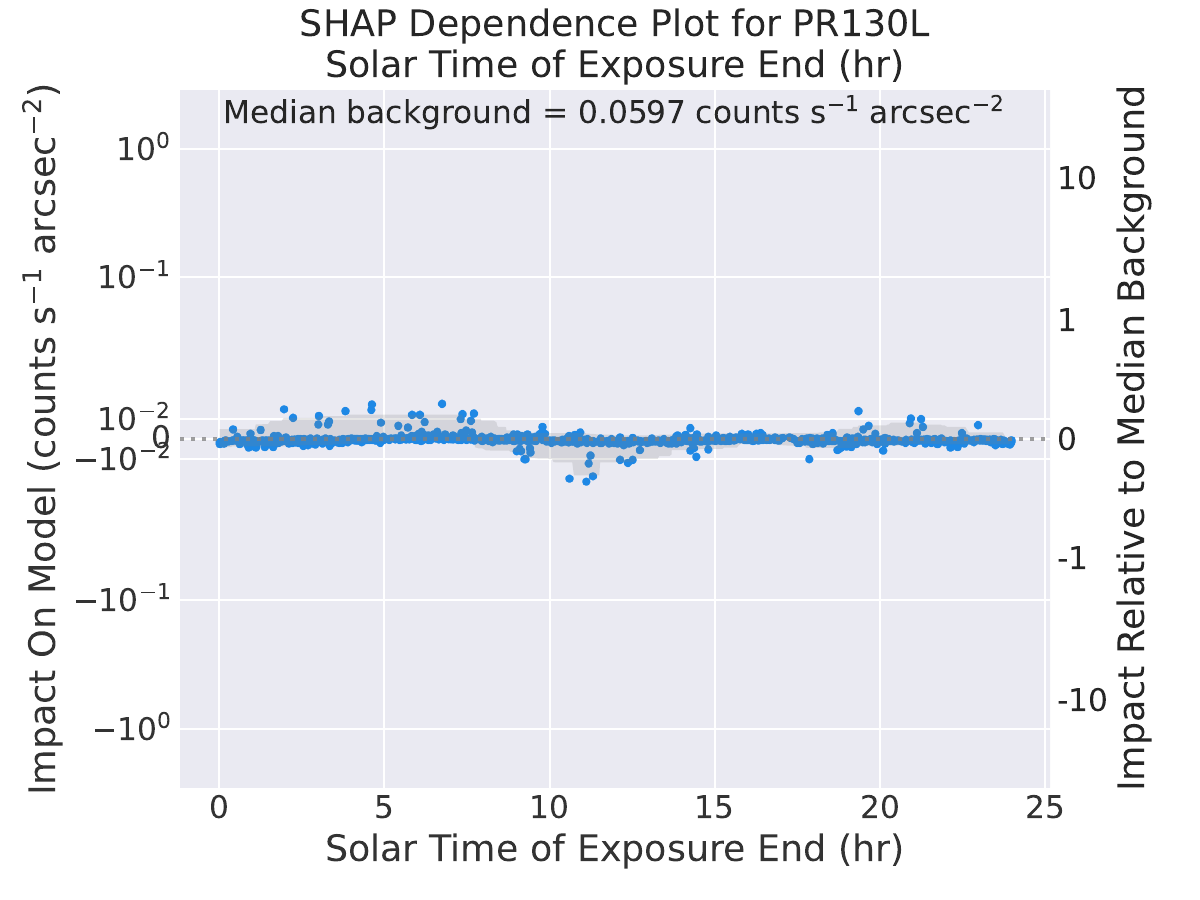}
\includegraphics[width=0.24\textwidth]{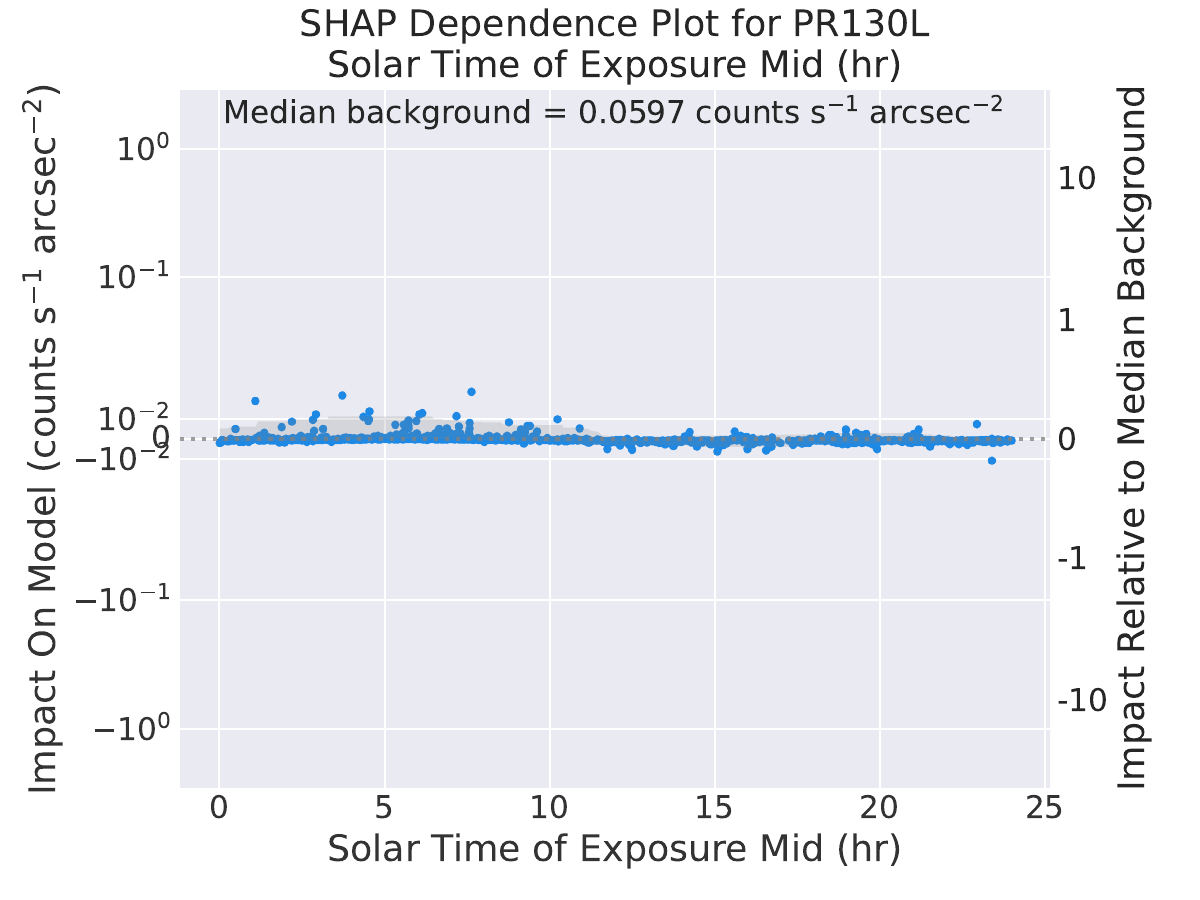}
\includegraphics[width=0.24\textwidth]{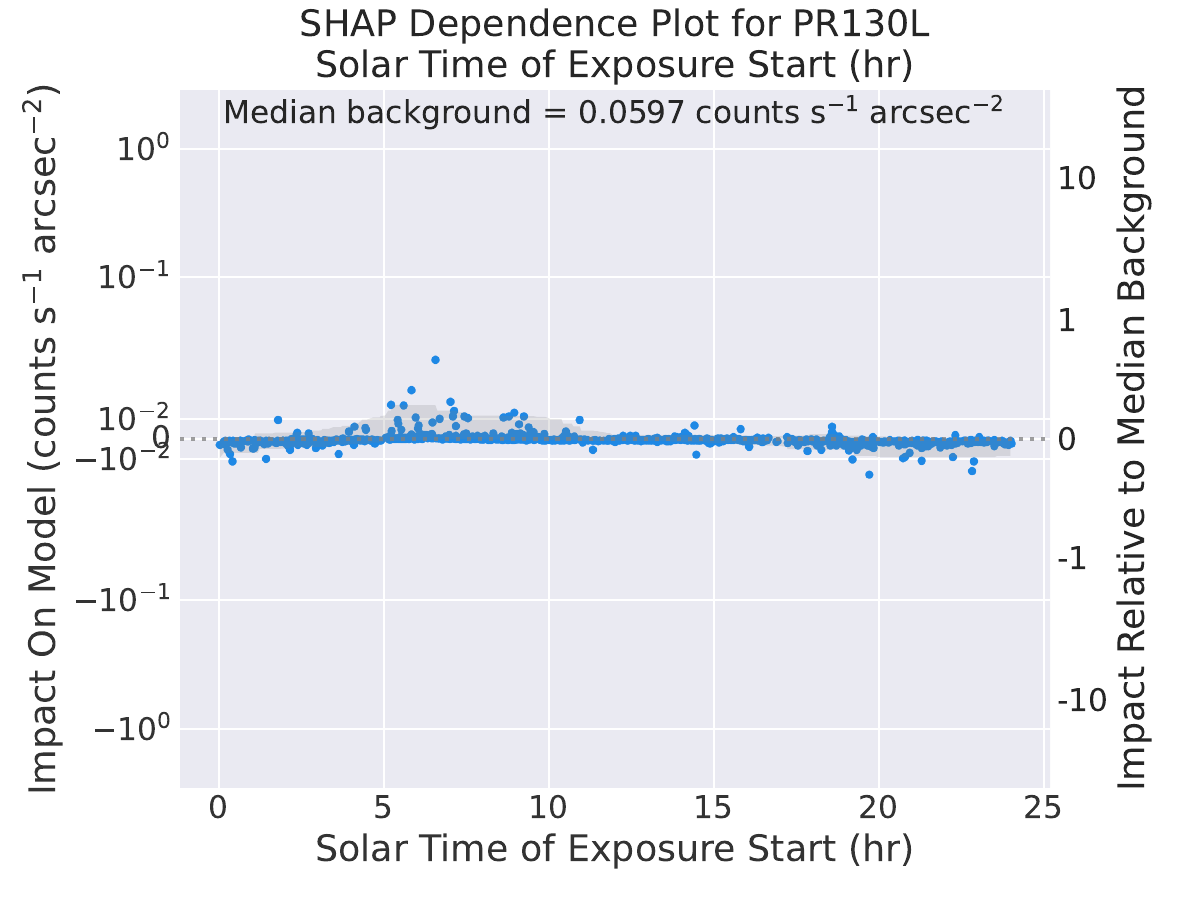}
\includegraphics[width=0.24\textwidth]{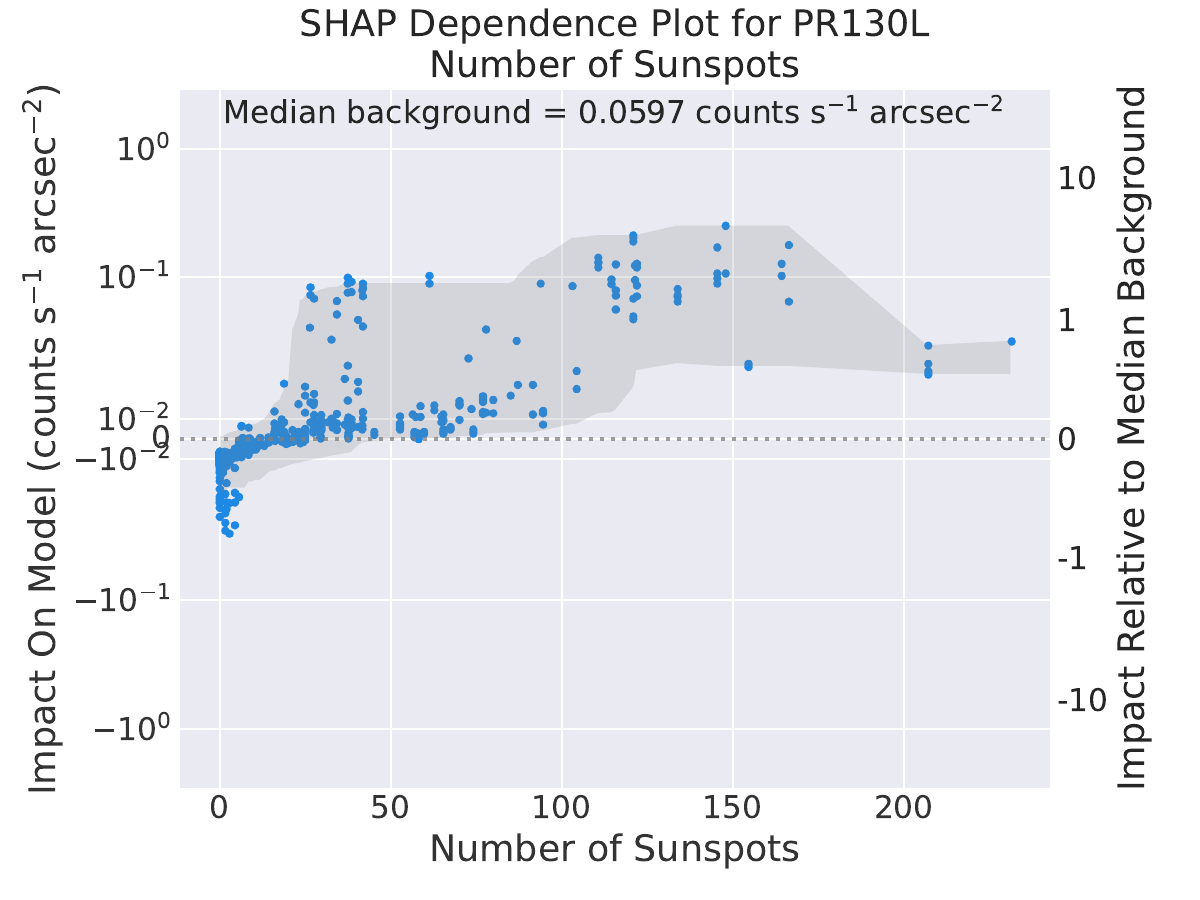}
\includegraphics[width=0.24\textwidth]{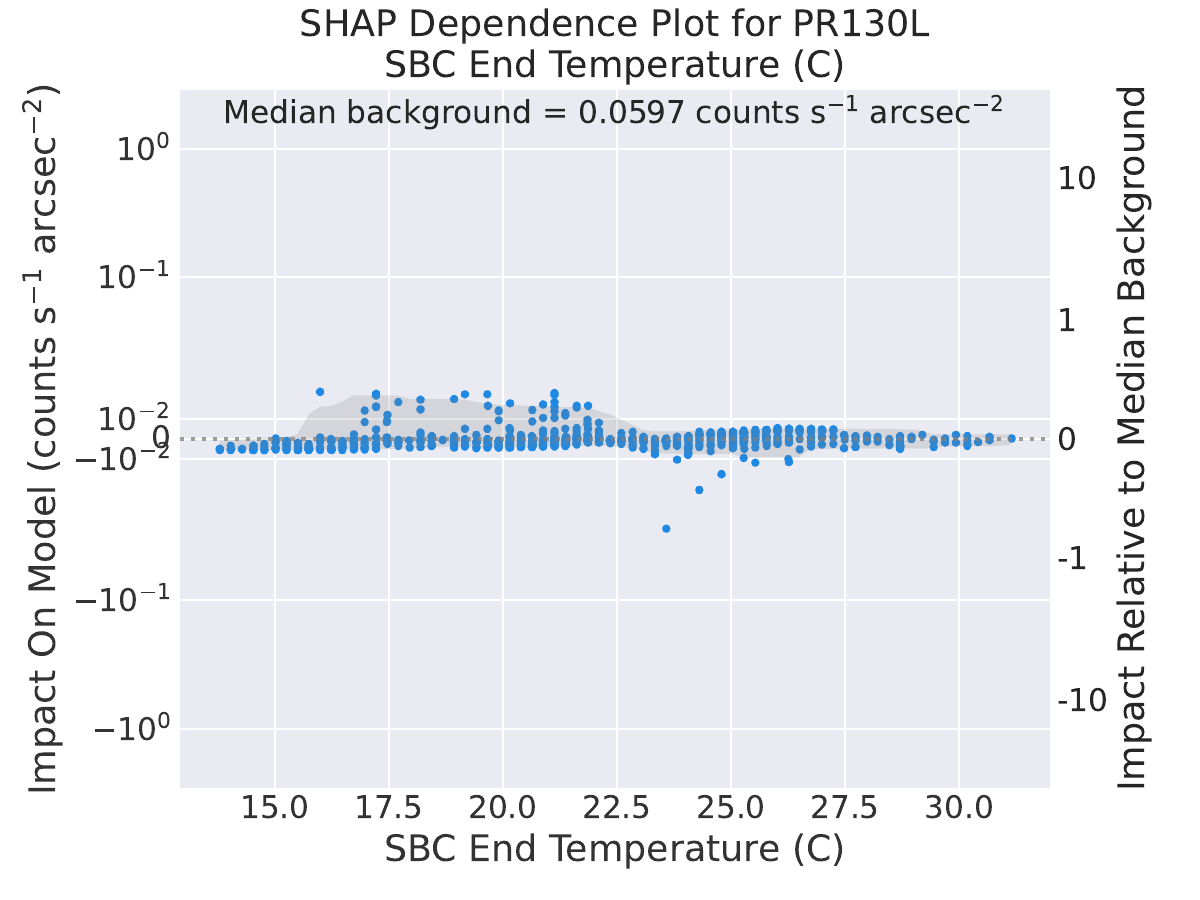}
\includegraphics[width=0.24\textwidth]{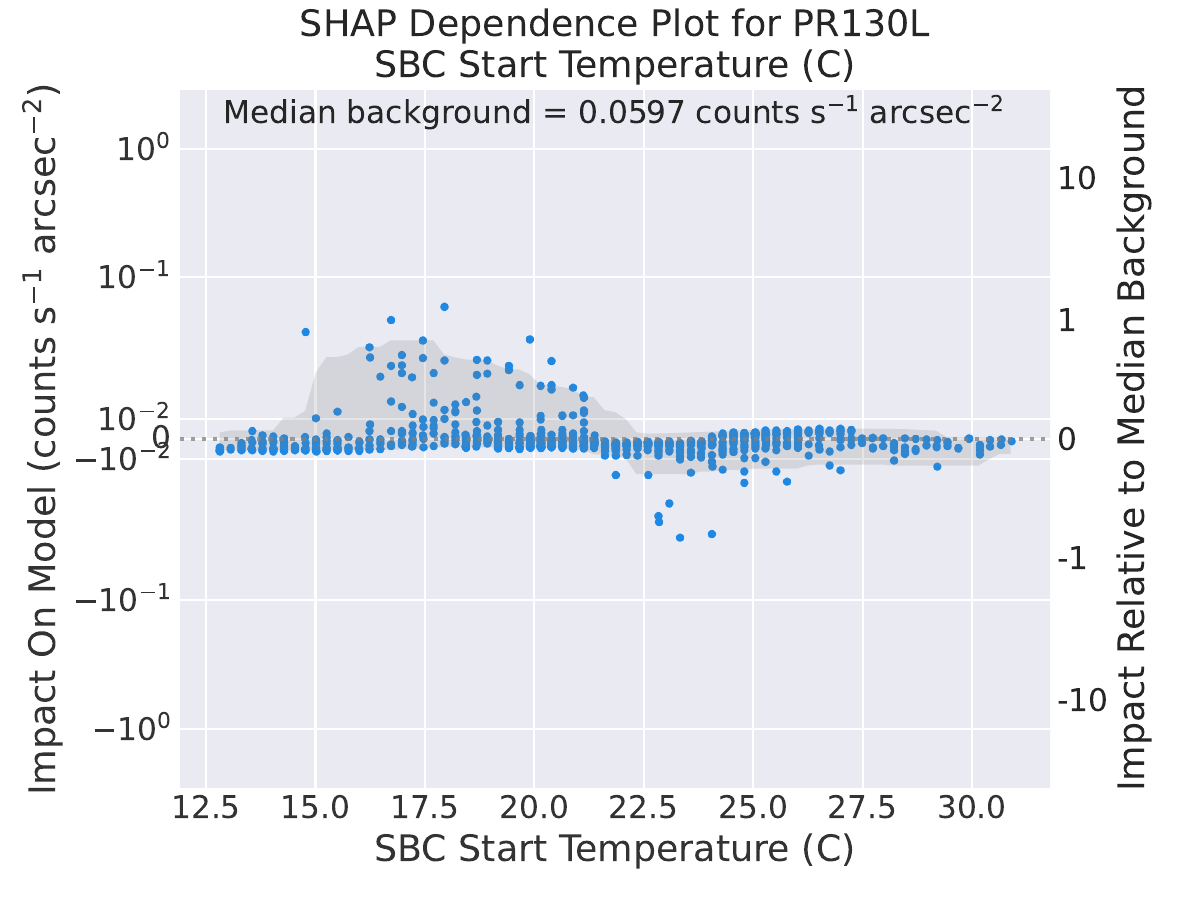}
\caption{SHAP dependence plots for QRF regression modeling of PR130L. Otherwise as per Figure~\ref{Fig:SHAP_Dependence_F115LP}.}
\label{Fig:SHAP_Dependence_PR130L}
\end{figure}

Here, in Figures~\ref{Fig:SHAP_Dependence_F115LP}, \ref{Fig:SHAP_Dependence_F122M}, \ref{Fig:SHAP_Dependence_F125LP}, \ref{Fig:SHAP_Dependence_F140LP}, \ref{Fig:SHAP_Dependence_F150LP}, \ref{Fig:SHAP_Dependence_F165LP}, \ref{Fig:SHAP_Dependence_PR110L}, and  \ref{Fig:SHAP_Dependence_PR130L} we provide the SHAP dependence plots for every SBC filter. 

\section{SHAP Beeswarm Plots} \label{AppendixSection:SHAP_Beeswarm_Plots}

\begin{figure}
\centering
\includegraphics[width=0.9\textwidth]{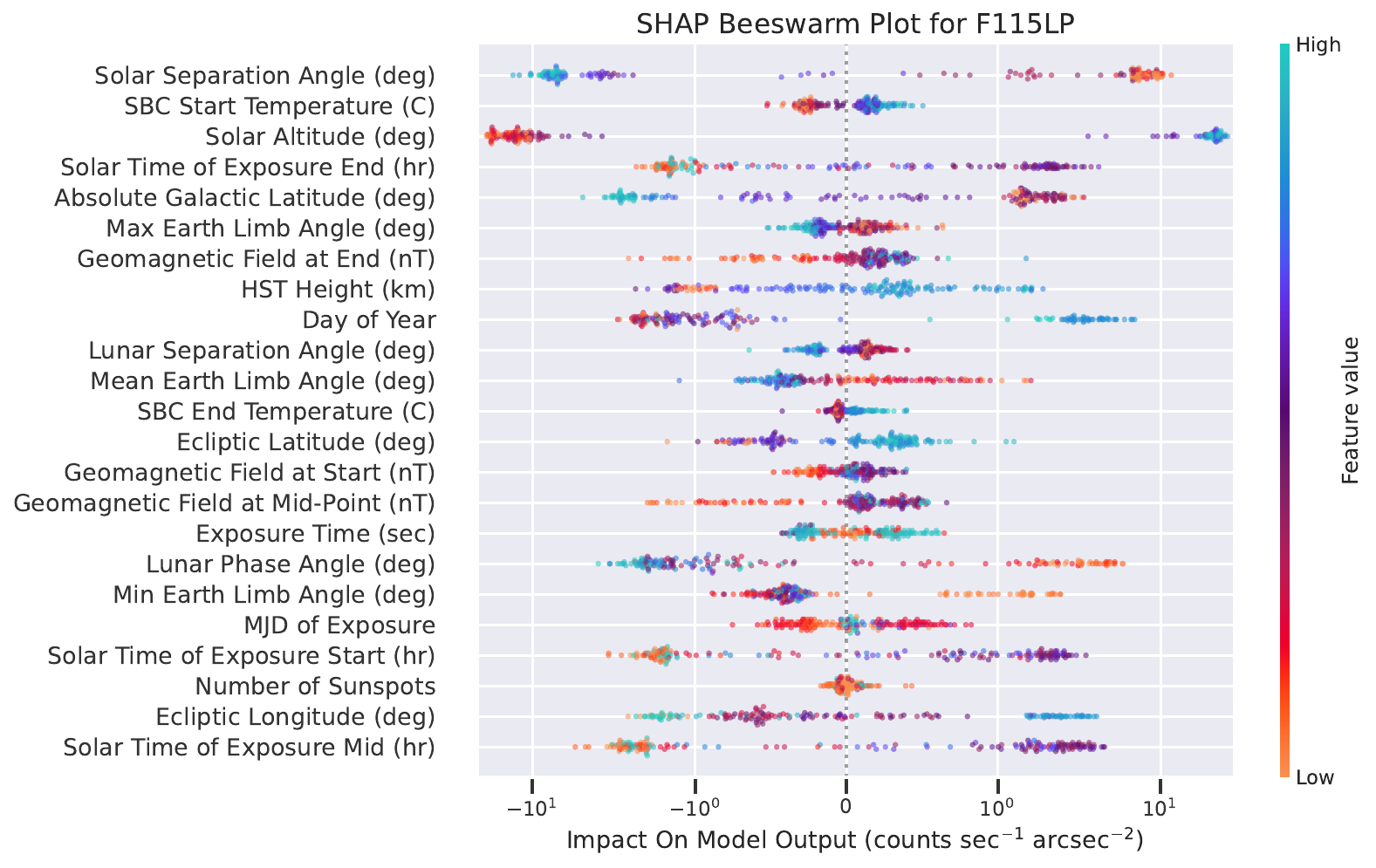}
\includegraphics[width=0.9\textwidth]{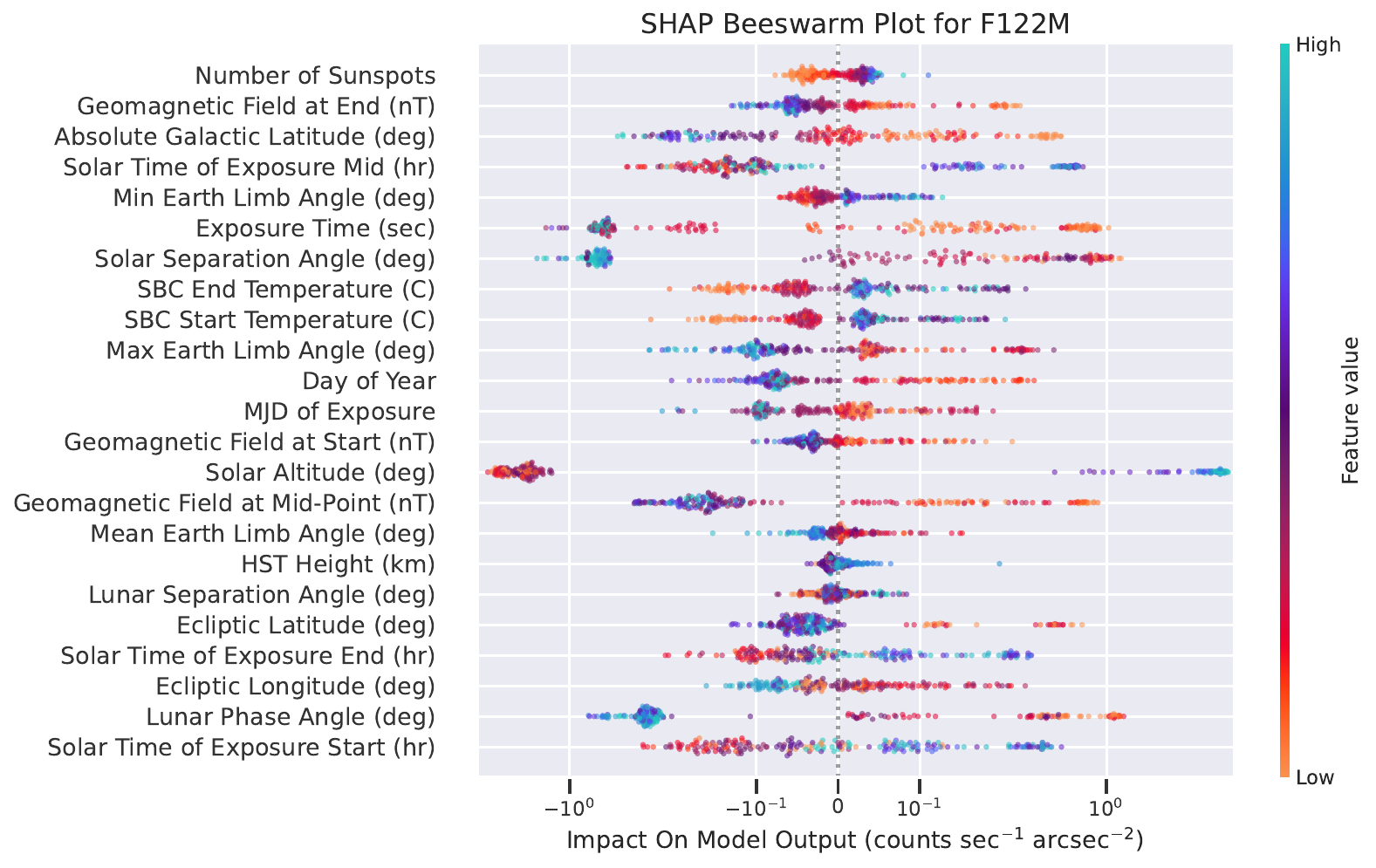}
\caption{SHAP beeswarm plots for QRF regression modeling of F115LP and F122M. Rows are ordered according to how tightly the parameter in question correlates with the marginal impact upon the model background prediction (with correlation calculated using the Kendell's tau correlation coefficient; \citealp{Kendall1990}). Note that a tighter of correlation does not necessary correspond to a larger impact on the model prediction; there are several parameters where the parameter value is very tightly correlated with the impact on the model prediction -- but where that impact is small.}
\label{Fig:SHAP_Beeswarm_F115LP_F122M}
\end{figure}

\begin{figure}
\centering
\includegraphics[width=0.9\textwidth]{SHAP_Plots_QuantileForestRegr/F125LP_SHAP_Beeswarm.pdf}
\includegraphics[width=0.9\textwidth]{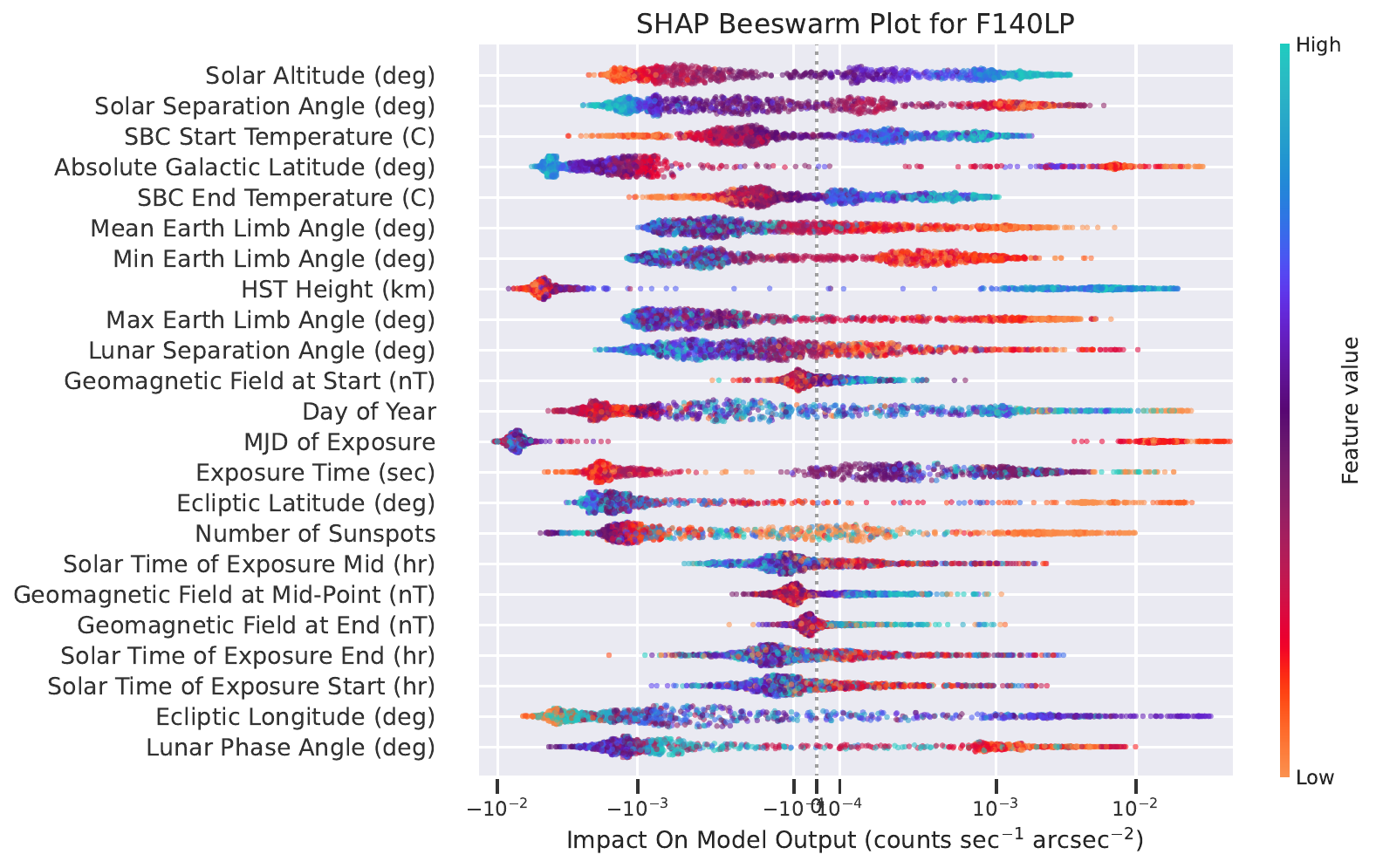}
\caption{As per Figure~\protect\ref{Fig:SHAP_Beeswarm_F115LP_F122M}, except for F125LP and F140LP.}
\label{Fig:SHAP_Beeswarm_F125LP_F140LP}
\end{figure}

\begin{figure}
\centering
\includegraphics[width=0.9\textwidth]{SHAP_Plots_QuantileForestRegr/F150LP_SHAP_Beeswarm.pdf}
\includegraphics[width=0.9\textwidth]{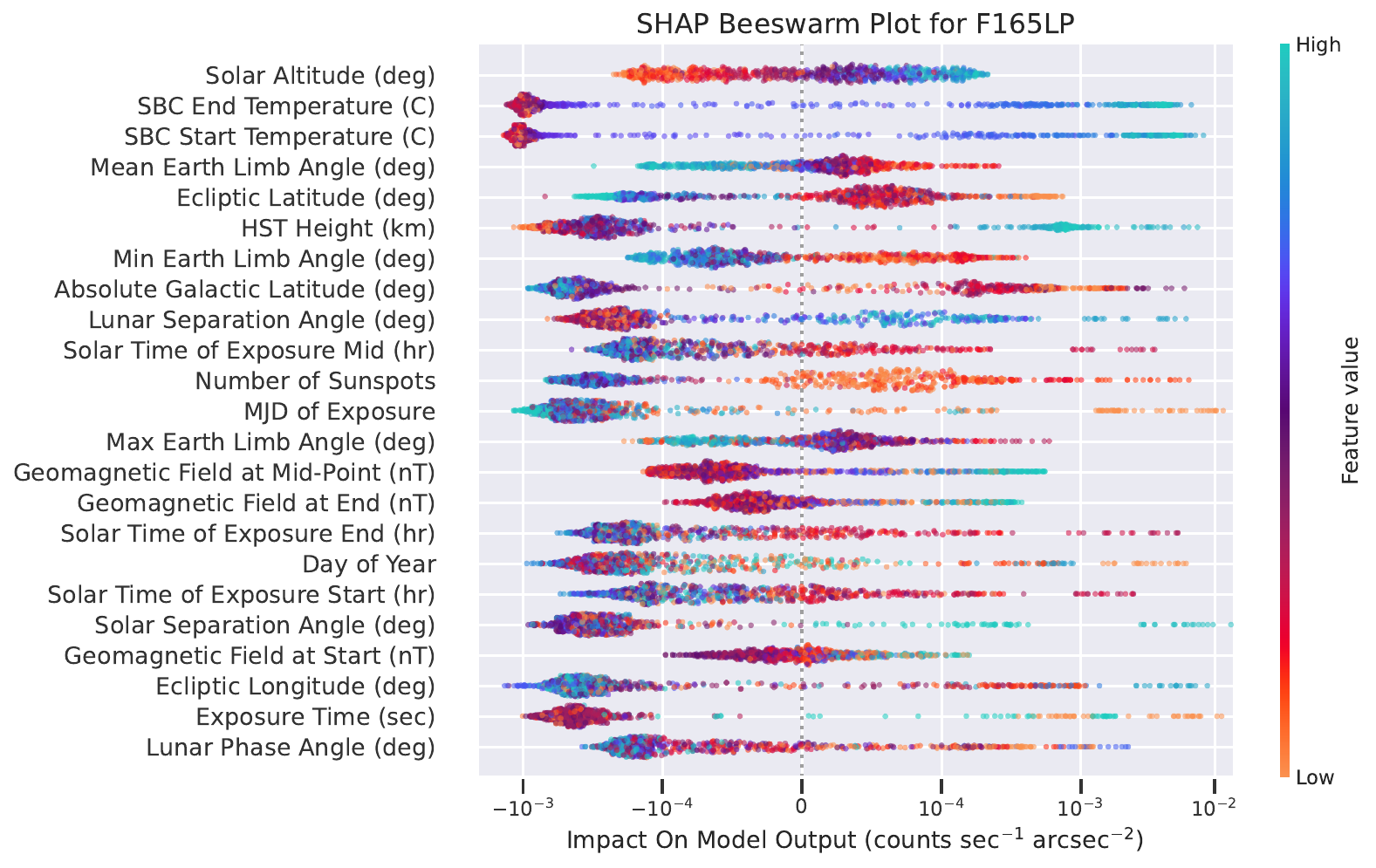}
\caption{As per Figure~\protect\ref{Fig:SHAP_Beeswarm_F115LP_F122M}, except for F125LP and F140LP.}
\label{Fig:SHAP_Beeswarm_F150LP_F165LP}
\end{figure}

\begin{figure}
\centering
\includegraphics[width=0.9\textwidth]{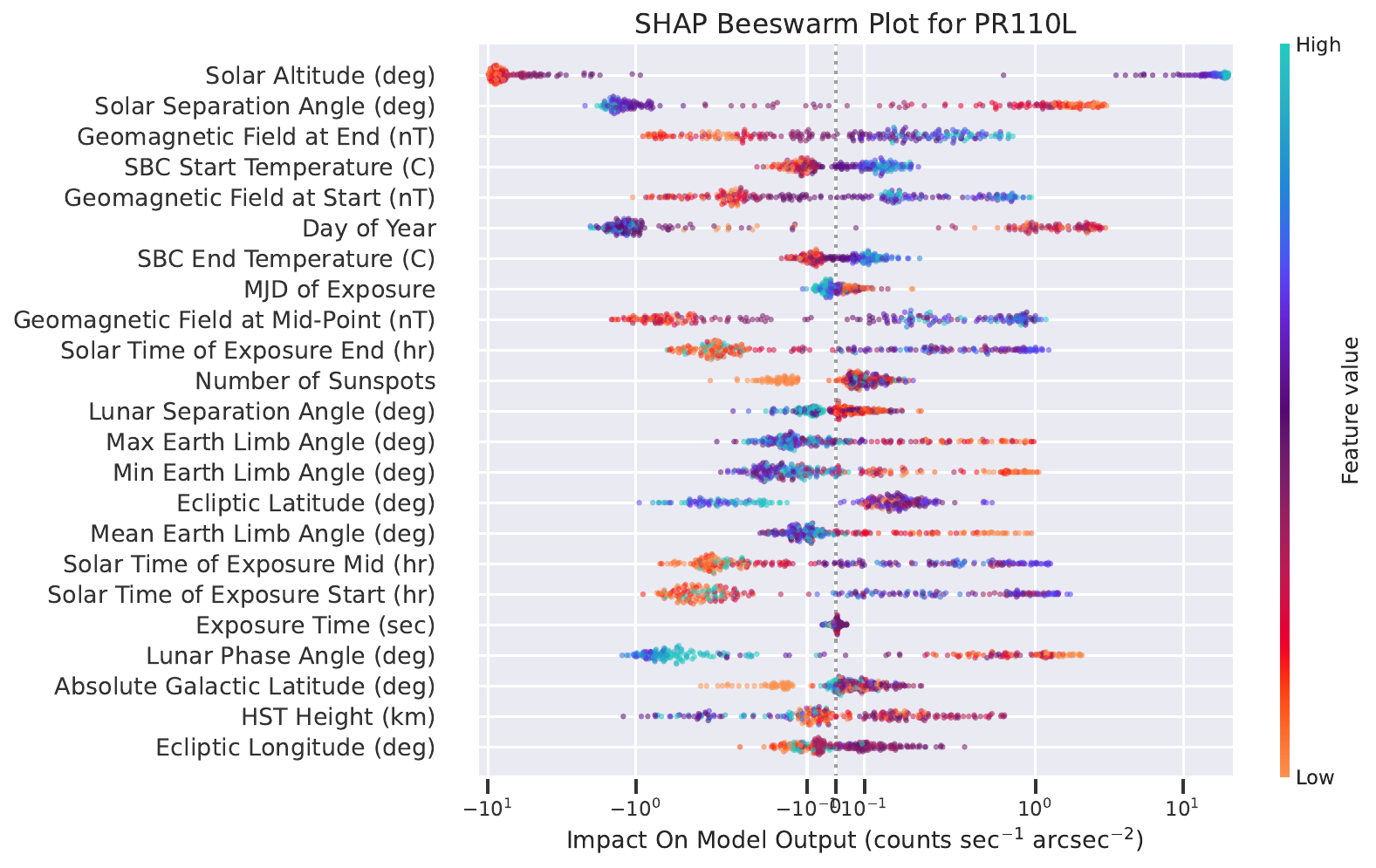}
\includegraphics[width=0.9\textwidth]{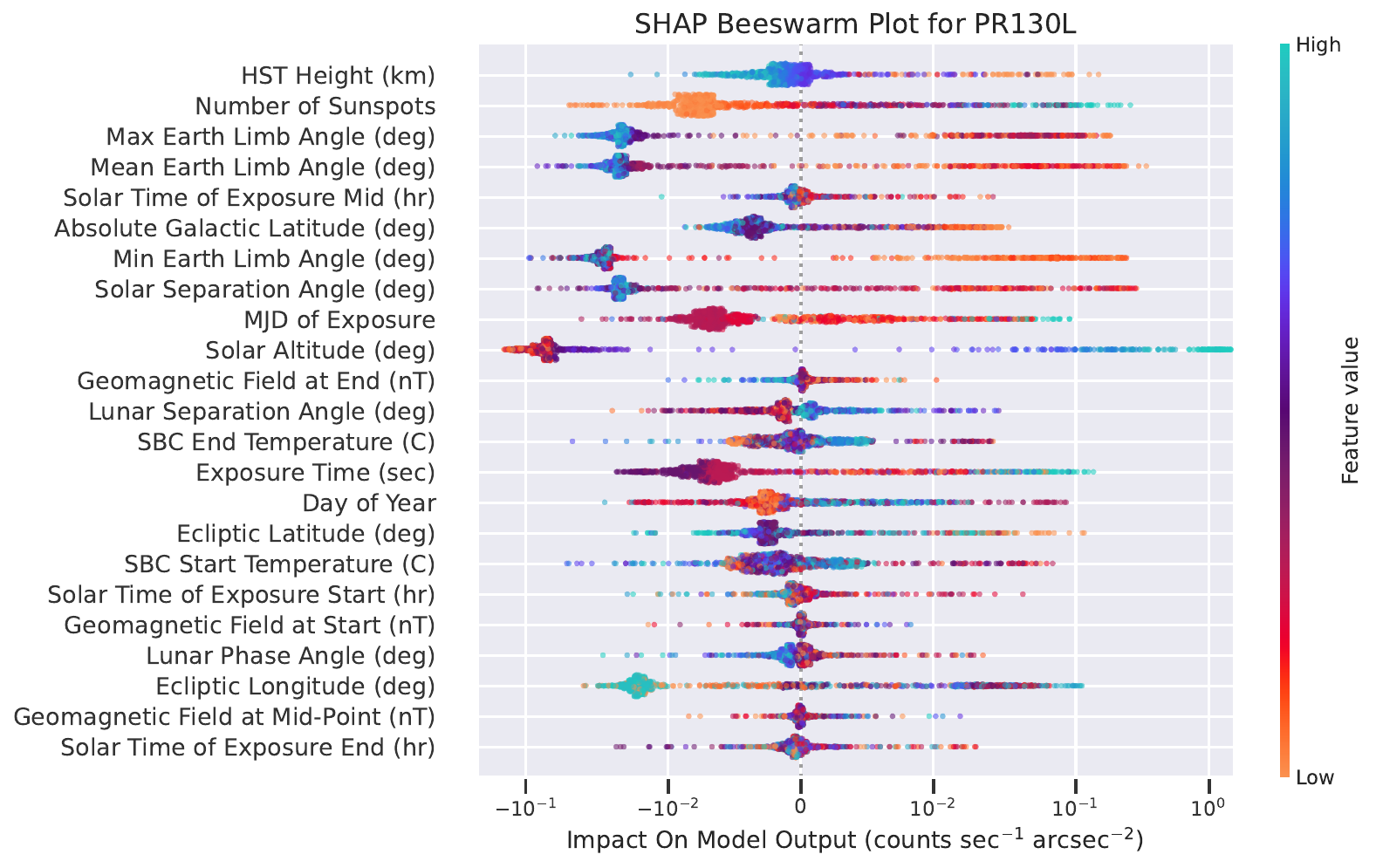}
\caption{As per Figure~\protect\ref{Fig:SHAP_Beeswarm_F115LP_F122M}, except for PR110L and PR130L.}
\label{Fig:SHAP_Beeswarm_PR110L_PR130L}
\end{figure}

Here, in Figures~\ref{Fig:SHAP_Beeswarm_F115LP_F122M}, \ref{Fig:SHAP_Beeswarm_F125LP_F140LP}, \ref{Fig:SHAP_Beeswarm_F150LP_F165LP}, and \ref{Fig:SHAP_Beeswarm_PR110L_PR130L}, we provide the SHAP beeswarm plot for every SBC filter. Note that the plots for F125LP and F150LP were already displayed in Figure~\ref{Fig:Example_Beeswarm_Plots} in Section~\ref{Subsubsection:SHAP_Beeswarm_Plots}, but are repeated here for completeness.

\newpage
{\normalsize 
\setlength{\bibsep}{1.0pt} 
\input{journalCommand.tex}
\bibliographystyle{mnras}
\bibliography{ChrisBib}}

\end{document}

%% file: journalCommand.tex
\def\ref@jnl#1{{\rmfamily #1}}%
\newcommand\aj{\ref@jnl{AJ}}%
\newcommand\araa{\ref@jnl{ARA\&A}}%
\newcommand\apj{\ref@jnl{ApJ}}%
\newcommand\apjl{\ref@jnl{ApJ}}%
\newcommand\apjs{\ref@jnl{ApJS}}%
\newcommand\ao{\ref@jnl{Appl.~Opt.}}%
\newcommand\apss{\ref@jnl{Ap\&SS}}%
\newcommand\aap{\ref@jnl{A\&A}}%
\newcommand\aapr{\ref@jnl{A\&A~Rev.}}%
\newcommand\aaps{\ref@jnl{A\&AS}}%
\newcommand\azh{\ref@jnl{AZh}}%
\newcommand\baas{\ref@jnl{BAAS}}%
\newcommand\jrasc{\ref@jnl{JRASC}}%
\newcommand\memras{\ref@jnl{MmRAS}}%
\newcommand\mnras{\ref@jnl{MNRAS}}%
\newcommand\pra{\ref@jnl{Phys.~Rev.~A}}%
\newcommand\prb{\ref@jnl{Phys.~Rev.~B}}%
\newcommand\prc{\ref@jnl{Phys.~Rev.~C}}%
\newcommand\prd{\ref@jnl{Phys.~Rev.~D}}%
\newcommand\pre{\ref@jnl{Phys.~Rev.~E}}%
\newcommand\prl{\ref@jnl{Phys.~Rev.~Lett.}}%
\newcommand\pasp{\ref@jnl{PASP}}%
\newcommand\pasj{\ref@jnl{PASJ}}%
\newcommand\qjras{\ref@jnl{QJRAS}}%
\newcommand\skytel{\ref@jnl{S\&T}}%
\newcommand\solphys{\ref@jnl{Sol.~Phys.}}%
\newcommand\sovast{\ref@jnl{Soviet~Ast.}}%
\newcommand\ssr{\ref@jnl{Space~Sci.~Rev.}}%
\newcommand\zap{\ref@jnl{ZAp}}%
\newcommand\nat{\ref@jnl{Nature}}%
\newcommand\iaucirc{\ref@jnl{IAU~Circ.}}%
\newcommand\aplett{\ref@jnl{Astrophys.~Lett.}}%
\newcommand\apspr{\ref@jnl{Astrophys.~Space~Phys.~Res.}}%
\newcommand\bain{\ref@jnl{Bull.~Astron.~Inst.~Netherlands}}%
\newcommand\fcp{\ref@jnl{Fund.~Cosmic~Phys.}}%
\newcommand\gca{\ref@jnl{Geochim.~Cosmochim.~Acta}}%
\newcommand\grl{\ref@jnl{Geophys.~Res.~Lett.}}%
\newcommand\jcp{\ref@jnl{J.~Chem.~Phys.}}%
\newcommand\jgr{\ref@jnl{J.~Geophys.~Res.}}%
\newcommand\jqsrt{\ref@jnl{J.~Quant.~Spec.~Radiat.~Transf.}}%
\newcommand\memsai{\ref@jnl{Mem.~Soc.~Astron.~Italiana}}%
\newcommand\nphysa{\ref@jnl{Nucl.~Phys.~A}}%
\newcommand\physrep{\ref@jnl{Phys.~Rep.}}%
\newcommand\physscr{\ref@jnl{Phys.~Scr}}%
\newcommand\planss{\ref@jnl{Planet.~Space~Sci.}}%
\newcommand\procspie{\ref@jnl{Proc.~SPIE}}%
\newcommand\pasa{\ref@jnl{PASA}}%